\documentclass[useAMS,fleqn,usenatbib]{mnras}

\usepackage{newtxtext}

\usepackage[T1]{fontenc}
\usepackage{ae,aecompl}

\usepackage{graphicx,epsfig,longtable}
\usepackage{amssymb}
\usepackage{amsmath}
\usepackage{tabulary} 
\usepackage{tabularx} 
\usepackage{makeidx}
\usepackage{threeparttable}
\usepackage{array}
\usepackage{xcolor}
\usepackage[normalem]{ulem}
\usepackage{etoolbox}
\makeatletter
\patchcmd\@combinedblfloats{\box\@outputbox}{\unvbox\@outputbox}{}{%
   \errmessage{\noexpand\@combinedblfloats could not be patched}%
}%
 \makeatother

\newcommand{\Msun}{$\mathrm{M}_{\odot}$}
\newcommand{\nucpha}{Nuclear Physics A}
\title[]
  {Improved leakage-equilibration-absorption scheme (\textsc{ILEAS}) for neutrino physics in compact object mergers}
\author[R. Ardevol-Pulpillo et. al.]
  {R. Ardevol-Pulpillo,$^1$\textsuperscript{,}$^2$ H.-T. Janka$^1$, O. Just$^3$ and A. Bauswein,$^4$\textsuperscript{,}$^5$ \\
  $^1$Max-Planck-Institut f\"ur Astrophysik, Postfach 1317, 85741 Garching, Germany\\
  $^2$Physics Department, Technische Universit\"at M\"unchen,  James-Franck-Str. 1, 85748 Garching, Germany\\  
  $^3$Astrophysical Big Bang Laboratory, RIKEN Cluster for Pioneering Research, 2-1 Hirosawa, Wako, Saitama 351-0198, Japan\\
  $^4$GSI Helmholtzzentrum f\"ur Schwerionenforschung, Planckstra\ss e 1, 64291 Darmstadt, Germany\\
  $^5$Heidelberger Institut f\"ur Theoretische Studien, Schloss-Wolfsbrunnenweg 35, 69118 Heidelberg, Germany\\
}
\date{Released 2018 Xxxxx XX}

\pubyear{2018}

\pagerange{\pageref{firstpage}--\pageref{lastpage}} \pubyear{2018}

\def\ga{\,\,\raise0.14em\hbox{$>$}\kern-0.76em\lower0.28em\hbox
{$\sim$}\,\,}
\def\la{\,\,\raise0.14em\hbox{$<$}\kern-0.76em\lower0.28em\hbox
{$\sim$}\,\,}
\def\Msun{$\mathrm{M}_{\odot}$}
\begin{document}

\label{firstpage}

\maketitle

\begin{abstract}
We present a new, computationally efficient, energy-integrated approximation 
for neutrino effects in hot and dense astrophysical environments such
as supernova cores and compact binary mergers and their remnants.
Our new method, termed ILEAS for Improved Leakage-Equilibration-Absorption
Scheme, improves the lepton-number and energy losses of traditional leakage
descriptions by a novel prescription of the diffusion time-scale based on
a detailed energy integral of the flux-limited diffusion equation. The leakage
module is supplemented by a neutrino-equilibration treatment that ensures
the proper evolution of the total lepton number and medium plus neutrino energies
as well as neutrino-pressure effects in the neutrino-trapping domain. Moreover,
we employ a simple and straightforwardly applicable ray-tracing algorithm for
including re-absorption of escaping neutrinos especially in the decoupling
layer and during the transition to semi-transparent conditions. ILEAS is 
implemented on a three-dimensional (3D) Cartesian grid with a minimum of
free and potentially case-dependent parameters and exploits the basic 
physics constraints that should be fulfilled in the neutrino-opaque and
free-streaming limits.
We discuss a suite of tests for stationary and time-dependent proto-neutron star
models and post-merger black-hole-torus configurations, for which 3D ILEAS 
results are demonstrated to agree with energy-dependent 1D and 2D two-moment (M1) 
neutrino transport on the level of 10--15 percent in basic neutrino properties.
This also holds for the radial profiles of the neutrino luminosities and of the
electron fraction. Even neutrino absorption maps around torus-like neutrino
sources are qualitatively similar without any fine-tuning, confirming that ILEAS can 
satisfactorily reproduce local losses and re-absorption of neutrinos as found in 
sophisticated transport calculations.
 \end{abstract}

\begin{keywords}
gravitational waves --- hydrodynamics --- nucleosynthesis --- neutrinos -- stars: neutron
\end{keywords}

\section{Introduction}

 Neutrinos play an important role in high-energy stellar astrophysics, more precisely in the context of the birth and death of neutron stars (NSs), among others. Already in the 1990s, the neutrino-induced shock revival was theorized as a mechanism for exploding core-collapse supernovae (SNe) (\citealt{1985ApJ...295...14B}, \citealt{1990RvMP...62..801B}, see \citealt{2012ARNPS..62..407J}, \citealt{2013RvMP...85..245B} and \citealt{2015PASA...32....9F} for recent reviews). After a successful explosion, neutrinos are essential to understand the long-term cooling of the new-born NS (e.g. \citealt{2010PhRvL.104y1101H}). At the other end of their lives, NSs in binaries with another compact object (CO), either a NS, a black hole (BH) or a white dwarf (WD), may be able to merge within a Hubble time. In such scenarios, NS matter reaches very high temperatures and densities, emitting copious amounts of neutrinos (see e.g. \citealt{2011LRR....14....6S}, \citealt{2012LRR....15....8F} and \citealt{2015IJMPD..2430012R} for reviews). Despite being dynamically only a secondary ingredient, neutrinos, by their emission and absorption, drive the neutron-to-proton ratio of the NS as well as the ejected matter. This aspect will determine the distribution of synthesized elements in the ejecta \citep{2015MNRAS.452.3894G, 2015MNRAS.448..541J, 2014ApJ...789L..39W, 2015PhRvD..91f4059S, 2015ApJ...815...82L, 2016PhRvD..93d4019F, 2016PhRvD..94l3016F, 2016MNRAS.460.3255R, 2017PhRvD..96l4005B, 2018CQGra..35c4001M, 2016CQGra..33r4002L} as well as the associated electromagnetic transient, known as kilonova, powered by the radioactive decay of neutron-rich elements (\citealt{1998ApJ...507L..59L, 2005astro.ph.10256K, 2010MNRAS.406.2650M, 2011ApJ...738L..32G, 2011ApJ...736L..21R, 2016ApJ...831..190H, 2016ApJ...829..110B, 2017PhRvD..96l4005B}, see \citealt{2017ARNPS..67..253T} and \citealt{2016ARNPS..66...23F,2017LRR....20....3M} for recent reviews on r-process in NS mergers and kilonovae, respectively). Neutrinos are also needed to understand the fate of the hypermassive NS (HMNS) remnants and the evolution of the surrounding torus (if any) \citep{2014MNRAS.443.3134P, 2014MNRAS.441.3444M, 2015PhRvD..91l4021F,2017ApJ...846..114F,2017PhRvD..96l3015W}. Merger remnants have been envisioned as the central engines of short gamma-ray bursts (sGRB), and neutrino pair-annihilation has been shown to deposit significant amounts of energy, which might help in powering an ultrarelativistic jet \citep{2016ApJ...816L..30J, 2017JPhG...44h4007P}. The recent detection of a NS merger though its associated signals in the form of gravitational waves (GW) and electromagnetic radiation (EM), as predicted by theoretical models, highlights the remarkable contributions of detailed numerical simulations to our understanding of the universe.
  
 The evolution of the neutrino phase space distribution function obeys the Boltzmann transport equation. In three spatial dimensions (3D), this becomes a six-dimensional, time-dependent problem for each neutrino species, which is considerably arduous to solve when no further approximations are applied to reduce its dimensionality (e.g. \citealt{1966AnPhy..37..487L}). Therefore, numerous schemes of varying complexity and accuracy have been developed to cope with this challenging task. 
 
 In the context of NS mergers, truncated moment schemes are the most sophisticated treatments successfully used. In such schemes, the angular dependence of the neutrino momentum distribution is removed by evolving a hierarchy of ("moment") equations, which are obtained from angular moment integration of the Boltzmann equation. Thus one introduces moments as angular integrals of the neutrino phase-space distribution function, i.e. neutrino energy density, flux density, pressure and higher-order moments. It is necessary, in order to close the set of equations, to find a way to express the highest employed moments, which are not evolved by but appear in the moment equations. The so-called M1 schemes (e.g. \citealt{2011PThPh.125.1255S}) time-integrate the evolution equations for the zeroth- and first-order moments, closing the system with an analytical relation expressing the highest employed moments as local functions of the evolved ones, and are used in grey (e.g. \citealt{2015PhRvD..91l4021F}) as well as energy-dependent versions (e.g. \citealt{2015MNRAS.453.3386J}). Despite their enormous current popularity for describing the transport of neutrinos, they suffer from the inability to properly handle crossing radiation beams, and thus have a tendency to overestimate the neutrino densities in the polar directions in NS-merger remnant simulations \citep{2015MNRAS.448..541J, 2018PhRvD..98f3007F}. These limitations call for alternative treatments of neutrino re-absorption, such as ray-tracing, in order to provide an accurate description of the ejecta composition, essential for predicting EM counterparts to NS mergers. Recently, an alternative Monte Carlo closure has been suggested by \cite{2018MNRAS.475.4186F}, but its applicability in merger simulations has not been demonstrated yet. Monte Carlo (MC) codes are still obviated from their direct use in full-scale merger simulations because of their tremendous computational costs and memory requirements, for which reason they have been set aside in favour of less expensive methods, except for some MC applications in snapshot calculations \citep{2015ApJ...813...38R, 2018PhRvD..98f3007F}. Ray-tracing algorithms, which solve the Boltzmann equation in one dimension, have also been employed in snapshot calculations of NS merger remnants by \cite{2015MNRAS.448..541J} in the Newtonian framework and, recently, by \cite{2018PhRvD..98j3014D} in a general-relativistic version, which also includes the effects of neutrino scattering in an approximative manner based on the existence of a precomputed M1 solution.
 
 Leakage schemes are a very popular and computationally simple approximation for the treatment of neutrino effects in NS merger simulations. They had been introduced first in Newtonian merger models as grey versions by \cite{1996A&A...311..532R, 1997A&A...319..122R, 1999A&A...344..573R, 2001A&A...380..544R} and, with a different handling of the energy integration, by \cite{2003MNRAS.342..673R, 2003MNRAS.345.1077R, 2012MNRAS.426.1940K, 2013MNRAS.430.2585R,2013RSPTA.37120272R}. More recently, leakage schemes have been used in many relativistic merger simulations, too (for example, \citealt{2011PhRvL.107e1102S,2011PhRvL.107u1101S,2012CQGra..29l4003K,2013ApJ...776...47D,2014PhRvD..90b4026F,2017CQGra..34d4002F,2014PhRvD..89j4029N,2015PhRvD..92d4045P,2016PhRvD..94d3003L,2016CQGra..33r4002L,2016PhRvD..94b4023B}). Moreover, leakage methods have been applied in evolution studies of NS merger remnants, including HMNSs \citep{2014MNRAS.443.3134P, 2015ApJ...813....2M,2018CQGra..35c4001M,2014MNRAS.441.3444M,2017MNRAS.472..904L} as well as BH-torus systems \citep{2007PThPh.118..257S,2015MNRAS.446..750F,2015MNRAS.449..390F}.

 The most basic version of leakage schemes accounts for neutrino energy and lepton-number losses by local source terms in the hydrodynamics equations, following the original implementations by \cite{1996A&A...311..532R} and \cite{2003MNRAS.342..673R}, possibly upgraded by gravitational redshift effects \citep{2010CQGra..27k4107S,2010CQGra..27k4103O,2013PhRvD..88f4009G}. In more advanced versions this basic functionality of leakage schemes is supplemented by treatments of a trapped neutrino component and by neutrino absorption. Such improvements have been accomplished by hybrid methods between leakage and M1 \citep{2012PTEP.2012aA304S, 2015PhRvD..91f4059S, 2016PhRvD..93l4046S,2017ApJ...846..114F,2017PhRvD..96l3012S,2018PhRvD..97b3009K} or `M0' (zeroth moment with a closure) \citep{2016MNRAS.460.3255R,2017ApJ...838L...2R,2017ApJ...842L..10R,2018ApJ...852L..29R,2018PhRvL.120k1101Z}. Alternatively, equilibrium and time-scale arguments have been used to parametrize the trapping physics, and complex propagation paths have been considered to connect the locations of neutrino production in a leakage treatment with the neutrinospheric decoupling region for describing neutrino absorption exterior to the trapping domain (see \citealt{2014MNRAS.443.3134P,2014A&A...568A..11P,2016ApJS..223...22P}).

 Comparisons by \cite{2015PhRvD..91l4021F, 2016PhRvD..93d4019F} reveal major differences between results with their grey M1-based SpEC code and `classical' leakage results for the first 10--15\,ms after the collision of compact binary stars. In contrast, \cite{2016ApJS..223...22P} report very good qualitative and partially quantitative agreement in key quantities when testing their Advanced Spectral Leakage (ASL) scheme against Boltzmann transport in the context of Newtonian, spherically symmetric (1D) hydrodynamic simulations of several 100\,ms of post-bounce accretion in core-collapse supernovae. However, the ASL code involves a variety of parameters that were calibrated on grounds of this considered problem. It is not obvious that the thus determined parameter values work equally well for a broader class of conditions. Moreover, also the axisymmetric (2D) and three-dimensional (3D) simulations of stellar core-collapse and post-bounce accretion, which \cite{2016ApJS..223...22P} applied their ASL code to, still contain a quasi-spherical, highly opaque neutrino source that accretes mass from the collapsing star at high rates (as in the 1D simulations). These tests are not conclusive with respect to the question how well the ASL code is able to perform in the merger case, where the remnant is rotationally deformed and not accreting. Radial profiles of the neutrino quantities for the transition from the high-opacity to the low-opacity regime and tests with non-spherical neutrino sources, which could facilitate such a judgement, are not available for upgraded leakage schemes in the literature.

 In this work we present a new implementation of an improved leakage treatment that does not only take into account local energy and lepton-number losses by neutrino emission, but it also accounts for the fact that neutrinos can equilibrate with matter in the optically thick regime and that they are still re-absorbed by matter when they propagate through optically thin regions. Our new method, which is termed Improved Leakage-Equilibration-Absorption Scheme (ILEAS), is designed to fulfil a number of requirements: (1) low algorithmic complexity in order to enable easy numerical realization; (2) proper and consistent reproduction of the correct physical behaviour of the neutrino-matter system at high optical depths; (3) description of the transition to the low-opacity regime with a minimum number of free parameters and ad hoc recipes of approximation; and (4) high computational efficiency that permits the calculation of large sets of merger simulations to explore the multidimensional parameter space (system masses and mass ratios, spins, orbital parameters, NS equations of state) that describes NS-NS/BH binaries. The computational efficiency is also facilitated by the fact that ILEAS, in contrast to transport calculations with explicit schemes, is not subject to any time-stepping constraints by the Courant-Friedrichs-Lewy condition (CFL).

 ILEAS is implemented on a 3D Cartesian grid and makes use of the grey description of the leakage loss terms applied by \cite{1996A&A...311..532R}. The greyness of the treatment benefits all of the mentioned requirements. Advancing beyond the original treatment by \cite{1996A&A...311..532R}, ILEAS introduces a new definition of the neutrino-loss time-scale based on the energy-integrated equation of flux-limited diffusion. This allows for a considerably improved description of the neutrino drain from regions of high optical depths. The effects of a trapped neutrino component are taken into account by considering neutrinos as part of an equilibrated neutrino-matter fluid in the trapping regime. Neutrino absorption in the transition to the optically thin limit is handled by a simplified ray-tracing method that adopts an analytical integration of the radiation attenuation along the ray paths of escaping neutrinos following \cite{2001A&A...368..527J}.

 To assess the quality of the ILEAS scheme, we consider different stages during the cooling evolution of a spherical proto-NS (PNS) and perform steady-state as well as time-dependent calculations (co-evolving the medium temperature and electron fraction on a fixed background density). We compare the leakage results for the neutrino emission with 1D neutrino transport results obtained with the ALCAR and VERTEX codes. Both of these codes are energy-dependent two-moment schemes employing an algebraic (M1) closure and a variable Eddington factor closure based on a solution of the Boltzmann equation, respectively. Moreover, we perform time-dependent calculations for the neutrino emission from optically thick (high-mass) as well as optically thin (low-mass) axisymmetric BH-accretion tori in direct comparison with ALCAR results. Our tests demonstrate very good compatibility between leakage and transport results (global quantities agree on the level of roughly 10 percent or better) with respect to radial luminosity profiles, neutrino luminosities evolving over periods of tens of milliseconds, mean energies, and spatial distributions of electron fraction and neutrino-energy absorption rates.

 Our paper is structured as follows. In section~\ref{sec:model} we describe the physical and algorithmic components of the ILEAS code and their numerical realization, in section~\ref{sec:tests} we present our set of neutrino transport tests for proto-NS and BH-torus models, as well as first demonstrations of the application of ILEAS in NS-NS merger models, and in section~\ref{sumary} we summarize our work. In the four following appendices~\ref{appendix:tdiffs}--\ref{appendix:nabs} we compare results for different definitions of the diffusion time-scale used in previous literature, present an overview of the neutrino opacities and source terms employed by ILEAS, discuss different versions of implementing the $\beta$-processes in the leakage treatment, and provide test results for an alternative method to compute neutrino-number re-absorption, respectively.
   
\section{Numerical description}\label{sec:model}

\subsection{Weak interactions with \textsc{ILEAS} in CFC relativistic hydrodynamics}\label{sec:model:coupling}

\begin{figure}
\begin{center}
\makebox[0pt][c]{%
\minipage{0.5\textwidth}
\includegraphics[width=\textwidth]{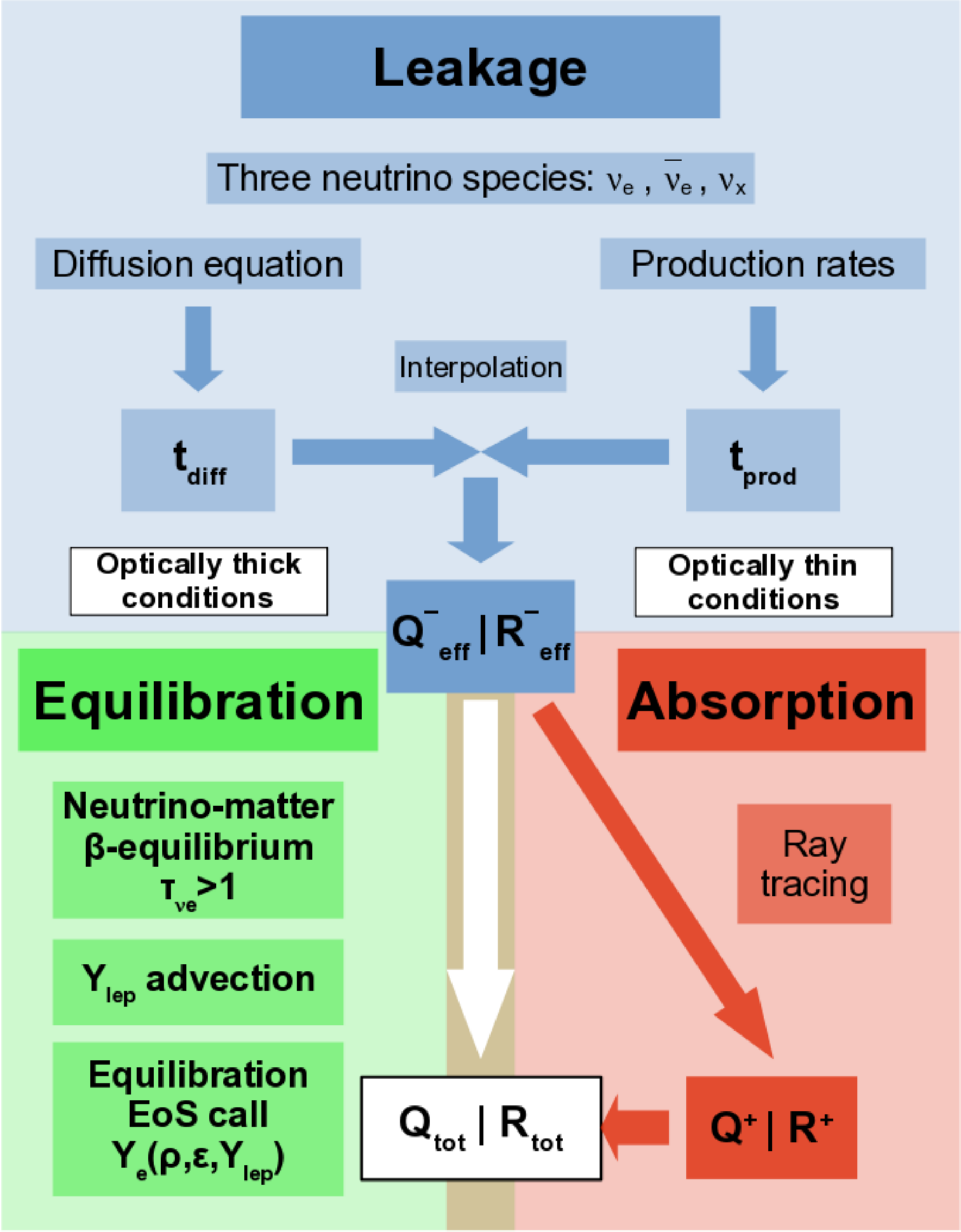}
\endminipage\hfill
}%
\caption{Elements of the different modules which compose our ILEAS scheme (leakage, equilibration and absorption) and their primary interdependences.} \label{schemechart}
\end{center}
\end{figure}

 We present a novel neutrino leakage scheme, \textsc{ILEAS}, that is capable of reproducing the fundamental aspects of the neutrino physics described by more sophisticated transport schemes at lower computational costs. The scheme calculates the energy and lepton number changes caused by weak interactions of three neutrino species: electron neutrinos, $\nu_e$, electron antineutrinos, $\bar{\nu}_e$, and heavy-lepton neutrinos, $\nu_x$ (which include $\mu$ and $\tau$ neutrinos and their antiparticles). Neutrinos are considered to be massless because their relevant mean energies are of order MeV, orders of magnitude larger than their rest mass ($<1$ eV). Neutrino flavour oscillations are ignored in our treatment. The full scheme is composed of three major modules which model different aspects of the transport of neutrinos, summarized in figure~\ref{schemechart}: the \textit{leakage}, the \textit{equilibration} and the \textit{absorption} modules. The leakage unit estimates the local number and energy loss rates associated with neutrinos which `leak' out of the system, as an interpolation between trapping and free streaming conditions. At high optical depths, neutrinos of all species are in equilibrium with matter, which we account for explicitly with our equilibration unit. This effect is ignored in most leakage schemes with few recent exceptions \citep{2010CQGra..27k4107S,2011PhRvL.107e1102S,2016ApJS..223...22P}, but was used as initial condition for nuclear network calculations \citep{2015MNRAS.452.3894G}. Finally, the absorption module computes the energy and number deposition rates due to interactions of the escaping neutrinos with the optically thin material, by means of a simple ray-tracing algorithm. 
 
 The leakage and absorption modules provide the neutrino cooling rates, $Q_{\nu_i}^{-}$, and heating rates, $Q_{\nu_i}^{+}$, respectively, for all three neutrino species. The total energy source term, which will enter the hydrodynamical evolution equations, can be calculated from them as
 \begin{equation}
  Q_{\mathrm{tot}}\ =\ \sum_{i=\nu_e,\bar{\nu}_e}{Q_{i}^{+}}-\sum_{i=\nu_e,\bar{\nu}_e,\nu_x}{Q_{i}^{-}}.\label{qtot}
 \end{equation}
 We remark that the first sum includes only contributions from $\nu_e$ and $\bar{\nu}_e$, namely charged-current absorptions on nucleons. We neglect here the effects of neutrino-antineutrino annihilation, which affect all flavours of neutrinos. However, they have no direct impact on the electron fraction and usually contribute little to the mass of ejected material (e.g. \citealt{2016ApJ...816L..30J, 2017JPhG...44h4007P, 2017ApJ...846..114F}). Moreover, they are difficult to treat accurately without detailed knowledge of the neutrino phase-space distribution \citep{2018PhRvD..98f3007F}.
 
 Similarly, the lepton change rates, $R_{\nu_i}^{-}$ and $R_{\nu_i}^{+}$ can be combined to the total (electron flavour) lepton change rate as,
 \begin{equation}
  R_{\mathrm{tot}}\ =\ R_{\nu_e}^{+}-R_{\bar{\nu}_e}^{+}-R_{\nu_e}^{-}+R_{\bar{\nu}_e}^{-}.\label{rtot}
 \end{equation} 
 Details for the calculation of the rates will be discussed in sections~\ref{sec:model:leakage} and~\ref{sec:model:abs}.
 
 Because we want to apply \textsc{ILEAS} in the context of NS mergers, we provide as an example the implementation of the source terms in the evolution equations of our conformally flat (CFC\footnote{Conformally Flat Condition \citep{1980grg1.conf...23I, 1996PhRvD..54.1317W}}), relativistic NS merger code. For a more detailed description of the complete scheme, we refer the reader to \cite{2007A&A...467..395O}. In the present section (\ref{sec:model:coupling}) we use the convention $c=G=1$.
 
 In the CFC approximation, the metric can be expressed as
 \begin{equation}
 \label{CFC3+1}
 \mathrm{d}s^2\ =\ (-\alpha^2\ +\ \beta_i\beta^i)\mathrm{d}t^2\ +\ 2\beta_i\mathrm{d}x^i\mathrm{d}t\ +\ \psi^4\delta_{ij}\mathrm{d}x^i\mathrm{d}x^j,
 \end{equation}
 where $\alpha$, $\beta_i$ and $\psi$ are the metric potentials, i.e. lapse, shift and conformal factor, respectively. 
 
 As in most hydrodynamic solvers, we define the conserved quantities, namely conserved rest-mass density, $\rho^{\ast}$, conserved specific momentum, $\hat{u}_i$, and conserved specific energy, $\tau$, as a function of their primitive counterparts, rest mass density, $\rho$, velocity, $v_i$, and specific internal energy, $\varepsilon$, via
 \begin{align}
 &\rho^{\ast}\ =\ \rho \alpha u^0 \psi^6 , \label{conserveda}\\
 &\hat{u}_i\ =\ Hu_i\ =\ H(v^i\ +\ \beta^i)\psi^4 u^0 , \label{conservedb}\\
 &\tau\ =\ H W\ -\ \frac{P}{\rho W}\ -\ \sqrt{1\ +\ \frac{\hat{u}_i \hat{u}_j \delta^{ij}}{\psi^4}} . \label{conservedc}
 \end{align}
 Here the Lorenz factor is defined as $W\ =\ \alpha u^0\ =\ \sqrt{1\ +\ \gamma^{ij} u_i u_j}$, with $\gamma^{ij}$ being the spatial components of the metric, $u^0$ and $u^i$ are the time and space components of the 4-velocity, $H$ represents the relativistic specific enthalpy, defined as $H\ =\ 1\ +\ P/\rho\ +\ \varepsilon$, $P$ is the fluid pressure and $\delta^{ij}$ is the Kronecker delta. We then write the relativistic Euler equations, where we include the neutrino source term defined in equation~\eqref{qtot}, $Q_{\mathrm{tot}}$, in the momentum and energy equations with the pertinent corrections,
 \begin{align} 
  \frac{\mathrm{d}}{\mathrm{d}t} \rho^{\ast}\ &=\ -\rho^{\ast}\partial_i v^i, \label{Hydroeqa} \\
  \frac{\mathrm{d}}{\mathrm{d}t} \hat{u}_i\ &=\ -\frac{1}{\rho^{\ast}}\alpha\psi^6\partial_iP-\alpha\hat{u}^0\partial_i\alpha+\hat{u}_j\partial_i\beta^j+\frac{2\hat{u}_k\hat{u}_k}{\psi^5\hat{u}^0}\partial_i\psi \nonumber \\
 &+\frac{Q_{\mathrm{tot}}\alpha\hat{u}_i}{\rho HW} , \label{Hydroeqb}\\
  \frac{\mathrm{d}}{\mathrm{d}t}\tau\ &=\ -\frac{\psi^6}{\rho^{\ast}}(v^i+\beta^i)\left(1-\frac{HW}{\omega}\right)(\partial_iP)-\psi^6\frac{P}{\rho^{\ast}}\partial_i(v^i+\beta^i) \nonumber \\
 &-6\psi^5\frac{P}{\rho^{\ast}}(v^i+\beta^i)(\partial_i\psi)-\frac{\hat{u}_i}{\psi^4}\left(1-\frac{HW}{\omega}\right)(\partial_i\alpha)\nonumber \\
 &+\frac{1}{\psi^4}\left(\frac{1}{HW}-\frac{1}{\omega}\right)\left[\hat{u}_i\hat{u}_j\partial_j\beta^i-\frac{1}{3}\hat{u}_i\hat{u}_i\partial_j\beta^j\right] \nonumber \\
 &+\frac{Q_{\mathrm{tot}}\alpha}{\rho}\left[1-\frac{\hat{u}_i\hat{u}_j\delta^{ij}}{\psi^4HW\omega}\right], \label{Hydroeqc}
 \end{align}
 where $\mathrm{d}/\mathrm{d}t=\partial_t+v^i\partial_i$ and $\omega=\sqrt{1+\left(\hat{u}_i\hat{u}_j\delta^{ij}/\psi^4\right)}$. 
 
 To close the system, one needs a microphysical equation of state (EoS) as a function of $\rho$, $\varepsilon$ and $Y_e$, representing the thermodynamics of the fluid. In the equilibration module, we treat the regions where neutrinos are trapped and in $\beta$-equilibrium with the medium in a specific way by redefining the specific energy density, $\varepsilon$, pressure, $P$, and specific enthalpy, $H$, to include the contributions from the combined fluid of matter plus trapped neutrinos. This means, that in order to close the set of evolution equations in those regions, we need to build an additional set of EoS tables which also incorporates the contributions from the neutrinos. 
 
 Without the inclusion of weak interactions, the net electron fraction, $Y_e$, is just advected with the fluid ($\mathrm{d}Y_e/\mathrm{d}t=0$). The leakage and absorption modules, however, provide a source term, $R_{\mathrm{tot}}$, as defined in equation~\eqref{rtot}, which enters the evolution equation of $Y_e$,
 \begin{equation}
  \frac{\mathrm{d}}{\mathrm{d}t} Y_e\ =\ \frac{R_{\mathrm{tot}}\alpha}{\mathcal{A}\rho W}. \label{yeadv}
 \end{equation} 
 where $\mathcal{A}$ is Avogadro's constant. Note that as the source terms obtained from \textsc{ILEAS} ($R_{\mathrm{tot}}$ and $Q_{\mathrm{tot}}$) are expressed in CGS units, $\mathcal{A}=1/m_b$, where $m_b$ is the atomic mass unit. To model the trapping conditions, we advect the trapped $\nu_e$ and $\bar{\nu}_e$ lepton fractions (equation~\ref{ynui}) in addition to the $Y_e$,
 \begin{align}
  &\frac{\mathrm{d}}{\mathrm{d}t} Y_{\nu_e}^{\mathrm{trap}}\ =\ 0, \label{leptonadvnue}\\
  &\frac{\mathrm{d}}{\mathrm{d}t} Y_{\bar{\nu}_e}^{\mathrm{trap}}\ =\ 0. \label{leptonadvanue}
 \end{align} 
 The final goal of this procedure is to obtain an updated \textit{trapped lepton fraction} at the end of every time-step, defined as
 \begin{equation}
 Y_{\mathrm{lep}}=Y_e+Y_{\nu_e}^{\mathrm{trap}}-Y_{\bar{\nu}_e}^{\mathrm{trap}}.\label{Ylep}
 \end{equation}
 We can then use this $Y_{\mathrm{lep}}$ in an \textit{equilibration step} to recover the new equilibrium values for $Y_e$, $Y_{\nu_e}^{\mathrm{trap}}$ and $Y_{\bar{\nu}_e}^{\mathrm{trap}}$. This requires the construction of a set of EoS tables which invert the dependence on $\rho$, $\varepsilon$ and $Y_{\mathrm{lep}}$ to obtain $Y_e$. We will expand the details on the equilibration module in section~\ref{sec:model:equil}.
 
\subsection{The neutrino leakage scheme}\label{sec:model:leakage}

\begin{figure*}
\begin{center}

\includegraphics[width=\textwidth]{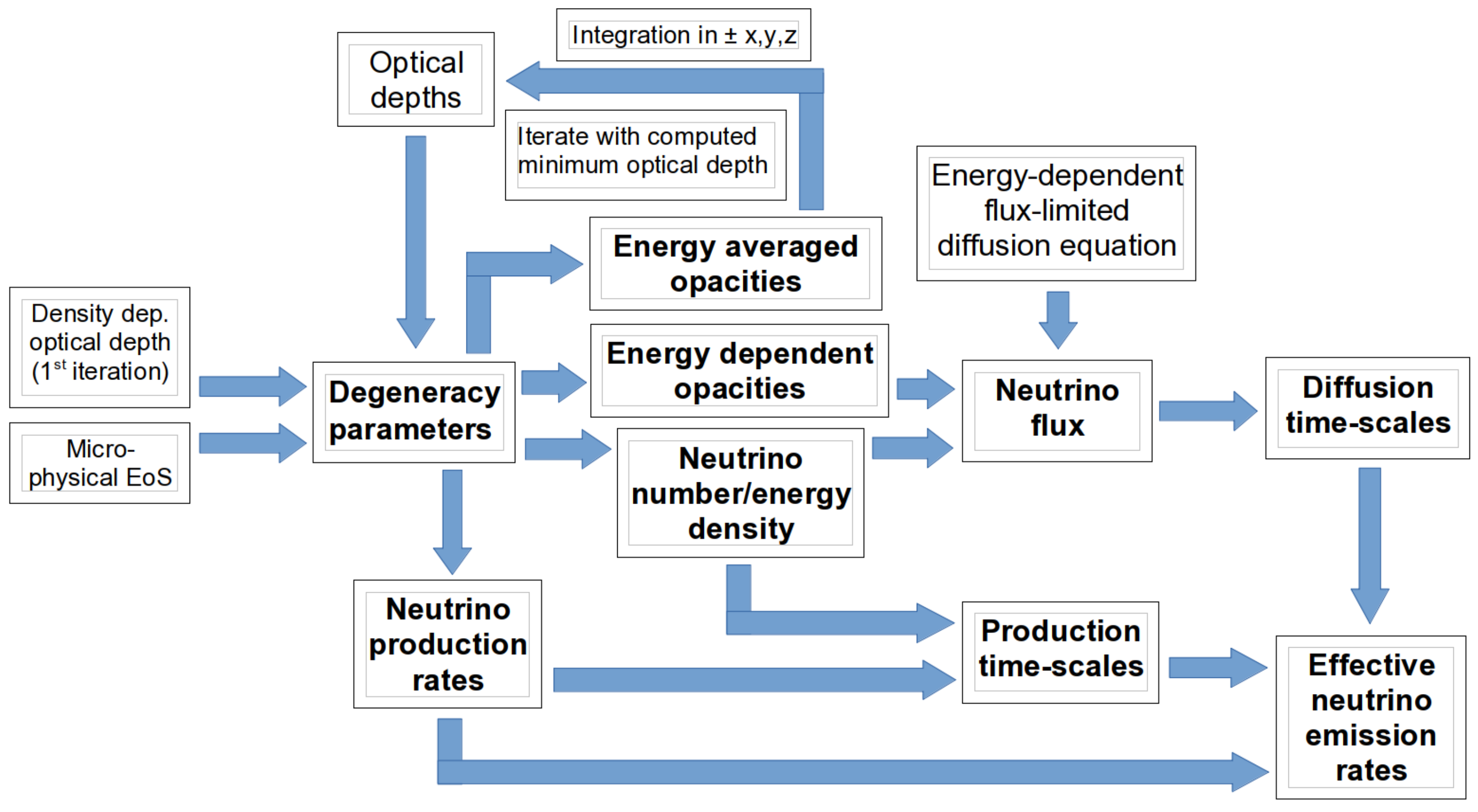}
\caption{Components of the leakage module and their mutual interdependences (not including neutrino re-absorption nor equilibration).}\label{leakagechart}
\end{center}
\end{figure*}

 The leakage part of our code is based on the archetypical leakage scheme from \cite{1996A&A...311..532R}. The essence of the model consists in the evaluation of the local \textit{effective} neutrino production rates, 
 \begin{equation}	
 R_{\nu_i}^{-}\ =\ R_{\nu_i}\gamma_{\nu_i,\mathrm{num}}^{\mathrm{eff}},\label{Reff}
 \end{equation}
 and 
 \begin{equation}
 Q_{\nu_i}^{-}\ =\ Q_{\nu_i}\gamma_{\nu_i,\mathrm{en}}^{\mathrm{eff}},\label{Qeff}
 \end{equation}
 where $R_{\nu_i}$ and $Q_{\nu_i}$ are the local neutrino production rates for number and energy, respectively, as defined in equations~\eqref{Rpur} and~\eqref{Qpur}. $\gamma_{\nu_i,\mathrm{num}}^{\mathrm{eff}}$ and $\gamma_{\nu_i,\mathrm{en}}^{\mathrm{eff}}$ are obtained by means of an interpolation between the relevant time-scales in the optically thick and optically thin conditions as
 \begin{equation}
   \gamma_{\nu_i}^{\mathrm{eff}}\ =\ \left(1+\frac{t_{\nu_i}^{\mathrm{diff}}}{t_{\nu_i}^{\mathrm{prod}}}\right)^{-1}.\label{gammaeff}
 \end{equation}
 Here $t_{\nu_i}^{\mathrm{diff}}$ is the diffusion time-scale, relevant in the diffusion/optically thick regime (see equations~\ref{tdiffn} and~\ref{tdiffe}), and $t_{\nu_i}^{\mathrm{prod}}$ is the production time-scale, relevant in the free streaming/optically thin regime (see equations~\ref{tprodn} and~\ref{tprode}). In figure~\ref{leakagechart} we provide a schematic depiction of the \textit{leakage} part of \textsc{ILEAS}.
 
 Although energy-dependent leakage schemes have been developed and successfully used, a grey approximation offers advantages in connection to our treatment of the equilibration regime, while keeping the scheme at a minimum with respect to computational cost, especially in the absorption module. Therefore, we employ spectrally averaged/integrated quantities for our calculations\footnote{Denoted with an overbar, when susceptible to confusion with their energy-dependent counterparts.} (see appendix~\ref{appendix:reac} for details). In particular, we carefully take into account the energy dependence in the calculation of the diffusion time-scale as will be explained in section~\ref{sec:model:tdiff}.
 
 We assume the neutrino spectrum to follow a Fermi-Dirac distribution with matter temperature, $T$, expressed in energy units,  
 \begin{equation}
  f(\epsilon;T,\eta_{\nu_i})\ =\ \frac{1}{1+\mathrm{e}^{(\epsilon/T-\eta_{\nu_i})}},\label{fermid}
 \end{equation}
 for neutrinos with energy $\epsilon$. The neutrino degeneracy parameter, $\eta_{\nu_i}\ =\ \mu_{\nu_i}/T$, (with $\mu_{\nu_i}$ being the neutrino chemical potential) is prescribed as an interpolation of the equilibrium degeneracy, $\eta_{\nu_i}^{{\mathrm{eq}}}$, at high optical depth and a vanishing value at low optical depth \citep{1996A&A...311..532R}:
 
 \begin{equation}
    \eta_{\nu_i}\ =\ \eta_{\nu_i}^{{\mathrm{eq}}}(1-\mathrm{e}^{-\tau_{\nu_i}}).\label{etanu}
 \end{equation}
  
 The equilibrium degeneracy of electron neutrinos obeys
 
 \begin{equation}
    \eta_{\nu_e}^{{\mathrm{eq}}}\ =\ \eta_{e}+\eta_{p}-\eta_n-Q/T,\label{betachem}
 \end{equation}

 where $\eta_e$ is the electron degeneracy (including rest mass), $\eta_p$ and $\eta_n$ are the proton and neutron degeneracies (without rest mass) and $Q=m_nc^2-m_pc^2=1.2935$MeV is the nucleon rest-mass energy difference. Electron antineutrinos are assumed to have an equilibrium degeneracy given by $\eta_{\bar{\nu}_e}^{{\mathrm{eq}}}=-\eta_{\nu_e}^{{\mathrm{eq}}}$, whereas the degeneracy of heavy-lepton neutrinos is considered to be zero, $\eta_{\nu_x}=0$. Ensuring a behaviour of $\eta_{\nu_i}$ compatible with the transition to low optical depth is essential when using microphysical EoSs, in order to avoid unphysical values of the Fermi integrals and their ratios. In the semi-transparent regime, however, when the neutrino phase-space distribution function begins to deviate from equilibrium, leakage schemes can only approximate $\eta_{\nu_i}$. In the case of \textsc{ILEAS} we use an interpolation, which can have a non-negligible impact on the neutrino luminosities in comparison to transport schemes.
 
 In equation~\eqref{etanu}, $\tau_{\nu_i}$ is the optical depth for neutrino species $\nu_i$, estimated as the \textit{minimum} optical depth calculated in the six Cartesian directions ($\pm x,\ \pm y,\ \pm z$) as 
 \begin{equation}
  \tau_{\nu_i}=\int_r^{\infty}{\bar{\kappa}_{\nu_i}^{j=1}(r')\mathrm{d}r'}.\label{tau}
 \end{equation}
 We point out to the reader that contrary to previous leakage schemes, ILEAS does not require the optical depth to determine the diffusion time-scales (see section~\ref{sec:model:tdiff}), but optical depths are used merely for the interpolation of the neutrino degeneracies (equation~\ref{etanu}) and the location of the neutrinospheres, which are used for the detailed treatment of equilibration (see section~\ref{sec:model:equil}) and absorption (see section~\ref{sec:model:abs}). Moreover, in section~\ref{sec:tests} we demonstrate the good agreement between the results obtained by ILEAS and two neutrino transport codes, particularly in a spherically symmetric PNS scenario, which is a rather problematic setup for a Cartesian grid. Therefore, we consider our geometric assumptions when calculating the optical depth to be sufficiently accurate.
 
 In equation~\eqref{tau}, the energy-averaged opacity, $\bar{\kappa}_{\nu_i}^{j=1}$, is defined as in equation~\eqref{kappa_total}. We consider as opacity sources the absorption of electron neutrinos and electron antineutrinos on neutrons and protons, respectively, and the scattering of all neutrino species on nucleons, alpha particles and heavy nuclei. We employ the same absorption opacities as in \cite{1996A&A...311..532R}, with the additional inclusion of stimulated absorption\footnote{Only in the calculation of $t_{\nu_i}^{\mathrm{diff}}$.} (neutrino blocking) and nucleon rest-mass corrections. The scattering opacities are also taken from the same source but with the nucleon blocking factors from \cite{1993ApJ...405..637M} (see appendix~\ref{appendix:reac:opaa} for details). Contrary to \cite{1996A&A...311..532R}, we do not assume matter to be fully dissociated and employ the nucleon number fractions from the EoS instead, in the computation of the nucleon blocking factors (equations~\ref{protonblk} and~\ref{neutronblk}). Since $\eta_{\nu_i}$ are necessary for the calculation of the opacities, one iteration step is performed by replacing $\tau_{\nu_i}$ in equation \eqref{etanu} by a function of the density, $f(\rho)=(10^{11}*\rho^{-1})^2$, with $\rho$ in $\mathrm{g*cm}^{-3}$. We find that there is no need for multiple iterations, as the results converge very quickly.

 \begin{table}
  \begin{center}
      \caption{Neutrino interactions implemented in our scheme. Appendix~\ref{appendix:reac} provides the formulation employed for each reaction. Note that the current version of ILEAS includes neutrino-pair production but not pair annihilation processes.}\label{table:nureac}  
    \begin{threeparttable}
     \begin{tabular}{lcc}
      \hline
      \hline
      \noalign{\vskip 2mm}  
      \parbox[c]{0.1\textwidth}{Name}        & \parbox[c]{0.1\textwidth}{Interaction}        & \parbox[c]{0.1\textwidth}{\centering $\nu$ species} \vspace{1mm}  \\ 
      \hline  
      \noalign{\vskip 2mm}  
      $\beta$-react. for $\nu_e$	          & $p+e^- \leftrightarrow n+\nu_e$				& $\nu_e$	   			\\  
      $\beta$-react. for $\bar{\nu}_e$  & $n+e^+ \leftrightarrow p+\bar{\nu}_e$     			& $\bar{\nu}_e$	   			\\                                                                                                
      $e^-e^+$ annihil.			              & $e^-+e^+ \rightarrow \nu_i+\bar{\nu}_i$    			& $\nu_e$, $\bar{\nu}_e$, $\nu_x$	\\                                                                                             
      Plasmon decay			                  & $\gamma_{\mathrm{trans}} \rightarrow \nu_i+\bar{\nu}_i$	& $\nu_e$, $\bar{\nu}_e$, $\nu_x$	\\                  
      N-N bremsstr.\tnote{$\dagger$}	  & $N+N \rightarrow N+N+\nu_i+\bar{\nu}_i$			& $\nu_x$	   			\\                                                                                    
      Nucleon scatt.			                  & $N+\nu_i\rightarrow N+\nu_i$					& $\nu_e$, $\bar{\nu}_e$, $\nu_x$	\\                                                                                             
      $\alpha$ part. scatt.		              & $\alpha+\nu_i\rightarrow \alpha+\nu_i$			& $\nu_e$, $\bar{\nu}_e$, $\nu_x$	\\   
      Nuclei scatt.			                      & $(A,Z)+\nu_i\rightarrow (A,Z)+\nu_i$				& $\nu_e$, $\bar{\nu}_e$, $\nu_x$	
		      \\  
      \noalign{\vskip 2mm}  
      \hline
      \hline
     \end{tabular}
     \begin{tablenotes}	
       \item[$\dagger$]  $N=p,n$
     \end{tablenotes}
    \end{threeparttable}	
  \end{center}
\end{table}
 
 The description of the nucleon blocking coefficients\footnote{We relabelled the final state blocking coefficient $\xi_N$ instead of $\eta_N$ \citep{1985ApJS...58..771B} to avoid confusion with degeneracy parameters.}, $\xi_{NN}$ (with $N$ being the nucleon type, either neutron, $n$, or proton, $p$), appearing in the absorption opacities and production rates (appendix~\ref{appendix:reac}) are implemented following \cite{1985ApJS...58..771B}, assuming the nucleons are ideal non-relativistic Fermi gases. Due to this approximation, using the nucleon chemical potentials from modern EoSs is inconsistent, because the corresponding chemical potentials contain corrections due to nucleon self-interaction potentials in a dense medium. In fact, it causes these blocking factors to become unphysical (either negative or bigger than unity) and unable to reproduce the non-degenerate limit. In order to avoid this undesirable behaviour, we calculate the chemical potentials by inverting the expressions for the number densities of free Fermi gases (see \citealt{Ramppthesis, 2002A&A...396..361R}). These chemical potentials are used only for the computation of the nucleon blocking factors.
 
 With the knowledge of the neutrino degeneracies, we can calculate the neutrino production rates, $R_{\nu_i}$ and $Q_{\nu_i}$, for number and energy, respectively (see appendix~\ref{appendix:reac}). The neutrino interactions included in this work are summarized in table~\ref{table:nureac}. Besides the production rates employed in \cite{1996A&A...311..532R}, we include nucleon-nucleon bremsstrahlung as a source for heavy-lepton neutrinos, which is the dominant production channels at high densities, and $\nu_e$ emission by electron captures on heavy nuclei\footnote{See appendix~\ref{appendix:reac:sourc} for details on the implementation of the bremsstrahlung rate and $\nu_e$ production rates by electron captures on nuclei.}. 
  
 At this stage, it is useful to define the energy-dependent neutrino number and energy density,
\begin{equation}
 E_{\nu_i}^j(\epsilon)\ =\ g_{\nu_i}\frac{4\pi}{(hc)^3}\epsilon^{2+j}f(\epsilon;T,\eta_{\nu_i}),\label{neutrinodensity}
\end{equation}
 where $j=0$ is for the number and $j=1$ for the energy density. Integrated over the neutrino spectrum, they become
 \begin{equation}
    \bar{E}_{\nu_i}^{j}\ =\ g_{\nu_i}\frac{4\pi}{(hc)^3}T^{3+j}F_{2+j}(\eta_{\nu_i}),\label{edenseq}
 \end{equation}
 where $F_k=\int_0^{\infty}{x^kf(x;T,\eta_{\nu_i})\mathrm{d}x}$ are the relativistic Fermi integrals of order $k$ and the multiplicity factor, $g_{\nu_i}$, is unity for $\nu_e$ and $\bar{\nu}_e$ and 4 for $\nu_x$. Now we can calculate the production time-scales for number and energy, $t_{\nu_i}^{\mathrm{prod}}$, as\footnote{Note the change in the notation with respect to \cite{1996A&A...311..532R}.}
 
 \begin{equation}
    t_{\nu_i,\mathrm{num}}^{\mathrm{prod}}\ =\ \frac{\bar{E}^{j=0}_{\nu_i}}{R_{\nu_i}},\label{tprodn}
 \end{equation}
 \begin{equation}
    t_{\nu_i,\mathrm{en}}^{\mathrm{prod}}\ =\ \frac{\bar{E}^{j=1}_{\nu_i}}{Q_{\nu_i}}.\label{tprode}
 \end{equation}
 
\subsubsection{The diffusion time-scale}\label{sec:model:tdiff}

 At high optical depth, neutrinos are trapped and slowly diffuse through the medium on a much longer time-scale than they are produced. A simple estimate of this time-scale is obtained when considering a random walk. The average distance a particle can travel in an optically thick medium can be approximated as 
 \begin{equation}
  d\ =\ \sqrt{N}\lambda,\label{randomwalk}
 \end{equation}
 where $N$ is the number of times a particle scatters and $\lambda$ the mean free path between scatterings. Assuming neutrinos travel at the speed of light, one can estimate the diffusion time-scale as
 \begin{equation}
  t_{\nu_i}^{\mathrm{diff}}\ \sim\ N\frac{\lambda}{c}\ =\ \frac{d^2}{\lambda c}.\label{randomwalktdiff}
 \end{equation}

 A similar expression can be obtained from the diffusion equation for the zeroth-order angular moments, ignoring neutrino source terms and considering a static background medium
\begin{equation}
 \frac{\partial E_{\nu_i}(\epsilon)}{\partial t}\ =\ -\boldsymbol\nabla\boldsymbol\cdotp\boldsymbol{F}_{\nu_i}(\epsilon),\label{Diffeq}
\end{equation}
 where $E_{\nu_i}(\epsilon)$ is representative of $E_{\nu_i}^{0,1}$ in the previous section. The neutrino flux $\boldsymbol{F}_{\nu_i}(\epsilon)$ can be obtained from Fick's law for the diffusion case as
\begin{equation}
 \boldsymbol{F}_{\nu_i}(\epsilon)\ =\ \frac{-c}{3\kappa_{\nu_i}(\epsilon)}\boldsymbol{\nabla}E_{\nu_i}(\epsilon).\label{Flux}
\end{equation}
 The factor 3 in the diffusion coefficient arises from the assumption of an isotropic neutrino distribution (see \citealt{Mihalas}). Now by a simple dimensional analysis, using $\kappa_{\nu_i}=1/\lambda_{\nu_i}$, one gets
\begin{equation}
 \frac{E_{\nu_i}}{t_{\nu_i}^{\mathrm{diff}}}\ =\ \frac{c\lambda_{\nu_i} E_{\nu_i}}{3d^2},\label{tdiffdim}
\end{equation}
 easily recovering the result of equation~\eqref{randomwalktdiff} (modulo a factor $1/3$).
 
 Previous leakage schemes made different assumptions about the length-scale $d$ in order to derive a numerical value for $t_{\nu_i}^{\mathrm{diff}}$, any of which gives a good order of magnitude approximation of neutrino losses. In appendix~\ref{appendix:tdiffs}, we analyse in detail some of these prescriptions and compare the corresponding results with those from more sophisticated transport calculations, as well as those obtained in the present paper. There, one can see that all such leakage approximations perform poorly when one is interested in reproducing the local neutrino losses of detailed transport calculations: most neutrinos are radiated from a narrow region close to the neutrinosphere, defined as the radius where the optical depth is $\tau_{\nu_i}=2/3$, and hardly any from the optically thick region in the deeper interior. In addition, the total luminosities can exceed those of a transport calculation by a factor of 2 or more. The reason for this behaviour is the simplistic dimensional analysis used to estimate the diffusion time-scale, which leads to a steep decrease of the time-scale with radius, inversely proportional to the mean free path, preventing the diffusion of neutrinos out from high optical depth and favouring the escape of those produced near the NS surface.
 
 To obtain a more accurate treatment, we evaluate numerically the spatial derivatives in equations~\eqref{Diffeq} and~\eqref{Flux} using five-point stencils in order to recover the divergence of the flux. Since neutrinos with different energies diffuse at different speeds, which leads to a significant impact on the spectrally averaged diffusion time-scale in the semi-transparent regime, we retain the energy dependence in the calculation of the flux. Integration to obtain the diffusion time-scale yields, 
 \begin{equation}
  t_{\nu_i,\mathrm{num}}^{\mathrm{diff}}=\ \frac{\bar{E}^{j=0}_{\nu_i}}{\boldsymbol{\nabla}\boldsymbol\cdotp\int_0^{\infty}{\frac{-c}{3\kappa_{\nu_i}(\epsilon)}\Lambda_{\nu_i}^{j=0} (\epsilon)\boldsymbol{\nabla}E^{j=0}_{\nu_i}(\epsilon)}\mathrm{d}\epsilon},\label{tdiffn}
\end{equation}
 \begin{equation}
  t_{\nu_i,\mathrm{en}}^{\mathrm{diff}}=\ \frac{\bar{E}^{j=1}_{\nu_i}}{\boldsymbol{\nabla}\boldsymbol\cdotp\int_0^{\infty}{\frac{-c}{3\kappa_{\nu_i}(\epsilon)}\Lambda_{\nu_i}^{j=1} (\epsilon)\boldsymbol{\nabla}E^{j=1}_{\nu_i}(\epsilon)}\mathrm{d}\epsilon},\label{tdiffe}
\end{equation}
 for number and energy diffusion respectively, where the energy-dependent total opacities, $\kappa_{\nu_i}(\epsilon)$, are calculated as in equation~\eqref{kappa_totale}. Due to the inclusion of rest-mass corrections in the computation of the absorption opacities (see equations~\ref{kappa_abse_nue} and~\ref{kappa_abse_anue}), we cannot rely on an analytical solution of the Fermi integrals. For the energy integration we employ 15 energy bins in a logarithmic spacing up to 400 MeV (with bin limits at 5.0, 6.4, 8.4, 11.2, 15.2, 20.7, 28.4, 39.2, 54.3, 75.5, 105.2, 146.7, 204.8, 286.1 and 400.0 MeV), which is the same grid employed by the M1 scheme \textsc{ALCAR} in the models discussed for comparison in section~\ref{sec:tests}.

 It is well known that in the (semi)transparent region diffusion becomes acausal because the flux diverges as $\lambda=1/\kappa\rightarrow\infty$. In order to ensure the correct limits we employ a flux limiter, $\Lambda_{\nu_i}(\epsilon)$, as successfully used in flux limited diffusion schemes (FLD) \citep{1975NYASA.262...54W, 1981ApJ...248..321L}. Because the differences between different flux limiters are effectively small, we use the canonical expression suggested by \cite{1975NYASA.262...54W}, retaining the energy dependence in order to ensure causality for each of the energy bins,
 \begin{equation}
  \Lambda_{\nu_i}^j(\epsilon)=\left(1+\frac{1}{3\kappa_{\nu_i}^*(\epsilon)}\frac{|\boldsymbol{\nabla}E^{j}_{\nu_i}(\epsilon)|}{E^{j}_{\nu_i}(\epsilon)}\right)^{-1}.\label{fluxlimiter}
 \end{equation}
 
 The divergence of the flux, in equations~\eqref{tdiffn} and~\eqref{tdiffe}, gives us information about the nature of a given region, either as a source from which neutrinos diffuse out ($\boldsymbol{\nabla}\boldsymbol\cdotp\boldsymbol{F}_{\nu_i}(\epsilon)>0$) or as a sink where neutrinos flow to ($\boldsymbol{\nabla}\boldsymbol\cdotp\boldsymbol{F}_{\nu_i}(\epsilon)<0$). Because the leakage model is constructed to approximate the local neutrino \textit{losses}, it cannot directly deal with sinks, which would translate to negative diffusion time-scales. Therefore, no net neutrino losses occur in such regions and, in concordance, we assume the diffusion time-scale to be infinite, quenching all local neutrino losses. At low optical depths, however, this approach does not make sense because radiation does not obey the physics of diffusion, but the free streaming limit where $t_{\nu_i}^{\mathrm{diff}}<t_{\nu_i}^{\mathrm{prod}}$ should be recovered. Accordingly, we set $t_{\nu_i}^{\mathrm{diff}}=\infty$ only inside the neutrinosphere, where the optical depth is $\tau_{\nu_i}>2/3$, and take its absolute value outside (which will always be smaller than $t_{\nu_i}^{\mathrm{prod}}$). In the same spirit, small regions (less than few grid cells) bounded by two sinks are treated as sinks as well, as neutrinos will diffuse to the neighbouring sinks and remain trapped. This final correction turns out to be necessary to avoid overestimated neutrino emission near the neutrinosphere in some of the PNS snapshots at later times.
 
 Including relativistic corrections \citep{2011PThPh.125.1255S} for an asymptotically flat space-time ($\mathrm{d}s^2=-\alpha^2\mathrm{d}t^2+\psi^4\delta_{ij}\mathrm{d}x^i\mathrm{d}x^j$, where $\alpha$ is the lapse function, $\psi$ the conformal factor and we take the shift vector, $\boldsymbol{\beta}$, to be negligible for simplicity), the diffusion time-scale becomes:
 \begin{equation}
  t_{\nu_i,j}^{\mathrm{diff}}=\ \frac{\psi^2 \bar{E}^{j}_{\nu_i}}{\boldsymbol{\nabla}\boldsymbol\cdotp\left(\alpha \psi^2 \int_0^{\infty}{\frac{-c}{3\kappa_{\nu_i}(\epsilon)}\Lambda_{\nu_i}^j (\epsilon)\boldsymbol{\nabla}E^{j}_{\nu_i}(\epsilon)}\mathrm{d}\epsilon\right)},\label{tdiffrel}
 \end{equation}
 for $j=0,1$.

\subsection{Neutrino absorption in optically thin matter}\label{sec:model:abs} 

At low optical depths, neutrinos decouple from matter and essentially stream away at the speed of light. However,  before free streaming is reached, a significant fraction of these neutrinos can be re-absorbed. This neutrino energy and number deposition in semi-transparent regions is crucial for many astrophysical phenomena, such as the shock revival in SNe, the ejecta composition in CO mergers or neutrino-driven winds from the remnants of either event. Attempts to reliably simulate any of those scenarios, therefore, require to account for neutrino absorption. The `standard' leakage approach only serves the purpose of estimating neutrino losses, but does not take care of re-absorption. Therefore, a complimentary absorption scheme is needed. We present a description here based on the 1D formulation of radiation attenuation by \cite{2001A&A...368..527J}, generalized to any 3D geometry by means of a simple ray-tracing algorithm.

We start with the premise that neutrinos are produced in the centre of a given cell and approximately escape in the direction of the local gradient of the neutrino energy density, $-\boldsymbol{\nabla}\bar{E}_{\nu_i}^{j=1}$, following a straight ray. This is a fair assumption in spherical symmetry, and although in complex geometries neutrinos will scatter and change direction along their way out (and also be subject to gravitational ray bending), it is a reasonable first-order approximation\footnote{\cite{2014MNRAS.443.3134P, 2014A&A...568A..11P} spent significant effort on designing recipes to construct radiation paths for their ray-tracing treatment. We refrain from adding complications to our code in this aspect, first to save computer time, second because our simple scheme works well in near-surface or low optical depth regions that dominate the neutrino emission and absorption (as proven in practise by our test results), and third because any complicated path definition will still remain an approximation whose general validity cannot be guaranteed without verification by comparison to detailed transport.}. We use a 3D slab formalism \citep{SlabMethod} to find all cells of our 3D Cartesian grid crossed by a given ray and estimate the deposited energy and number as a function of the distance traversed in the cell, reducing the escaping luminosity accordingly. 

We note that in the following we use the \textit{ray coordinate}, $s$, to denote positions along the rays, whereas the Cartesian coordinates, $\boldsymbol x$, are used to define the position (\textit{centre}) of a grid cell. Therefore, for a ray emitted by a cell at $\boldsymbol x_1$, the origin of its ray coordinate, $s=0$, corresponds to the Cartesian coordinates of the emitting cell. The path traversed by a ray crossing an absorbing grid cell at $\boldsymbol x_2$ is then characterized by $s^-_2$ and $s^+_2$, representing the positions where the ray intersects with the boundaries of such a cell. We also include re-absorption of neutrinos produced by the emitting cell itself, along a path from the centre of the cell, $s^-_1=0$, to its boundary, $s^+_1$.

The luminosity produced by a cell, $\boldsymbol x_1$, is generally calculated in leakage schemes as
\begin{equation}
 L_{\nu_i}^{\mathrm{ray}}(\boldsymbol x_1)\ \approx\ Q_{\nu_i}^{-}(\psi^6V)_{\mathrm{cell,em}},\label{lumapprox}
\end{equation}
including metric corrections to the volume of the emitting cell. Following equation~(72) from \cite{2001A&A...368..527J}, the amount of energy per unit volume deposited in an absorbing cell at $\boldsymbol x_2$, by a ray travelling along a path from $s^-_2$ to $s^+_2$, is determined by
\begin{align}
 &\left[L_{\nu_i}^{\mathrm{ray}}(s^+_2)-L_{\nu_i}^{\mathrm{ray}}(s^-_2) \right] (\psi^{6} V)_{\mathrm{cell,abs}}^{-1}\ =\  \nonumber \\ 
 &L_{\nu_i}^{\mathrm{ray}}(s^-_2)\left[1-\mathrm{exp}\left(-\int_{s^-_2}^{s^+_2}{\frac{\bar{\kappa}_{\nu_i,\mathrm{a}}(\boldsymbol x_2)}{\langle\chi_{\nu_i}(\boldsymbol x_2)\rangle}\mathrm{d}s'}\right)\right](\psi^{6} V)_{\mathrm{cell,abs}}^{-1},\label{lumiabs}
\end{align}
where $L_{\nu_i}^{\mathrm{ray}}(s^-_2)$ is the incoming luminosity, $L_{\nu_i}^{\mathrm{ray}}(s^+_2)$ the outgoing luminosity, $\bar{\kappa}_{\nu_i,\mathrm{a}}$ the spectrally averaged absorption opacity (equations~\ref{kappa_abs_nue} and~\ref{kappa_abs_anue}) and $\langle\chi_{\nu_i}\rangle$ is the energy-averaged flux factor (see below). We note that the luminosity attenuation is calculated along the ray coordinate, $s$, as a function of the distance traversed, whereas all the thermodynamical and metric variables are assumed to be homogeneous within each cell ($a(s^-_j)=a(s^+_j)=a(\boldsymbol x_j)$).

The luminosity arriving at the absorbing cell, $L_{\nu_i}^{\mathrm{ray}}(s^-_2)$, relates to the luminosity produced in the cell from which the ray originates, $L_{\nu_i}^{\mathrm{ray}}(s^-_1)=L_{\nu_i}^{\mathrm{ray}}(\boldsymbol x_1)$, by
\begin{equation}
  L_{\nu_i}^{\mathrm{ray}}(s^-_2)\ =\ L_{\nu_i}^{\mathrm{ray}}(s^-_1)\mathrm{exp}\left(-\int_{s^-_1}^{s^-_2}{\frac{\bar{\kappa}_{\nu_i,\mathrm{a}}(\boldsymbol x_j)}{\langle\chi_{\nu_i}(\boldsymbol x_j)\rangle}\mathrm{d}s'}\right)\frac{\alpha(\boldsymbol x_1)^2}{\alpha(\boldsymbol x_2)^2} \label{Rel}
\end{equation}
where $j$ runs over the cells intersected by the ray prior to reaching $s^-_2$ (see equation \ref{lumiabs}). We include gravitational redshift of the luminosities between the emitting and absorbing cells following \cite{2010CQGra..27k4103O}, but we omit Doppler effects for simplicity.

In equation~\eqref{lumiabs} $\langle\chi_{\nu_i}\rangle$ is defined as the ratio of the neutrino flux to the neutrino energy density times the speed of light. In leakage schemes, however, there is no notion of local neutrino number/energy density outside the diffusive regime, and therefore, an approximate expression is required. In free streaming conditions, the average neutrino flux factor, $\langle\chi_{\nu_i}\rangle$, approaches the value of $1$ as the radiation becomes forward peaked far away from the source, while at high optical depths $\langle\chi_{\nu_i}\rangle$ becomes very small. Its exact behaviour between both extremes, however, remains strongly dependent on the geometry of the neutrino emitting object. In the case of a spherical cooling PNS (e.g. \citealt{1991A&A...244..378J}), $\langle\chi_{\nu_i}\rangle$ is known to be about $1/4$ at the neutrinosphere, and we adopt for such a case the interpolation suggested by \cite{2010CQGra..27k4103O}, $\langle\chi_{\nu_i}\rangle_{\mathrm{PNS}}^{-1}\ =\ 4.275\tau_{\nu_i}+1.15$. For more complex geometries, such as a BH-torus system or a binary NS merger, more sophisticated models for the streaming factor, which encode the geometric effects, would be necessary. However, we take the aforementioned linear interpolation to be sufficiently good for the present work, as shown in the tested scenarios in section~\ref{sec:tests}. 

We obtain the absorption rate, $Q_{\nu_i}^+$, of a given cell with Cartesian coordinate $\boldsymbol x_2$, from the superposition of all the rays crossing this cell depositing energy according to equation \eqref{lumiabs}, and for homogeneous conditions in the cell, as 
\begin{align}
  Q_{\nu_i}^+(\boldsymbol x_2)\ &=\ \gamma_{\nu_i,\mathrm{en}}^{\mathrm{eff}}\boldsymbol\cdotp\sum_{\mathrm{rays}}{\frac{L_{\nu_i}^{\mathrm{ray}}(s^-)}{(\psi^{6} V)_{\mathrm{cell,abs}}}}\boldsymbol\nonumber \\ 
  &\cdotp{\left[1-\mathrm{exp}\left({\frac{-\bar{\kappa}_{\nu_i,\mathrm{a}}(\boldsymbol x_2)(s^{+}-s^{-})_{\mathrm{ray}}}{\langle\chi_{\nu_i}(\boldsymbol x_2)\rangle}}\right)\right]},\label{Q+}
\end{align}
 where $(s^{+}-s^{-})_{\mathrm{ray}}$ delimits the path the each ray travels inside the cell. The factor $\gamma_{\nu_i,\mathrm{en}}^{\mathrm{eff}}$ ensures that the absorption is mainly applied in the optically thin regime (see equation~\ref{gammaeff} for the definition of $\gamma_{\nu_i,\mathrm{en}}^{\mathrm{eff}}$) because our leakage scheme treats neutrino losses as effective emission (emission minus absorption) in the optically thick region. The mean absorption opacity of a given cell, $\bar{\kappa}_{\nu_i,\mathrm{a}}$, is calculated as in equations~\eqref{kappa_abs_nue} and~\eqref{kappa_abs_anue} with the corresponding spectrum of the neutrinos as seen by the fluid at a given location (see below). 
 
 In the framework of the leakage scheme, neutrinos are assumed to instantaneously leak out of the system, which would imply that they carry their production spectrum along the ray. Physically, however, neutrinos slowly diffuse out of the hot NS, thermalizing with the medium in the process, until the optical depth becomes small enough for them to freely stream away. In order to account for this behaviour, we determine the \textit{spectrum} of neutrinos (produced at $s_1$ and irradiating the fluid at $s_2$) by differentiating between the three following cases: 
 \begin{itemize}
  \item Neutrinos emitted anywhere are re-absorbed by cells which are \textit{inside} of the neutrinosphere ($\tau_{\nu_i,{s_2}}>2/3$) with a Fermi spectrum characterized by the local matter temperature and neutrino degeneracy (equation~\ref{etanu}) of the re-absorption cell (thermal spectrum, $f(\epsilon;T_{s_2},\eta_{\nu_i,{s_2}})$).
  
  \item Neutrinos emitted from \textit{inside} of the neutrinosphere ($\tau_{\nu_i,{s_1}}>2/3$) are re-absorbed by cells \textit{outside} of the neutrinosphere ($\tau_{\nu_i,{s_2}}<2/3$) with a Fermi spectrum characterized by the matter temperature and neutrino degeneracy of the last cell crossed by the ray which is inside of the neutrinosphere (neutrinospheric spectrum, $f(\epsilon;T_{(\tau_{\nu_i}=2/3)},\eta_{\nu_i,(\tau_{\nu_i}=2/3)})$).
  
  \item Neutrinos produced \textit{outside} of the neutrinosphere ($\tau_{\nu_i,{s_1}}<2/3$) are re-absorbed by cells also \textit{outside} of the neutrinosphere ($\tau_{\nu_i,{s_2}}<2/3$) with the same Fermi spectrum with which they were produced, i.e. characterized by the matter temperature and neutrino degeneracy of the production cell (production spectrum, $f(\epsilon;T_{s_1},\eta_{\nu_i,{s_1}})$).
 \end{itemize}

As in other grey absorption schemes (e.g. \citealt{2010CQGra..27k4103O}), we then estimate the lepton number deposition as
\begin{equation}
  R_{\nu_i}^+\ =\ \frac{Q_{\nu_i}^+}{\bar{\epsilon}_{\nu_i}}.\label{R+}
\end{equation}
We calculate the mean neutrino energy, $\bar{\epsilon}_{\nu_i}$, of neutrinos being absorbed in $\beta$-processes by considering Fermi spectra:
\begin{align}
 \bar{\epsilon}_{\nu_i}\ =\ T\frac{F_5(\eta_{\nu_i})}{F_4(\eta_{\nu_i})}\frac{\alpha(\tilde{s}_1)}{\alpha(s_2)},\label{epsilon_nusphere}
\end{align}
where the matter temperature and the neutrino degeneracies are consistently taken as above for each of the described cases. The redshift, $\alpha(\tilde{s}_1)/\alpha(s_2)$, is then applied only for absorption on cells outside of the neutrinosphere and only between the neutrinosphere (if the neutrinos are produced inside, i.e. $\tau_{\nu_i,s_1}>2/3$), $\tilde{s}_1=s_{(\tau_{\nu_i}=2/3)}$, or the production cell (if they are produced outside), $\tilde{s}_1=s_1$, and the absorbing cell, $s_2$. Neutrinos absorbed inside of the neutrinosphere locally thermalize with matter, thus eliminating any trace of the prior neutrino spectrum\footnote{Effectively, for redshift effects, neutrinos absorbed inside of the neutrinosphere are seen as if they were emitted at the same location where absorption occurs.}. 

As an alternative approach, one can calculate the neutrino-number re-absorption rates, $R_{\nu_i}^+$, by the same procedure described for the treatment of the neutrino energy re-absorption, i.e. independently of the neutrino-energy absorption rates. Following equation~\eqref{Q+}, but using the neutrino-number luminosities and spectrally-averaged opacities over the neutrino number spectra, defined in equations~\eqref{kappa_abs_nue} and~\eqref{kappa_abs_anue} (with $j=0$), we obtain directly the neutrino-number absorption rates. The comparison between the results obtained by both approaches on a PNS snapshot calculation, see appendix~\ref{appendix:nabs}, does not reveal any significant differences in the mean energies of radiated neutrinos (see equation~\ref{eleak}), thus demonstrating the robustness of our treatment to methodical variations in details.

We employ a Gaussian smoothing filter with standard deviation $\sigma=1$ over the absorption rates in the three-dimensional spatial domain. This ensures the conservation of the total absorption rates and mitigates the drawbacks of employing a limited number of rays\footnote{One ray per cell of the Cartesian grid.} in a ray-tracing approach. Thus smoothing out high local rates over neighbouring spatial points moderately boosts the computational performance of the scheme.

\subsection{Neutrino equilibration in optically thick matter}\label{sec:model:equil}

 At the typical densities and temperatures achieved during NS mergers or SNe, a part of the neutrinos is expected to remain trapped in optically thick conditions. Under such circumstances, they will achieve local beta equilibrium with the surrounding matter within a very short time, carrying lepton number, and contributing to the energy and pressure of the stellar medium.
  
\begin{figure}
  \centering
\begin{center}
\makebox[0pt][c]{%
\minipage{0.5\textwidth}
\includegraphics[width=\textwidth]{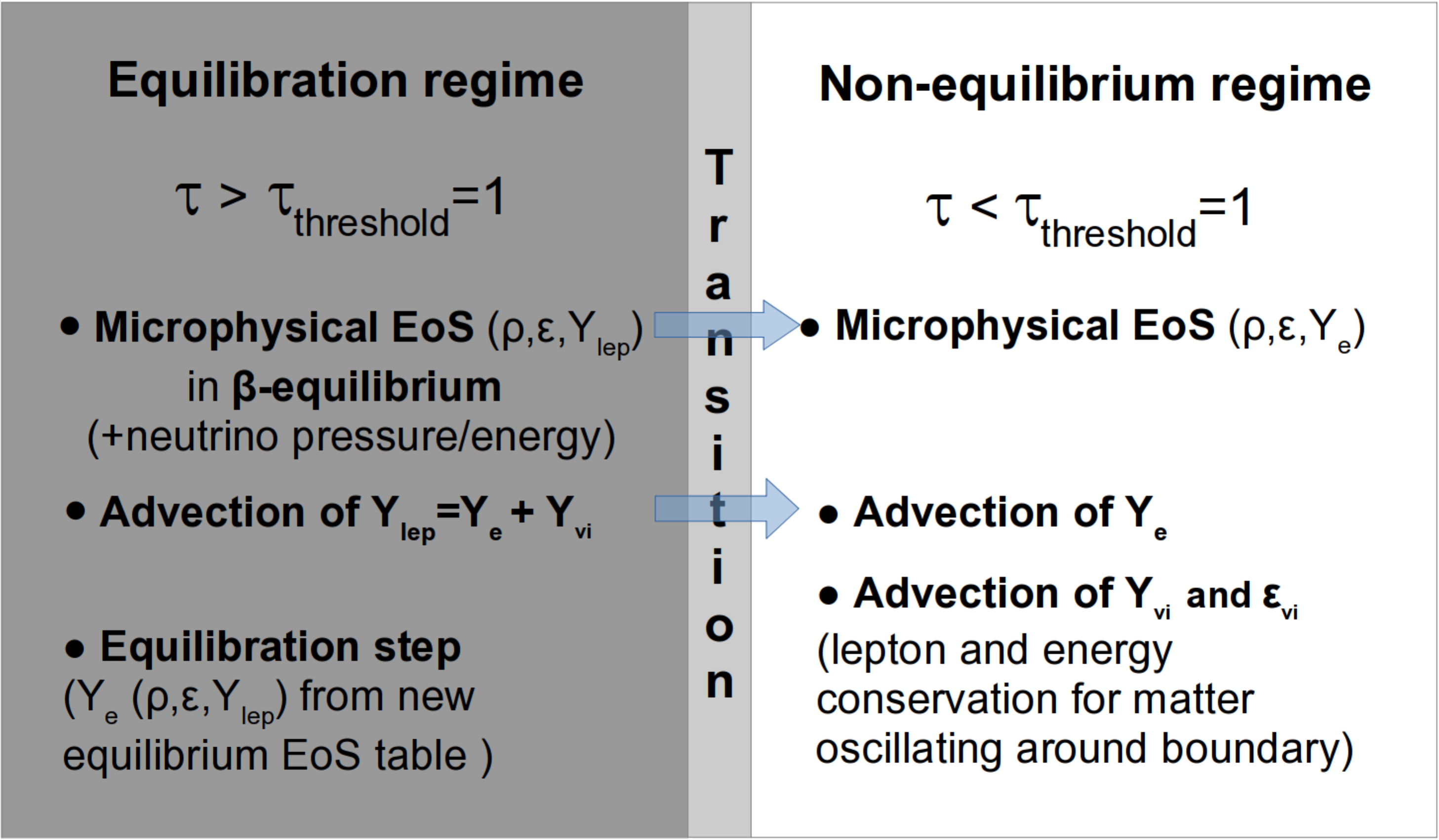}
\endminipage\hfill
}%
\caption{Diagram illustrating the functioning of the neutrino equilibration treatment. We use three overlapping equilibration regions, one for each neutrino species.} \label{equilibrationchart}
\end{center}
\end{figure}

 In order to account for this important effect, we developed an equilibration scheme which ensures that the fluid remains in beta equilibrium with the trapped neutrinos in the optically thick regime, by two measures. First, we employ a set of EoS tables which include the contributions of trapped neutrinos to the specific internal energy and pressure of the medium, to be used for the hydrodynamical evolution of the system. Second, we perform an \textit{equilibration step} after each hydro step, `reshuffling' the trapped leptons and recovering the equilibrium values of the corresponding thermodynamical quantities. This last step requires the advection of the trapped lepton fraction $Y_{\mathrm{lep}}$, which can be expressed as the sum of the individual trapped neutrino fractions, $Y_{\nu_i}^{\mathrm{trap}}$, and the electron fraction, $Y_e$, as we described in section~\ref{sec:model:coupling} (see equations~\ref{yeadv}-\ref{leptonadvanue} there). 
 
\begin{table}
  \begin{center}
      \caption{Neutrino equilibration regions included in \textsc{ILEAS}, discriminated by the trapped neutrino species in each of them.}\label{table:equilregions}  
    \begin{threeparttable}
     \begin{tabular}{lc}
      \hline
      \hline
      \noalign{\vskip 2mm}  
      \parbox[c]{0.1\textwidth}{Equilibration region}        & \parbox[c]{0.1\textwidth}{\centering Trapped $\nu$ species}    \vspace{1mm}  \\ 
      \hline   
      \noalign{\vskip 2mm}                                                                                                                       
      1		& $\nu_e,\ \bar{\nu}_e,\ \nu_x$		\\                                                                               
      2		& $\nu_e,\ \bar{\nu}_e$			\\                                                                        
      3		& $\nu_e,\ \nu_x$			\\                                                                        
      4		& $\bar{\nu}_e,\ \nu_x$			\\                                                                        
      5		& $\nu_e$				\\                                                                        
      6		& $\bar{\nu}_e$				\\                                                                        
      7		& $\nu_x$				\\                                                                 
      8		& none					\\  
      \noalign{\vskip 2mm}  
      \hline
      \hline
     \end{tabular}
    \end{threeparttable}	
  \end{center}
\end{table}

 We treat each of the three different neutrino species independently, describing overlapping equilibration regions, which requires us to build an additional EoS table for each possible combination of trapped species. This amounts to a total of eight different possibilities, listed in table~\ref{table:equilregions}. Even though we opted for the most general implementation of the neutrino equilibration regions, it is also possible to reduce the number of different equilibration zones by assuming a hierarchy in the minimum densities for which neutrinos of the different species remain trapped. In most relevant astrophysical scenarios, $\nu_x$ will decouple from matter at higher densities than the other two species, followed by $\bar{\nu}_e$, and finally $\nu_e$ at lower densities. Therefore, a simpler equilibration treatment could be achieved with only the inclusion of regions 1, 2, 5, and 8, yet capturing all important physical effects under most circumstances.
 
 Our new EoS tables provide all necessary thermodynamical quantities as functions of density, $\rho$, specific internal energy (including the stellar plasma and the corresponding trapped neutrino contribution), $\varepsilon$, and the trapped lepton fraction, $Y_{\mathrm{lep}}$, defined as in equation~\eqref{Ylep}. Therefore, for each of the equilibration regions listed in table~\ref{table:equilregions} we build an EoS table including only the contributions of the corresponding trapped neutrinos which we use only in the corresponding equilibration region. We remind the reader that only $\nu_e$ and $\bar{\nu}_e$ contribute to the trapped electron-lepton fraction in their respective equilibration regions, because $\nu_x$ do not carry electron flavour. Moreover, since $\nu_x$ are produced in pairs and are treated equally with respect to neutrinos and anti-neutrinos, we assume that muon and tauon numbers do not build up in the stellar environment. The thermodynamical quantities we need to obtain from the EoS call are 
 
\begin{itemize}
   \item the total pressure, which includes the contributions from medium and trapped neutrinos, $P$,
   \item the temperature, $T$, 
   \item the chemical potentials, $\mu_n$, $\mu_p$ and $\mu_e$, 
   \item the individual lepton fractions, $Y_e$, $Y_{\nu_e}^{\mathrm{trap}}$ and $Y_{\bar{\nu}_e}^{\mathrm{trap}}$ ,
   \item and the individual specific neutrino energy contributions, $\varepsilon_{\nu_e}$, $\varepsilon_{\bar{\nu}_e}$ and $\varepsilon_{\nu_x}$. 
\end{itemize}
  
 These two last sets of quantities, the individual lepton fractions and the individual specific neutrino energies, are relevant for the treatment of the boundaries of each equilibration region, as will be detailed below.
 
 The neutrino contribution to the specific internal energy of the fluid can be calculated from the neutrino equilibrium energy density of a given neutrino species, $\bar{E}_{\nu_i}^{1}$ (equation~\ref{edenseq} with $j=1$), as
 \begin{equation}
  \varepsilon_{\nu_i}\ =\ \frac{\bar{E}_{\nu_i}^{1}}{\rho c^2}.\label{specifice}
 \end{equation}
 The neutrino fraction, $Y_{\nu_i}^{\mathrm{trap}}$, can be obtained from the neutrino equilibrium number density, $\bar{E}_{\nu_i}^{0}$ (equation~\ref{edenseq} with $j=0$), as
 \begin{equation}
  Y_{\nu_i}^{\mathrm{trap}}\ =\ \frac{\bar{E}_{\nu_i}^{0}}{\rho \mathcal{A}},\label{ynui}
 \end{equation} 
 where $\mathcal{A}$ is Avogadro's constant. The Fermi integrals for the equilibrium energy density of $\nu_x\bar{\nu}_x$ pairs are computed by the analytical expression from \cite{1978ApJ...224..631B}, whereas for $\nu_e$ and $\bar{\nu}_{e}$ we analytically approximate the Fermi integrals in equation~\eqref{edenseq} (with $j=0$ for number and $j=1$ for energy) following \cite{1978A&A....67..185T}. Then, one can calculate the pressure of each neutrino species as
 \begin{equation}
  P_{\nu_i}\ =\ \frac{1}{3}\bar{E}_{\nu_i}^{1}.\label{nupressure}
 \end{equation}  
 
 We apply our equilibration treatment for a given neutrino species, $\nu_i$, down to optical depths $\tau_{\nu_i}\geq1$. At lower optical depths, the deviations from the equilibrium energy density become significant (>20 per cent), and thus the assumption of beta equilibrium is not suitable. 
 
 Each equilibration region listed in table~\ref{table:equilregions} employs a different EoS table, which depends on the composition of matter in that region via $Y_{\mathrm{lep}}$ and the fluid (stellar medium plus trapped neutrinos) specific internal energy, $\varepsilon$. During the dynamical evolution of a system, SPH particles or grid cells will switch between the different equilibration regions. In order to ensure energy conservation of material crossing these boundaries, we add or subtract from that cell's or particle's $\varepsilon$, the corresponding neutrino contribution. This requires the recovery of the neutrino specific energy component, $\varepsilon_{\nu_i}$ from the EoS tables at every time step. Because outside of a given equilibration region, we assume that $\nu_i$ are not in equilibrium with matter but do not leave the system immediately if marginally outside of their equilibration region, we simply advect $\varepsilon_{\nu_i}$, i.e. $\varepsilon_{\nu_i}$ is only updated \textit{inside} the $\nu_i$ equilibration region. Similarly, we advect the individual $Y_{\nu_i}^{\mathrm{trap}}$ not exclusively in the corresponding trapping region, but in the whole domain (see equations~\ref{leptonadvnue} and~\ref{leptonadvanue}). This advection serves the purpose of avoiding non-physical energy and lepton losses by material oscillating around a given boundary. When matter flows inside an equilibration region, its advected $Y_{\nu_i}^{\mathrm{trap}}$ and $\varepsilon_{\nu_i}$ will contribute again to the fluid's total lepton fraction and specific internal energy, respectively. This boundary treatment is sufficiently good under the assumption that material re-entering a given regime spent too little time outside to experience significant losses of its neutrino content. Since hardly ever material re-enters the equilibrium domain from having been far outside, this treatment is sufficiently good. We remind the reader again that neutrino losses from the stellar medium are accounted for by the leakage module, both in the trapping and free-streaming domains. In figure~\ref{equilibrationchart} we summarize the equilibration module for a given equilibration region and the transition from this domain to its neighbouring regions, where equilibrium is not fulfilled.
 
 \subsection{Extraction of neutrino properties from \textsc{ILEAS}}\label{sec:model:postproc}
 
 It is often desirable to extract some relevant neutrino-related quantities from numerical simulations, to be used for post-processing, in nucleosynthesis calculations or to treat neutrino oscillations, to predict the detectability of a signal by neutrino detectors or just for diagnostics. In the present section, we describe how we calculate the neutrino luminosities and the radiated mean neutrino energies in \textsc{ILEAS}.

Given the neutrino loss and absorption rates, the net neutrino ``luminosities'' that reach distant observers (in the rest frame of the source's centre of mass) can be approximately written (neglecting Doppler effects, gravitational ray bending, time retardation and the shift vector) as\footnote{Note that the quantity denoted here as ``luminosity'', $L_{\nu_i}$, is actually the volume-integrated number or energy-loss rate, which should be distinguished from the viewing-angle dependent luminosities that can be measured by external observers.}
\begin{equation}
 L_{\nu_i}^{\mathrm{en}}(\boldsymbol x_{\mathrm{obs}})\ =\ {\int{(Q_{\nu_i}^-(\boldsymbol x)-Q_{\nu_i}^+(\boldsymbol x))\frac{\alpha(\boldsymbol x)^2}{\alpha(\boldsymbol x_{\mathrm{obs}})^2}\psi(\boldsymbol x)^{6}\mathrm{d}V}},\label{lumien}
\end{equation}
for energy, and
\begin{equation}
 L_{\nu_i}^{\mathrm{num}}(\boldsymbol x_{\mathrm{obs}})\ =\ {\int{(R_{\nu_i}^-(\boldsymbol x)-R_{\nu_i}^+(\boldsymbol x))\frac{\alpha(\boldsymbol x)}{\alpha(\boldsymbol x_{\mathrm{obs}})}\psi(\boldsymbol x)^{6}\mathrm{d}V}},\label{luminum}
\end{equation}
for number, where, $\boldsymbol x$ and $\boldsymbol x_{\mathrm{obs}}$ are the positions of the emitting/absorbing cell and of the observer respectively, and the integral runs over our grid domain. Trivially, for an observer at rest at an infinite distance, $\alpha(\boldsymbol x_{\mathrm{obs}})=\alpha(\infty)=1$.

The natural way to estimate the mean neutrino energies in the leakage framework is simply by the ratio of net energy and number luminosities,
\begin{equation}
 \langle\epsilon_{\nu_i}^{\mathrm{leak}}\rangle(\boldsymbol x_{\mathrm{obs}})\ =\ \frac{L_{\nu_i}^{\mathrm{en}}(\boldsymbol x_{\mathrm{obs}})}{L_{\nu_i}^{\mathrm{num}}(\boldsymbol x_{\mathrm{obs}})}.\label{eleak}
\end{equation}
This approach, however, does not yield very good agreement with transport results for the mean energies of radiated neutrinos (see section~\ref{sec:tests:snapPNS}), because it bears the deficiency mentioned earlier, namely that it ignores the thermalization of neutrinos produced at high optical depths on their way out of the star. As detailed in section~\ref{sec:model:abs}, in our absorption module we do not follow this leakage ansatz, but instead work with the local neutrino spectra inside the neutrinosphere, $\tau_{\nu_i}>2/3$, and assume that neutrinos in the optically thin region ($\tau_{\nu_i}<2/3$) carry either their neutrinospheric spectra, if produced in the optically thick region, or their production ones. 

In order to provide a more meaningful value for the radiated mean neutrino energies, we make the following approximations in a post processing step. First, we differentiate the optically thick and optically thin regimes, as introduced earlier, separated by the neutrinosphere at $\tau_{\nu_i}=2/3$. The mean energy for neutrinos produced in the optically thin regime is calculated in the fashion of leakage schemes, but accounting independently for the absorption of energy \textit{and number}, as
\begin{align}&\langle\epsilon_{\nu_i}^{\mathrm{thin}}\rangle(\boldsymbol x_{\mathrm{obs}})\ =\ \nonumber \\
 &\frac{Q_{\nu_i}^-(\boldsymbol x_1)\mathrm{exp}\left(-2\int_{s^-_1}^{s^+_1}{{\bar{\kappa}_{\nu_i,\mathrm{a}}^{j=1}(\boldsymbol x_1)}/{\langle\chi_{\nu_i}(\boldsymbol x_1)\rangle}\mathrm{d}s'}\right)\alpha(\boldsymbol x_1)}{R_{\nu_i}^-(\boldsymbol x_1)\mathrm{exp}\left(-2\int_{s^-_1}^{s^+_1}{{\bar{\kappa}_{\nu_i,\mathrm{a}}^{j=0}(\boldsymbol x_1)}/{\langle\chi_{\nu_i}(\boldsymbol x_1)\rangle}\mathrm{d}s'}\right)\alpha(\boldsymbol x_{\mathrm{obs}})}.\label{ethin}
\end{align}
Here the spectrally averaged opacities for energy and number absorption, $\bar{\kappa}_{\nu_i,\mathrm{a}}^{j=1}$ and $\bar{\kappa}_{\nu_i,\mathrm{a}}^{j=0}$, are calculated as in equations~\eqref{kappa_abs_nue} and~\eqref{kappa_abs_anue} with the neutrino production spectrum, $f(\epsilon;T_{s_1},\eta_{\nu_i,{s_1}})$, of the emitting cell. We remind the reader that $s$ is the ray coordinate, as used in section~\ref{sec:model:abs}, and the ray origin, $s^-_1$, corresponds to the centre of the cell in the Cartesian coordinate $\boldsymbol x_1$. Because we want to evaluate the mean energies as observable from outside of our domain, we include the gravitational redshift from the production cell to an observer positioned at $\boldsymbol x_{\mathrm{obs}}$. Given the steep density distribution typical of the environments of PNSs or HMNSs, it is a fairly accurate approximation that most neutrinos will be re-absorbed near their emission location. In this spirit, we approximate the path in the line integral in equation~\eqref{ethin} by the total distance the ray would travel if it crossed the whole production cell, which we consider as a proxy for the absorption along the whole outgoing ray\footnote{Hence the factor 2 before the integrals in equation \eqref{ethin}.}. Note that in the absorption module (section~\ref{sec:model:abs}), we only take the path from the centre to the edge of the cell for self-absorption of a production cell, but follow the whole paths of outgoing rays.

Furthermore, we define an absorption correction factor of the mean energy  for neutrinos coming from inside the neutrinosphere,
\begin{equation}
 c_{\mathrm{abs}}\ =\ \left[\frac{\mathrm{exp}\left(-\int_{s^-}^{s^+}{{\bar{\kappa}_{\nu_i,\mathrm{a}}^\mathrm{j=1}(\boldsymbol x)}/{\langle\chi_{\nu_i}(\boldsymbol x)\rangle}\mathrm{d}s'}\right)}{\mathrm{exp}\left(-\int_{s^-}^{s^+}{{\bar{\kappa}_{\nu_i,\mathrm{a}}^\mathrm{j=0}(\boldsymbol x)}/{\langle\chi_{\nu_i}(\boldsymbol x)\rangle}\mathrm{d}s'}\right)}\right]_{(\tau_{\nu_i}=2/3)}.\label{corrfacabs}
\end{equation}
This correction factor is evaluated in the cells immediately outside of the neutrinosphere, using their local equilibrium neutrino spectrum. Rays escaping from inside the neutrinosphere and crossing such cells, will then have their mean energies corrected by means of $c_{\mathrm{abs}}$, representing the whole absorption outside the neutrinosphere. Each ray emerging from the optically thick regime will thus contribute to the final average with a mean energy
\begin{equation}
 \langle\epsilon_{\nu_i}^{\mathrm{thick}}\rangle(\boldsymbol x_{\mathrm{obs}})\ =\ \left[c_{\mathrm{abs}}T\frac{F_3(\eta_{\nu_i})}{F_2(\eta_{\nu_i})}\right]_{(\tau_{\nu_i}=2/3)}\boldsymbol\cdotp\frac{\alpha(\boldsymbol x_{(\tau_{\nu_i}=2/3)})}{\alpha(\boldsymbol x_{\mathrm{obs}})},\label{ethick}
\end{equation}
calculated where the ray crosses the neutrinosphere, and including the aforementioned correction factor. 

Finally, we obtain the total radiated mean neutrino energy by means of a weighted average of all rays, using the neutrino energy luminosities \textit{leaving} the corresponding cells either at the neutrinosphere for the optically thick rays or from production cells in the optically thin region:
\begin{align}
 \langle\epsilon_{\nu_i}^{\mathrm{tot}}\rangle\ =\ &\frac{\sum\limits_{k|\tau_{\nu_i}>2/3}{\langle\epsilon_{\nu_i,k}^{\mathrm{thick}}\rangle\boldsymbol\cdotp \Delta L_{\nu_i,k}(\boldsymbol x)}}{\sum\limits_{k\in V}{\Delta L_{\nu_i,k}(\boldsymbol x)}}+\nonumber \\
 &\frac{\sum\limits_{k|\tau_{\nu_i}<2/3}{\langle\epsilon_{\nu_i,k}^{\mathrm{thin}}\rangle\boldsymbol\cdotp \Delta L_{\nu_i,k}(\boldsymbol x)}}{\sum\limits_{k\in V}{\Delta L_{\nu_i,k}(\boldsymbol x)}}\label{etot}
\end{align}
Here the summations in the numerator go over all rays $k$ which are emitted from cells inside ($\tau_{\nu_i}>2/3$) or outside ($\tau_{\nu_i}<2/3$) the neutrinosphere, and the one in the denominator over the whole volume.

\section{\texorpdfstring{Astrophysical test applications: \\
    cooling PNS and BH-torus systems}{Lg}}\label{sec:tests}
    
All the \textsc{ILEAS} calculations presented in this work (section~\ref{sec:tests}) were performed on a three-dimensional Cartesian grid with $\sim$0.7~km of resolution in all three coordinate directions. This grid expands $\sim$100~km in all six Cartesian directions ($\pm x,\ \pm y,\ \pm z$) from the centre-of-mass of the system, covering the astrophysical objects and their immediate surroundings. The same grid is also employed for the NS merger simulations, providing full coverage of the late inspiral phase as well as the initial merger remnant and the absorption-dominated regions along polar directions.

In this section, we differentiate two kinds of test applications for our \textsc{ILEAS} scheme: snapshot calculations and time evolution. For the snapshot calculations we apply \textsc{ILEAS} on a snapshot of the hydrodynamical and thermodynamical data obtained from a simulation performed with \textsc{ALCAR} or \textsc{VERTEX} neutrino transport codes. After a short relaxation (see details below), we compare the results obtained by \textsc{ILEAS} and the corresponding neutrino transport code in the whole spatial domain. For the time evolution tests on the other hand, we will constrain ourselves on a comparison of the spatially-integrated results obtained by \textsc{ILEAS} and \textsc{ALCAR} as a function of time, with both schemes evolving the temperature and electron fraction of the matter background starting from the same original snapshot.

In order to test the performance of our scheme in time-dependent systems and to relax the thermodynamical background, we have coupled \textsc{ILEAS} to a simple time evolution scheme. As we want to focus on the neutrino effects, we only evolve the temperature (via the internal energy density of the fluid, $E_{\mathrm{fluid}}$) and the electron fraction, keeping the matter density fixed and ignoring the velocity terms\footnote{Radial velocities are small for the PNS case and not very high for the BH-torus systems.}. Not evolving the density allow to test the neutrino treatment independently from hydrodynamics.

We can calculate the changes in the electron fraction from the rate equation as
\begin{align}
 \left(\frac{\mathrm{d}Y_e}{\mathrm{d}t}\right)_{\mathrm{source}}\ =\ \frac{R_{\mathrm{tot}}}{\mathcal{A}\rho},\label{Yeevol}
\end{align}
where $R_{\mathrm{tot}}$ is given in equation~\eqref{rtot} and $\mathcal{A}$ is the Avogadro constant. For the fluid energy density, following the first law of thermodynamics for a quasi-static system with fixed density, we can express its evolution equation as
\begin{align}
 \left(\frac{\mathrm{d}E_{\mathrm{fluid}}}{\mathrm{d}t}\right)_{\mathrm{source}}\ =\ Q_{\mathrm{tot}},\label{uevol}
\end{align}
where $Q_{\mathrm{tot}}$ is given in equation~\eqref{qtot}. We solve these simple equations explicitly with a forward integration, allowing for changes on either quantity of up to 2 per cent in a single time-step. Then, we only need to call the EoS to obtain the temperature from the energy density, matter density and $Y_e$ (via bisection) and then the chemical potentials, which we use in the next leakage step. We also include equilibration as described in this paper.

We initialize the system by computing the fluid energy density and chemical potentials from the EoS using the density, temperature and electron fraction from the initial snapshot. Then we calculate the boundaries of the equilibration regions and set the initial trapped lepton fraction inside of each region to include the corresponding trapped neutrino components, as specified in table~\ref{table:equilregions}.

For the snapshot calculation tests, we only evolve the system for around 5~ms until a stationary (\textit{relaxed}) state of the thermodynamical conditions is reached. This prevents our results from being contaminated by an initial transient caused by the replacement of the neutrino treatment from \textsc{ALCAR} or \textsc{VERTEX} to \textsc{ILEAS}. The \textsc{ILEAS} results after this relaxation are then compared to the \textsc{ALCAR}/\textsc{VERTEX} data used at the time of mapping. In contrast, in time-evolution tests, calculations with \textsc{ILEAS} and \textsc{ALCAR} are performed in parallel over longer periods of time.

 \subsection{Snapshot calculations of a cooling proto-neutron star}\label{sec:tests:snapPNS}

\begin{figure*}
\begin{center}
\makebox[0pt][c]{%
\minipage{0.5\textwidth}
\includegraphics[width=\textwidth]{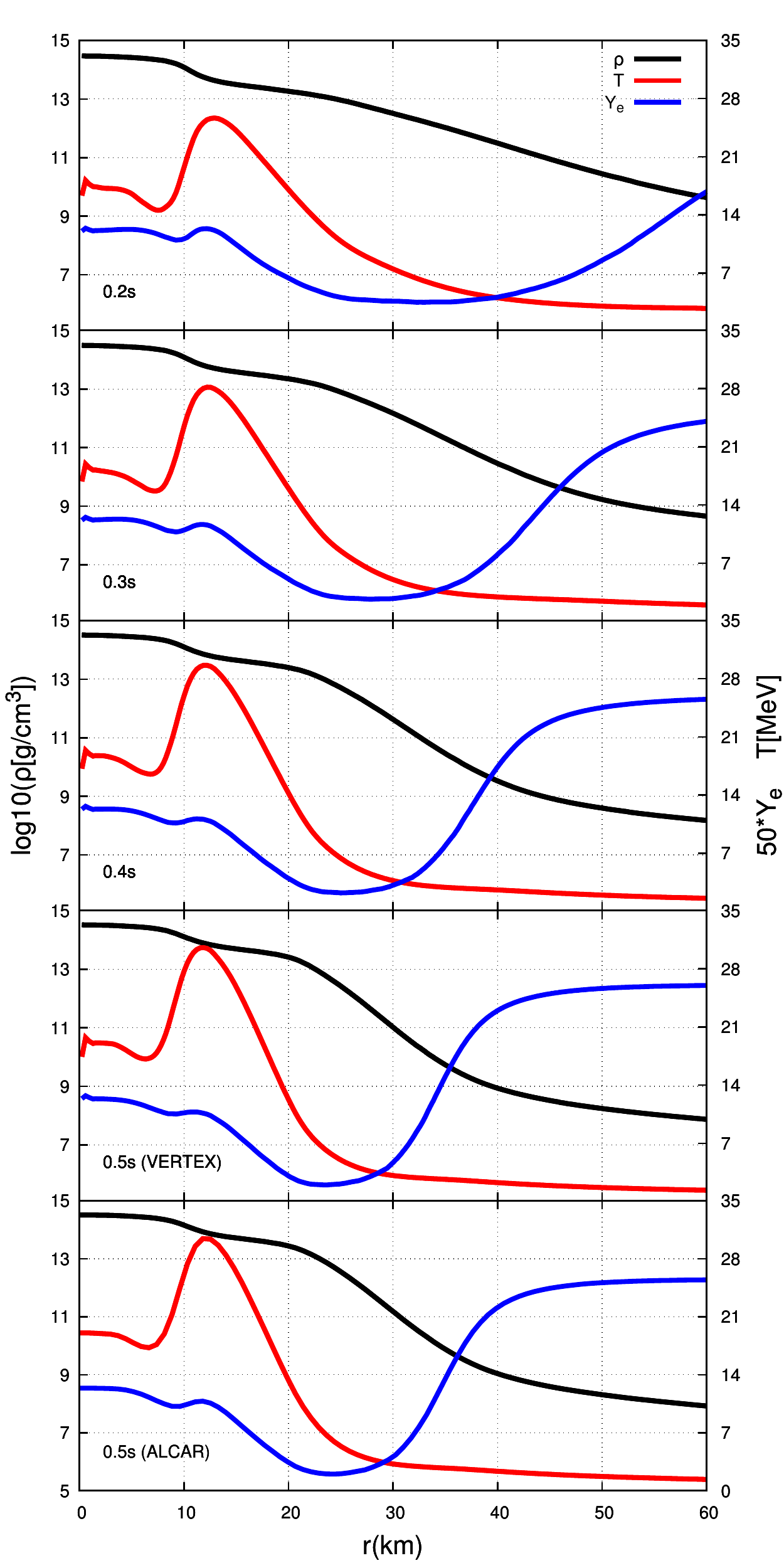}
\endminipage\hfill
\hspace{0.1cm}
\minipage{0.5\textwidth}
\includegraphics[width=\textwidth]{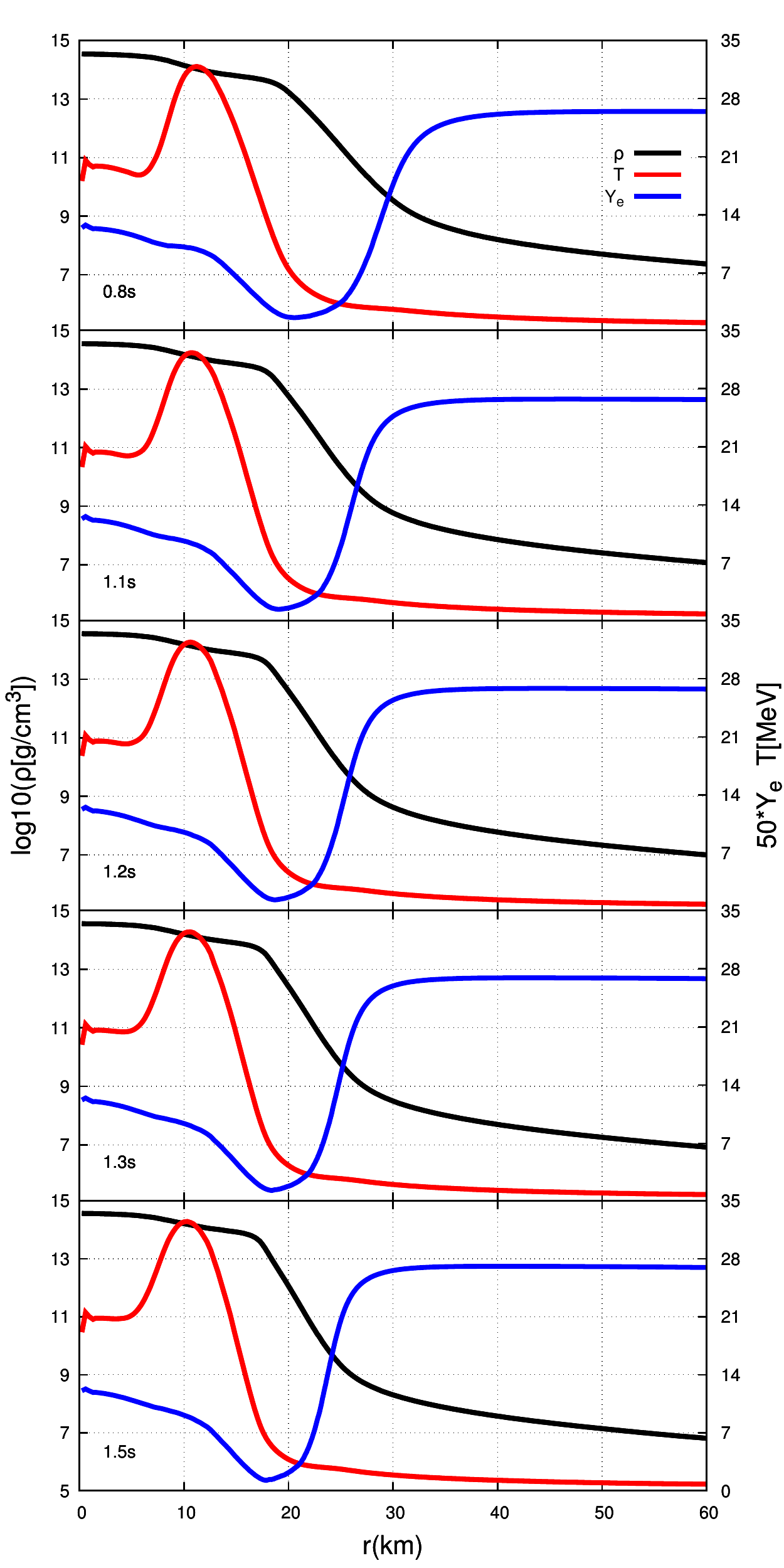}
\endminipage\hfill
}%
\caption{Hydrodynamical/thermodynamical profiles (density black, temperature red and electron fraction blue) of the different PNS snapshots employed in section~\ref{sec:tests:snapPNS}. Post-bounce times of the profiles are specified in the left lower corner of the panels.}\label{Hydro}
\end{center}
\end{figure*}

\begin{table}
  \begin{center}
      \caption{Neutrino interactions employed in section~\ref{sec:tests} for our tests with the \textsc{ILEAS} code in comparison to \textsc{ALCAR} and \textsc{VERTEX} calculations ($N=p,n$). Note that for the BH-Torus models in section~\ref{sec:tests:snapBHT} we neglected $\nu_x$ altogether. We point out that the current version of ILEAS includes neutrino-pair production but not pair annihilation processes.}\label{table:nureaccomp}  
    \begin{threeparttable}
     \begin{tabular}{lcc}
      \hline
      \hline
      \noalign{\vskip 2mm}  
      \parbox[c]{0.1\textwidth}{Name}        & \parbox[c]{0.1\textwidth}{Interaction}        & \parbox[c]{0.1\textwidth}{\centering $\nu$ species} \vspace{1mm}  \\ 
      \hline   
      \noalign{\vskip 2mm}                                                                                                                       
      $\beta$-react. for $\nu_e$	               & $p+e^- \leftrightarrow n+\nu_e$			            & $\nu_e$	   			\\  
      $\beta$-react. for $\bar{\nu}_e$	       & $n+e^+ \leftrightarrow p+\bar{\nu}_e$     		& $\bar{\nu}_e$	   			\\                                                                                                
      $e^-e^+$ annihil.			                   & $e^-+e^+ \rightarrow \nu_i+\bar{\nu}_i$    	& $\nu_x$				\\                  
      N-N bremsstr.\tnote{$\dagger$}		   & $N+N \rightarrow N+N+\nu_i+\bar{\nu}_i$		& $\nu_x$	   			\\                                                                             
      Nucleon scatt.			                           & $N+\nu_i\rightarrow N+\nu_i$				            & $\nu_e$, $\bar{\nu}_e$, $\nu_x$	\\  
      \noalign{\vskip 2mm}  
      \hline
      \hline
     \end{tabular}
     \begin{tablenotes}	
       \item[$\dagger$]  $N=p,n$
     \end{tablenotes}
    \end{threeparttable}	
  \end{center}
\end{table}

\begin{figure*}
\begin{center}
\makebox[0pt][c]{%
\minipage{\textwidth}
\includegraphics[width=\textwidth]{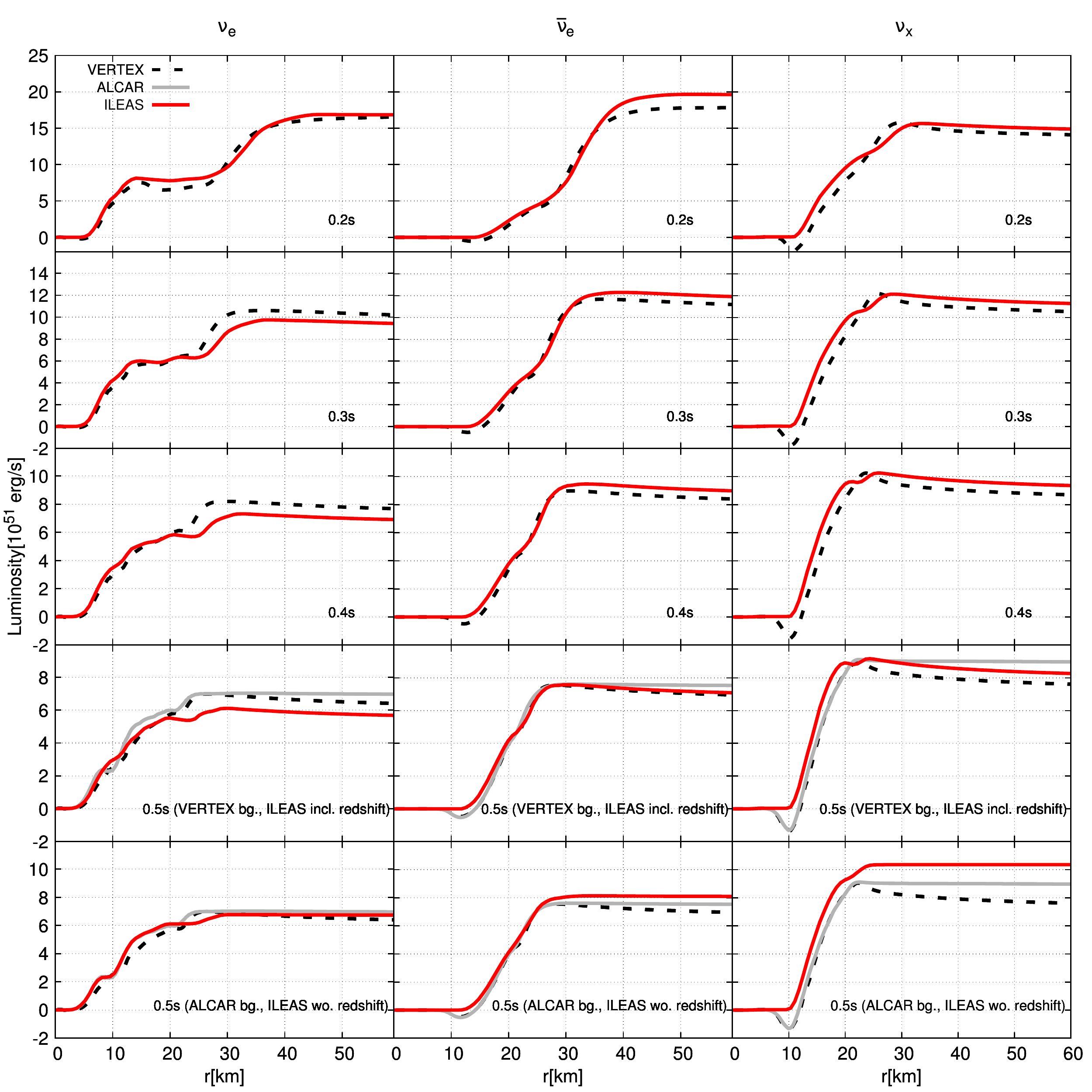}
\endminipage\hfill
}%
\caption{Radial neutrino luminosity profiles obtained by \textsc{ILEAS} on a quasi-stationary background for $\nu_e$, $\bar{\nu}_e$ and one representative of $\nu_x$, as measured in the local frame at each radius \textit{r}, compared to the results of more sophisticated transport schemes. Snapshots are taken from the results of PNS cooling simulations by {\protect\cite{2010PhRvL.104y1101H}} at different post-bounce times, which are noted in the lower right corner of each panel. The temperature and $Y_e$ profiles were relaxed for 5~ms with \textsc{ILEAS} to obtain stationary results. The \textsc{VERTEX} results shown are the original luminosities from the source model, and for the case at 0.5~s, we also show the results obtained by the M1 scheme \textsc{ALCAR} {\protect\citep{2015MNRAS.453.3386J}} for the same snapshot. The bottom row of plots shows the results of \textsc{ILEAS} obtained on the background adopted from \textsc{ALCAR} instead of directly using the output of the \textsc{VERTEX} simulation as in the row above. We caution the reader that the \textsc{ALCAR} luminosities, as well as the ones obtained by \textsc{ILEAS} on the \textsc{ALCAR} background (bottom row), do not include gravitational redshift.} \label{PNSSnapshots1}
\end{center}
\end{figure*}

\begin{figure*}
\begin{center}
\makebox[0pt][c]{%
\minipage{\textwidth}
\includegraphics[width=\textwidth]{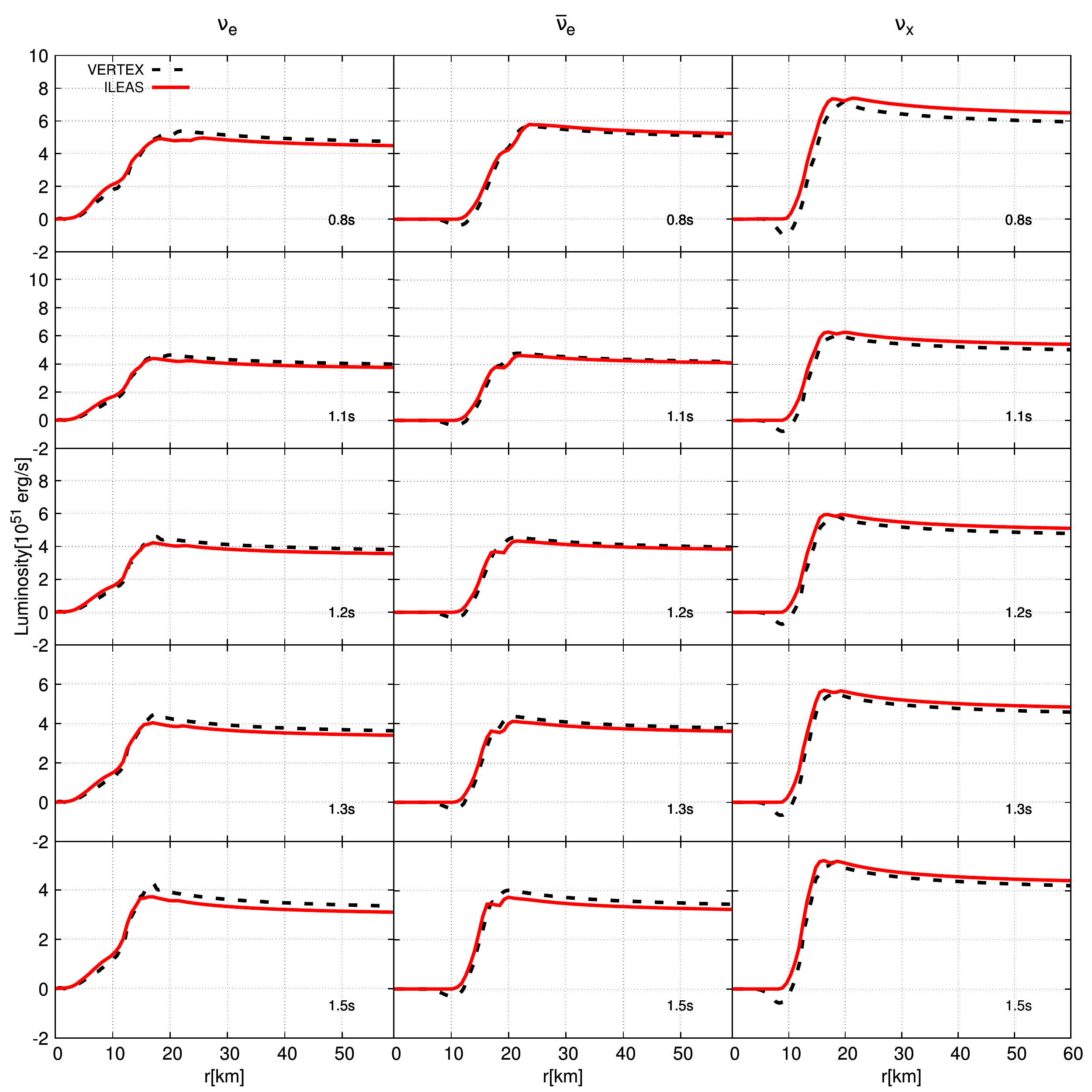}
\endminipage\hfill
}%
\caption{Radial neutrino luminosity profiles obtained by \textsc{ILEAS} on a quasi-stationary background for $\nu_e$, $\bar{\nu}_e$ and one representative of $\nu_x$, as measured in the local frame at each radius \textit{r}, compared to the results of more sophisticated transport schemes. Snapshots are taken from the results of PNS cooling simulations by {\protect\cite{2010PhRvL.104y1101H}} at different post-bounce times, which are noted in the lower right corner of each panel. The temperature and $Y_e$ profiles were relaxed for 5~ms with \textsc{ILEAS} to obtain stationary results. The \textsc{VERTEX} results shown are the original luminosities from the source model.} \label{PNSSnapshots2}
\end{center}
\end{figure*}

\begin{figure}
\begin{center}
\makebox[0pt][c]{%
\minipage{0.5\textwidth}
\includegraphics[width=\textwidth]{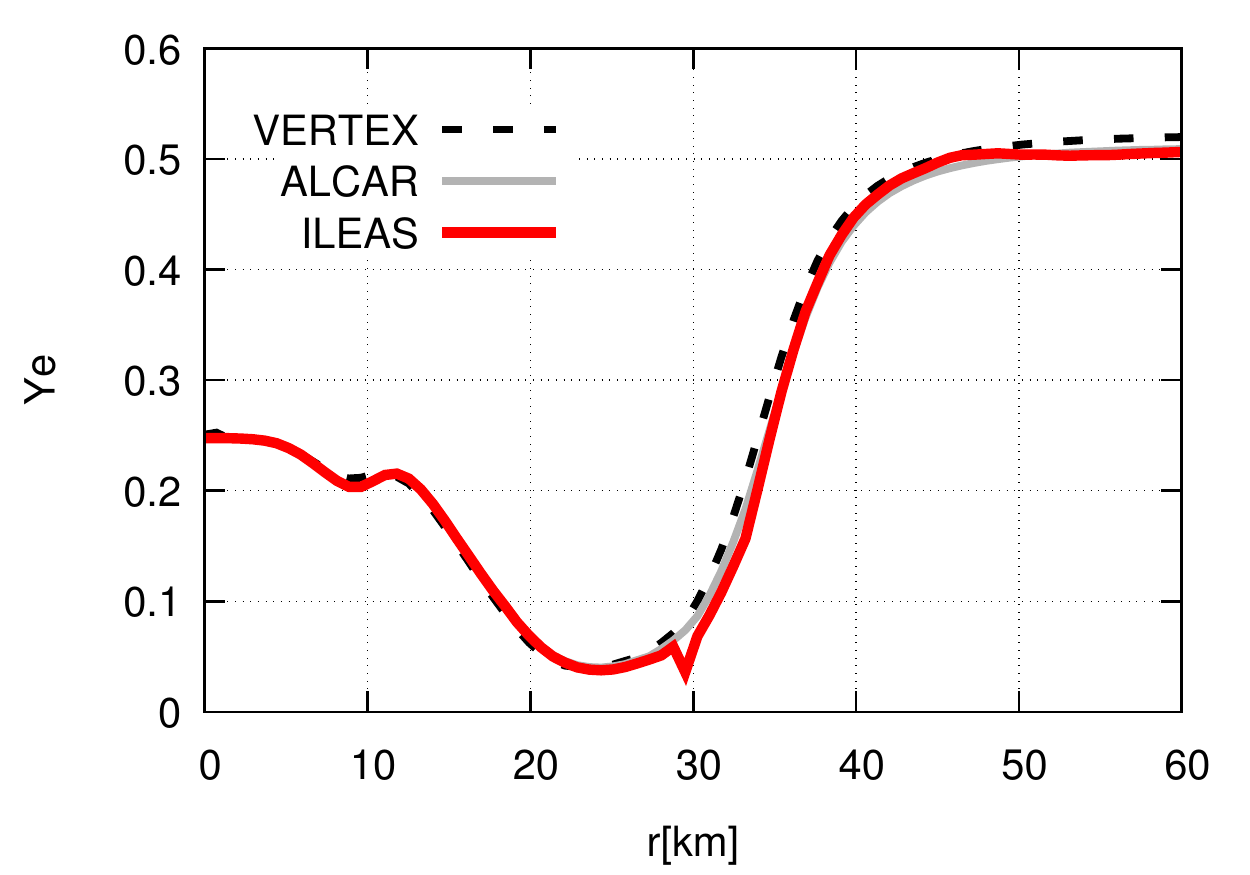}
\endminipage\hfill
}%
\caption{Electron fraction profile after 5~ms of relaxation obtained by \textsc{ILEAS} applied to the 0.5~s PNS snapshot from \textsc{ALCAR}. We show for comparison also the $Y_e$ profiles of the \textsc{VERTEX} and \textsc{ALCAR} simulations.} \label{ye05}
\end{center}
\end{figure}

In order to asses the quality of our new \textsc{ILEAS} code, we need to test it against more sophisticated transport schemes and in different regimes. Given that our ultimate goal is the application of \textsc{ILEAS} in the context of NS mergers, cooling PNSs present a relevant test scenario. During the explosion of a massive star in a SN, its core contracts to high densities and temperatures, giving birth to a young NS. The hot, dense interior of such a newly formed PNS is a perfect representation of an optically thick regime where the diffusion treatment can be tested. Additionally, the star is surrounded by a less dense envelope, where absorption of the neutrinos emitted from the NSs surface will apply. In between, the transition region around the neutrinosphere poses the most challenging conditions for treatments based on an interpolation of diffusive and free streaming regimes, such as in our \textsc{ILEAS} method. 

We apply our scheme to several snapshots from a hydrodynamical simulation performed by \cite{2010PhRvL.104y1101H}, who used the 1D version of the \textsc{PROMETHEUS}-\textsc{VERTEX} code with energy-dependent two-moment neutrino transport including Boltzmann-closure. We take the hydrodynamical and thermodynamical data (density, temperature and $Y_e$) at different times post-bounce from the model labelled Sr (reduced opacities), and map it to our 3D Cartesian grid using a standard trilinear interpolation\footnote{This cooling PNS is the remnant of an 8.8~\Msun\ (zero-age main sequence stellar mass) electron-capture SN.}. The motivation behind the chosen model is the similarity of our opacities and production reactions with the ones included in the original setup. Figure~\ref{Hydro} shows the density, temperature and electron fraction profiles of the corresponding snapshots.

For the sake of more detailed comparisons, we also employed the energy-dependent M1 scheme \textsc{ALCAR} \citep{2015MNRAS.453.3386J} to calculate the neutrino luminosities for one of the snapshots. Starting the evolution from an earlier timestep (0.4~s post-bounce) of the \textsc{VERTEX} simulation and evolving it hydrodynamically for 0.1~s, \textsc{ALCAR} was able to reproduce the results of \textsc{VERTEX} at 0.5~s with remarkable accuracy. We use this evolved \textsc{ALCAR} background (at 0.5~s post-bounce) for our direct, detailed comparison of the results obtained by \textsc{ILEAS} and \textsc{ALCAR}. The neutrino interactions employed by both schemes for the tests are summarized in table~\ref{table:nureaccomp}. We must point out that the prescriptions for $\nu_x$ production rates (pair processes and bremsstrahlung) differ between both codes, therefore a bigger disagreement is to be expected in the luminosities of these heavy-lepton neutrinos (see \citealt{2002A&A...396..361R} and appendix~\ref{appendix:reac} for the exact definitions of the rates employed by \textsc{ALCAR} and \textsc{ILEAS}, respectively). Moreover, some differences will unavoidably arise from the fact that \textsc{ILEAS} is in essence a grey scheme while \textsc{ALCAR} is fully energy-dependent. Finally, \textsc{ILEAS} is implemented on a 3D Cartesian grid, whereas \textsc{ALCAR} uses a spherical (polar) grid, offering advantages with respect to resolving radial gradients.

As a foreword to the comparison, it is important to note that there are still some noteworthy differences between \textsc{ALCAR} and the standard formulation of \textsc{ILEAS} in the derivation of the neutrino production rates. The former calculates the rates for $\beta$-production of $\nu_e$ and $\bar{\nu}_e$ from a formulation that ensures detailed balance, based on blocking-corrected absorption opacities, $\kappa^*$, defined in equations~\eqref{kappa_abs_nue_star} and~\eqref{kappa_abs_anue_star}, following \cite{2002A&A...396..361R}. On the other hand, \textsc{ILEAS} employs the emissivity, $j_{\nu_i}$, defined as in equations~\eqref{emissivitynue} and~\eqref{emissivityanue} \citep{1985ApJS...58..771B}, to compute the rates. In appendix~\ref{appendix:rates} we show the derivation of the rates in both schemes, and provide a detailed comparison of the effects of each prescription on the neutrino luminosities with \textsc{ILEAS}. In order to enable a more accurate comparison, \textsc{ILEAS} employs the prescription of the $\beta$-production rates from \textsc{ALCAR} in the results shown in this section, also using the same energy binning as described in section~\ref{sec:model:tdiff}. 

Figures~\ref{PNSSnapshots1} and~\ref{PNSSnapshots2} show the luminosity profiles of each neutrino species obtained by \textsc{ILEAS} for the selected time snapshots from the \textsc{VERTEX} simulation, in comparison to the original transport results. In the bottom panels of figure~\ref{PNSSnapshots1} we present the results obtained on the background evolved with \textsc{ALCAR}, where the results obtained by both transport codes are also plotted for comparison. Note that in this panel, for a better comparison with \textsc{ALCAR}, we do not include redshift in the calculations with \textsc{ILEAS}. In order to obtain the results presented in this section, we have relaxed the background using \textsc{ILEAS} to adjust the temperature and electron fraction to their new steady-state values (see above at the beginning of section~\ref{sec:tests:timeevol}). After a brief transient of a few ms, all quantities settle into a quasi-stationary state. We will discuss the details of the scheme employed for relaxation as well as for the longer time evolution of one of these snapshots in section~\ref{sec:tests:timeevol}.

In all the tested snapshots, from 0.2~s to 1.5~s post-bounce, \textsc{ILEAS} is able to reproduce the transport results for $\nu_i$ and $\bar{\nu}_i$ with better than 10 per cent accuracy. The slightly bigger discrepancies for $\nu_x$ are very likely associated with the different prescriptions of nucleon bremsstrahlung employed by the different codes. Moreover, \textsc{ILEAS} as a leakage scheme calculates only neutrino losses and, therefore, it is unable to model the negative neutrino fluxes observed with transport treatments at $\sim$10 km. The negative fluxes are a consequence of the local temperature maximum at about 12~km, which leads to a net neutrino diffusion flux directed towards the centre of the PNS, i.e. neutrinos in this region diffuse inward. Because \textsc{ILEAS} is unable to reproduce such an effect by construction, the diffusion time-scale in those regions, which would become negative, is set to infinity, preventing any leakage of neutrinos out of the star\footnote{See section~\ref{sec:model:tdiff} for details on our treatment of negative diffusion time-scales.}.

The performance of \textsc{ILEAS} in the optically thick region is remarkable, especially for the \textsc{ALCAR} background, in which case both codes use exactly the same opacities for $\nu_e$ and $\bar{\nu}_e$. The good agreement arises from the definition of our diffusion time-scale. This effectively translates in a local source term calculated as $Q_{\nu_i}^{-} \simeq -\int_0^{\infty}{\boldsymbol{\nabla}\boldsymbol\cdotp\boldsymbol{F}_{\nu_i}\mathrm{d}\epsilon}$, which, in the case of quasi-stationarity, $\partial E_{\nu_i}(\epsilon)/\partial t\ \sim\ 0$, is essentially the same result as with \textsc{ALCAR}. As we approach the semi-transparent region, however, the results start to differ slightly due to the deviations from $\beta$-equilibrium of the neutrino spectrum, which we approximated using our interpolation of the neutrino degeneracies (equation~\ref{etanu}). This is one of the most delicate aspects of our scheme, as the diffusion time-scale depends sensitively on the neutrino spectrum, which cannot be properly determined by a leakage method. Finally, in the optically thin regime, our 3D absorption model successfully captures the essential features of energy and lepton-number deposition in the PNS envelope. This is visible from a very good agreement of the relaxed $Y_e(r)$ (figure~\ref{ye05}) profiles obtained with \textsc{ILEAS} and \textsc{ALCAR/VERTEX}, respectively. Furthermore, the $\nu_e$ and $\bar{\nu}_e$ luminosity profiles (figures~\ref{PNSSnapshots1} and \ref{PNSSnapshots2}) reproduce the transport results with great accuracy, while the profiles in appendix~\ref{appendix:tdiffs} which use the same definition of the diffusion time-scale but do not include neutrino re-absorption, show a clear overproduction of both neutrino species (figure~\ref{Tdiffcomparison}, Model~7). In section~\ref{sec:tests:snapBHT} we will discuss in further detail the features of our 3D absorption scheme in the context of a BH-torus system.

It is interesting to note the small differences in the relaxed electron fraction profile. Figure~\ref{ye05} shows the original profile from the 0.5~s PNS snapshot evolved by \textsc{VERTEX} in comparison to the profiles obtained by \textsc{ALCAR}, and the one further relaxed using \textsc{ILEAS}. It catches the eye that there is a slight but systematic shift of the rising flank of the $Y_e$ ``trough'', which is located close to the PNS surface, to slightly larger radii for the \textsc{ILEAS} model. In fact, this effect is generic because of the poor ability of any leakage scheme to accurately model the semi-transparent regime regardless of the absorption or equilibration parts. However, we emphasize that our implementation of the diffusion time-scale in \textsc{ILEAS} performs extremely well also in this respect compared to other schemes presented in the literature (e.g. \citealt{2016ApJS..223...22P}), as can be seen by our test results obtained with other definitions of the diffusion time-scale, summarized in appendix~\ref{appendix:tdiffs}. Tentatively, the remaining moderate overestimation of the loss of neutrino-lepton number from a narrow layer around the neutrinosphere could be mitigated by a further improved handling of the neutrino spectrum out of equilibrium.

Table~\ref{table:results} lists a summary of the luminosities and mean energies of the three neutrino species, as seen by a local observer in the rest frame of the neutrino source at the edge of our grid, $\sim$100~km, obtained by \textsc{ILEAS} for all our tested conditions, in comparison to the original results obtained by the corresponding transport codes. All \textsc{ILEAS} results are extracted after a few milliseconds of relaxation, employing the formulations described in section~\ref{sec:model:postproc}. As mentioned earlier, the neutrino luminosities obtained by \textsc{ILEAS} for all tested PNS snapshots provide a very good approximation of the luminosities obtained by the transport calculations. 

The mean neutrino energies calculated in the leakage approach, however, exhibit a greater disagreement with the transport results, especially at later times of the PNS evolution (table~\ref{table:results}). For $\nu_e$ the leakage mean energies are increasingly lower compared to the M1 results (up to $\sim$4~MeV at 1.5 s), whereas the $\bar{\nu}_e$ show the opposite trend, but to a smaller extent (up to $\sim$3~MeV). This energy discrepancy does not significantly improve when we compute our neutrino number and energy absorption rates independently of each other as detailed in appendix~\ref{appendix:nabs}. This systematic and consistent disagreement with transport results is probably linked to the approximative treatment of the semi-transparent regime by the flux-limiting approach. Figure~\ref{nabslumis} in appendix~\ref{appendix:nabs} reveals that, particularly in the case of $\nu_e$, a dominant component of the neutrino-number luminosity is emitted from the semi-transparent region (right below the neutrinosphere), whereas the neutrino-energy luminosity comes from deeper inside the NS. For this reason, the calculation of the neutrino-number luminosity is more sensitive to the approximations applied in our flux-limiting prescription for the diffusion time-scale, thus impacting the neutrino mean energies computed by the leakage method.

To offer an alternative measure of the radiated mean energies, which is also more compatible with the mean energies of neutrinos used in our absorption module, we provide the approximate diagnostic mean energies defined by equation~\eqref{etot}. We find that, as expected, these post-processed energies considerably improve our mean energy estimates for $\nu_e$, with just moderate corrections for $\bar{\nu}_e$, providing an agreement better than typically $\sim$15 per cent ($\sim$1.5~MeV difference in the worst case). The larger differences observed in the $\nu_x$ mean neutrino energies stem from the different prescriptions of bremsstrahlung employed by \textsc{ILEAS}, \textsc{ALCAR}, and \textsc{VERTEX}.

 \subsection{Snapshot calculations: black hole-torus system}\label{sec:tests:snapBHT}

\begin{figure}
\begin{center}
\makebox[0pt][c]{%
\minipage{0.51\textwidth}
\includegraphics[width=0.97\textwidth]{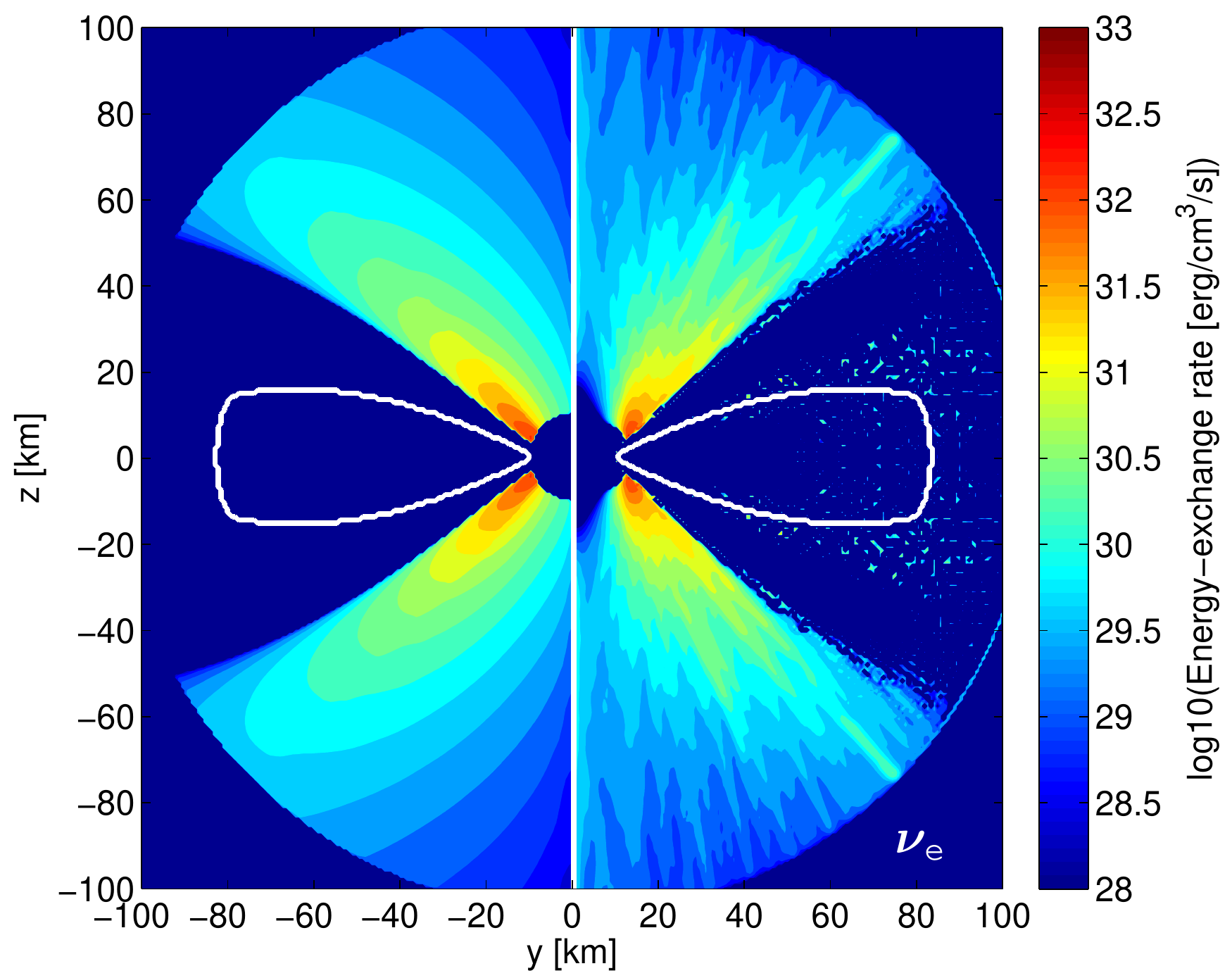}
\endminipage\hfill
}%
\\
\makebox[0pt][c]{%
\minipage{0.51\textwidth}
\includegraphics[width=0.97\textwidth]{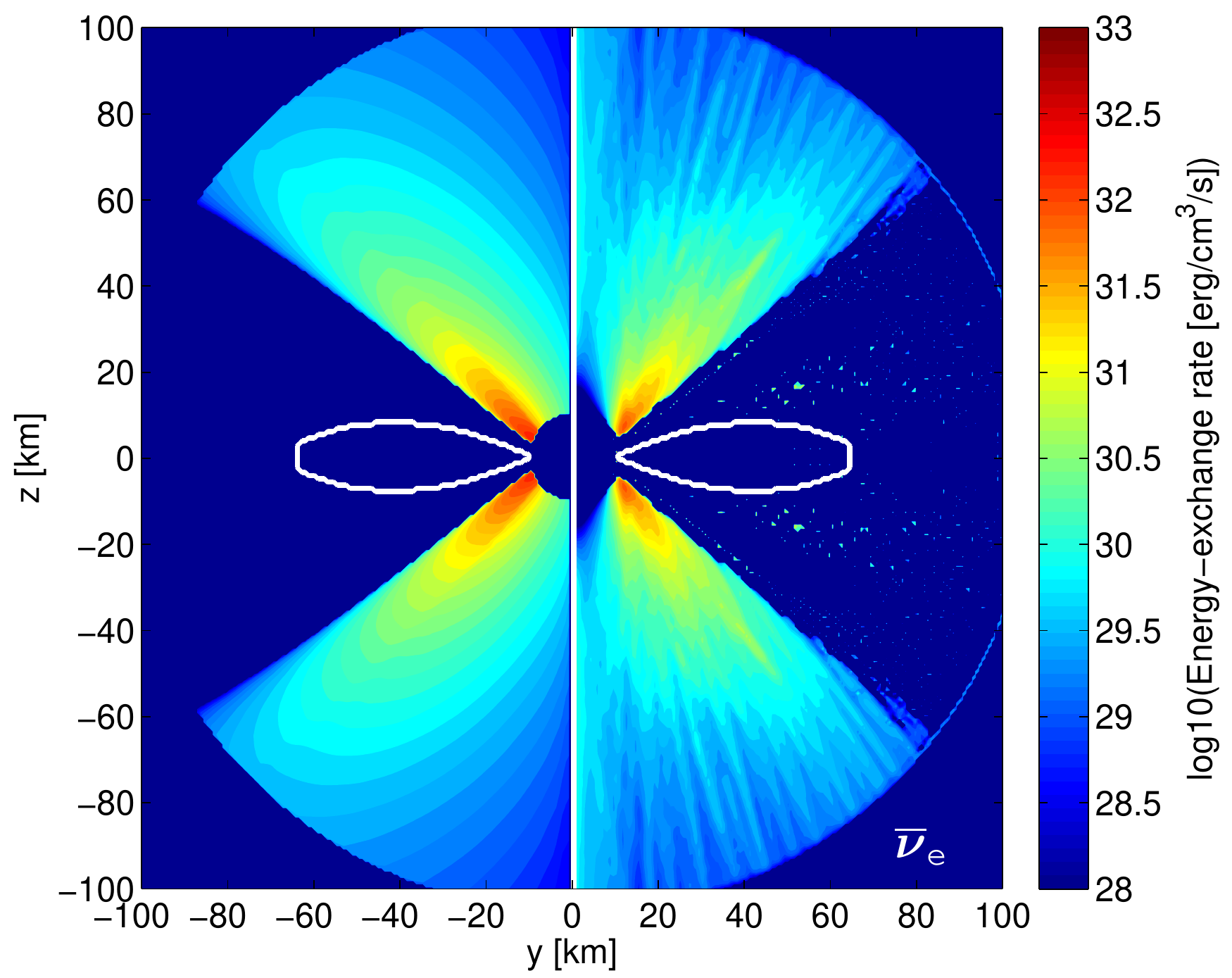}
\endminipage\hfill
}%
\\
\makebox[0pt][c]{%
\minipage{0.51\textwidth}
\includegraphics[width=0.98\textwidth]{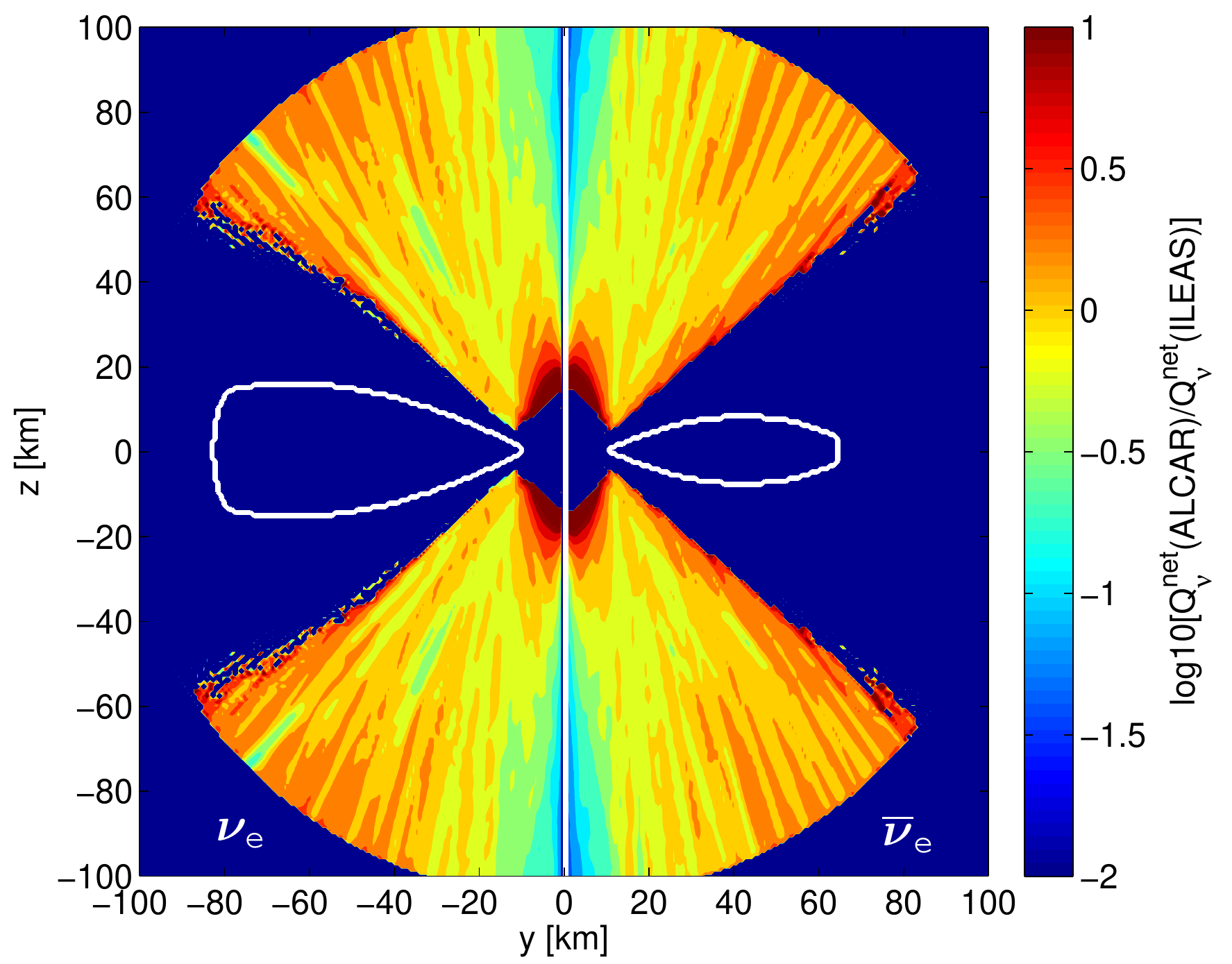}
\endminipage\hfill
}%
\caption{Results of our neutrino absorption scheme for an 0.3~\Msun\ torus around a 3~\Msun\ BH, top plot for $\nu_e$, middle plot for $\bar{\nu}_e$. Colour coding displays the net energy-exchange rate $Q_{\nu_i}^{\mathrm{net}}=Q_{\nu_i}^{+}-Q_{\nu_i}^{-}$ ($\nu_i=\nu_e,\bar{\nu}_e$) in the absorption-dominated region, where this net rate is positive. The left half-panel of each of these plots shows the results obtained by the \textsc{ALCAR} scheme, while the right ones are the results with \textsc{ILEAS}. The white contours depict the neutrinospheres, where $\tau_{\nu_i}=2/3$. The bottom plot displays the ratio of the net energy-exchange rates between \textsc{ALCAR} and \textsc{ILEAS} in regions where both rates are positive (left panel for $\nu_e$, right panel for $\bar{\nu}_e$).} \label{BHTorusAbs1}
\end{center}
\end{figure}
\begin{table*}
  \centering
  \caption{Neutrino luminosities and mean energies obtained by \textsc{ILEAS} applied to several snapshots of a PNS cooling simulation at different times and two BH-torus models, in comparison to the results from transport calculations with different codes. All leakage quantities are computed as described in section~\ref{sec:model:postproc}. \textit{Leakage mean energies} provide the mean energies calculated by equation~\eqref{eleak}, while \textit{mean energies for diagnostics} are obtained via equation~\eqref{etot}. All values are taken for a local observer in the rest frame of the source at the edge of the grid (100 km). $\nu_x$ luminosities refer to a single representative of the four species of heavy-lepton neutrinos.}\label{table:results}
    \begin{tabular}{lccccccc}
      \hline
      \hline
      \noalign{\vskip 2mm}  
      Model        & $\nu$-species & \parbox[c]{1.6cm}{\centering Transport luminosity ($10^{51}\ \mathrm{erg}\boldsymbol\cdotp\mathrm{s}^{-1}$)} &  \parbox[c]{1.6cm}{\centering Leakage luminosity ($10^{51}\ \mathrm{erg}\boldsymbol\cdotp\mathrm{s}^{-1}$)}  & \parbox[c]{1.6cm}{\centering Transport mean energy (MeV)} & \parbox[c]{1.6cm}{\centering Leakage mean energy (MeV)} & \parbox[c]{1.6cm}{\centering Mean energy for diagnostics (MeV)} & \parbox[c]{1.6cm}{\centering Transport code} \vspace{1mm}  \\ 
      \hline 
      \noalign{\vskip 2mm}                                                                                                                                                                                                                
      PNS 0.2~s        & $\nu_e$     	    & 16.2      & 16.4       & 9.72        & 7.85       & 11.04   & \textsc{VERTEX} \\ 
      PNS 0.2~s        & $\bar{\nu}_e$    & 17.3      & 19.1       & 12.42       & 12.43      & 12.72  & \textsc{VERTEX} \\ 
      PNS 0.2~s        & $\nu_x$             & 13.7      & 14.5       & 14.32       & 21.38      & - 	       & \textsc{VERTEX} \\                                                                                                                 
      PNS 0.3~s        & $\nu_e$     	    & 9.8        & 9.1        & 9.43        & 7.20       & 10.65    & \textsc{VERTEX} \\ 
      PNS 0.3~s        & $\bar{\nu}_e$    & 10.8      & 11.5       & 12.18       & 11.98      & 12.27  & \textsc{VERTEX} \\ 
      PNS 0.3~s        & $\nu_x$             & 10.2      & 11.0       & 13.80       & 20.14      & - 	      & \textsc{VERTEX} \\                                                                                                               
      PNS 0.4~s        & $\nu_e$     	    & 7.4        & 6.7        & 9.31        & 6.96       & 10.52  & \textsc{VERTEX} \\ 
      PNS 0.4~s        & $\bar{\nu}_e$    & 8.1        & 8.7        & 12.00       & 11.88      & 12.30  & \textsc{VERTEX} \\ 
      PNS 0.4~s        & $\nu_x$             & 8.4        & 9.1        & 13.51       & 19.19      & - 	     & \textsc{VERTEX} \\                                                                                                                            
      PNS 0.5~s        & $\nu_e$     	    & 6.2        & 5.5        & 9.26        & 7.11       & 10.58  & \textsc{VERTEX} \\
      PNS 0.5~s        & $\bar{\nu}_e$    & 6.7        & 6.9        & 11.86       & 11.32      & 12.28  & \textsc{VERTEX} \\ 
      PNS 0.5~s        & $\nu_x$             & 7.3        & 8.0        & 13.33       & 18.75      & - 	     & \textsc{VERTEX} \\                                                                                                                         
      PNS 0.5~s        & $\nu_e$     	    & 7.0        & 6.7        & 9.93        & 7.95       & 11.62  & \textsc{ALCAR}  \\
      PNS 0.5~s        & $\bar{\nu}_e$    & 7.6        & 8.1        & 13.32       & 12.62      & 13.13  & \textsc{ALCAR}  \\
      PNS 0.5~s        & $\nu_x$             & 9.0        & 10.4       & 15.67       & 21.46      & - 	     & \textsc{ALCAR}  \\                                                                                                                         
      PNS 0.8~s        & $\nu_e$     	    & 4.6        & 4.4        & 9.24        & 7.42       & 10.44  & \textsc{VERTEX} \\
      PNS 0.8~s        & $\bar{\nu}_e$    & 4.9        & 5.1        & 11.64       & 11.39      & 12.45  & \textsc{VERTEX} \\
      PNS 0.8~s        & $\nu_x$             & 5.7        & 6.3        & 13.02       & 18.07      & - 	     & \textsc{VERTEX} \\                                                                                                                
      PNS 1.1~s        & $\nu_e$     	    & 3.8        & 3.6        & 9.24        & 6.56       & 10.10  & \textsc{VERTEX} \\ 
      PNS 1.1~s        & $\bar{\nu}_e$    & 4.0        & 4.0        & 11.45       & 12.54      & 12.64  & \textsc{VERTEX} \\ 
      PNS 1.1~s        & $\nu_x$             & 4.9        & 5.3        & 12.76       & 17.48      & - 	     & \textsc{VERTEX} \\                                                                                                                          
      PNS 1.2~s        & $\nu_e$     	    & 3.7        & 3.5        & 9.24        & 6.16       & 10.02  & \textsc{VERTEX} \\
      PNS 1.2~s        & $\bar{\nu}_e$    & 3.8        & 3.7        & 11.43       & 12.98      & 12.73  & \textsc{VERTEX} \\
      PNS 1.2~s        & $\nu_x$             & 4.7        & 5.0        & 12.69       & 17.17      & - 	      & \textsc{VERTEX} \\                                                                                                                
      PNS 1.3~s        & $\nu_e$     	    & 3.5        & 3.3        & 9.24        & 5.79       & 10.08  & \textsc{VERTEX} \\ 
      PNS 1.3~s        & $\bar{\nu}_e$    & 3.6        & 3.5        & 11.38       & 13.46      & 12.85  & \textsc{VERTEX} \\ 
      PNS 1.3~s        & $\nu_x$             & 4.4        & 4.7        & 12.61       & 16.89      & - 	     & \textsc{VERTEX} \\                                                                                                                
      PNS 1.5~s        & $\nu_e$     	    & 3.2        & 3.0        & 9.22        & 5.18       & 10.01  & \textsc{VERTEX} \\ 
      PNS 1.5~s        & $\bar{\nu}_e$    & 3.3        & 3.2        & 11.27       & 14.26      & 12.88  & \textsc{VERTEX} \\ 
      PNS 1.5~s        & $\nu_x$             & 4.1        & 4.3        & 12.43       & 16.46      & - 	  & \textsc{VERTEX} \\                                                                                                                           
      BH-torus 0.3~\Msun & $\nu_e$     	     & 23.3       & 21.5       & 12.13       & 12.66      & 14.19  & \textsc{ALCAR}  \\
      BH-torus 0.3~\Msun & $\bar{\nu}_e$  & 18.4       & 16.7       & 14.97       & 15.89      & 17.16  & \textsc{ALCAR}  \\                                                                                                                     
      BH-torus 0.1~\Msun & $\nu_e$     	     & 6.5        & 6.5        & 12.02       & 12.69      & 14.85  & \textsc{ALCAR}  \\
      BH-torus 0.1~\Msun & $\bar{\nu}_e$  & 5.2        & 4.8        & 14.20       & 14.50      & 16.28  & \textsc{ALCAR}  \\  
      \noalign{\vskip 2mm}                                                                                                                       
      \hline
      \hline
  \end{tabular}
\end{table*}

 \begin{figure*}
\begin{center}
\makebox[0pt][c]{%
\minipage{0.51\textwidth}
\includegraphics[width=\textwidth]{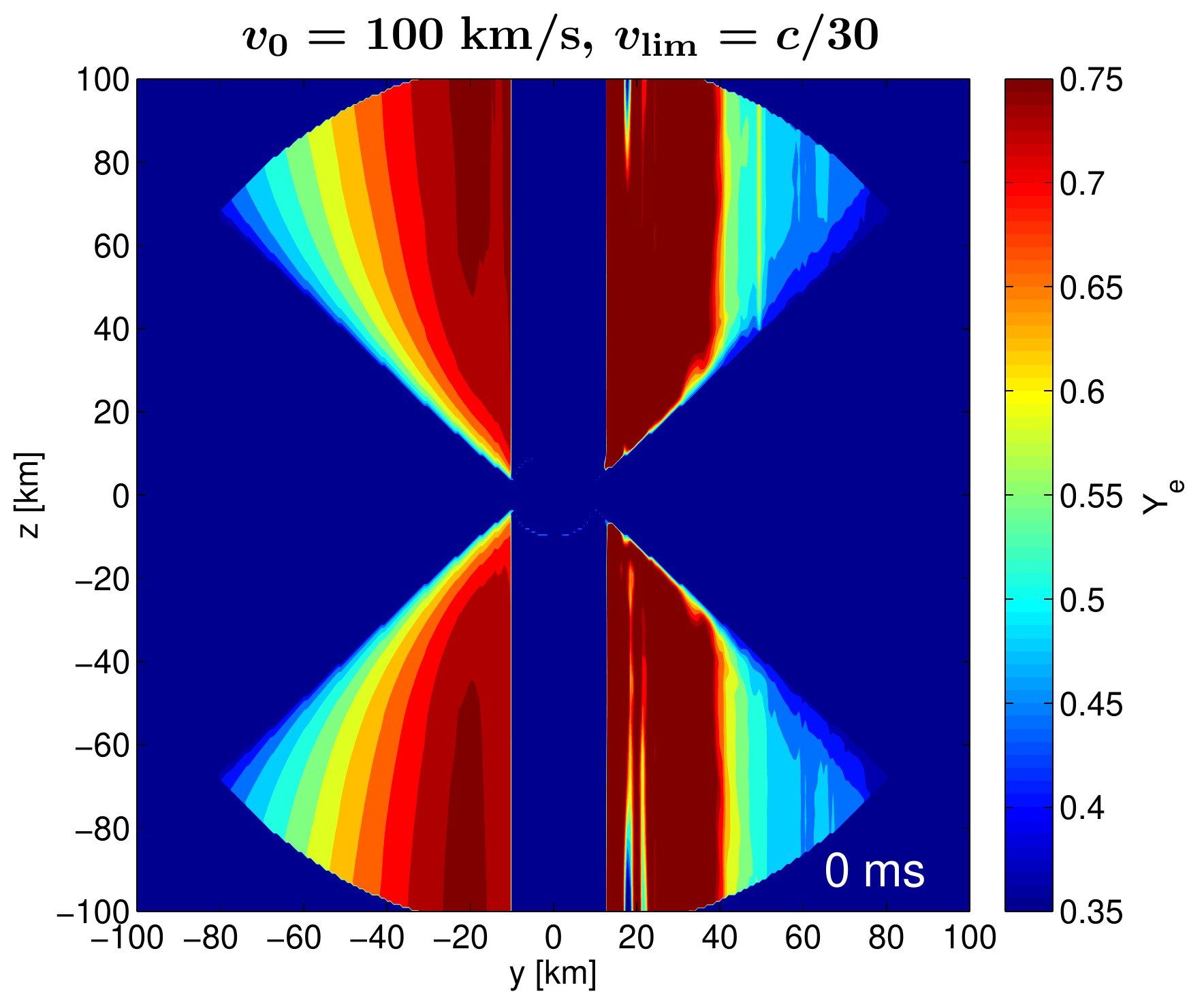}
\endminipage\hfill
\hspace{-0.1cm}
\minipage{0.51\textwidth}
\includegraphics[width=\textwidth]{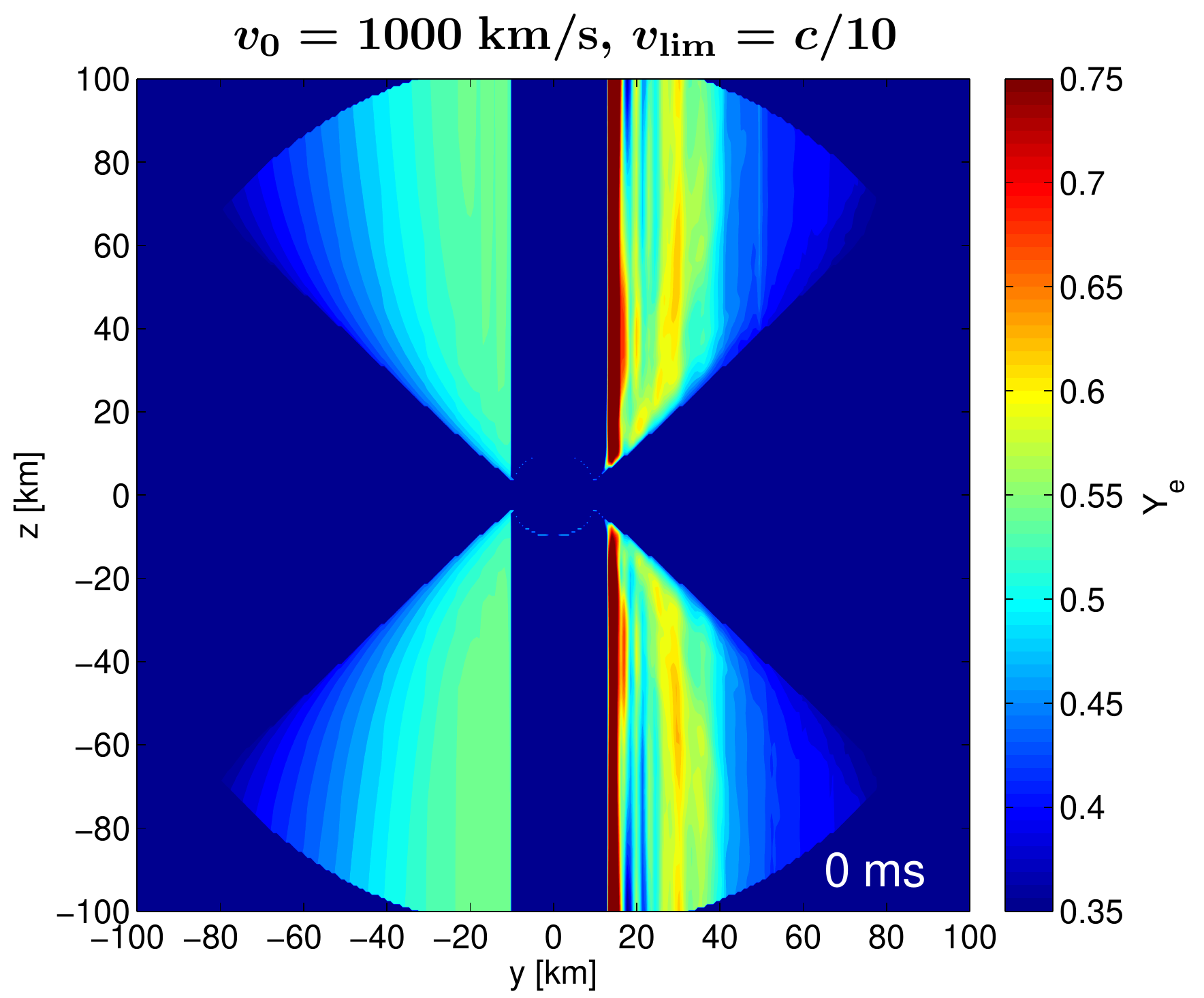}
\endminipage\hfill
}%
\\
\makebox[0pt][c]{%
\minipage{0.51\textwidth}
\includegraphics[width=\textwidth]{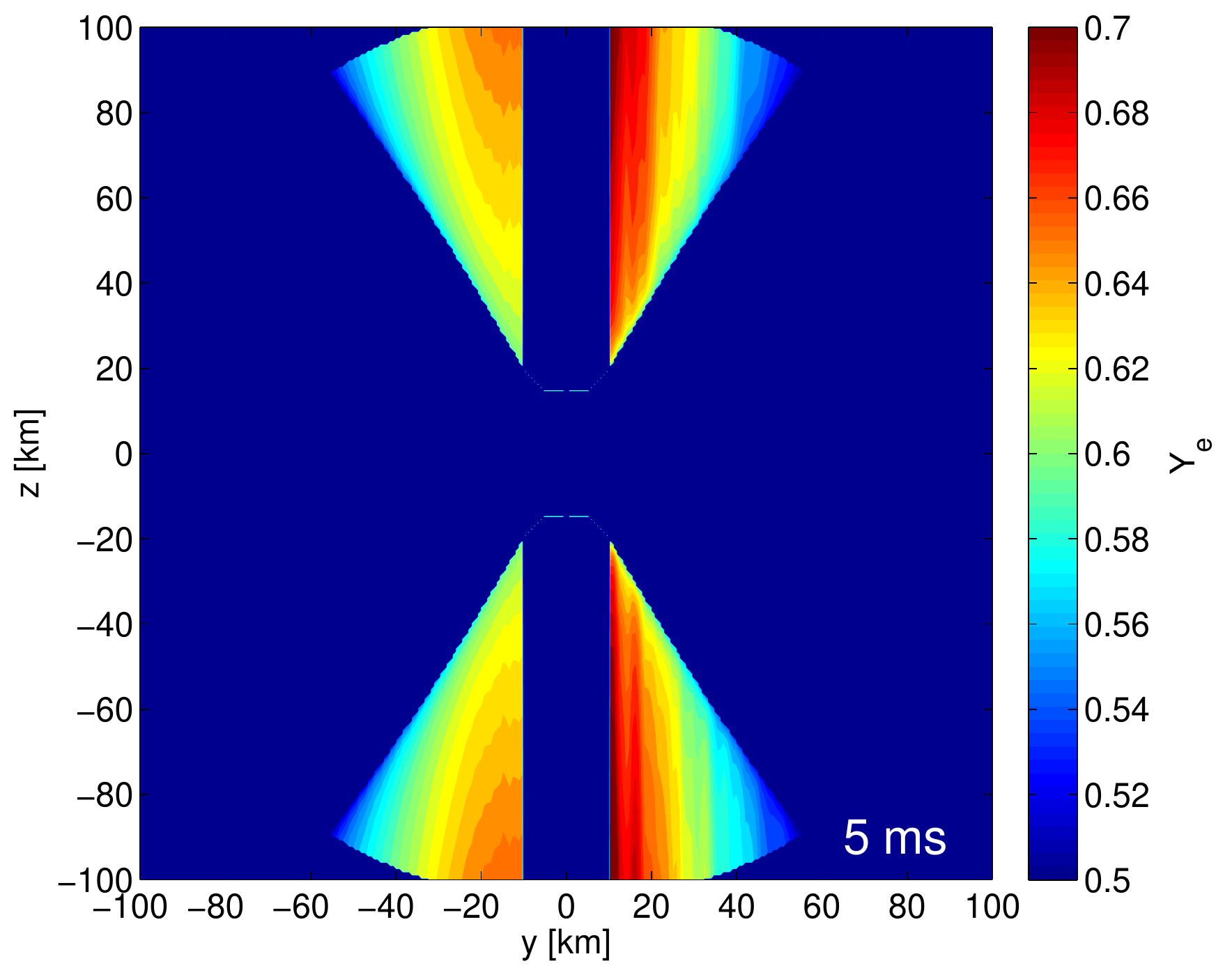}
\endminipage\hfill
\hspace{-0.1cm}
\minipage{0.51\textwidth}
\includegraphics[width=\textwidth]{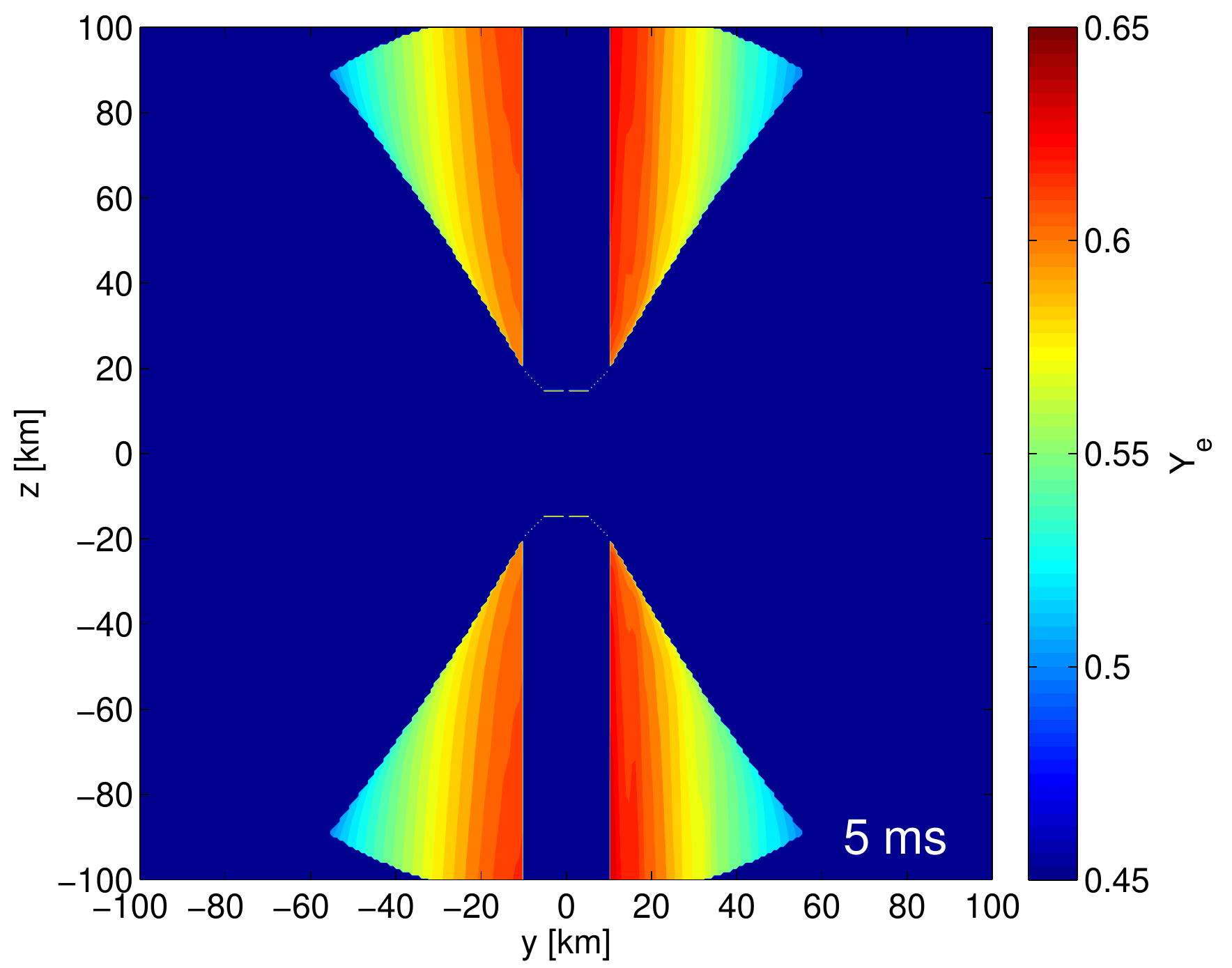}
\endminipage\hfill
}%
\caption{Evolution of the $Y_e$ of a prescribed outflow launched from the torus (initial torus mass 0.3~\Msun) surrounding a BH (3~\Msun) as it travels along a straight path in the z-direction, exposed to the neutrino rates calculated by \textsc{ALCAR} (left half-panels) and \textsc{ILEAS} (right half-panels). The plots in the top row display the results obtained for the same BH-torus snapshot also used in figure~\ref{BHTorusAbs1}, while those in the bottom row show the results for a snapshot obtained after 5~ms of evolution with \textsc{ALCAR} and \textsc{ILEAS}, respectively (see section~\ref{sec:tests:timeevol}). The two plots in the left column provide the results obtained for a slow wind, with an initial outflow velocity $v_0=100$~km/s and a limiting velocity $v_{\mathrm{lim}}=c/30$, whereas the ones in the right-column provide the results obtained by a fast wind, with $v_0=1000$~km/s and $v_{\mathrm{lim}}=c/10$  (see text for details).} \label{yeevol}
\end{center}
\end{figure*}

In order to assess the performance of our scheme on a possible remnant of a CO merger, we calculate the neutrino luminosities for two different BH-torus systems evolved previously using the \textsc{ALCAR} code \citep{2015MNRAS.453.3386J}. Both models are composed of a 3~\Msun~BH surrounded by a torus of NS debris: a thin torus with 0.1~\Msun~and a thicker one with 0.3~\Msun, respectively. The neutrino reactions employed for these cases are the same as for the PNS (table~\ref{table:nureaccomp}), except for heavy-lepton neutrinos, which are switched off in both calculations because of their minor relevance for this setup. The results obtained by \textsc{ALCAR} and \textsc{ILEAS} presented in this section do not include the effects of redshift.

In table~\ref{table:results} we also include the neutrino luminosities and mean energies for $\nu_e$ and $\bar{\nu}_e$ as obtained by \textsc{ILEAS}, applied to the two BH-torus systems. Because tori are optically thinner than PNSs, their cooling time-scale is much shorter, and the temperature can change considerably during the relaxation of the background. We took this into account by providing the results of both \textsc{ALCAR} and \textsc{ILEAS} after 3~ms of evolution starting from the original snapshots. Even though we also provide the mean energies calculated by equation~\eqref{etot}, the ones obtained by the leakage approximation via equation~\eqref{eleak} should be more accurate in the case of BH-torus systems, for two simple reasons. First, in the BH-torus models considered in this work, matter becomes optically thin during the relaxation (the optical depth is  $\tau_{\nu_i}<2/3$ almost everywhere after a few milliseconds of evolution) or optically thick material encloses a very small volume, so that the leakage ansatz, namely that neutrinos stream away with the mean energy obtained from their local production, is a reasonable approximation. Second, the gradients in the hydrodynamical and thermodynamical quantities are considerably flatter than in the PNS case. Therefore, the reasoning that most absorption occurs in the production cell, which is employed to estimate the mean energies in equation~\eqref{etot}, is a less accurate approximation. Because the leakage mean energies employ a more accurate description of absorption in optically thin regions, which are the far dominant conditions in the tori, we advise the reader to consider the leakage mean energies for any diagnostic analysis or comparison.

Figure~\ref{BHTorusAbs1} shows the performance of our absorption scheme on the snapshot of a thick torus (initial torus mass 0.3~\Msun) around a 3~\Msun\ BH. Despite the ray patterns caused by the ray-tracing approach, the qualitative resemblance in the top and middle plots between \textsc{ALCAR} (left half-panels) and \textsc{ILEAS} (right half-panels) is remarkable. Moreover, the bottom panel in figure~\ref{BHTorusAbs1} displays the ratio of the net energy-exchange rates obtained by \textsc{ALCAR} and \textsc{ILEAS} in the absorption-dominated regions, which also highlights the overall quantitatively satisfactory agreement between both schemes, within a factor of $\sim$2 accuracy. We refrain from performing a comparison of the rates immediately above the BH and in the close vicinity of the z-axis, because a consistent treatment of general relativistic and special relativistic effects would be needed to describe the influence of the BH or ultrarelativistic GRB jets. Furthermore, \cite{2018PhRvD..98f3007F}, for example, compared Monte Carlo results and M1 results in the context of a HMNS surrounded by a torus and pointed out that the inexact M1 closure strongly overestimates the number density in the polar regions, by $\sim$50 per cent for $\nu_e$ and $\bar{\nu}_e$, which leads to significant boosting of the absorption rates by charged-current reactions and excess heating. \cite{2015MNRAS.448..541J} also reported similar behaviour when comparing BH-torus calculations with a ray-tracing Boltzmann solver against their M1 results. Therefore, a detailed quantitative comparison between \textsc{ILEAS} and M1 results in the vicinity of the polar axis could be misleading.

For a more direct assessment of the impact of differences in the absorption rates between \textsc{ALCAR} and \textsc{ILEAS} on possible outflows originating from the torus, we performed another test. For this purpose we defined parametrized outflows in the polar directions and compared the evolution of $Y_e$ under the influence of the emission and absorption rates from \textsc{ALCAR} and \textsc{ILEAS}.

In a steady-state situation, the evolution of the electron fraction of an outflow can be approximated by \citep{1996ApJ...472..440M}
 \begin{equation}
  \frac{\mathrm{d}}{\mathrm{d}t} Y_e\ =\ v(z)\frac{\mathrm{d}}{\mathrm{d}z} Y_e\ =\ \frac{R_{\mathrm{tot}}(z)}{\mathcal{A}\rho(z)}, \label{dyez}
 \end{equation} 
with the total lepton-number exchange rate, $R_{\mathrm{tot}}$, defined as in equation~\eqref{rtot}. While $\rho(z)$ is adopted from the hydrodynamic solution of the ALCAR calculation, we assume the unbound material to move in the z-direction with the parametrized velocity, $v(z)$,
\begin{equation}
v(z)\ =\ \mathrm{min}\left[v_{\mathrm{lim}}\left(\frac{z-z_0}{\Delta z}\right)+v_0,v_{\mathrm{lim}}\right],\label{vz}
\end{equation}
accelerating along the $z$-direction to a terminal velocity $v_{\mathrm{lim}}$. In equation~\eqref{vz}, $\Delta z$ determines the length-scale over which the velocity reaches its limiting value. We set it to 30~km for the presented tests. $z_0$ defines the position of the surface from which the outflow is launched. We locate this surface at the position where the \textsc{ALCAR} net lepton-number-exchange rates, $R_{\nu_i}^{\mathrm{net}}=R_{\nu_i}^{+}-R_{\nu_i}^{-}$, of $\nu_e$ as well as $\bar{\nu}_e$, become absorption-dominated. We then use exactly the same surface location to launch the outflow in both models, in one case employing the \textsc{ALCAR} rates and in the other one employing the \textsc{ILEAS} rates. The last free parameter of this toy model is $v_0$, which determines the initial velocity of the outflow. In order to assess the influence of the parameters $v_0$ and $v_{\mathrm{lim}}$, which determine the duration of time the ejecta remain in the near-surface region, where the absorption rates are higher, we test two different sets of values corresponding to a slow wind, $v_0=100$~km/s and $v_{\mathrm{lim}}=c/30$, and a fast wind, $v_0=1000$~km/s and $v_{\mathrm{lim}}=c/10$ (see left and right columns in figure~\ref{yeevol}, respectively).

Integrating equation~\eqref{dyez}, the $Y_e$ of the outflow at a given position is simply determined by
 \begin{equation}
  Y_e(z)\ =\ Y_e(z_0)+\int_{z_0}^z{\frac{R_{\mathrm{tot}}(z')}{\mathcal{A}\rho(z') v(z')}\mathrm{d}z'}.\label{yez}
 \end{equation} 

 Figure~\ref{yeevol} shows the evolution of the $Y_e$ as described by equation~\eqref{yez} along the paths travelled by the outflow in the polar directions. The left half of each panel illustrates the results obtained when employing the rates calculated by \textsc{ALCAR}, whereas the right half shows the results with the \textsc{ILEAS} rates. The two wind descriptions (slow wind in the left column of plots, fast wind in the right), agree qualitatively for the snapshot at 0~ms (top row) to the extent that $Y_e>0.5$ is achieved in the same spatial region close to the polar axis. There differences in $Y_e$ are on the level of $\sim$20 per cent for the slow wind, showing the limitations of our scheme, and of $\sim$10 per cent for the fast wind ($\sim$0.55 for \textsc{ALCAR} compared to $\sim$0.5--0.6 for \textsc{ILEAS}), with local variations due to the ray patterns associated with ILEAS. The regions where $Y_e<0.5$ display even better quantitative agreement within $\sim$5 per cent. We warn the reader that, as we already pointed out, \cite{2015MNRAS.448..541J} and \cite{2018PhRvD..98f3007F} found possible deficiencies in the calculation of the absorption rates by M1 schemes in the polar region when contrasting with ray-tracing and MC results, respectively, and thus any comparison in such regions should be taken with caution. Furthermore, as we will see in section~\ref{sec:tests:timeevol}, the long-term evolution of the BH-torus model with \textsc{ILEAS} produces a transient at the beginning of the simulation as the system relaxes to a new quasi-steady state (figure~\ref{BHTtimedep}), which could affect the discussed results. For this reason, we also provide in figure~\ref{yeevol} (bottom row) the results of outflow calculations applied on snapshots obtained after 5~ms of evolution with \textsc{ALCAR} and \textsc{ILEAS}, respectively (see section~\ref{sec:tests:timeevol}, figure~\ref{BHTtimedep}). At these later times, the evolution of the composition of the ejecta predicted by \textsc{ALCAR} and \textsc{ILEAS} agrees very well, especially in the fast wind scenario.
  
It must be noted that \textsc{ILEAS} assumes the flux factor to follow the simple interpolation $\langle\chi_{\nu_i}\rangle_{\mathrm{PNS}}^{-1}\ =\ 4.275\tau_{\nu_i}+1.15$, suggested by \cite{2010CQGra..27k4103O}. This is an acceptable approximation for the case of a cooling PNS, but fails to capture the geometry of the BH-torus system. More sophisticated prescriptions of the flux factor, which account for geometric effects, would certainly improve the accuracy of the absorption scheme in the area around the inner edge of the torus. This also concerns the results for $Y_e$ in outflows and is likely to reduce differences between ILEAS and transport calculations, in particular also for slow winds. However, such improvement is beyond the scope of this work, and we consider the obtained results with the presented approximations as satisfactory. 

\subsection{Time evolution of a proto-neutron star and two black hole-torus systems}\label{sec:tests:timeevol}

\begin{figure}
\begin{center}
\makebox[0pt][c]{%
\minipage{0.5\textwidth}
\includegraphics[width=\textwidth]{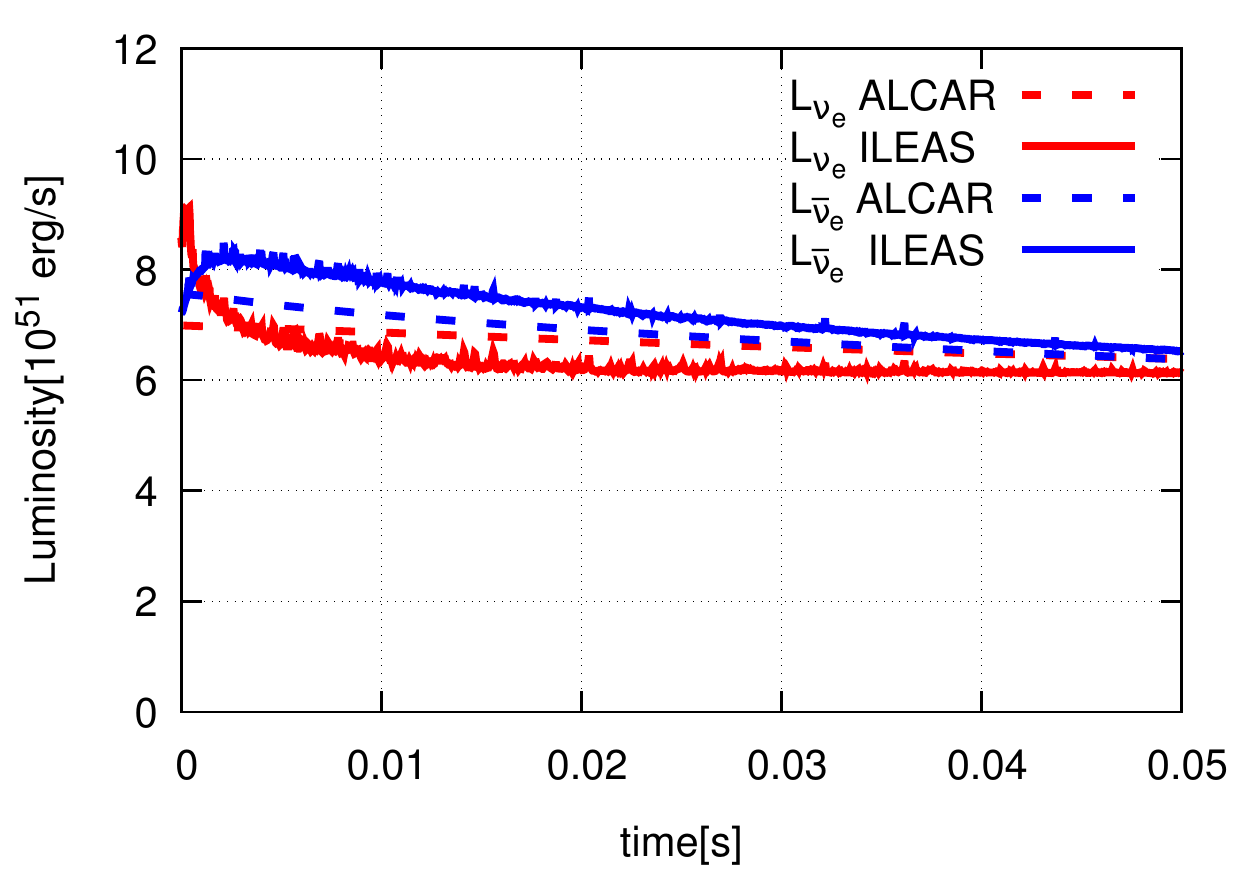}
\endminipage\hfill
}%
\\
\makebox[0pt][c]{%
\minipage{0.5\textwidth}
\includegraphics[width=\textwidth]{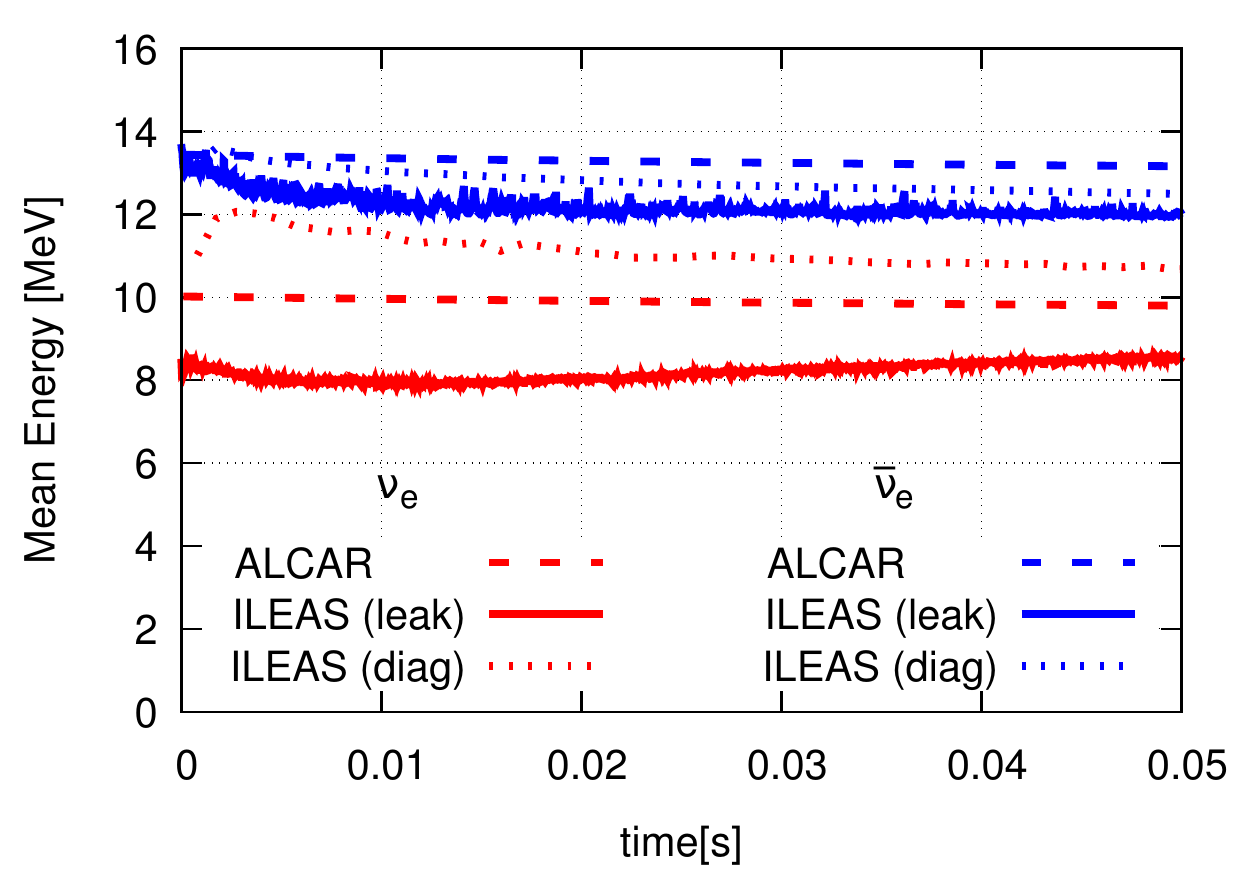}
\endminipage\hfill
}%
\caption{Top panel: time evolution of the neutrino luminosities obtained by \textsc{ILEAS} (solid lines) and \textsc{ALCAR} (dashed lines) produced by a cooling PNS with fixed density background, but evolved $T$ and $Y_e$ profiles. Bottom panel: time evolution of the mean neutrino energies obtained by \textsc{ALCAR} (dashed lines) and \textsc{ILEAS} using the leakage approach (equation~\ref{eleak}, solid lines) or the one for diagnostics (equation~\ref{etot}, dotted lines) for the same PNS model. The luminosities and mean energies are computed for a local observer in the centre-of-mass frame of the neutrino source at the edge of our grid (100~km).  The starting time corresponds to the \textsc{ALCAR} snapshot discussed in section~\ref{sec:tests:snapPNS} (0.5~s post-bounce).} \label{PNStimedep}
\end{center}
\end{figure}

\begin{figure*}
\begin{center}
\makebox[0pt][c]{%
\minipage{0.5\textwidth}
\includegraphics[width=\textwidth]{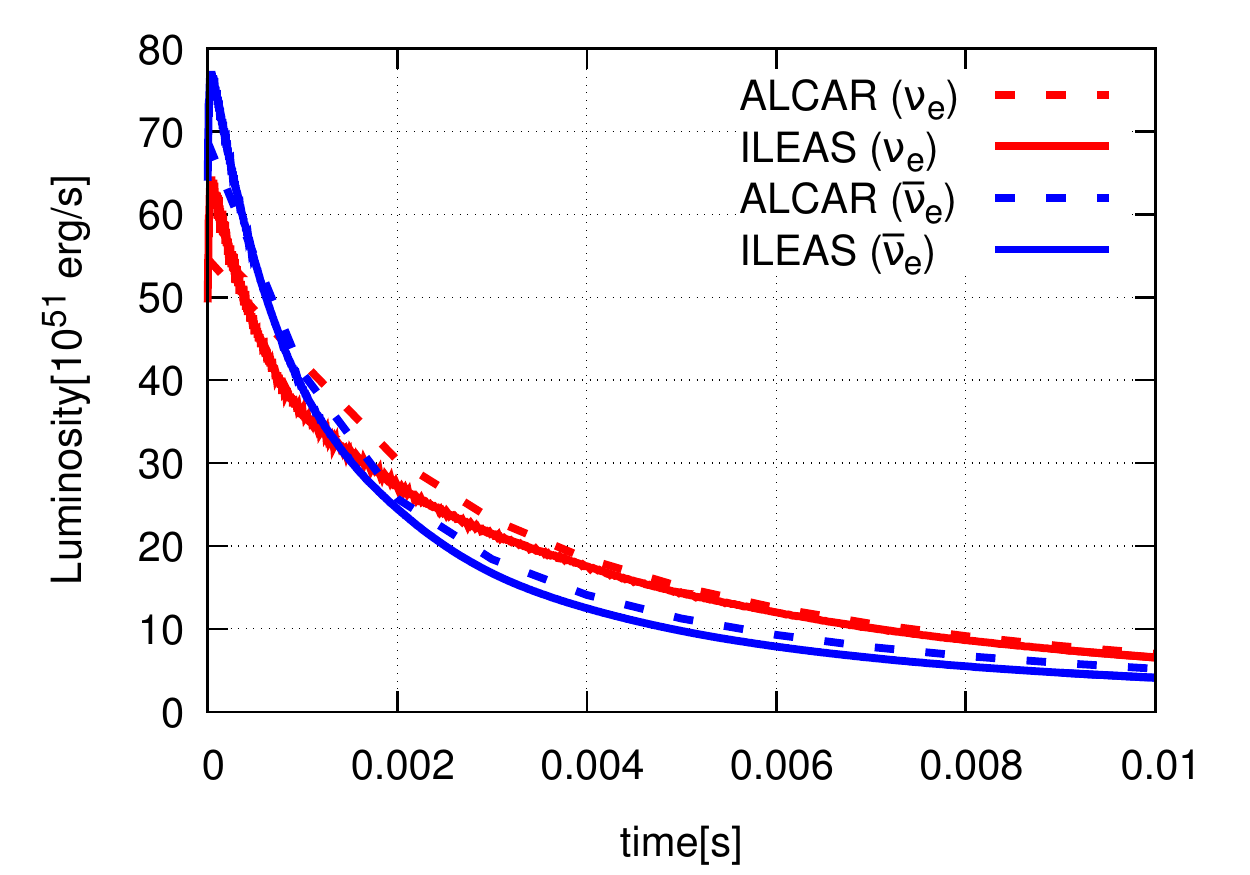}
\endminipage\hfill
\minipage{0.5\textwidth}
\includegraphics[width=\textwidth]{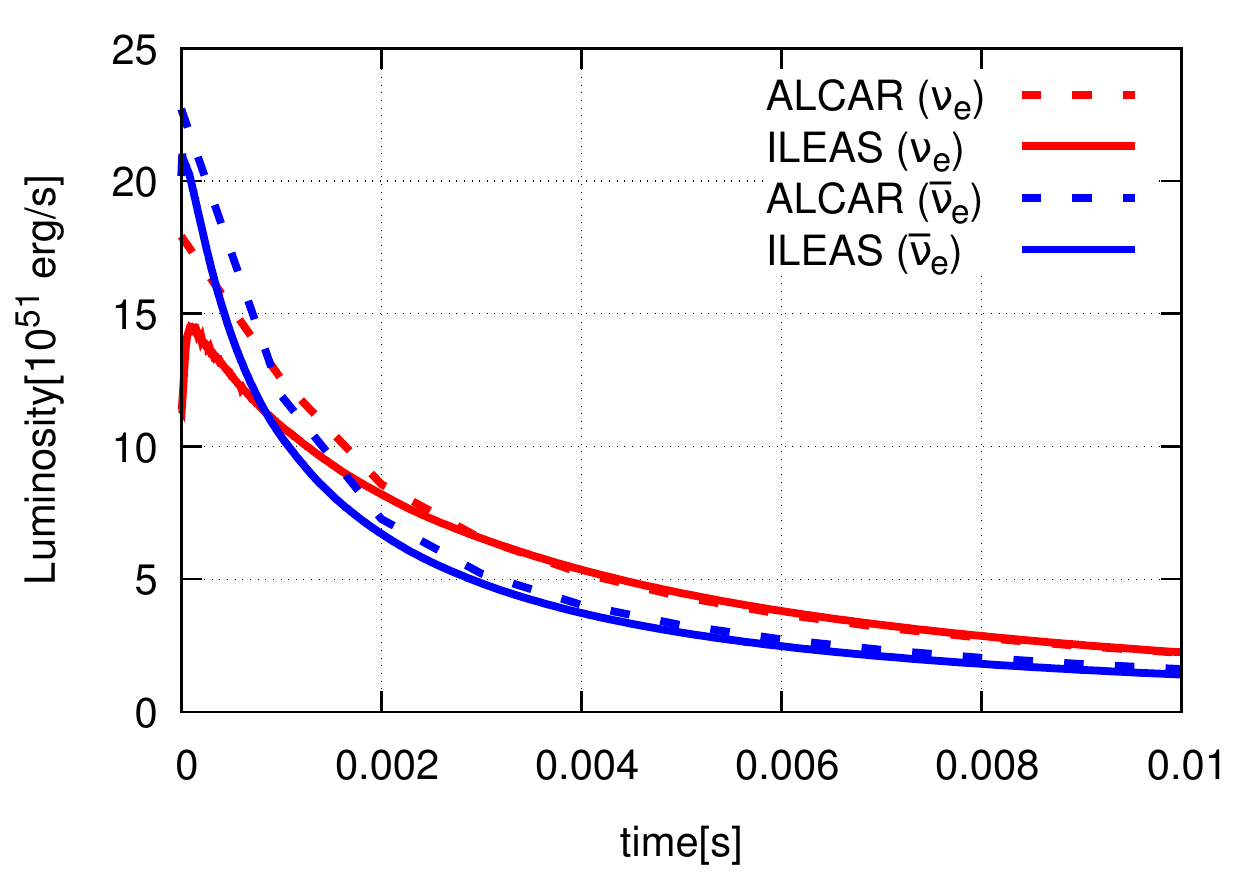}
\endminipage\hfill
}%
\\
\makebox[0pt][c]{%
\minipage{0.5\textwidth}
\includegraphics[width=\textwidth]{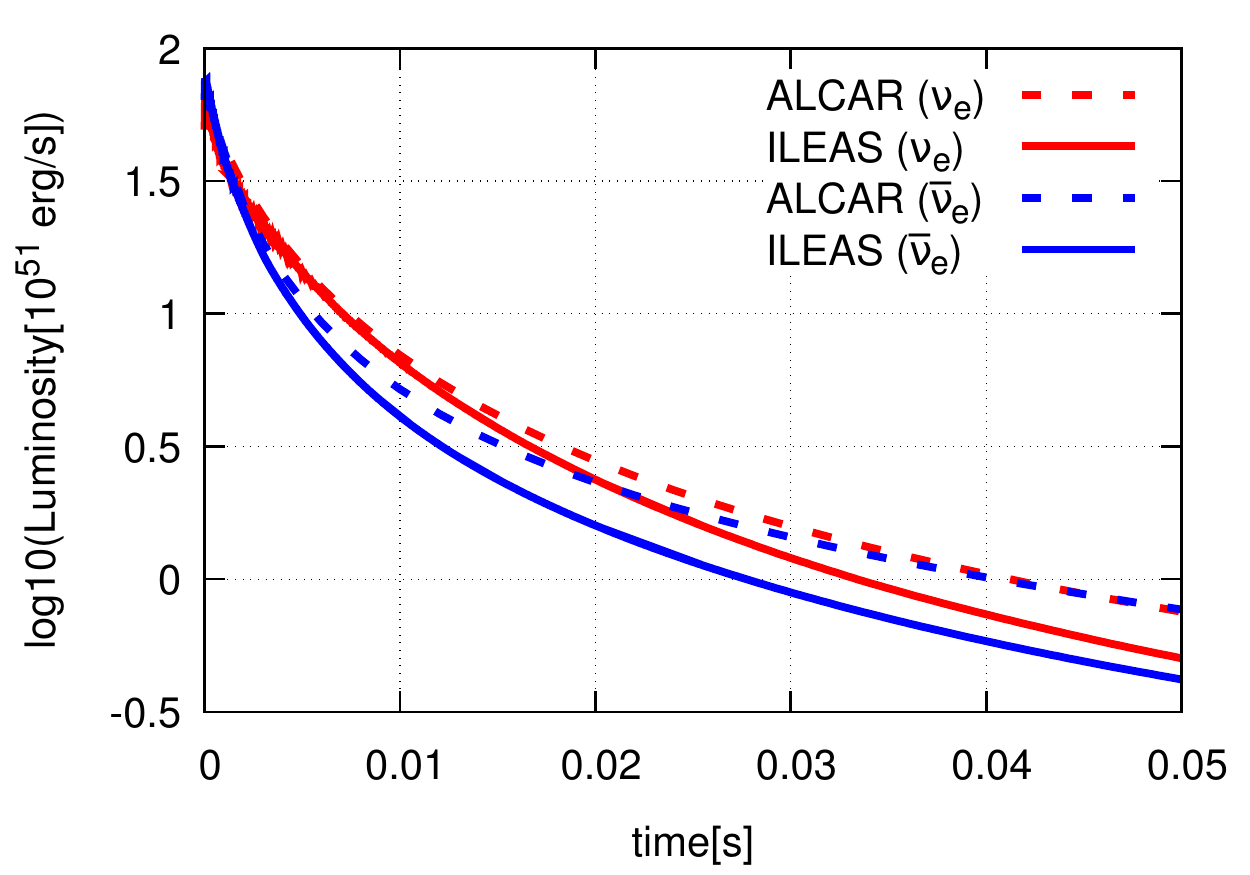}
\endminipage\hfill
\minipage{0.5\textwidth}
\includegraphics[width=\textwidth]{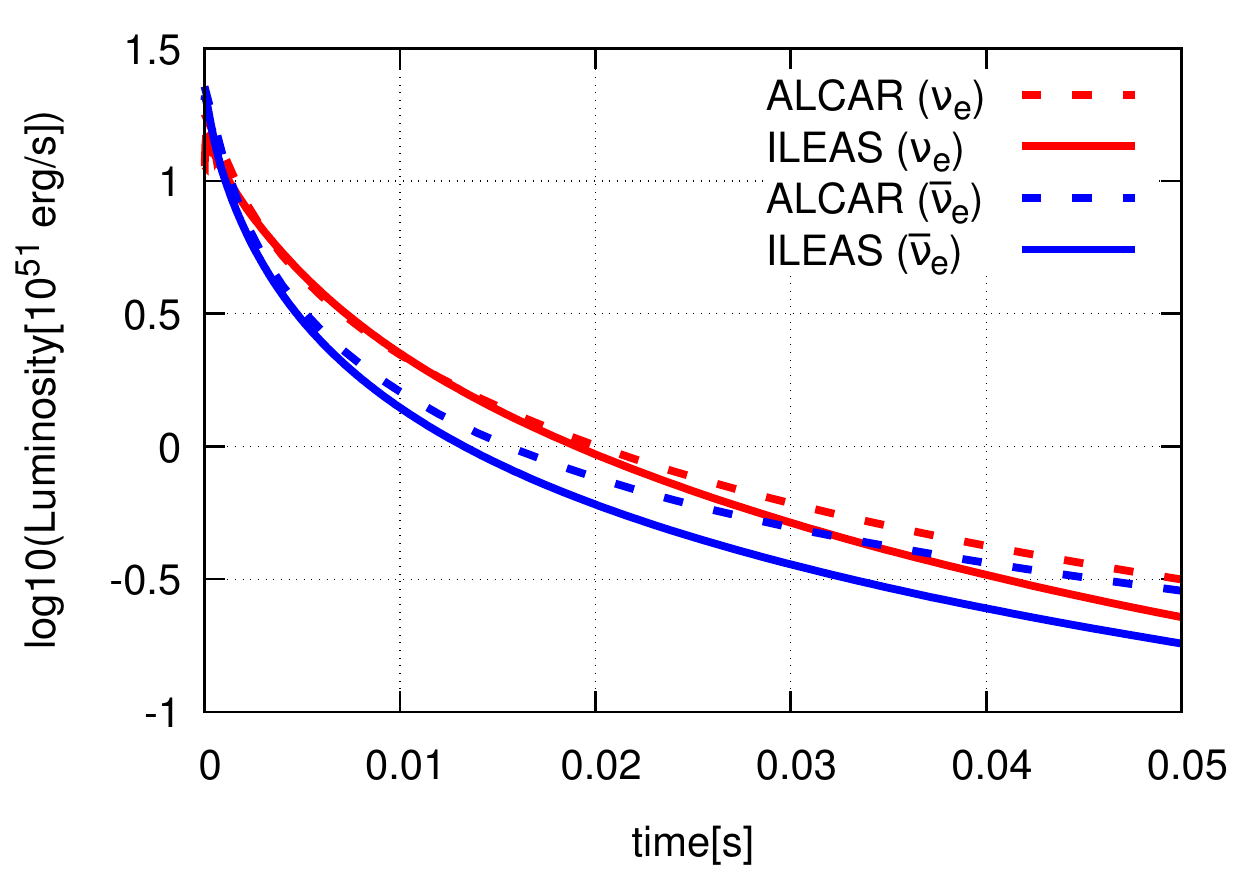}
\endminipage\hfill
}%
\\
\makebox[0pt][c]{%
\minipage{0.5\textwidth}
\includegraphics[width=\textwidth]{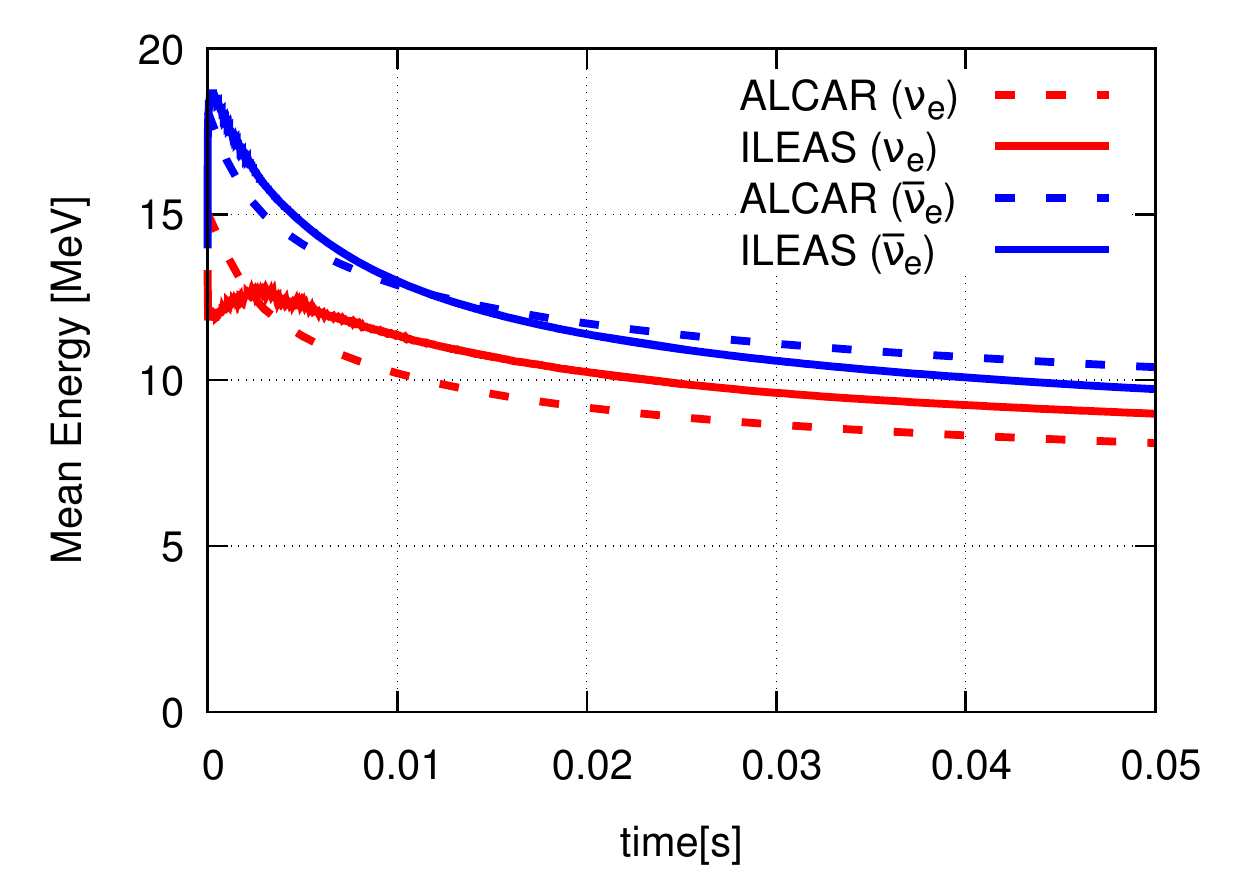}
\endminipage\hfill
\minipage{0.5\textwidth}
\includegraphics[width=\textwidth]{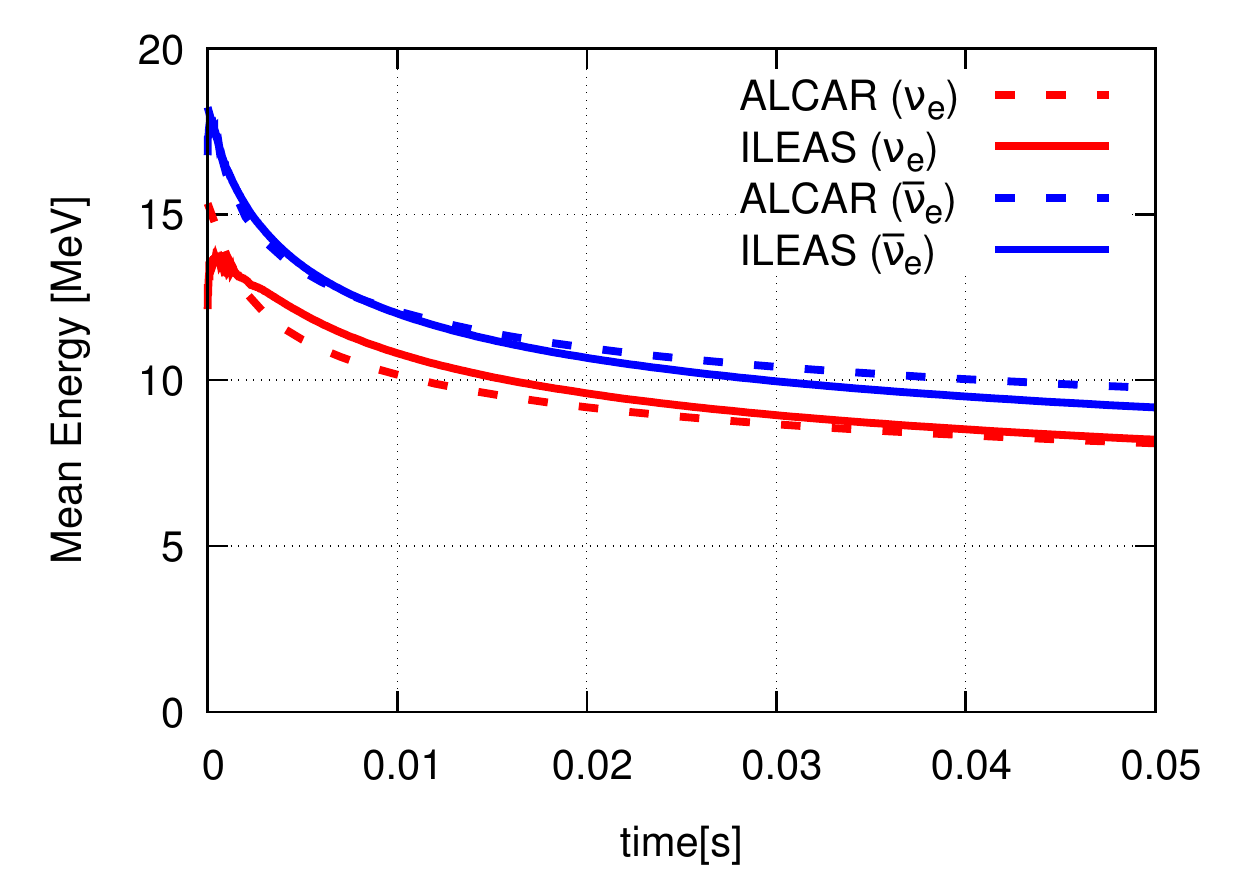}
\endminipage\hfill
}%
\caption{Top and middle panels: time evolution of the neutrino luminosities obtained by \textsc{ILEAS} (solid lines)  and \textsc{ALCAR} (dashed lines) produced by a thick BH-torus (left panels, initial torus mass 0.3~\Msun around a 3~\Msun\ BH) and by a thin BH-torus (right panels, initial torus mass 0.1~\Msun around a 3~\Msun\ BH) with fixed density background, but evolved $T$ and $Y_e$ distributions. The top panels show the first 10~ms of evolution, while the middle ones show the full duration of the simulations with the luminosities on a logarithmic scale for better visibility. Bottom panels: time evolution of the mean neutrino energies obtained by \textsc{ALCAR} (dashed lines) and \textsc{ILEAS} (equation~\ref{eleak}, solid lines) for the same BH-torus model. The luminosities and mean energies are computed for a local observer in the centre-of-mass frame of the neutrino source at the edge of our grid (100~km).} \label{BHTtimedep}
\end{center}
\end{figure*}

In this section, we discuss the time evolution starting from the \textsc{ALCAR} PNS snapshot at 0.5~s post-bounce described in section~\ref{sec:tests:snapPNS} and the two BH-torus models from section~\ref{sec:tests:snapBHT}. As detailed at the beginning of section~\ref{sec:tests}, we evolve the electron fraction and the matter temperature through the internal energy density of the fluid via equations~\eqref{uevol} and \eqref{Yeevol}, respectively. 

Figure~\ref{PNStimedep} displays the time evolution of the neutrino luminosities for the PNS snapshot. The time axis starts at the time of the original snapshot, where the evolution is started. After a brief transient of a few milliseconds, the electron fraction and the temperature relax to their equilibrium values, and the system slowly evolves in a quasi-steady state. Thus the results listed in table~\ref{table:results} correspond to the results of the plot at 5~ms. \textsc{ILEAS} is capable of reproducing the results obtained by \textsc{ALCAR} with $\sim$10 per cent accuracy throughout the 50~ms simulated. In figure~\ref{PNStimedep} we also provide the evolution of the mean neutrino energies obtained by \textsc{ALCAR} and \textsc{ILEAS}, using both equations~\eqref{eleak} and~\eqref{etot} described in section~\ref{sec:model:postproc}, for the same PNS model. We find an agreement within 1--2~MeV between both codes, with a tendency of improvement at later times and better results for the diagnostic mean neutrino energies as given by equation~\eqref{etot}.

Similarly, figure~\ref{BHTtimedep} displays the time evolution of the neutrino luminosities for the two BH-torus models. It is important to note that in this case the unavoidable transient which occurs when switching on \textsc{ILEAS}, proceeds to swiftly cool the optically thin disc before a stationary state can be reached. The natural consequence is, therefore, that the \textsc{ILEAS} luminosities become smaller than those obtained by \textsc{ALCAR}, whose background remains hotter. Nevertheless, for most of the time the results obtained by \textsc{ILEAS} agree to less than 10 per cent with the ones obtained by \textsc{ALCAR}. The bottom panels in figure~\ref{BHTtimedep} display the evolution of the mean neutrino energies obtained by \textsc{ALCAR} and \textsc{ILEAS} (using equation~\ref{eleak}) for the same two BH-torus models. As for the PNS case, our results agree within $\sim$1~MeV or, for the thin torus case, even better than 1 MeV during most of the simulation. Note that the results listed in table~\ref{table:results} correspond to the those in the plots from figure~\ref{BHTtimedep} measured at 3~ms.

\subsection{Performance of ILEAS in numerical simulations of binary neutron star mergers}\label{sec:tests:NSNS}

 \begin{figure*}
\begin{center}
\makebox[0pt][c]{%
\minipage{0.51\textwidth}
\includegraphics[width=\textwidth]{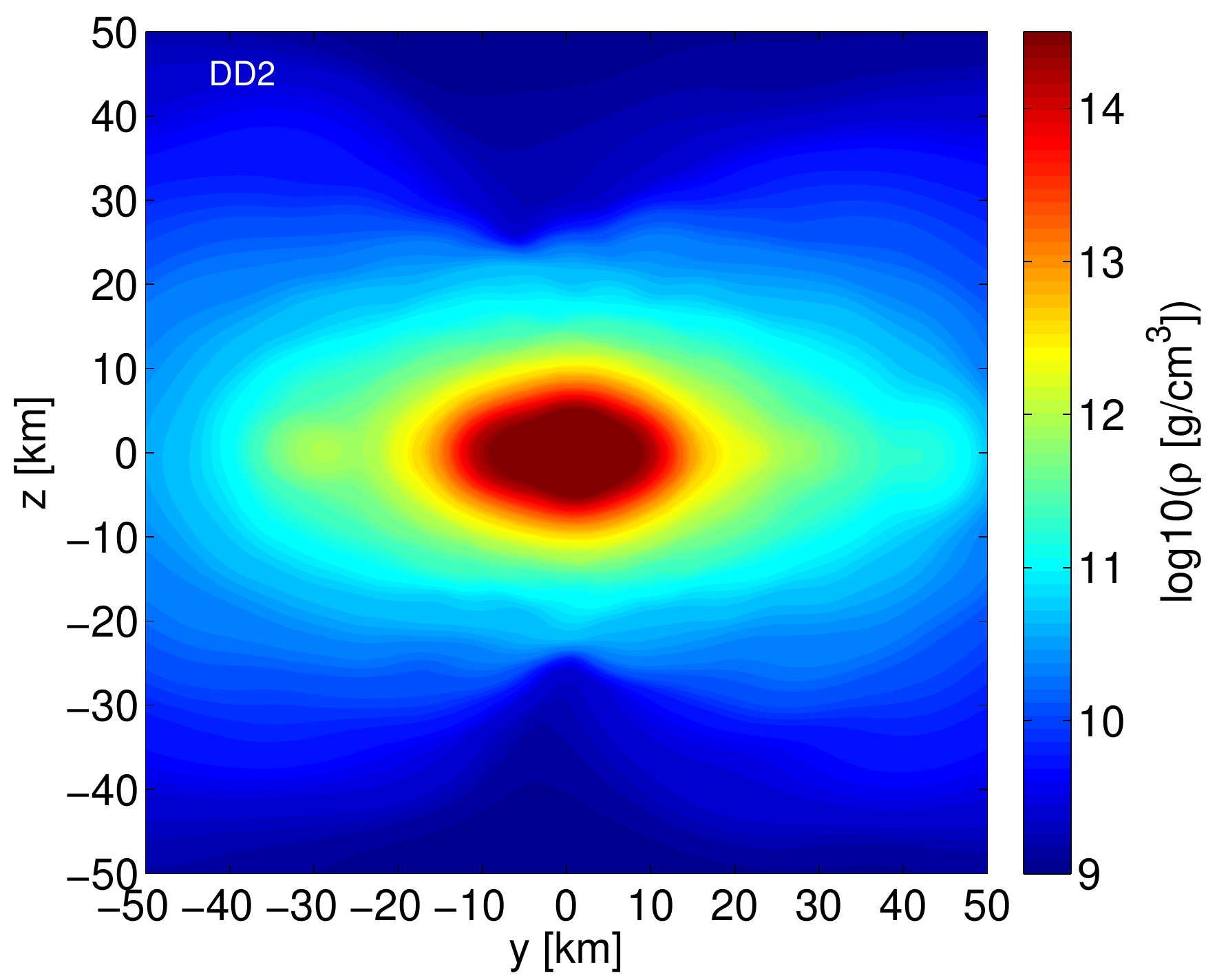}
\endminipage\hfill
\hspace{-0.1cm}
\minipage{0.51\textwidth}
\includegraphics[width=\textwidth]{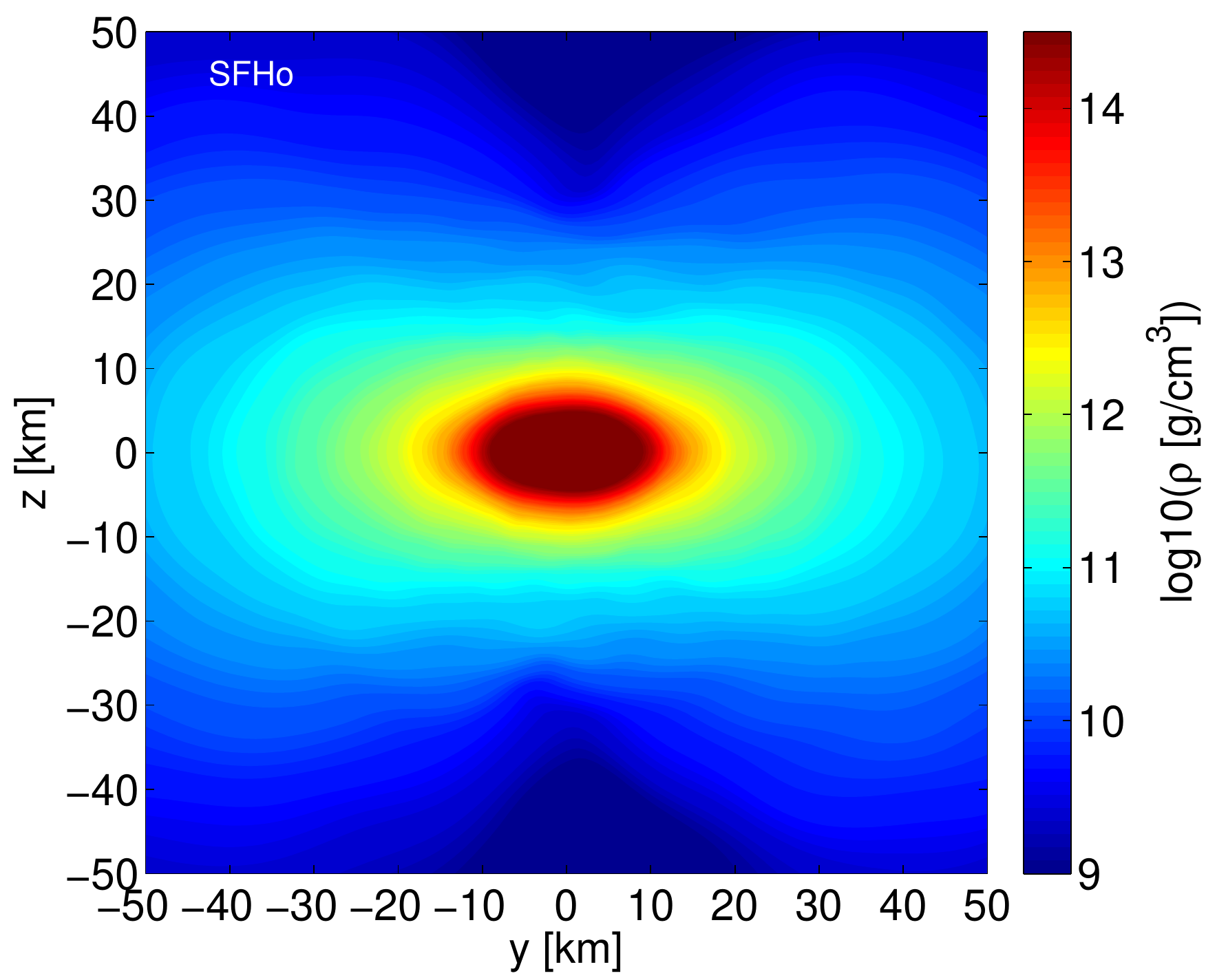}
\endminipage\hfill
}%
\\
\makebox[0pt][c]{%
\minipage{0.51\textwidth}
\includegraphics[width=\textwidth]{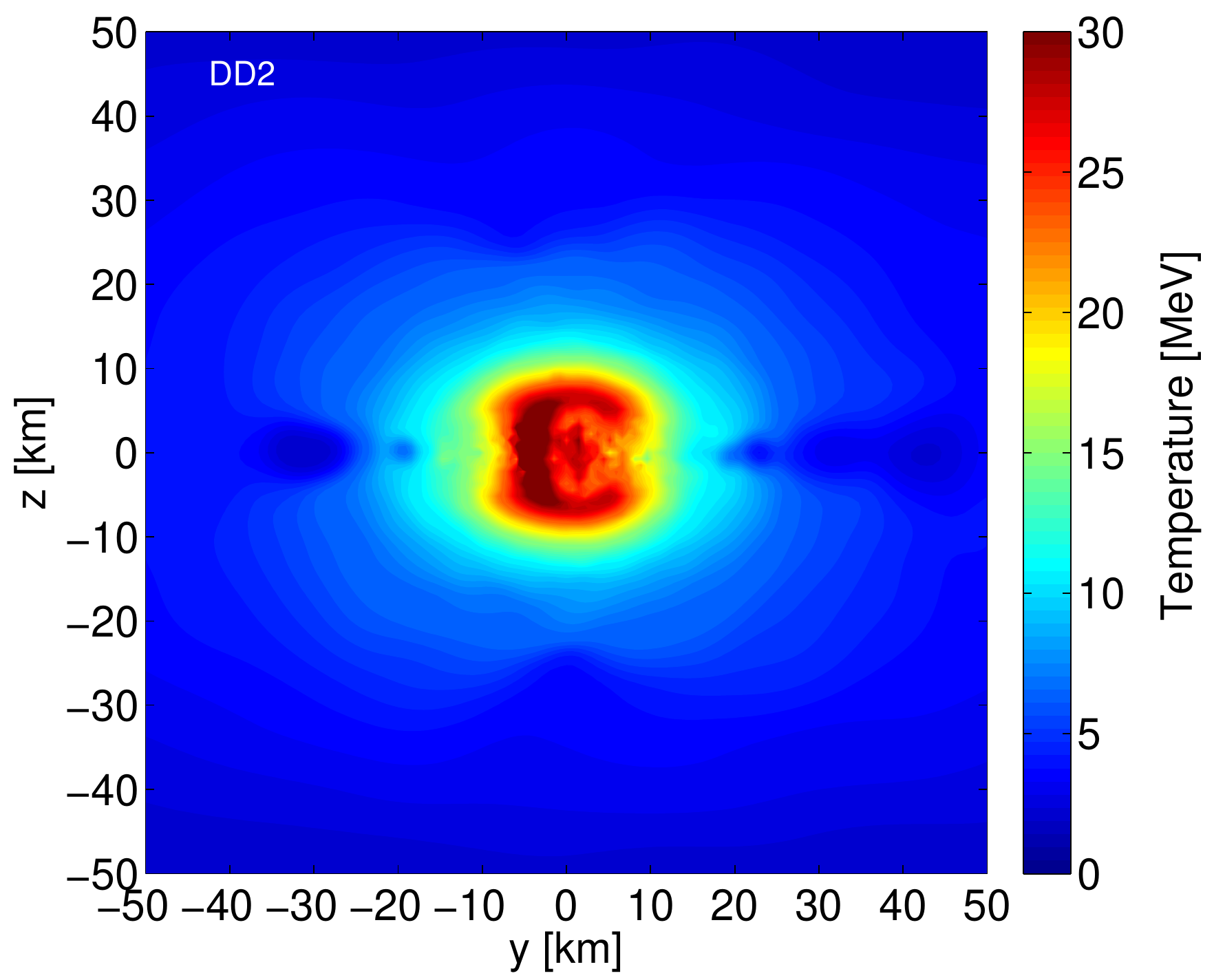}
\endminipage\hfill
\hspace{-0.1cm}
\minipage{0.51\textwidth}
\includegraphics[width=\textwidth]{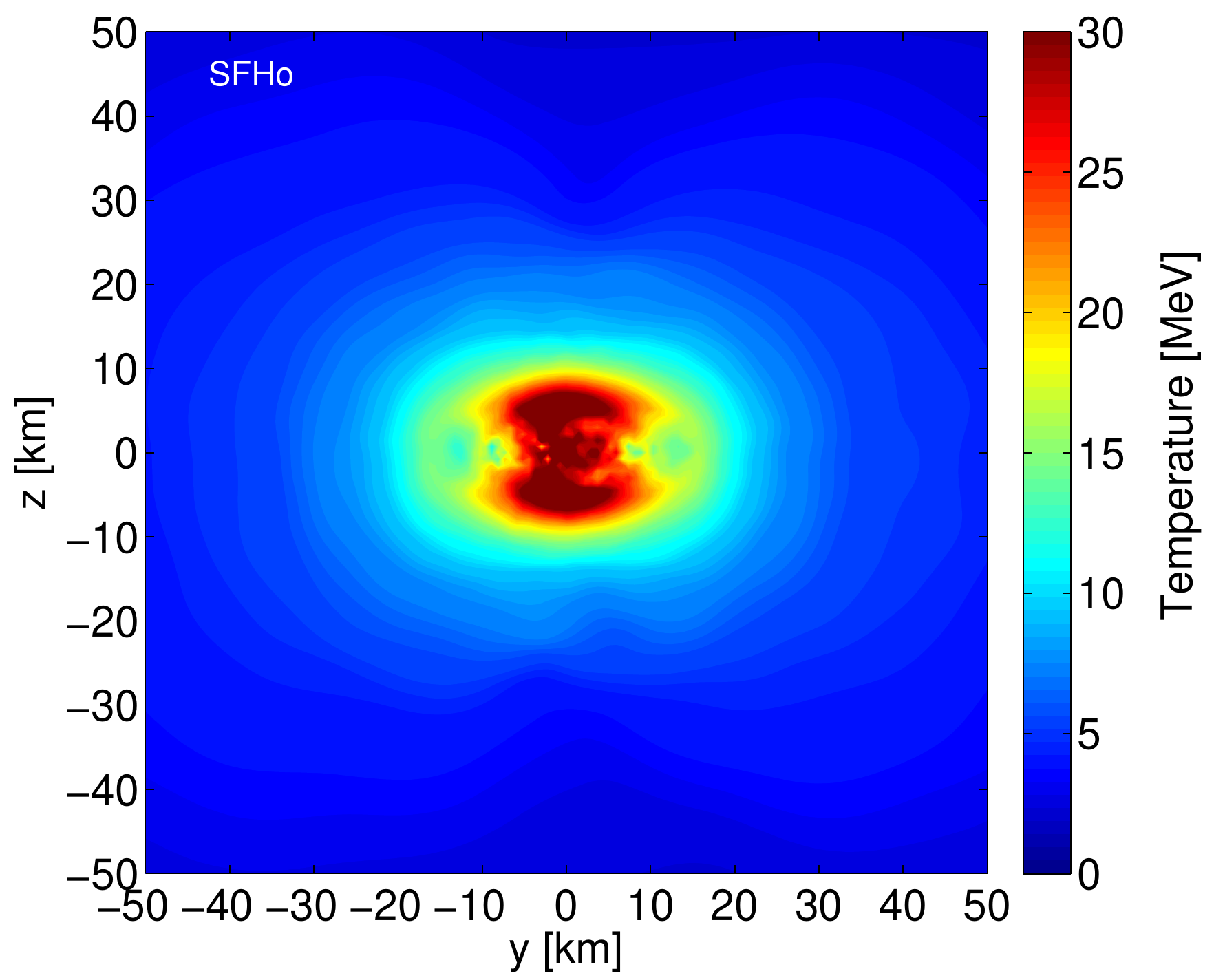}
\endminipage\hfill
}%
\\
\makebox[0pt][c]{%
\minipage{0.51\textwidth}
\includegraphics[width=\textwidth]{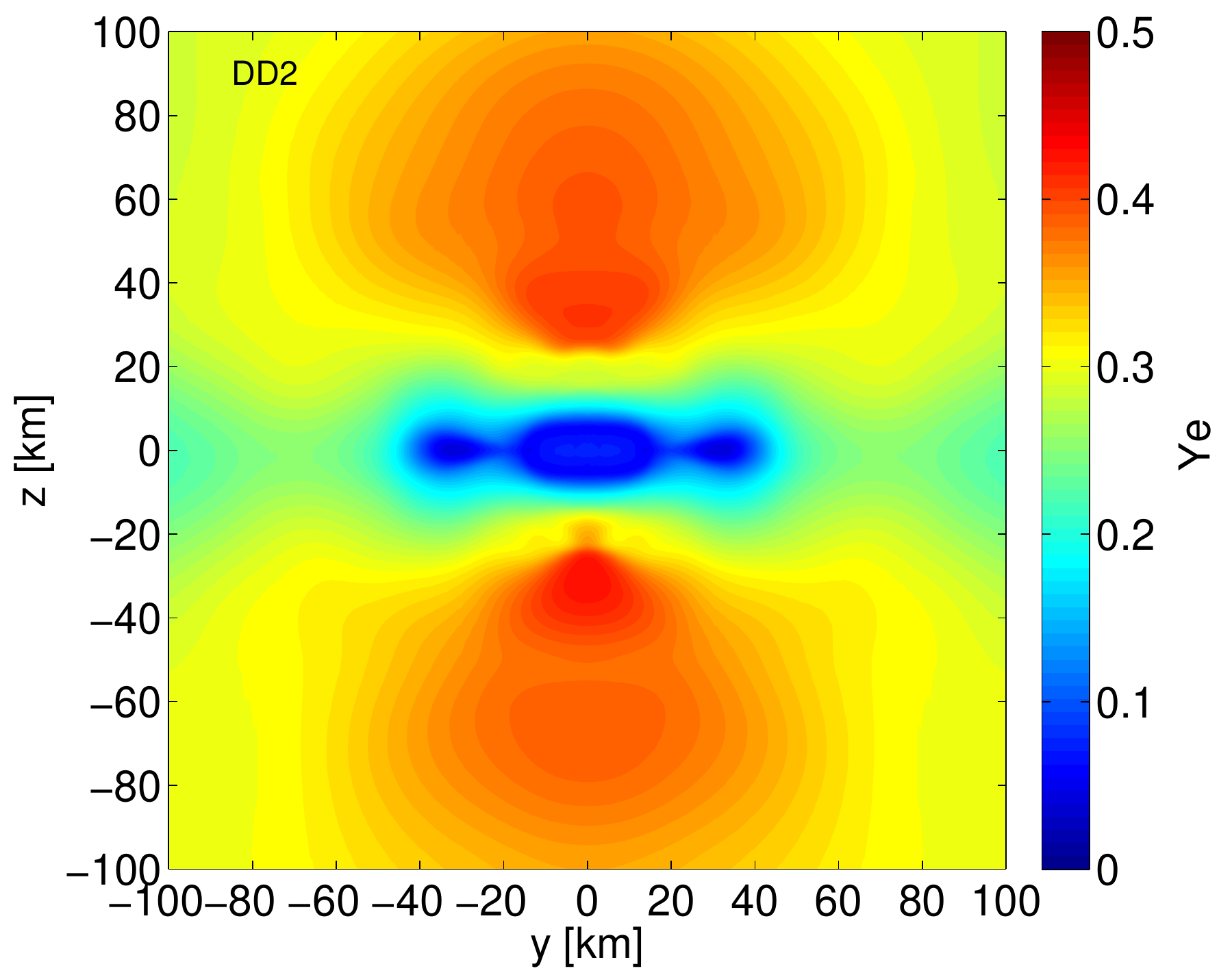}
\endminipage\hfill
\hspace{-0.1cm}
\minipage{0.51\textwidth}
\includegraphics[width=\textwidth]{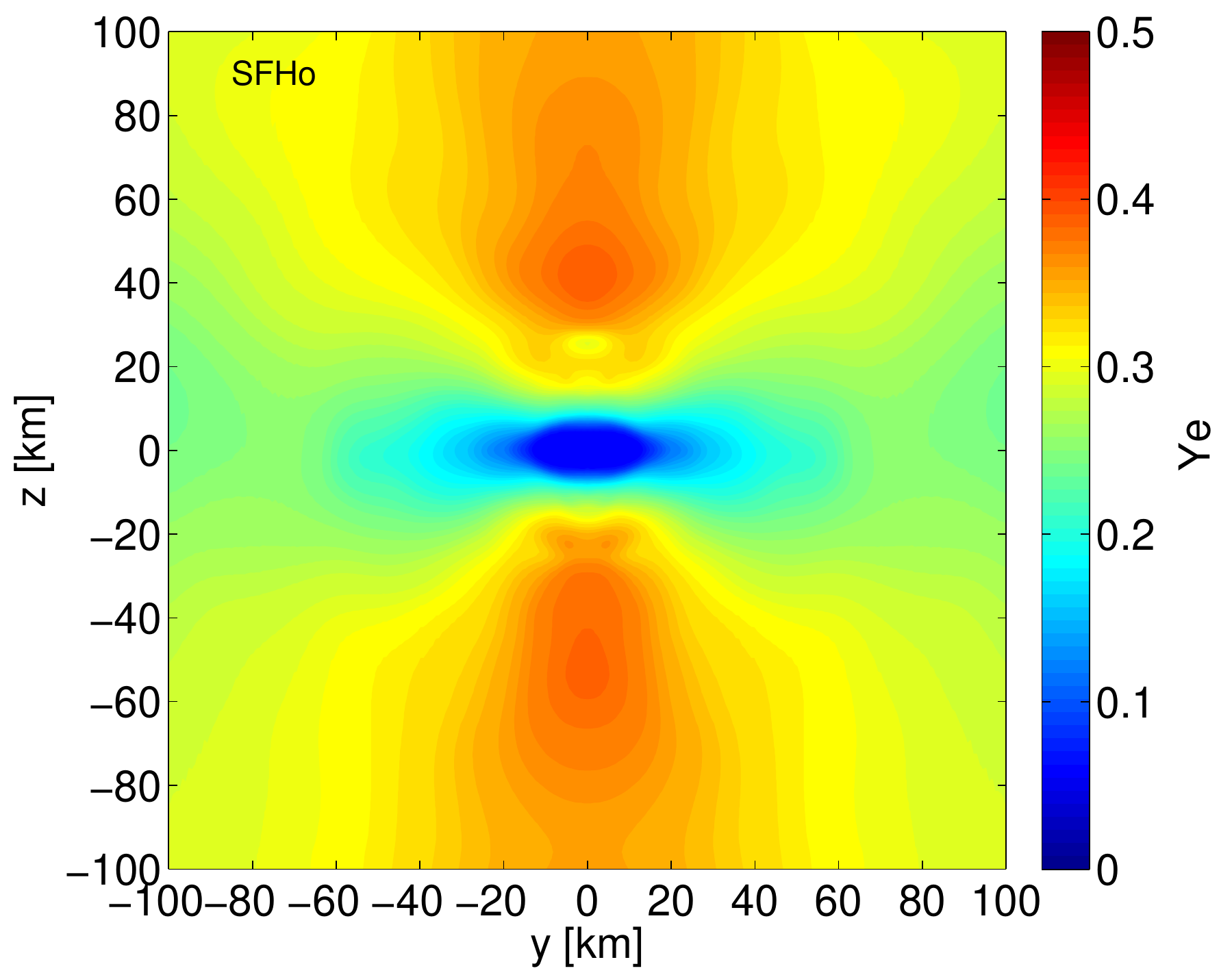}
\endminipage\hfill
}%
\caption{Vertical slices of the remnant of two binary NS mergers showing the density (top row), temperature (middle row) and electron fraction (bottom row). The slices were taken 5~ms after merger (defined from the moment when the lapse function reaches its first minimum) from two symmetric simulations with initial $M_{\mathrm{NS}}=1.35$~\Msun, using the DD2 EoS (left column) and the SFHo EoS (right column). We caution the reader that the bottom row of panels displays the angular averages of the electron fraction, as well as spatial ranges on the axes different from the two other rows.} \label{NSNShydro}
\end{center}
\end{figure*}

 \begin{figure*}
\begin{center}
\makebox[0pt][c]{%
\minipage{0.51\textwidth}
\includegraphics[width=\textwidth]{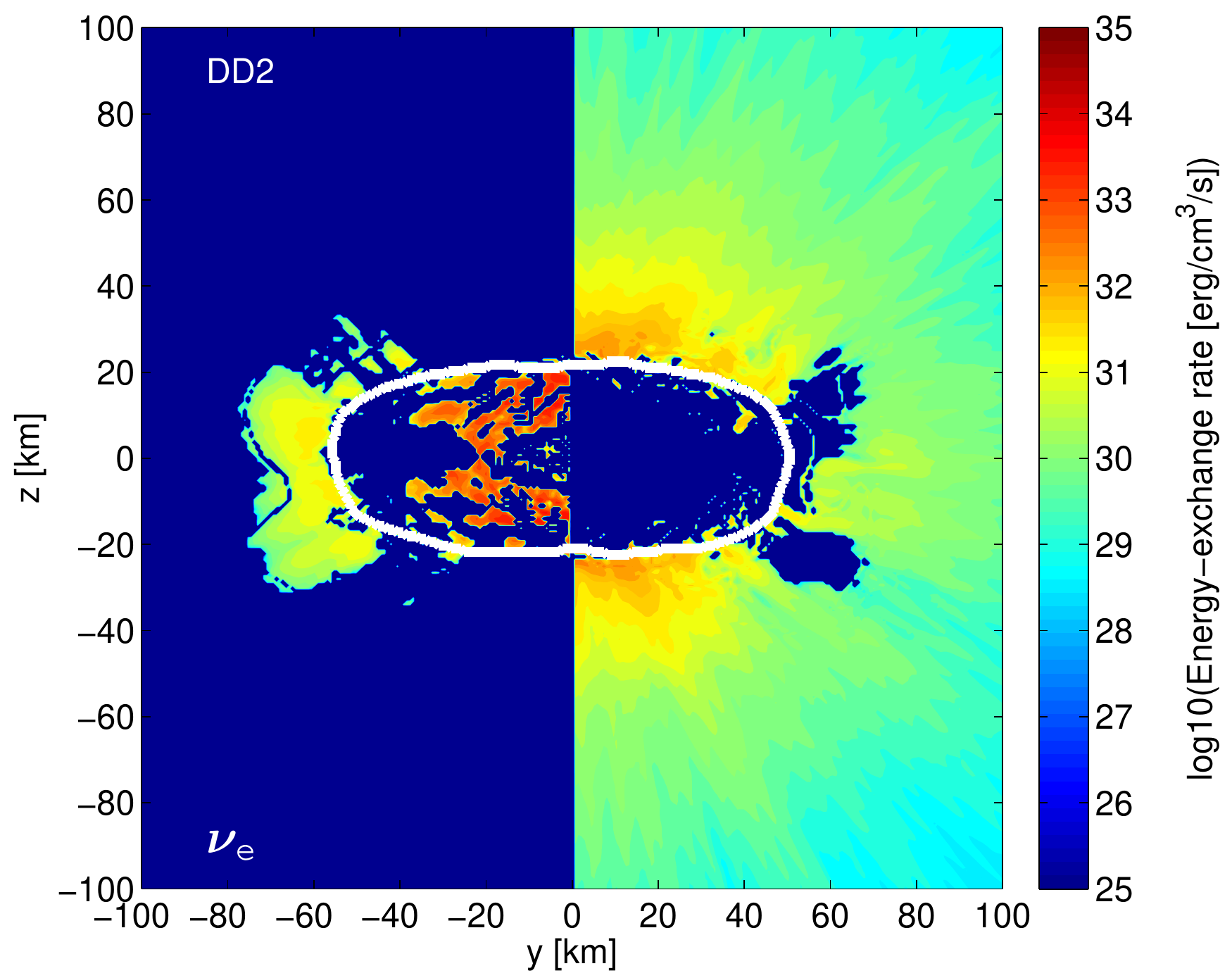}
\endminipage\hfill
\hspace{-0.1cm}
\minipage{0.51\textwidth}
\includegraphics[width=\textwidth]{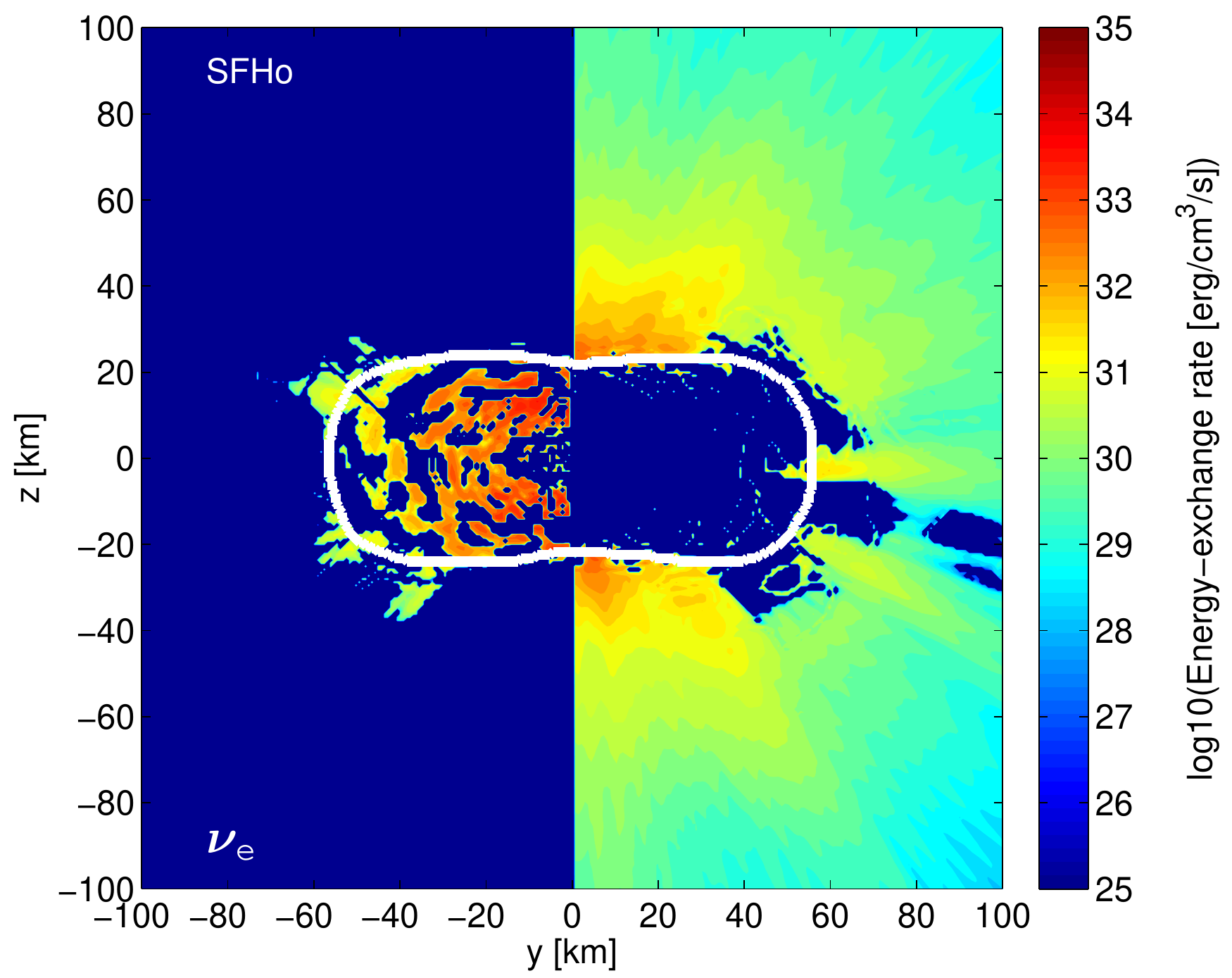}
\endminipage\hfill
}%
\\
\makebox[0pt][c]{%
\minipage{0.51\textwidth}
\includegraphics[width=\textwidth]{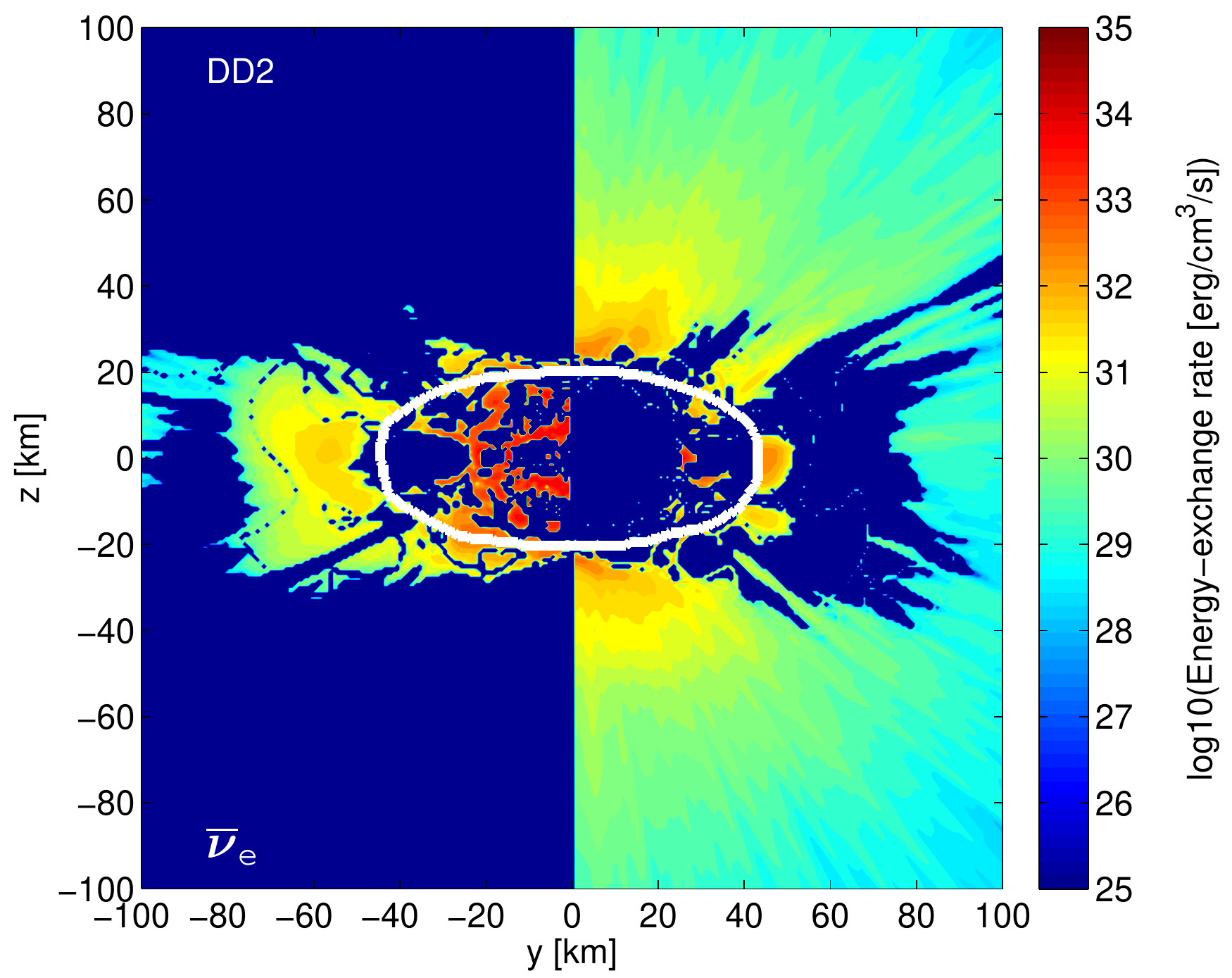}
\endminipage\hfill
\hspace{-0.1cm}
\minipage{0.51\textwidth}
\includegraphics[width=\textwidth]{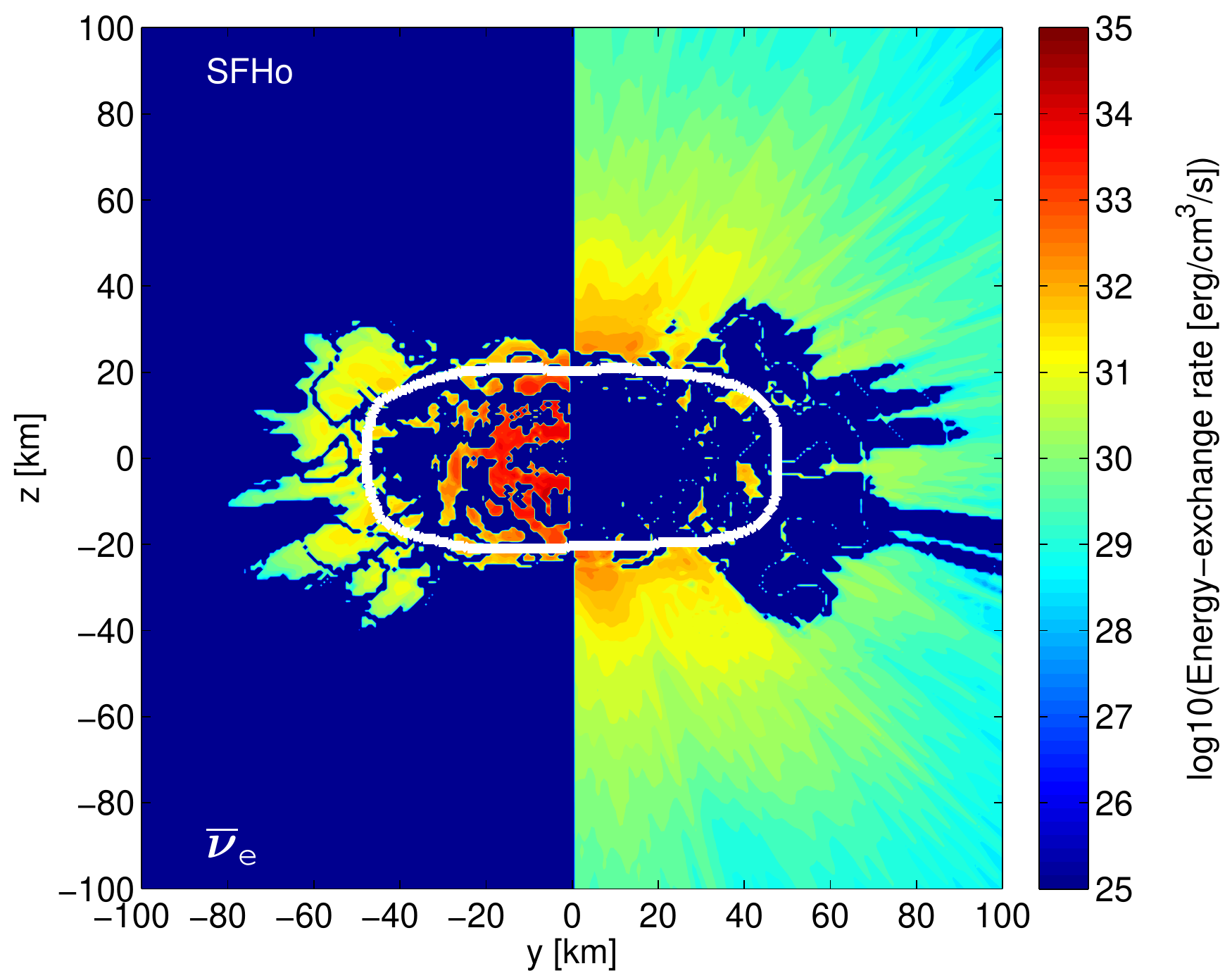}
\endminipage\hfill
}%
\\
\makebox[0pt][c]{%
\minipage{0.51\textwidth}
\includegraphics[width=\textwidth]{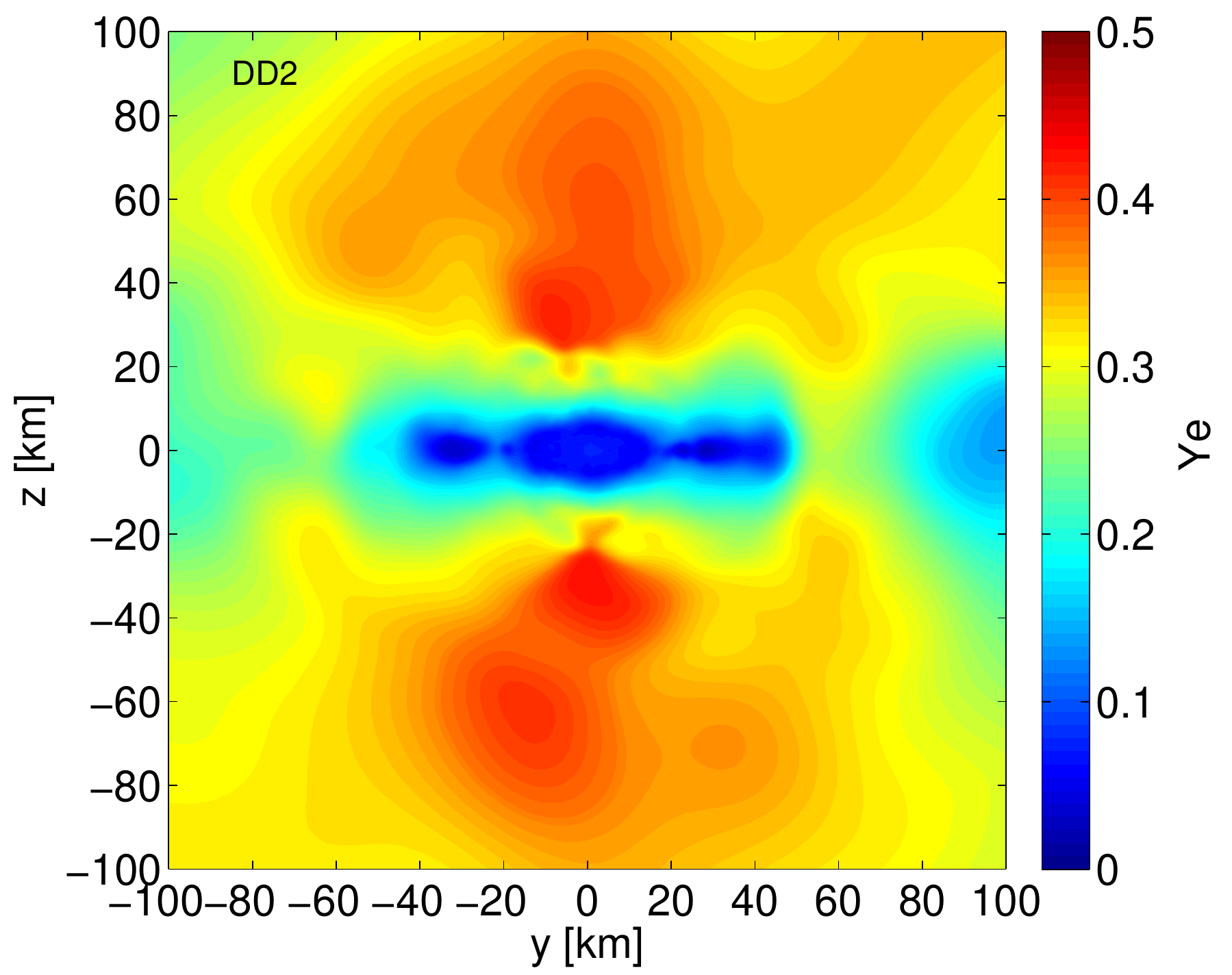}
\endminipage\hfill
\hspace{-0.1cm}
\minipage{0.51\textwidth}
\includegraphics[width=\textwidth]{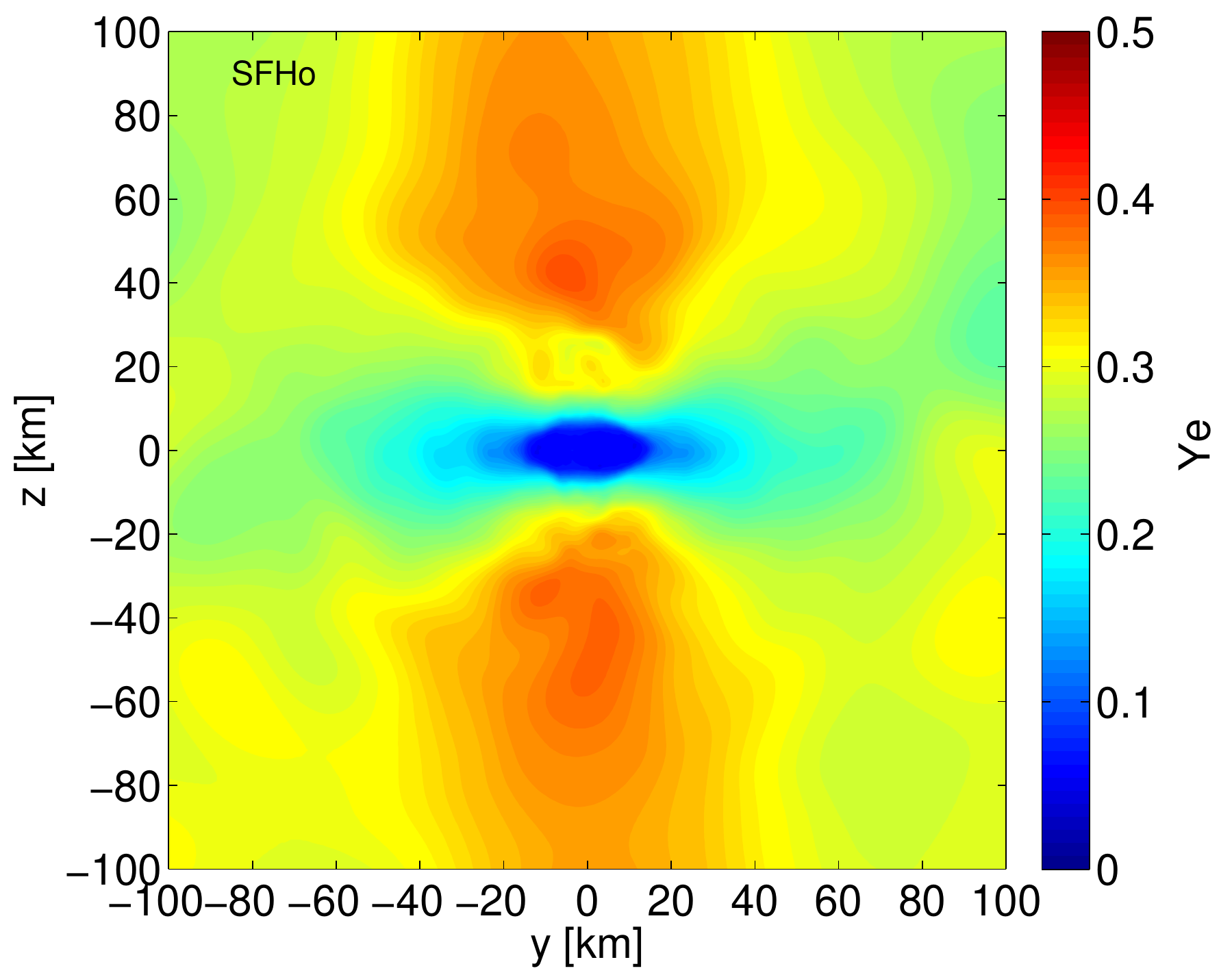}
\endminipage\hfill
}%
\caption{Vertical slices of the two NS merger remnants listed in table~\ref{simulationsejecta} showing the energy-exchange rates $Q_{\nu_i}^{\mathrm{net}}=Q_{\nu_i}^{+}-Q_{\nu_i}^{-}$ of $\nu_i=\nu_e$ (top row of plots) and $\nu_i=\bar{\nu}_e$ (middle row of plots) at 5~ms after merger (defined from the moment when the lapse function reaches its first minimum). The left column of plots shows the results from a NS-NS merger simulation using the DD2 EoS, whereas the right column of plots displays the results of a simulation with the SFHo EoS. The left half-panel of each plot displays colour-coded the regions where neutrino emission dominates over absorption, while the right half-panels display the absorption-dominated regions in the other hemisphere. We caution the reader that these are vertical slices from three-dimensional simulations, and thus there is no symmetry between the left and right half-panels. The white contours in the upper four plots delineate the location of the neutrinosphere ($\tau_{\nu_i}=2/3$) of the corresponding neutrino species. For reasons of direct comparison we also show in the bottom row of plots the instantaneous $Y_e$ distribution in the cross-sectional slices.} \label{NSNSabsem}
\end{center}
\end{figure*}

\begin{figure}
\begin{center}
\makebox[0pt][c]{%
\minipage{0.5\textwidth}
\includegraphics[width=\textwidth]{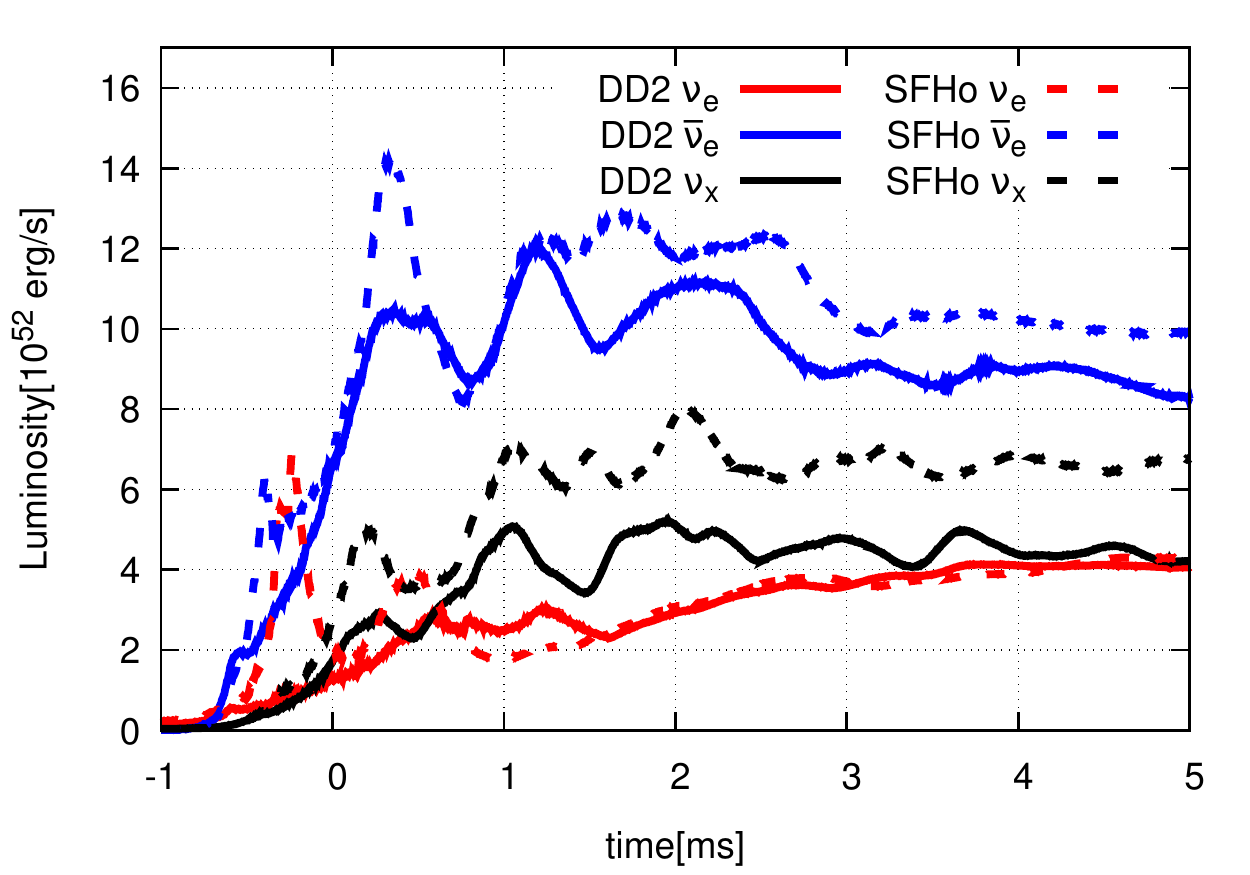}
\endminipage\hfill
}%
\\
\makebox[0pt][c]{%
\minipage{0.5\textwidth}
\includegraphics[width=\textwidth]{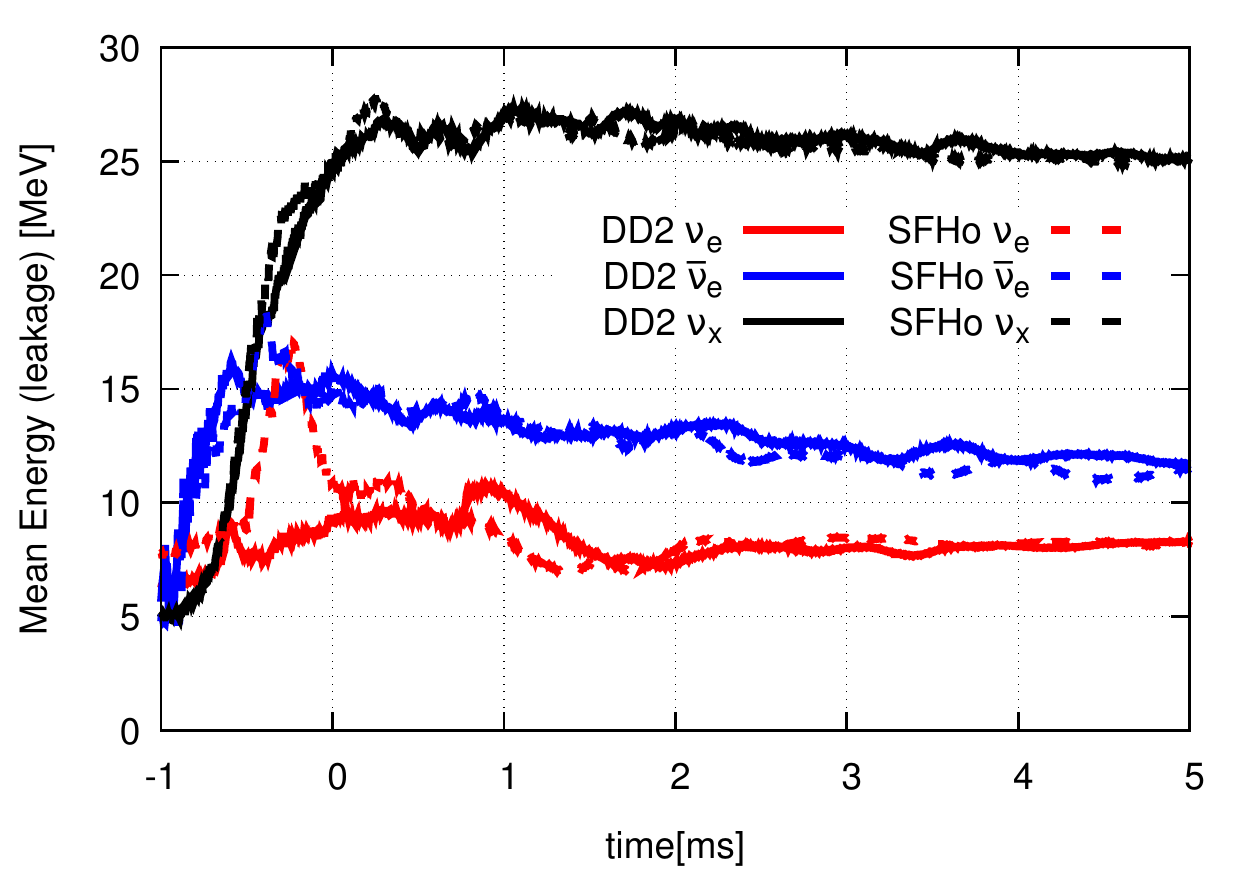}
\endminipage\hfill
}%
\\
\makebox[0pt][c]{%
\minipage{0.5\textwidth}
\includegraphics[width=\textwidth]{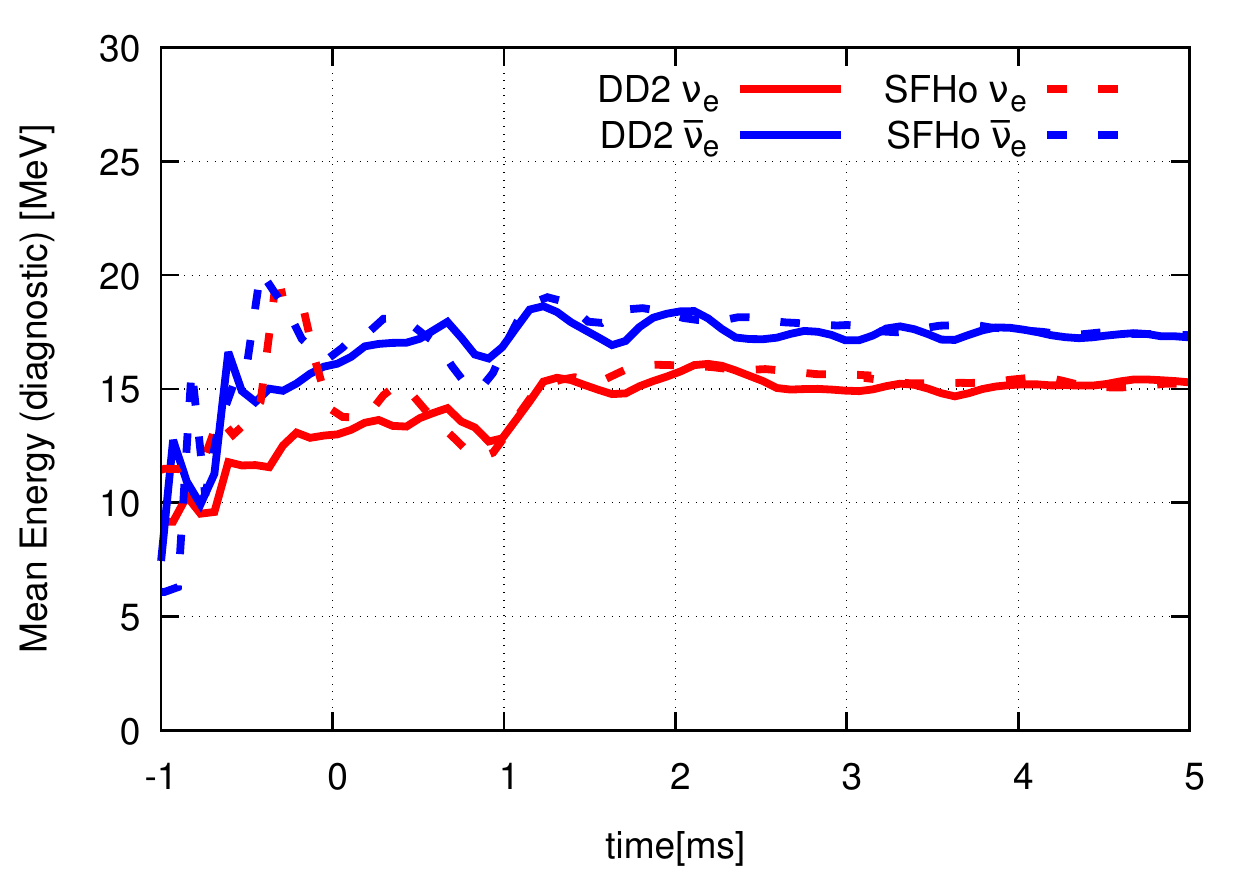}
\endminipage\hfill
}%
\caption{Top panel: neutrino luminosities from all neutrino species ($\nu_e$, $\bar{\nu}_e$ and one representative of $\nu_x$) obtained by \textsc{ILEAS} in the two numerical simulations of NS mergers listed in table~\ref{simulationsejecta} performed with the SPH-CFC code described in {\protect\cite{2007A&A...467..395O}}. Middle panel: mean neutrino energies of all neutrino species, calculated in the leakage approach (equation~\ref{eleak}) for the same NS merger models. Bottom panel: mean neutrino energies of electron-type neutrinos calculated for diagnostics (equation~\ref{etot}). The luminosities and mean energies are computed for an observer at infinity and the time is measured with respect to the first minimum of the lapse function.} \label{lumimeaneNSNS}
\end{center}
\end{figure}

\begin{figure}
  \centering
\begin{center}
\makebox[0pt][c]{%
\minipage{0.5\textwidth}
\includegraphics[width=\textwidth]{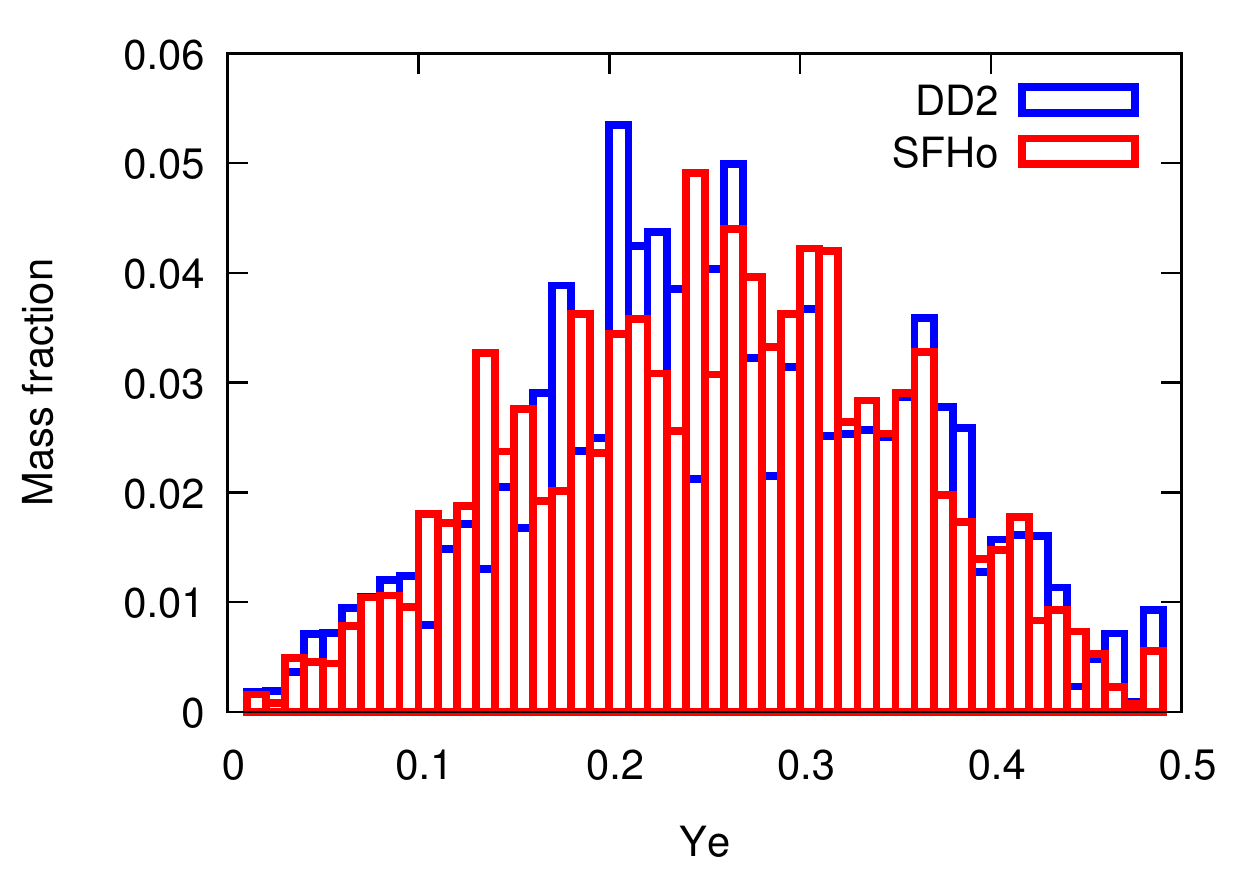}
\endminipage\hfill
}%
\caption{Histogram showing the electron fraction (vs. mass fraction) of the material ejected in the two symmetric NS merger simulations listed in table~\ref{simulationsejecta}. The ejecta properties were measured 5~ms after merger (defined from the moment when the lapse function reaches its first minimum). The total ejecta masses and average ejecta $Y_e$ obtained in the reported models are summarized in table~\ref{simulationsejecta}.} \label{yehist}
\end{center}
\end{figure}

In order to demonstrate the capabilities of the neutrino scheme described in this work, we present the results of two numerical simulations of the merger of two NSs including the effects of weak interactions, despite the unavailability of reference solutions that could allow us to exploit these simulations as tests for our neutrino scheme. For such merger simulations, \textsc{ILEAS} was coupled to our SPH-CFC code \citep{2007A&A...467..395O, 2013ApJ...773...78B} as described in section~\ref{sec:model}. The two models included in this section are evolved from a symmetric initial binary configuration, with two irrotational $1.35$~\Msun\ NSs with the DD2 \citep{2010PhRvC..81a5803T, 2010NuPhA.837..210H} and SFHo \citep{Steiner2013a} EoS, respectively.

Because \textsc{ILEAS} is implemented on a Cartesian grid instead of being implemented on the SPH formulation, it is necessary to map the relevant quantities evolved by our SPH-CFC merger code ($\rho$, temperature, $Y_e$ and the gravitational potentials $\alpha$ and $\psi$) onto the grid. This has been shown to be an ambiguous procedure, with various options described in the literature (see e.g \citealt{2007PASA...24..159P, 2018arXiv180303652R}). We found that the most suitable approach consistent in all the phases of the NS merger evolution is the \textit{normalized} rendering method described in \cite{2007PASA...24..159P}, but weighted with the conserved rest mass density, $\rho^*$.

For rendering a given quantity $A$ from SPH particles ($A_j$) to our grid ($\langle A\rangle(\boldsymbol{r})$) we apply
\begin{align}
 \langle A\rangle(\boldsymbol{r})\ =\ \frac{\sum_j{\frac{m_j}{\rho^*_j}A_j\rho^*_jW(|\boldsymbol{r}-\boldsymbol{r}_j|,h_j)}}{\sum_j{\frac{m_j}{\rho^*_j}\rho^*_jW(|\boldsymbol{r}-\boldsymbol{r}_j|,h_j)}},\label{renderrho}
\end{align}
where the summation runs over all the particles whose kernel overlaps with a grid point, $\boldsymbol{r}$. Following standard SPH nomenclature, $m_j$ is the mass of a given SPH particle $j$ and $W(|\boldsymbol{r}-\boldsymbol{r}_j|,h_j)$ the SPH kernel, with $|\boldsymbol{r}-\boldsymbol{r}_j|$ being the distance between the centre of the SPH particle and the grid point and $h_j$ the smoothing length. For the NS merger models presented in this section we employ a spherically symmetric cubic spline kernel (see \citealt{2007A&A...467..395O} for details on the SPH implementation). The mapping of the neutrino source terms from \textsc{ILEAS}' Cartesian grid back to the SPH particles is performed with a trilinear interpolation.

A direct comparison with results employing \textsc{ALCAR} or \textsc{VERTEX} neutrino transport in the context of NS mergers is currently not possible, as neither of these neutrino treatments are ready for use in our SPH-CFC code. Nevertheless, numerical simulations of the same astrophysical set-up are available in the literature (e.g. \citealt{2015PhRvD..91f4059S, 2016PhRvD..93l4046S, 2015PhRvD..92d4045P}), which allows us to qualitatively compare our results to those obtained by other groups employing different schemes. It is important to remark, however, that substantial quantitative differences are to be expected already from the diverse hydrodynamical solvers. Particularly, the high numerical viscosity of SPH can influence the neutrino-relevant temperature profile of the HMNS. Note however the overall good agreement between SPH and grid-based codes as far as bulk properties like the ejecta mass are concerned (see for instance \citealt{2013ApJ...773...78B} and \citealt{2013PhRvD..88d4026H}). Moreover, differences are likely to result from resolution differences and to occur already between the different grid-based codes, but no detailed comparisons exist in the literature. The functional dependence of neutrino interactions on high powers of the temperature makes it very difficult if not impossible to determine the exact origin of variations in neutrino-related quantities between different codes, caused either by the hydro-/thermodynamical evolution or the neutrino treatment. 

In figure~\ref{NSNShydro} we provide vertical slices of the relevant hydro-/thermodynamical quantities ($\rho$, temperature and $Y_e$) for both simulated models at 5~ms post-merger. The times are measured with respect to the first minimum of the lapse function. \cite{2016PhRvD..93l4046S} and \cite{2015PhRvD..92d4045P} provide similar plots for the same model setups, albeit at different post-merger times. In the first row of plots, one can clearly see the structure of the merger remnant: a deformed rotating HMNS surrounded by a torus of merger debris. The central object retains the initial, low $Y_e$ (see bottom row in figure~\ref{NSNShydro}) and is surrounded by a dense, neutron-rich inner disk ($Y_e<0.2$) together with an extended outer disk of higher electron-fraction material ($0.2<Y_e<0.3$). The polar regions are filled by material with even higher electron fraction ($Y_e\gtrsim0.3$).

 \begin{table}
  \begin{center}
      \caption{List of NS merger models presented in this work, with their initial setup (NS masses and EoS) and the properties of the ejected material (mass and average $Y_e$) extracted at 5~ms post-merger. (The merger time is defined from the moment when the lapse function reaches its first minimum.)}\label{simulationsejecta}  
    \begin{threeparttable}
     \begin{tabular}{lccc}
      \hline
      \hline
      \noalign{\vskip 2mm}  
      \parbox[c]{0.1\textwidth}{EoS}        & \parbox[c]{0.1\textwidth}{\centering NS masses (\Msun)}        & \parbox[c]{0.1\textwidth}{\centering Ejecta mass ($10^{-3}$~\Msun)}        & \parbox[c]{0.1\textwidth}{\centering Ejecta $\langle Y_e\rangle$}         \vspace{1mm}  \\ 
      \hline    
      \noalign{\vskip 2mm}                                                                                                                         
      DD2			& 1.35-1.35		& 1.8		& 0.26	   	\\  
      SFHo			& 1.35-1.35		& 3.2		& 0.26	   	\\  
      \noalign{\vskip 2mm}   
      \hline
      \hline
     \end{tabular}
    \end{threeparttable}	
  \end{center}
\end{table}

The top and middle rows of plots in figure~\ref{NSNSabsem} display the emission-dominated (left-hand panels) and absorption-dominated (right-hand panels) regions of the post-merger remnants (also in sectional planes of the 3D distribution perpendicular to the orbital plane at 5~ms post-merger). The blue regions of the left-hand panels which lie inside the neutrinosphere (delimited with a white contour) represent regions where the diffusion time-scale becomes negative. In such regions, the diffusion time-scale employed by \textsc{ILEAS} is set to infinity, quenching local neutrino losses (see section~\ref{sec:model:tdiff}). Outside the contour, neutrino re-absorption mostly dominates over emission, except in some regions of the torus surrounding the HMNS. A vast number of neutrinos are re-absorbed right above and below the central object (right-hand panels of the top and middle rows in figure~\ref{NSNSabsem}), substantially increasing the electron fraction in such regions (see bottom panels in figure~\ref{NSNShydro} and figure~\ref{NSNSabsem}). This creates a baryon-polluted high-opacity region in the polar directions, possibly obstructing the formation of jet-like structures (e.g. \citealt{2016ApJ...816L..30J}). 

The top panel in figure~\ref{lumimeaneNSNS} displays the time evolution of the neutrino luminosities of the three neutrino species for both merger models.  \textsc{ILEAS} is able to qualitatively reproduce the results obtained by other NS-NS merger simulations which also include neutrinos, such as the hierarchy of the neutrino luminosities ($L_{\bar{\nu}_e}>L_{\nu_x}\gtrsim L_{\nu_e}$) or the dependence on the EoS. Softer EoSs yield more compact NSs, which collide more violently, thus producing hotter merger remnants that emit higher neutrino luminosities. Quantitatively however, the luminosities displayed in figure~\ref{lumimeaneNSNS} are a factor $2-4$ lower than the ones reported by \cite{2016PhRvD..93l4046S} and \cite{2015PhRvD..92d4045P}. Here it is necessary to point out that the results presented in these two papers \citep{2015PhRvD..92d4045P, 2016PhRvD..93l4046S} also show a disagreement of up to a factor 2 between each other. Nevertheless, as mentioned before, it is impossible to identify the exact origin of the differences (hydro-/thermodynamical evolution or neutrino treatment) without a detailed comparison between the different schemes, testing and comparing the hydro and neutrino transport methods separately.

The \textit{leakage} neutrino mean energies (equation~\ref{eleak}) obtained with ILEAS for both models (see middle panel of figure~\ref{lumimeaneNSNS}) are comparable with the approximate values reported by \cite{2016PhRvD..93l4046S} which also employed a leakage+absorption scheme (although our results are $2-3$~MeV lower). The similarity of the neutrino mean energies of $\nu_e$ and $\bar{\nu}_e$ for both EoS we observe in figure~\ref{lumimeaneNSNS} was already present in the results found by \cite{2016PhRvD..93d4019F} when simulating the merger of two $1.2$~\Msun\ NSs with an M1 scheme to treat neutrino effects. Moreover, the neutrino mean energies they describe are $\sim5$~MeV higher than the ones found by \cite{2016PhRvD..93l4046S} but in agreement with our neutrino mean energies \textit{for diagnostics} (equation~\ref{etot}, see bottom panel of figure~\ref{lumimeaneNSNS}) within $\pm3$~MeV.

Following \cite{2013ApJ...773...78B}, we consider as \textit{ejected material} those SPH particles which are gravitationally \textit{unbound}, i.e. which fulfil the condition $\varepsilon_{\mathrm{stationary}}>1$. Here $\varepsilon_{\mathrm{stationary}}$ is derived from the hydrodynamical equations described in section~\ref{sec:model:coupling} assuming a stationary metric and neglecting the pressure forces \citep{2002PhRvD..65j3005O},
\begin{equation}
 \varepsilon_{\mathrm{stationary}}\ =\ v^i \hat{u}_i+\frac{\epsilon}{u^0}+\frac{1}{u^0}, \label{ustationary}
\end{equation}
with $v_i$, the fluid velocity, $\hat{u}_i$, the conserved specific momentum, $\epsilon$, the specific internal energy and $u^0$, the time-component of the 4-velocity, all defined as in section~\ref{sec:model:coupling}. At infinity equation~\ref{ustationary} reduces to the Newtonian expression of the total energy of a fluid element. In table~\ref{simulationsejecta} we summarize the ejecta properties of the two NS merger models discussed in this section.

In figure~\ref{yehist} we plot the $Y_e$ distribution of the ejected material. In agreement with some of the previous results in the literature \citep{2015PhRvD..91f4059S, 2016PhRvD..93l4046S}, we find material distributed all the way to $Y_e=0.5$, with most material concentrated between $Y_e=0.1$ and $Y_e=0.4$. Therefore, high-mass r-process elements are expected to be synthesized in the dynamical ejecta of NS-NS mergers, as well as some contribution of first-peak r-process elements from the higher $Y_e$ tail of the distribution. We must note, however, that other groups with different neutrino schemes found the ejecta to be more neutron-rich, reaching maximum $Y_e$ values of only 0.3 (e.g. \citealt{2015PhRvD..92d4045P}, who, however, do not include neutrino re-absorption effects). 

Our numerical scheme operates on one supercomputer node. With 32 CPU cores, which is the maximum number of cores currently available per node, and OpenMP parallelization, 5~ms of evolution from the simulations presented in this section took 9 days to compute (wall clock time). During our numerical simulations, 90 per cent of the evolution step time is spent in \textsc{ILEAS}. The absorption module takes 75 per cent of this time while the rest is spent on the leakage part, mainly in the calculation of the diffusion time-scales. The amount of time spent by the equilibration module is negligible. The performance scales very well with the number of CPU cores. The computing time required for the 5~ms of evolution is almost doubled ($\sim16$~days) if the number of CPU cores is halved. Each individual NS was modelled by 175000 SPH particles and the grid employed is the same as that applied in the other tests presented in this work (see beginning of section~\ref{sec:tests}).

\section{Summary}\label{sumary}

The detection of a NS-NS merger using GW interferometers and its associated EM transients \citep{PhysRevLett.119.161101,2017ApJ...848L..12A} 
has opened the doors to a new era of multi-messenger astronomy. The vivid discussions about the nature of the observed kilonova (e.g. \citealt{2017ApJ...848L..17C, 2017ApJ...848L..18N, 2017ApJ...848L..19C, 2017Natur.551...80K, 2017Natur.551...75S, 2017Natur.551...67P, 2017ApJ...850L..37P, 2018MNRAS.481.3423W}) has brought to light the need of a reliable understanding not only of the composition of the ejected material, but also its dependence on the direction of ejection. At the same time, the error-bars associated with the measured NS masses, together with the underlying uncertainty of the EoS of NS matter, call for the exploration of a wide distribution of initial conditions for numerical simulations. In order to reconcile these two needs we have introduced \textsc{ILEAS}, an improved leakage scheme which accounts for the basic physical effects of neutrino transport at a moderate computational cost. \textsc{ILEAS} is ideal for exploring wide parameter spaces in three dimensions, where $\sim$10 per cent of accuracy is enough to capture the essential impact of weak interactions.

Leakage models have been used to emulate neutrino losses in the context of NS 
mergers since the 1990's \citep{1996A&A...311..532R}, but little work has been 
devoted to assess their accuracy (exceptions are 
\citealt{2016PhRvD..93d4019F,2016ApJS..223...22P}). 
In fact, we show that, in their standard formulations, leakage schemes have a 
tendency to overproduce neutrinos in the region close to the neutrinosphere, 
which could lead to numerical artefacts in near-surface regions and incorrect 
estimates of the ejecta composition. Moreover, the traditional leakage schemes 
were only a simple ansatz to estimate the local neutrino losses, ignoring other 
important physical effects inherent to neutrino transport, such as equilibration 
or re-absorption. In the recent years, truncated moment schemes have been 
developed and successfully used in the context of NS mergers, providing a 
more sophisticated alternative to leakage schemes. These approaches, however, 
also possess disadvantages and shortcomings of their own, such as problems 
with crossing flows (see, e.g., \citealt{2018PhRvD..98f3007F}) or the need of more computational resources, in particular when combined with a ray-tracing scheme for computing high-resolution neutrino distributions. While not a proper neutrino transport 
scheme, \textsc{ILEAS} is able to capture all the aforementioned physical 
effects, yet retaining the simple and inexpensive aspects of leakage schemes. 
Its improved diffusion time-scale, obtained directly from the flux-limited 
diffusion equation, provides a much better estimate of the neutrino losses 
in optically thick regions. This is reinforced by the inclusion of equilibration: 
the equilibration step ensures the recovery of the correct lepton fractions in the 
$\beta$-equilibrium regime, and the EoS also includes the energy and pressure 
contributions of the trapped neutrinos. Finally, by means of a simple 
multi-dimensional ray-tracing algorithm, we account for the re-absorption in 
optically thin conditions of neutrinos leaking out from the system. In order 
to keep the absorption module computationally efficient, we decided to resort 
to a grey approximation for \textsc{ILEAS}. However, our spectral calculation 
of the diffusive flux allows us to approximately capture the energy-dependent 
decoupling of neutrinos from matter in the integrated diffusion time-scale. 
Our results show that, despite the inherent approximations, \textsc{ILEAS} 
is sufficiently good to reproduce the results of more sophisticated transport 
schemes on the level of 10 per cent, locally and globally.

Motivated by its future application in the context of NS mergers, we tested the 
performance of \textsc{ILEAS} by comparison to available simulations 
representing some of the typical conditions encountered during NS mergers. 
We presented the results obtained with \textsc{ILEAS} applied on 3D mappings 
of several PNS cooling snapshots from the 1D \textsc{VERTEX-PROMETHEUS} simulation 
(Sr) with energy-dependent neutrino transport performed by \cite{2010PhRvL.104y1101H}. 
For all tested snapshots, ranging from 0.2~s until 1.5~s post-bounce, \textsc{ILEAS} 
was able, after a short relaxation of the medium (i.e., after evolving temperature
and $Y_e$ for 5\,ms), to reproduce not only the total 
\textsc{VERTEX} luminosities, but also the complete radial luminosity profiles 
within $\sim$10 per cent accuracy. In order to provide a more detailed comparison, 
we also tested \textsc{ILEAS} on a snapshot obtained from the evolution of the 
same PNS performed by \textsc{ALCAR} \citep{2015MNRAS.453.3386J}, which includes 
an energy-dependent M1 transport solver and exactly the same neutrino reactions 
as \textsc{ILEAS} for $\nu_e$ and $\bar{\nu}_e$. As with the \textsc{VERTEX} 
cases, \textsc{ILEAS} reached an agreement within 10 per cent accuracy with the 
\textsc{ALCAR} results. Furthermore, we evolved the temperature and $Y_e$ of the 
\textsc{ALCAR} snapshot (keeping the density fixed) for 50~ms with both 
\textsc{ALCAR} and \textsc{ILEAS}, and the good agreement was maintained 
throughout the simulation. 

As possible remnants of CO mergers, BH-torus systems provide a useful scenario 
for testing the performance of our scheme in the low optical depth limit. 
Snapshot calculations allowed us to attest the capability of our absorption 
treatment to capture the qualitative features of neutrino absorption in 
comparison with the results attained by \textsc{ALCAR}. Furthermore, we evolved 
two BH-torus models for 50~ms using \textsc{ALCAR} and \textsc{ILEAS}, in the 
same fashion as the PNS snapshot. In spite of the initial over-cooling caused 
by the unavoidable transient produced when switching on \textsc{ILEAS}, the 
neutrino luminosities of both models preserved an agreement of $\sim$10 per cent. 

In conclusion, \textsc{ILEAS} has been shown to reproduce within $\sim$10 per cent 
accuracy basic results of more sophisticated transport schemes also in 
multi-dimensional scenarios. Albeit not as accurate as full-fledged 3D transport, 
\textsc{ILEAS} includes all the relevant physical effects of neutrino transport 
and surpasses the quality of previous, conventional leakage schemes while retaining 
most of their efficiency and simplicity. These features make \textsc{ILEAS} an 
appropriate description of neutrinos for numerical simulations of NS mergers, 
where the relatively short evolution time-scales may not require a full-scale 
3D neutrino transport to still obtain a consistent picture of the composition 
evolution of merger medium and ejecta. The exploration of the vast parameter 
space of possible binary configurations demands computationally efficient but 
sufficiently accurate codes. \textsc{ILEAS} is intended to serve this requirement.

In section~\ref{sec:tests:NSNS} we presented first NS-NS merger simulations for 
two different EoSs employing \textsc{ILEAS} for the neutrino treatment. The models 
demonstrate the feasibility, numerical stability and basic agreement with results 
available in the literature, although detailed comparisons of the neutrino quantities 
are not possible because of the combination of different neutrino treatments and 
different hydrodynamics schemes.

Nevertheless, we caution the reader that \textsc{ILEAS} cannot be perceived as 
a perfect replacement for a full neutrino transport scheme, especially not in 
situations where the transfer of energy and leptons from one location to another 
by neutrino diffusion is crucial to describe the interior evolution of an object,
as, for example, in the case of the long-time neutrino-cooling of PNSs. 
\textsc{ILEAS} is a suitable alternative to neutrino transport codes 
particularly in conditions where the dynamical changes of the system happen on 
a time-scale shorter than or comparable to the neutrino diffusion time-scale. 
Examples are the rapid evolution of two merging NSs or the cooling of 
(semi-)transparent tori around merger remnants. Finally, for its use in 
consistent GR simulations, \textsc{ILEAS} needs to be extended to include 
effects such as Doppler shift, time retardation and general relativistic 
gravitational ray bending, which were not essential in the test calculations 
presented in this work. In future versions, \textsc{ILEAS} will also be supplemented by a module 
describing the effects of neutrino-antineutrino annihilation.

A computationally considerably more expensive alternative to ILEAS is 
advertised by \cite{2018PhRvD..98f3007F}, who plan to upgrade their grey M1 scheme
by an Eddington tensor closure obtained from a Monte Carlo (MC) solution of
the Boltzmann equation exterior to regions of high optical depths, instead
of using the analytical closure relations applied so far. Whether such
a hybrid code yields stable results with acceptable numerical ``noise''
for affordable low-resolution MC calculations will have to be demonstrated.
For reasons of accuracy, \cite{2018PhRvD..98f3007F} recommend to evolve the
neutrino number densities as well as the neutrino energy densities in order
to obtain reasonably accurate local estimates of the average neutrino energies.
In order to evaluate errors of M1 results with analytical closure
relations, they performed a time-dependent calculation over 4.5\,ms with 
their relativistic MC solver for a HMNS as a representative remnant of a NS-NS merger, using
the time-dependent fluid quantities from the M1 radiation-hydrodynamics run
and not feeding back the MC results into fluid or M1 transport solutions.

They found relative differences of 10--30 per cent in the average neutrino
energies between the M1 and the MC transport results and concluded that
this implies that the absorption and scattering opacities can be off by 
$\sim$30 per cent up to close to a factor of 2, dependent on positions closer to
the polar axis (i.e.\ the rotation axis of the remnant) or farther away from
it. Moreover, because of artificial shocks associated with the use
of a non-linear, algebraic closure relation, they diagnosed that the M1 code
accumulates neutrinos close to the polar axis, leading to an excess of the
neutrino density in the polar regions by about 50 per cent for $\nu_e$ and $\bar\nu_e$
and by nearly a factor of 2 for $\nu_x$. The $\nu\bar\nu$ pair-annihilation 
rate above the poles of the HMNS is underestimated by factors of 2--3 by the M1 description. 
Similar results had already been reported in the context 
of BH-torus systems by \cite{2015MNRAS.448..541J} (in the appendix there), who also found 
an overestimation of the neutrino densities in the polar regions when comparing their 
energy-dependent M1 scheme, ALCAR, with a ray-tracing Boltzmann solution. 
However, in the tested BH-torus scenario, the results by \cite{2015MNRAS.448..541J} imply an \textit{over}-estimation of the pair-annihilation rate around the polar axis. While the spectral and opacity differences may be handled better by fully
energy dependent (and considerably more complex and costly) transport codes
such as ALCAR, the overestimated number densities and underestimated 
pair-annihilation rates have to await their cure through a replacement of
the analytic closure by a Boltzmann transport solution, possibly based on
MC results.

In view of the considerable error margins associated with a grey M1 
approximation and considering the high computational demands of future 
hybrid schemes, our ILEAS method constitutes itself as an interesting 
option for the next generation of NS-NS/BH merger simulations surveying the 
huge multi-dimensional parameter space of possibilities. ILEAS is not
only computationally very efficient but also appears to be competitive 
concerning its accuracy compared to other forefront developments
of neutrino transport treatments for CO mergers and their remnants. The
work by \cite{2018PhRvD..98f3007F} underlines that M1 solutions have their own
shortcomings when applied to the highly aspherical environments of merger
remnants. For this reason our tests with BH-torus systems, comparing ILEAS
to M1 results from the ALCAR code, cannot be considered as finally
conclusive regarding the accuracy of ILEAS. Direct comparisons of ILEAS
and MC results would be desirable.

\section*{Acknowledgements}

The authors are grateful to N. Rahman for helpful discussions. The project was supported by the Deutsche Forschungsgemeinschaft through Sonderforschungsbereich SFB-1258 ``Neutrinos and Dark Matter in Astro- and Particle Physics (NDM)'' and the Excellence Cluster Universe (EXC 153; http://www.universe-cluster.de/) and by the European Research Council through grant ERC-AdG No. 341157-COCO2CASA. Calculations were performed on Hydra and Draco of the Max Planck Computing and Data Facility (MPCDF). OJ acknowledges support by the Special Postdoctoral Researchers (SPDR) program and iTHEMS cluster of RIKEN. AB acknowledges support by the Klaus Tschira Foundation, the European Research Council (ERC) under the European Union's Horizon 2020 research and innovation programme under grant agreement No. 759253 and by the Sonderforschungsbereich SFB 881 ``The Milky Way System'' (subproject A10) of the German Research Foundation (DFG).

\bibliographystyle{mnras}
\bibliography{Bibliography}

\begin{thebibliography}{}
\makeatletter
\relax
\def\mn@urlcharsother{\let\do\@makeother \do\$\do\&\do\#\do\^\do\_\do\%\do\~}
\def\mn@doi{\begingroup\mn@urlcharsother \@ifnextchar [ {\mn@doi@}
  {\mn@doi@[]}}
\def\mn@doi@[#1]#2{\def\@tempa{#1}\ifx\@tempa\@empty \href
  {http://dx.doi.org/#2} {doi:#2}\else \href {http://dx.doi.org/#2} {#1}\fi
  \endgroup}
\def\mn@eprint#1#2{\mn@eprint@#1:#2::\@nil}
\def\mn@eprint@arXiv#1{\href {http://arxiv.org/abs/#1} {{\tt arXiv:#1}}}
\def\mn@eprint@dblp#1{\href {http://dblp.uni-trier.de/rec/bibtex/#1.xml}
  {dblp:#1}}
\def\mn@eprint@#1:#2:#3:#4\@nil{\def\@tempa {#1}\def\@tempb {#2}\def\@tempc
  {#3}\ifx \@tempc \@empty \let \@tempc \@tempb \let \@tempb \@tempa \fi \ifx
  \@tempb \@empty \def\@tempb {arXiv}\fi \@ifundefined
  {mn@eprint@\@tempb}{\@tempb:\@tempc}{\expandafter \expandafter \csname
  mn@eprint@\@tempb\endcsname \expandafter{\@tempc}}}

\bibitem[\protect\citeauthoryear{Abbott et~al.,}{Abbott
  et~al.}{2017a}]{PhysRevLett.119.161101}
Abbott B.~P.,  et~al., 2017a, \mn@doi [Phys. Rev. Lett.]
  {10.1103/PhysRevLett.119.161101}, 119, 161101

\bibitem[\protect\citeauthoryear{{Abbott} et~al.,}{{Abbott}
  et~al.}{2017b}]{2017ApJ...848L..12A}
{Abbott} B.~P.,  et~al., 2017b, \mn@doi [\apjl] {10.3847/2041-8213/aa91c9},
  \href {http://adsabs.harvard.edu/abs/2017ApJ...848L..12A} {848, L12}

\bibitem[\protect\citeauthoryear{{Barnes}, {Kasen}, {Wu}  \&
  {Mart{\'{\i}}nez-Pinedo}}{{Barnes} et~al.}{2016}]{2016ApJ...829..110B}
{Barnes} J.,  {Kasen} D.,  {Wu} M.-R.,   {Mart{\'{\i}}nez-Pinedo} G.,  2016,
  \mn@doi [\apj] {10.3847/0004-637X/829/2/110}, \href
  {http://adsabs.harvard.edu/abs/2016ApJ...829..110B} {829, 110}

\bibitem[\protect\citeauthoryear{{Bauswein}, {Goriely}  \& {Janka}}{{Bauswein}
  et~al.}{2013}]{2013ApJ...773...78B}
{Bauswein} A.,  {Goriely} S.,   {Janka} H.-T.,  2013, \mn@doi [\apj]
  {10.1088/0004-637X/773/1/78}, \href
  {http://adsabs.harvard.edu/abs/2013ApJ...773...78B} {773, 78}

\bibitem[\protect\citeauthoryear{{Bernuzzi}, {Radice}, {Ott}, {Roberts},
  {M{\"o}sta}  \& {Galeazzi}}{{Bernuzzi} et~al.}{2016}]{2016PhRvD..94b4023B}
{Bernuzzi} S.,  {Radice} D.,  {Ott} C.~D.,  {Roberts} L.~F.,  {M{\"o}sta} P.,
  {Galeazzi} F.,  2016, \mn@doi [\prd] {10.1103/PhysRevD.94.024023}, \href
  {http://adsabs.harvard.edu/abs/2016PhRvD..94b4023B} {94, 024023}

\bibitem[\protect\citeauthoryear{{Bethe}}{{Bethe}}{1990}]{1990RvMP...62..801B}
{Bethe} H.~A.,  1990, \mn@doi [Reviews of Modern Physics]
  {10.1103/RevModPhys.62.801}, \href
  {http://adsabs.harvard.edu/abs/1990RvMP...62..801B} {62, 801}

\bibitem[\protect\citeauthoryear{{Bethe} \& {Wilson}}{{Bethe} \&
  {Wilson}}{1985}]{1985ApJ...295...14B}
{Bethe} H.~A.,  {Wilson} J.~R.,  1985, \mn@doi [\apj] {10.1086/163343}, \href
  {http://adsabs.harvard.edu/abs/1985ApJ...295...14B} {295, 14}

\bibitem[\protect\citeauthoryear{{Bludman} \& {van Riper}}{{Bludman} \& {van
  Riper}}{1978}]{1978ApJ...224..631B}
{Bludman} S.~A.,  {van Riper} K.~A.,  1978, \mn@doi [\apj] {10.1086/156412},
  \href {http://adsabs.harvard.edu/abs/1978ApJ...224..631B} {224, 631}

\bibitem[\protect\citeauthoryear{{Bovard}, {Martin}, {Guercilena}, {Arcones},
  {Rezzolla}  \& {Korobkin}}{{Bovard} et~al.}{2017}]{2017PhRvD..96l4005B}
{Bovard} L.,  {Martin} D.,  {Guercilena} F.,  {Arcones} A.,  {Rezzolla} L.,
  {Korobkin} O.,  2017, \mn@doi [\prd] {10.1103/PhysRevD.96.124005}, \href
  {http://adsabs.harvard.edu/abs/2017PhRvD..96l4005B} {96, 124005}

\bibitem[\protect\citeauthoryear{{Bruenn}}{{Bruenn}}{1985}]{1985ApJS...58..771B}
{Bruenn} S.~W.,  1985, \mn@doi [\apjs] {10.1086/191056}, \href
  {http://adsabs.harvard.edu/abs/1985ApJS...58..771B} {58, 771}

\bibitem[\protect\citeauthoryear{{Burrows}}{{Burrows}}{2013}]{2013RvMP...85..245B}
{Burrows} A.,  2013, \mn@doi [Reviews of Modern Physics]
  {10.1103/RevModPhys.85.245}, \href
  {http://adsabs.harvard.edu/abs/2013RvMP...85..245B} {85, 245}

\bibitem[\protect\citeauthoryear{{Burrows}, {Reddy}  \& {Thompson}}{{Burrows}
  et~al.}{2006}]{2006NuPhA.777..356B}
{Burrows} A.,  {Reddy} S.,   {Thompson} T.~A.,  2006, \mn@doi [Nuclear Physics
  A] {10.1016/j.nuclphysa.2004.06.012}, \href
  {http://adsabs.harvard.edu/abs/2006NuPhA.777..356B} {777, 356}

\bibitem[\protect\citeauthoryear{{Chornock} et~al.,}{{Chornock}
  et~al.}{2017}]{2017ApJ...848L..19C}
{Chornock} R.,  et~al., 2017, \mn@doi [\apjl] {10.3847/2041-8213/aa905c}, \href
  {http://adsabs.harvard.edu/abs/2017ApJ...848L..19C} {848, L19}

\bibitem[\protect\citeauthoryear{{Cooperstein}, {van den Horn}  \&
  {Baron}}{{Cooperstein} et~al.}{1986}]{1986ApJ...309..653C}
{Cooperstein} J.,  {van den Horn} L.~J.,   {Baron} E.~A.,  1986, \mn@doi [\apj]
  {10.1086/164633}, \href {http://adsabs.harvard.edu/abs/1986ApJ...309..653C}
  {309, 653}

\bibitem[\protect\citeauthoryear{{Cooperstein}, {van den Horn}  \&
  {Baron}}{{Cooperstein} et~al.}{1987}]{1987ApJ...321L.129C}
{Cooperstein} J.,  {van den Horn} L.~J.,   {Baron} E.,  1987, \mn@doi [\apjl]
  {10.1086/185019}, \href {http://adsabs.harvard.edu/abs/1987ApJ...321L.129C}
  {321, L129}

\bibitem[\protect\citeauthoryear{{Cowperthwaite} et~al.,}{{Cowperthwaite}
  et~al.}{2017}]{2017ApJ...848L..17C}
{Cowperthwaite} P.~S.,  et~al., 2017, \mn@doi [\apjl]
  {10.3847/2041-8213/aa8fc7}, \href
  {http://adsabs.harvard.edu/abs/2017ApJ...848L..17C} {848, L17}

\bibitem[\protect\citeauthoryear{{Deaton} et~al.,}{{Deaton}
  et~al.}{2013}]{2013ApJ...776...47D}
{Deaton} M.~B.,  et~al., 2013, \mn@doi [\apj] {10.1088/0004-637X/776/1/47},
  \href {http://adsabs.harvard.edu/abs/2013ApJ...776...47D} {776, 47}

\bibitem[\protect\citeauthoryear{{Deaton}, {O'Connor}, {Zhu}, {Bohn}, {Jesse},
  {Foucart}, {Duez}  \& {McLaughlin}}{{Deaton}
  et~al.}{2018}]{2018PhRvD..98j3014D}
{Deaton} M.~B.,  {O'Connor} E.,  {Zhu} Y.~L.,  {Bohn} A.,  {Jesse} J.,
  {Foucart} F.,  {Duez} M.~D.,   {McLaughlin} G.~C.,  2018, \mn@doi [\prd]
  {10.1103/PhysRevD.98.103014}, \href
  {http://adsabs.harvard.edu/abs/2018PhRvD..98j3014D} {98, 103014}

\bibitem[\protect\citeauthoryear{{Faber} \& {Rasio}}{{Faber} \&
  {Rasio}}{2012}]{2012LRR....15....8F}
{Faber} J.~A.,  {Rasio} F.~A.,  2012, \mn@doi [Living Reviews in Relativity]
  {10.12942/lrr-2012-8}, \href
  {http://adsabs.harvard.edu/abs/2012LRR....15....8F} {15, 8}

\bibitem[\protect\citeauthoryear{{Fern{\'a}ndez} \& {Metzger}}{{Fern{\'a}ndez}
  \& {Metzger}}{2016}]{2016ARNPS..66...23F}
{Fern{\'a}ndez} R.,  {Metzger} B.~D.,  2016, \mn@doi [Annual Review of Nuclear
  and Particle Science] {10.1146/annurev-nucl-102115-044819}, \href
  {http://adsabs.harvard.edu/abs/2016ARNPS..66...23F} {66, 23}

\bibitem[\protect\citeauthoryear{{Fern{\'a}ndez}, {Kasen}, {Metzger}  \&
  {Quataert}}{{Fern{\'a}ndez} et~al.}{2015a}]{2015MNRAS.446..750F}
{Fern{\'a}ndez} R.,  {Kasen} D.,  {Metzger} B.~D.,   {Quataert} E.,  2015a,
  \mn@doi [\mnras] {10.1093/mnras/stu2112}, \href
  {http://adsabs.harvard.edu/abs/2015MNRAS.446..750F} {446, 750}

\bibitem[\protect\citeauthoryear{{Fern{\'a}ndez}, {Quataert}, {Schwab}, {Kasen}
   \& {Rosswog}}{{Fern{\'a}ndez} et~al.}{2015b}]{2015MNRAS.449..390F}
{Fern{\'a}ndez} R.,  {Quataert} E.,  {Schwab} J.,  {Kasen} D.,   {Rosswog} S.,
  2015b, \mn@doi [\mnras] {10.1093/mnras/stv238}, \href
  {http://adsabs.harvard.edu/abs/2015MNRAS.449..390F} {449, 390}

\bibitem[\protect\citeauthoryear{{Foglizzo} et~al.,}{{Foglizzo}
  et~al.}{2015}]{2015PASA...32....9F}
{Foglizzo} T.,  et~al., 2015, \mn@doi [\pasa] {10.1017/pasa.2015.9}, \href
  {http://adsabs.harvard.edu/abs/2015PASA...32....9F} {32, e009}

\bibitem[\protect\citeauthoryear{{Foucart}}{{Foucart}}{2018}]{2018MNRAS.475.4186F}
{Foucart} F.,  2018, \mn@doi [\mnras] {10.1093/mnras/sty108}, \href
  {http://adsabs.harvard.edu/abs/2018MNRAS.475.4186F} {475, 4186}

\bibitem[\protect\citeauthoryear{{Foucart} et~al.,}{{Foucart}
  et~al.}{2014}]{2014PhRvD..90b4026F}
{Foucart} F.,  et~al., 2014, \mn@doi [\prd] {10.1103/PhysRevD.90.024026}, \href
  {http://adsabs.harvard.edu/abs/2014PhRvD..90b4026F} {90, 024026}

\bibitem[\protect\citeauthoryear{{Foucart} et~al.,}{{Foucart}
  et~al.}{2015}]{2015PhRvD..91l4021F}
{Foucart} F.,  et~al., 2015, \mn@doi [\prd] {10.1103/PhysRevD.91.124021}, \href
  {http://adsabs.harvard.edu/abs/2015PhRvD..91l4021F} {91, 124021}

\bibitem[\protect\citeauthoryear{{Foucart} et~al.,}{{Foucart}
  et~al.}{2016a}]{2016PhRvD..93d4019F}
{Foucart} F.,  et~al., 2016a, \mn@doi [\prd] {10.1103/PhysRevD.93.044019},
  \href {http://adsabs.harvard.edu/abs/2016PhRvD..93d4019F} {93, 044019}

\bibitem[\protect\citeauthoryear{{Foucart}, {O'Connor}, {Roberts}, {Kidder},
  {Pfeiffer}  \& {Scheel}}{{Foucart} et~al.}{2016b}]{2016PhRvD..94l3016F}
{Foucart} F.,  {O'Connor} E.,  {Roberts} L.,  {Kidder} L.~E.,  {Pfeiffer}
  H.~P.,   {Scheel} M.~A.,  2016b, \mn@doi [\prd] {10.1103/PhysRevD.94.123016},
  \href {http://adsabs.harvard.edu/abs/2016PhRvD..94l3016F} {94, 123016}

\bibitem[\protect\citeauthoryear{{Foucart} et~al.,}{{Foucart}
  et~al.}{2017}]{2017CQGra..34d4002F}
{Foucart} F.,  et~al., 2017, \mn@doi [Classical and Quantum Gravity]
  {10.1088/1361-6382/aa573b}, \href
  {http://adsabs.harvard.edu/abs/2017CQGra..34d4002F} {34, 044002}

\bibitem[\protect\citeauthoryear{{Foucart}, {Duez}, {Kidder}, {Nguyen},
  {Pfeiffer}  \& {Scheel}}{{Foucart} et~al.}{2018}]{2018PhRvD..98f3007F}
{Foucart} F.,  {Duez} M.~D.,  {Kidder} L.~E.,  {Nguyen} R.,  {Pfeiffer} H.~P.,
   {Scheel} M.~A.,  2018, \mn@doi [\prd] {10.1103/PhysRevD.98.063007}, \href
  {http://adsabs.harvard.edu/abs/2018PhRvD..98f3007F} {98, 063007}

\bibitem[\protect\citeauthoryear{{Fujibayashi}, {Sekiguchi}, {Kiuchi}  \&
  {Shibata}}{{Fujibayashi} et~al.}{2017}]{2017ApJ...846..114F}
{Fujibayashi} S.,  {Sekiguchi} Y.,  {Kiuchi} K.,   {Shibata} M.,  2017, \mn@doi
  [\apj] {10.3847/1538-4357/aa8039}, \href
  {http://adsabs.harvard.edu/abs/2017ApJ...846..114F} {846, 114}

\bibitem[\protect\citeauthoryear{{Galeazzi}, {Kastaun}, {Rezzolla}  \&
  {Font}}{{Galeazzi} et~al.}{2013}]{2013PhRvD..88f4009G}
{Galeazzi} F.,  {Kastaun} W.,  {Rezzolla} L.,   {Font} J.~A.,  2013, \mn@doi
  [\prd] {10.1103/PhysRevD.88.064009}, \href
  {http://adsabs.harvard.edu/abs/2013PhRvD..88f4009G} {88, 064009}

\bibitem[\protect\citeauthoryear{{Goriely}, {Bauswein}  \& {Janka}}{{Goriely}
  et~al.}{2011}]{2011ApJ...738L..32G}
{Goriely} S.,  {Bauswein} A.,   {Janka} H.-T.,  2011, \mn@doi [\apjl]
  {10.1088/2041-8205/738/2/L32}, \href
  {http://adsabs.harvard.edu/abs/2011ApJ...738L..32G} {738, L32}

\bibitem[\protect\citeauthoryear{{Goriely}, {Bauswein}, {Just}, {Pllumbi}  \&
  {Janka}}{{Goriely} et~al.}{2015}]{2015MNRAS.452.3894G}
{Goriely} S.,  {Bauswein} A.,  {Just} O.,  {Pllumbi} E.,   {Janka} H.-T.,
  2015, \mn@doi [\mnras] {10.1093/mnras/stv1526}, \href
  {http://adsabs.harvard.edu/abs/2015MNRAS.452.3894G} {452, 3894}

\bibitem[\protect\citeauthoryear{{Hannestad} \& {Raffelt}}{{Hannestad} \&
  {Raffelt}}{1998}]{1998ApJ...507..339H}
{Hannestad} S.,  {Raffelt} G.,  1998, \mn@doi [\apj] {10.1086/306303}, \href
  {http://adsabs.harvard.edu/abs/1998ApJ...507..339H} {507, 339}

\bibitem[\protect\citeauthoryear{{Hecht}}{{Hecht}}{1989}]{Hechtthesis}
{Hecht} T.,  1989, Master's thesis, Technische Universit{\"a}t M{\"u}nchen

\bibitem[\protect\citeauthoryear{{Hempel} \& {Schaffner-Bielich}}{{Hempel} \&
  {Schaffner-Bielich}}{2010}]{2010NuPhA.837..210H}
{Hempel} M.,  {Schaffner-Bielich} J.,  2010, \mn@doi [\nucpha]
  {10.1016/j.nuclphysa.2010.02.010}, \href
  {http://adsabs.harvard.edu/abs/2010NuPhA.837..210H} {837, 210}

\bibitem[\protect\citeauthoryear{{Hotokezaka}, {Kiuchi}, {Kyutoku},
  {Muranushi}, {Sekiguchi}, {Shibata}  \& {Taniguchi}}{{Hotokezaka}
  et~al.}{2013}]{2013PhRvD..88d4026H}
{Hotokezaka} K.,  {Kiuchi} K.,  {Kyutoku} K.,  {Muranushi} T.,  {Sekiguchi}
  Y.-i.,  {Shibata} M.,   {Taniguchi} K.,  2013, \mn@doi [\prd]
  {10.1103/PhysRevD.88.044026}, \href
  {http://adsabs.harvard.edu/abs/2013PhRvD..88d4026H} {88, 044026}

\bibitem[\protect\citeauthoryear{{Hotokezaka}, {Nissanke}, {Hallinan}, {Lazio},
  {Nakar}  \& {Piran}}{{Hotokezaka} et~al.}{2016}]{2016ApJ...831..190H}
{Hotokezaka} K.,  {Nissanke} S.,  {Hallinan} G.,  {Lazio} T.~J.~W.,  {Nakar}
  E.,   {Piran} T.,  2016, \mn@doi [\apj] {10.3847/0004-637X/831/2/190}, \href
  {http://adsabs.harvard.edu/abs/2016ApJ...831..190H} {831, 190}

\bibitem[\protect\citeauthoryear{{H{\"u}depohl}, {M{\"u}ller}, {Janka}, {Marek}
   \& {Raffelt}}{{H{\"u}depohl} et~al.}{2010}]{2010PhRvL.104y1101H}
{H{\"u}depohl} L.,  {M{\"u}ller} B.,  {Janka} H.-T.,  {Marek} A.,   {Raffelt}
  G.~G.,  2010, \mn@doi [Physical Review Letters]
  {10.1103/PhysRevLett.104.251101}, \href
  {http://adsabs.harvard.edu/abs/2010PhRvL.104y1101H} {104, 251101}

\bibitem[\protect\citeauthoryear{{Isenberg} \& {Nester}}{{Isenberg} \&
  {Nester}}{1980}]{1980grg1.conf...23I}
{Isenberg} J.,  {Nester} J.,  1980, in {Held} A.,  ed., ~ Vol. 1, General
  Relativity and Gravitation. Vol. 1. One hundred years after the birth of
  Albert Einstein. Edited by A. Held. New York, NY: Plenum Press, p. 23, 1980.
  p.~23

\bibitem[\protect\citeauthoryear{{Janka}}{{Janka}}{1991}]{1991A&A...244..378J}
{Janka} H.-T.,  1991, \aap, \href
  {http://adsabs.harvard.edu/abs/1991A%26A...244..378J} {244, 378}

\bibitem[\protect\citeauthoryear{{Janka}}{{Janka}}{2001}]{2001A&A...368..527J}
{Janka} H.-T.,  2001, \mn@doi [\aap] {10.1051/0004-6361:20010012}, \href
  {http://adsabs.harvard.edu/abs/2001A%26A...368..527J} {368, 527}

\bibitem[\protect\citeauthoryear{{Janka}}{{Janka}}{2012}]{2012ARNPS..62..407J}
{Janka} H.-T.,  2012, \mn@doi [Annual Review of Nuclear and Particle Science]
  {10.1146/annurev-nucl-102711-094901}, \href
  {http://adsabs.harvard.edu/abs/2012ARNPS..62..407J} {62, 407}

\bibitem[\protect\citeauthoryear{{Just}, {Bauswein}, {Pulpillo}, {Goriely}  \&
  {Janka}}{{Just} et~al.}{2015a}]{2015MNRAS.448..541J}
{Just} O.,  {Bauswein} A.,  {Pulpillo} R.~A.,  {Goriely} S.,   {Janka} H.-T.,
  2015a, \mn@doi [\mnras] {10.1093/mnras/stv009}, \href
  {http://adsabs.harvard.edu/abs/2015MNRAS.448..541J} {448, 541}

\bibitem[\protect\citeauthoryear{{Just}, {Obergaulinger}  \& {Janka}}{{Just}
  et~al.}{2015b}]{2015MNRAS.453.3386J}
{Just} O.,  {Obergaulinger} M.,   {Janka} H.-T.,  2015b, \mn@doi [\mnras]
  {10.1093/mnras/stv1892}, \href
  {http://adsabs.harvard.edu/abs/2015MNRAS.453.3386J} {453, 3386}

\bibitem[\protect\citeauthoryear{{Just}, {Obergaulinger}, {Janka}, {Bauswein}
  \& {Schwarz}}{{Just} et~al.}{2016}]{2016ApJ...816L..30J}
{Just} O.,  {Obergaulinger} M.,  {Janka} H.-T.,  {Bauswein} A.,   {Schwarz} N.,
   2016, \mn@doi [\apjl] {10.3847/2041-8205/816/2/L30}, \href
  {http://adsabs.harvard.edu/abs/2016ApJ...816L..30J} {816, L30}

\bibitem[\protect\citeauthoryear{{Kasen}, {Metzger}, {Barnes}, {Quataert}  \&
  {Ramirez-Ruiz}}{{Kasen} et~al.}{2017}]{2017Natur.551...80K}
{Kasen} D.,  {Metzger} B.,  {Barnes} J.,  {Quataert} E.,   {Ramirez-Ruiz} E.,
  2017, \mn@doi [\nat] {10.1038/nature24453}, \href
  {http://adsabs.harvard.edu/abs/2017Natur.551...80K} {551, 80}

\bibitem[\protect\citeauthoryear{{Kay} \& {Kajiya}}{{Kay} \&
  {Kajiya}}{1986}]{SlabMethod}
{Kay} T.~L.,  {Kajiya} J.~T.,  1986, Siggraph'86, Association for Computing
  Machinery, \href
  {https://www.google.de/url?sa=t&rct=j&q=&esrc=s&source=web&cd=1&ved=0ahUKEwi2idHnlcHXAhULEVAKHfs2Cx8QFggoMAA&url=http%3A%2F%2Fciteseerx.ist.psu.edu%2Fviewdoc%2Fdownload%3Fdoi%3D10.1.1.124.4731%26rep%3Drep1%26type%3Dpdf&usg=AOvVaw02RxEnNEKJdQBBFM9aT9fj}
  {20}

\bibitem[\protect\citeauthoryear{{Kiuchi}, {Sekiguchi}, {Kyutoku}  \&
  {Shibata}}{{Kiuchi} et~al.}{2012}]{2012CQGra..29l4003K}
{Kiuchi} K.,  {Sekiguchi} Y.,  {Kyutoku} K.,   {Shibata} M.,  2012, \mn@doi
  [Classical and Quantum Gravity] {10.1088/0264-9381/29/12/124003}, \href
  {http://adsabs.harvard.edu/abs/2012CQGra..29l4003K} {29, 124003}

\bibitem[\protect\citeauthoryear{{Korobkin}, {Rosswog}, {Arcones}  \&
  {Winteler}}{{Korobkin} et~al.}{2012}]{2012MNRAS.426.1940K}
{Korobkin} O.,  {Rosswog} S.,  {Arcones} A.,   {Winteler} C.,  2012, \mn@doi
  [\mnras] {10.1111/j.1365-2966.2012.21859.x}, \href
  {http://adsabs.harvard.edu/abs/2012MNRAS.426.1940K} {426, 1940}

\bibitem[\protect\citeauthoryear{{Kulkarni}}{{Kulkarni}}{2005}]{2005astro.ph.10256K}
{Kulkarni} S.~R.,  2005, ArXiv Astrophysics e-prints, \href
  {http://adsabs.harvard.edu/abs/2005astro.ph.10256K} {}

\bibitem[\protect\citeauthoryear{{Kyutoku}, {Kiuchi}, {Sekiguchi}, {Shibata}
  \& {Taniguchi}}{{Kyutoku} et~al.}{2018}]{2018PhRvD..97b3009K}
{Kyutoku} K.,  {Kiuchi} K.,  {Sekiguchi} Y.,  {Shibata} M.,   {Taniguchi} K.,
  2018, \mn@doi [\prd] {10.1103/PhysRevD.97.023009}, \href
  {http://adsabs.harvard.edu/abs/2018PhRvD..97b3009K} {97, 023009}

\bibitem[\protect\citeauthoryear{{Lehner}, {Liebling}, {Palenzuela},
  {Caballero}, {O'Connor}, {Anderson}  \& {Neilsen}}{{Lehner}
  et~al.}{2016a}]{2016CQGra..33r4002L}
{Lehner} L.,  {Liebling} S.~L.,  {Palenzuela} C.,  {Caballero} O.~L.,
  {O'Connor} E.,  {Anderson} M.,   {Neilsen} D.,  2016a, \mn@doi [Classical and
  Quantum Gravity] {10.1088/0264-9381/33/18/184002}, \href
  {http://adsabs.harvard.edu/abs/2016CQGra..33r4002L} {33, 184002}

\bibitem[\protect\citeauthoryear{{Lehner}, {Liebling}, {Palenzuela}  \&
  {Motl}}{{Lehner} et~al.}{2016b}]{2016PhRvD..94d3003L}
{Lehner} L.,  {Liebling} S.~L.,  {Palenzuela} C.,   {Motl} P.~M.,  2016b,
  \mn@doi [\prd] {10.1103/PhysRevD.94.043003}, \href
  {http://adsabs.harvard.edu/abs/2016PhRvD..94d3003L} {94, 043003}

\bibitem[\protect\citeauthoryear{{Levermore} \& {Pomraning}}{{Levermore} \&
  {Pomraning}}{1981}]{1981ApJ...248..321L}
{Levermore} C.~D.,  {Pomraning} G.~C.,  1981, \mn@doi [\apj] {10.1086/159157},
  \href {http://adsabs.harvard.edu/abs/1981ApJ...248..321L} {248, 321}

\bibitem[\protect\citeauthoryear{{Li} \& {Paczy{\'n}ski}}{{Li} \&
  {Paczy{\'n}ski}}{1998}]{1998ApJ...507L..59L}
{Li} L.-X.,  {Paczy{\'n}ski} B.,  1998, \mn@doi [\apjl] {10.1086/311680}, \href
  {http://adsabs.harvard.edu/abs/1998ApJ...507L..59L} {507, L59}

\bibitem[\protect\citeauthoryear{{Lindquist}}{{Lindquist}}{1966}]{1966AnPhy..37..487L}
{Lindquist} R.~W.,  1966, \mn@doi [Annals of Physics]
  {10.1016/0003-4916(66)90207-7}, \href
  {http://adsabs.harvard.edu/abs/1966AnPhy..37..487L} {37, 487}

\bibitem[\protect\citeauthoryear{{Lippuner} \& {Roberts}}{{Lippuner} \&
  {Roberts}}{2015}]{2015ApJ...815...82L}
{Lippuner} J.,  {Roberts} L.~F.,  2015, \mn@doi [\apj]
  {10.1088/0004-637X/815/2/82}, \href
  {http://adsabs.harvard.edu/abs/2015ApJ...815...82L} {815, 82}

\bibitem[\protect\citeauthoryear{{Lippuner}, {Fern{\'a}ndez}, {Roberts},
  {Foucart}, {Kasen}, {Metzger}  \& {Ott}}{{Lippuner}
  et~al.}{2017}]{2017MNRAS.472..904L}
{Lippuner} J.,  {Fern{\'a}ndez} R.,  {Roberts} L.~F.,  {Foucart} F.,  {Kasen}
  D.,  {Metzger} B.~D.,   {Ott} C.~D.,  2017, \mn@doi [\mnras]
  {10.1093/mnras/stx1987}, \href
  {http://adsabs.harvard.edu/abs/2017MNRAS.472..904L} {472, 904}

\bibitem[\protect\citeauthoryear{{Martin}, {Perego}, {Arcones}, {Thielemann},
  {Korobkin}  \& {Rosswog}}{{Martin} et~al.}{2015}]{2015ApJ...813....2M}
{Martin} D.,  {Perego} A.,  {Arcones} A.,  {Thielemann} F.-K.,  {Korobkin} O.,
   {Rosswog} S.,  2015, \mn@doi [\apj] {10.1088/0004-637X/813/1/2}, \href
  {http://adsabs.harvard.edu/abs/2015ApJ...813....2M} {813, 2}

\bibitem[\protect\citeauthoryear{{Martin}, {Perego}, {Kastaun}  \&
  {Arcones}}{{Martin} et~al.}{2018}]{2018CQGra..35c4001M}
{Martin} D.,  {Perego} A.,  {Kastaun} W.,   {Arcones} A.,  2018, \mn@doi
  [Classical and Quantum Gravity] {10.1088/1361-6382/aa9f5a}, \href
  {http://adsabs.harvard.edu/abs/2018CQGra..35c4001M} {35, 034001}

\bibitem[\protect\citeauthoryear{{Mart{\'{\i}}nez-Pinedo}, {Fischer}, {Lohs}
  \& {Huther}}{{Mart{\'{\i}}nez-Pinedo} et~al.}{2012}]{2012PhRvL.109y1104M}
{Mart{\'{\i}}nez-Pinedo} G.,  {Fischer} T.,  {Lohs} A.,   {Huther} L.,  2012,
  \mn@doi [Physical Review Letters] {10.1103/PhysRevLett.109.251104}, \href
  {http://adsabs.harvard.edu/abs/2012PhRvL.109y1104M} {109, 251104}

\bibitem[\protect\citeauthoryear{{McLaughlin}, {Fuller}  \&
  {Wilson}}{{McLaughlin} et~al.}{1996}]{1996ApJ...472..440M}
{McLaughlin} G.~C.,  {Fuller} G.~M.,   {Wilson} J.~R.,  1996, \mn@doi [\apj]
  {10.1086/178077}, \href {http://adsabs.harvard.edu/abs/1996ApJ...472..440M}
  {472, 440}

\bibitem[\protect\citeauthoryear{{Metzger}}{{Metzger}}{2017}]{2017LRR....20....3M}
{Metzger} B.~D.,  2017, \mn@doi [Living Reviews in Relativity]
  {10.1007/s41114-017-0006-z}, \href
  {http://adsabs.harvard.edu/abs/2017LRR....20....3M} {20, 3}

\bibitem[\protect\citeauthoryear{{Metzger} \& {Fern{\'a}ndez}}{{Metzger} \&
  {Fern{\'a}ndez}}{2014}]{2014MNRAS.441.3444M}
{Metzger} B.~D.,  {Fern{\'a}ndez} R.,  2014, \mn@doi [\mnras]
  {10.1093/mnras/stu802}, \href
  {http://adsabs.harvard.edu/abs/2014MNRAS.441.3444M} {441, 3444}

\bibitem[\protect\citeauthoryear{{Metzger} et~al.,}{{Metzger}
  et~al.}{2010}]{2010MNRAS.406.2650M}
{Metzger} B.~D.,  et~al., 2010, \mn@doi [\mnras]
  {10.1111/j.1365-2966.2010.16864.x}, \href
  {http://adsabs.harvard.edu/abs/2010MNRAS.406.2650M} {406, 2650}

\bibitem[\protect\citeauthoryear{{Mezzacappa} \& {Bruenn}}{{Mezzacappa} \&
  {Bruenn}}{1993}]{1993ApJ...405..637M}
{Mezzacappa} A.,  {Bruenn} S.~W.,  1993, \mn@doi [\apj] {10.1086/172394}, \href
  {http://adsabs.harvard.edu/abs/1993ApJ...405..637M} {405, 637}

\bibitem[\protect\citeauthoryear{{Mihalas} \& {Weibel-Mihalas}}{{Mihalas} \&
  {Weibel-Mihalas}}{1984}]{Mihalas}
{Mihalas} D.,  {Weibel-Mihalas} B.,  1984, {Foundations of Radiation
  Hydrodynamics}.
Oxford University Press

\bibitem[\protect\citeauthoryear{{Neilsen}, {Liebling}, {Anderson}, {Lehner},
  {O'Connor}  \& {Palenzuela}}{{Neilsen} et~al.}{2014}]{2014PhRvD..89j4029N}
{Neilsen} D.,  {Liebling} S.~L.,  {Anderson} M.,  {Lehner} L.,  {O'Connor} E.,
   {Palenzuela} C.,  2014, \mn@doi [\prd] {10.1103/PhysRevD.89.104029}, \href
  {http://adsabs.harvard.edu/abs/2014PhRvD..89j4029N} {89, 104029}

\bibitem[\protect\citeauthoryear{{Nicholl} et~al.,}{{Nicholl}
  et~al.}{2017}]{2017ApJ...848L..18N}
{Nicholl} M.,  et~al., 2017, \mn@doi [\apjl] {10.3847/2041-8213/aa9029}, \href
  {http://adsabs.harvard.edu/abs/2017ApJ...848L..18N} {848, L18}

\bibitem[\protect\citeauthoryear{{O'Connor} \& {Ott}}{{O'Connor} \&
  {Ott}}{2010}]{2010CQGra..27k4103O}
{O'Connor} E.,  {Ott} C.~D.,  2010, \mn@doi [Classical and Quantum Gravity]
  {10.1088/0264-9381/27/11/114103}, \href
  {http://adsabs.harvard.edu/abs/2010CQGra..27k4103O} {27, 114103}

\bibitem[\protect\citeauthoryear{{Oechslin}, {Rosswog}  \&
  {Thielemann}}{{Oechslin} et~al.}{2002}]{2002PhRvD..65j3005O}
{Oechslin} R.,  {Rosswog} S.,   {Thielemann} F.-K.,  2002, \mn@doi [\prd]
  {10.1103/PhysRevD.65.103005}, \href
  {http://adsabs.harvard.edu/abs/2002PhRvD..65j3005O} {65, 103005}

\bibitem[\protect\citeauthoryear{{Oechslin}, {Janka}  \& {Marek}}{{Oechslin}
  et~al.}{2007}]{2007A&A...467..395O}
{Oechslin} R.,  {Janka} H.-T.,   {Marek} A.,  2007, \mn@doi [\aap]
  {10.1051/0004-6361:20066682}, \href
  {http://adsabs.harvard.edu/abs/2007A%26A...467..395O} {467, 395}

\bibitem[\protect\citeauthoryear{{Palenzuela}, {Liebling}, {Neilsen}, {Lehner},
  {Caballero}, {O'Connor}  \& {Anderson}}{{Palenzuela}
  et~al.}{2015}]{2015PhRvD..92d4045P}
{Palenzuela} C.,  {Liebling} S.~L.,  {Neilsen} D.,  {Lehner} L.,  {Caballero}
  O.~L.,  {O'Connor} E.,   {Anderson} M.,  2015, \mn@doi [\prd]
  {10.1103/PhysRevD.92.044045}, \href
  {http://adsabs.harvard.edu/abs/2015PhRvD..92d4045P} {92, 044045}

\bibitem[\protect\citeauthoryear{{Perego}, {Rosswog}, {Cabez{\'o}n},
  {Korobkin}, {K{\"a}ppeli}, {Arcones}  \& {Liebend{\"o}rfer}}{{Perego}
  et~al.}{2014a}]{2014MNRAS.443.3134P}
{Perego} A.,  {Rosswog} S.,  {Cabez{\'o}n} R.~M.,  {Korobkin} O.,
  {K{\"a}ppeli} R.,  {Arcones} A.,   {Liebend{\"o}rfer} M.,  2014a, \mn@doi
  [\mnras] {10.1093/mnras/stu1352}, \href
  {http://adsabs.harvard.edu/abs/2014MNRAS.443.3134P} {443, 3134}

\bibitem[\protect\citeauthoryear{{Perego}, {Gafton}, {Cabez{\'o}n}, {Rosswog}
  \& {Liebend{\"o}rfer}}{{Perego} et~al.}{2014b}]{2014A&A...568A..11P}
{Perego} A.,  {Gafton} E.,  {Cabez{\'o}n} R.,  {Rosswog} S.,
  {Liebend{\"o}rfer} M.,  2014b, \mn@doi [\aap] {10.1051/0004-6361/201423755},
  \href {http://adsabs.harvard.edu/abs/2014A%26A...568A..11P} {568, A11}

\bibitem[\protect\citeauthoryear{{Perego}, {Cabez{\'o}n}  \&
  {K{\"a}ppeli}}{{Perego} et~al.}{2016}]{2016ApJS..223...22P}
{Perego} A.,  {Cabez{\'o}n} R.~M.,   {K{\"a}ppeli} R.,  2016, \mn@doi [\apjs]
  {10.3847/0067-0049/223/2/22}, \href
  {http://adsabs.harvard.edu/abs/2016ApJS..223...22P} {223, 22}

\bibitem[\protect\citeauthoryear{{Perego}, {Yasin}  \& {Arcones}}{{Perego}
  et~al.}{2017a}]{2017JPhG...44h4007P}
{Perego} A.,  {Yasin} H.,   {Arcones} A.,  2017a, \mn@doi [Journal of Physics G
  Nuclear Physics] {10.1088/1361-6471/aa7bdc}, \href
  {http://adsabs.harvard.edu/abs/2017JPhG...44h4007P} {44, 084007}

\bibitem[\protect\citeauthoryear{{Perego}, {Radice}  \& {Bernuzzi}}{{Perego}
  et~al.}{2017b}]{2017ApJ...850L..37P}
{Perego} A.,  {Radice} D.,   {Bernuzzi} S.,  2017b, \mn@doi [\apjl]
  {10.3847/2041-8213/aa9ab9}, \href
  {http://adsabs.harvard.edu/abs/2017ApJ...850L..37P} {850, L37}

\bibitem[\protect\citeauthoryear{{Pian} et~al.,}{{Pian}
  et~al.}{2017}]{2017Natur.551...67P}
{Pian} E.,  et~al., 2017, \mn@doi [\nat] {10.1038/nature24298}, \href
  {http://adsabs.harvard.edu/abs/2017Natur.551...67P} {551, 67}

\bibitem[\protect\citeauthoryear{{Price}}{{Price}}{2007}]{2007PASA...24..159P}
{Price} D.~J.,  2007, \mn@doi [\pasa] {10.1071/AS07022}, \href
  {http://adsabs.harvard.edu/abs/2007PASA...24..159P} {24, 159}

\bibitem[\protect\citeauthoryear{{Radice}}{{Radice}}{2017}]{2017ApJ...838L...2R}
{Radice} D.,  2017, \mn@doi [\apjl] {10.3847/2041-8213/aa6483}, \href
  {http://adsabs.harvard.edu/abs/2017ApJ...838L...2R} {838, L2}

\bibitem[\protect\citeauthoryear{{Radice}, {Galeazzi}, {Lippuner}, {Roberts},
  {Ott}  \& {Rezzolla}}{{Radice} et~al.}{2016}]{2016MNRAS.460.3255R}
{Radice} D.,  {Galeazzi} F.,  {Lippuner} J.,  {Roberts} L.~F.,  {Ott} C.~D.,
  {Rezzolla} L.,  2016, \mn@doi [\mnras] {10.1093/mnras/stw1227}, \href
  {http://adsabs.harvard.edu/abs/2016MNRAS.460.3255R} {460, 3255}

\bibitem[\protect\citeauthoryear{{Radice}, {Bernuzzi}, {Del Pozzo}, {Roberts}
  \& {Ott}}{{Radice} et~al.}{2017}]{2017ApJ...842L..10R}
{Radice} D.,  {Bernuzzi} S.,  {Del Pozzo} W.,  {Roberts} L.~F.,   {Ott} C.~D.,
  2017, \mn@doi [\apjl] {10.3847/2041-8213/aa775f}, \href
  {http://adsabs.harvard.edu/abs/2017ApJ...842L..10R} {842, L10}

\bibitem[\protect\citeauthoryear{{Radice}, {Perego}, {Zappa}  \&
  {Bernuzzi}}{{Radice} et~al.}{2018}]{2018ApJ...852L..29R}
{Radice} D.,  {Perego} A.,  {Zappa} F.,   {Bernuzzi} S.,  2018, \mn@doi [\apjl]
  {10.3847/2041-8213/aaa402}, \href
  {http://adsabs.harvard.edu/abs/2018ApJ...852L..29R} {852, L29}

\bibitem[\protect\citeauthoryear{{Rampp}}{{Rampp}}{2000}]{Ramppthesis}
{Rampp} M.,  2000, PhD thesis, Technische Universit{\"a}t M{\"u}nchen

\bibitem[\protect\citeauthoryear{{Rampp} \& {Janka}}{{Rampp} \&
  {Janka}}{2002}]{2002A&A...396..361R}
{Rampp} M.,  {Janka} H.-T.,  2002, \mn@doi [\aap] {10.1051/0004-6361:20021398},
  \href {http://adsabs.harvard.edu/abs/2002A%26A...396..361R} {396, 361}

\bibitem[\protect\citeauthoryear{{Reddy}, {Prakash}  \& {Lattimer}}{{Reddy}
  et~al.}{1998}]{1998PhRvD..58a3009R}
{Reddy} S.,  {Prakash} M.,   {Lattimer} J.~M.,  1998, \mn@doi [\prd]
  {10.1103/PhysRevD.58.013009}, \href
  {http://adsabs.harvard.edu/abs/1998PhRvD..58a3009R} {58, 013009}

\bibitem[\protect\citeauthoryear{{Richers}, {Kasen}, {O'Connor},
  {Fern{\'a}ndez}  \& {Ott}}{{Richers} et~al.}{2015}]{2015ApJ...813...38R}
{Richers} S.,  {Kasen} D.,  {O'Connor} E.,  {Fern{\'a}ndez} R.,   {Ott} C.~D.,
  2015, \mn@doi [\apj] {10.1088/0004-637X/813/1/38}, \href
  {http://adsabs.harvard.edu/abs/2015ApJ...813...38R} {813, 38}

\bibitem[\protect\citeauthoryear{{Roberts}}{{Roberts}}{2012}]{2012ApJ...755..126R}
{Roberts} L.~F.,  2012, \mn@doi [\apj] {10.1088/0004-637X/755/2/126}, \href
  {http://adsabs.harvard.edu/abs/2012ApJ...755..126R} {755, 126}

\bibitem[\protect\citeauthoryear{{Roberts}, {Kasen}, {Lee}  \&
  {Ramirez-Ruiz}}{{Roberts} et~al.}{2011}]{2011ApJ...736L..21R}
{Roberts} L.~F.,  {Kasen} D.,  {Lee} W.~H.,   {Ramirez-Ruiz} E.,  2011, \mn@doi
  [\apjl] {10.1088/2041-8205/736/1/L21}, \href
  {http://adsabs.harvard.edu/abs/2011ApJ...736L..21R} {736, L21}

\bibitem[\protect\citeauthoryear{{Roberts}, {Reddy}  \& {Shen}}{{Roberts}
  et~al.}{2012}]{2012PhRvC..86f5803R}
{Roberts} L.~F.,  {Reddy} S.,   {Shen} G.,  2012, \mn@doi [\prc]
  {10.1103/PhysRevC.86.065803}, \href
  {http://adsabs.harvard.edu/abs/2012PhRvC..86f5803R} {86, 065803}

\bibitem[\protect\citeauthoryear{{Rosswog}}{{Rosswog}}{2013}]{2013RSPTA.37120272R}
{Rosswog} S.,  2013, \mn@doi [Philosophical Transactions of the Royal Society
  of London Series A] {10.1098/rsta.2012.0272}, \href
  {http://adsabs.harvard.edu/abs/2013RSPTA.37120272R} {371, 20120272}

\bibitem[\protect\citeauthoryear{{Rosswog}}{{Rosswog}}{2015}]{2015IJMPD..2430012R}
{Rosswog} S.,  2015, \mn@doi [International Journal of Modern Physics D]
  {10.1142/S0218271815300128}, \href
  {http://adsabs.harvard.edu/abs/2015IJMPD..2430012R} {24, 1530012}

\bibitem[\protect\citeauthoryear{{Rosswog} \& {Liebend{\"o}rfer}}{{Rosswog} \&
  {Liebend{\"o}rfer}}{2003}]{2003MNRAS.342..673R}
{Rosswog} S.,  {Liebend{\"o}rfer} M.,  2003, \mn@doi [\mnras]
  {10.1046/j.1365-8711.2003.06579.x}, \href
  {http://adsabs.harvard.edu/abs/2003MNRAS.342..673R} {342, 673}

\bibitem[\protect\citeauthoryear{{Rosswog}, {Ramirez-Ruiz}  \&
  {Davies}}{{Rosswog} et~al.}{2003}]{2003MNRAS.345.1077R}
{Rosswog} S.,  {Ramirez-Ruiz} E.,   {Davies} M.~B.,  2003, \mn@doi [\mnras]
  {10.1046/j.1365-2966.2003.07032.x}, \href
  {http://adsabs.harvard.edu/abs/2003MNRAS.345.1077R} {345, 1077}

\bibitem[\protect\citeauthoryear{{Rosswog}, {Piran}  \& {Nakar}}{{Rosswog}
  et~al.}{2013}]{2013MNRAS.430.2585R}
{Rosswog} S.,  {Piran} T.,   {Nakar} E.,  2013, \mn@doi [\mnras]
  {10.1093/mnras/sts708}, \href
  {http://adsabs.harvard.edu/abs/2013MNRAS.430.2585R} {430, 2585}

\bibitem[\protect\citeauthoryear{{R{\"o}ttgers} \& {Arth}}{{R{\"o}ttgers} \&
  {Arth}}{2018}]{2018arXiv180303652R}
{R{\"o}ttgers} B.,  {Arth} A.,  2018, preprint, \href
  {http://adsabs.harvard.edu/abs/2018arXiv180303652R} {} (\mn@eprint {arXiv}
  {1803.03652})

\bibitem[\protect\citeauthoryear{{Ruffert} \& {Janka}}{{Ruffert} \&
  {Janka}}{1999}]{1999A&A...344..573R}
{Ruffert} M.,  {Janka} H.-T.,  1999, \aap, \href
  {http://adsabs.harvard.edu/abs/1999A%26A...344..573R} {344, 573}

\bibitem[\protect\citeauthoryear{{Ruffert} \& {Janka}}{{Ruffert} \&
  {Janka}}{2001}]{2001A&A...380..544R}
{Ruffert} M.,  {Janka} H.-T.,  2001, \mn@doi [\aap]
  {10.1051/0004-6361:20011453}, \href
  {http://adsabs.harvard.edu/abs/2001A%26A...380..544R} {380, 544}

\bibitem[\protect\citeauthoryear{{Ruffert}, {Janka}  \& {Schaefer}}{{Ruffert}
  et~al.}{1996}]{1996A&A...311..532R}
{Ruffert} M.,  {Janka} H.-T.,   {Schaefer} G.,  1996, \aap, \href
  {http://adsabs.harvard.edu/abs/1996A%26A...311..532R} {311, 532}

\bibitem[\protect\citeauthoryear{{Ruffert}, {Janka}, {Takahashi}  \&
  {Schaefer}}{{Ruffert} et~al.}{1997}]{1997A&A...319..122R}
{Ruffert} M.,  {Janka} H.-T.,  {Takahashi} K.,   {Schaefer} G.,  1997, \aap,
  \href {http://adsabs.harvard.edu/abs/1997A%26A...319..122R} {319, 122}

\bibitem[\protect\citeauthoryear{{Sekiguchi}}{{Sekiguchi}}{2010}]{2010CQGra..27k4107S}
{Sekiguchi} Y.,  2010, \mn@doi [Classical and Quantum Gravity]
  {10.1088/0264-9381/27/11/114107}, \href
  {http://adsabs.harvard.edu/abs/2010CQGra..27k4107S} {27, 114107}

\bibitem[\protect\citeauthoryear{{Sekiguchi}, {Kiuchi}, {Kyutoku}  \&
  {Shibata}}{{Sekiguchi} et~al.}{2011a}]{2011PhRvL.107e1102S}
{Sekiguchi} Y.,  {Kiuchi} K.,  {Kyutoku} K.,   {Shibata} M.,  2011a, \mn@doi
  [Physical Review Letters] {10.1103/PhysRevLett.107.051102}, \href
  {http://adsabs.harvard.edu/abs/2011PhRvL.107e1102S} {107, 051102}

\bibitem[\protect\citeauthoryear{{Sekiguchi}, {Kiuchi}, {Kyutoku}  \&
  {Shibata}}{{Sekiguchi} et~al.}{2011b}]{2011PhRvL.107u1101S}
{Sekiguchi} Y.,  {Kiuchi} K.,  {Kyutoku} K.,   {Shibata} M.,  2011b, \mn@doi
  [Physical Review Letters] {10.1103/PhysRevLett.107.211101}, \href
  {http://adsabs.harvard.edu/abs/2011PhRvL.107u1101S} {107, 211101}

\bibitem[\protect\citeauthoryear{{Sekiguchi}, {Kiuchi}, {Kyutoku}  \&
  {Shibata}}{{Sekiguchi} et~al.}{2012}]{2012PTEP.2012aA304S}
{Sekiguchi} Y.,  {Kiuchi} K.,  {Kyutoku} K.,   {Shibata} M.,  2012, \mn@doi
  [Progress of Theoretical and Experimental Physics] {10.1093/ptep/pts011},
  \href {http://adsabs.harvard.edu/abs/2012PTEP.2012aA304S} {2012, 01A304}

\bibitem[\protect\citeauthoryear{{Sekiguchi}, {Kiuchi}, {Kyutoku}  \&
  {Shibata}}{{Sekiguchi} et~al.}{2015}]{2015PhRvD..91f4059S}
{Sekiguchi} Y.,  {Kiuchi} K.,  {Kyutoku} K.,   {Shibata} M.,  2015, \mn@doi
  [\prd] {10.1103/PhysRevD.91.064059}, \href
  {http://adsabs.harvard.edu/abs/2015PhRvD..91f4059S} {91, 064059}

\bibitem[\protect\citeauthoryear{{Sekiguchi}, {Kiuchi}, {Kyutoku}, {Shibata}
  \& {Taniguchi}}{{Sekiguchi} et~al.}{2016}]{2016PhRvD..93l4046S}
{Sekiguchi} Y.,  {Kiuchi} K.,  {Kyutoku} K.,  {Shibata} M.,   {Taniguchi} K.,
  2016, \mn@doi [\prd] {10.1103/PhysRevD.93.124046}, \href
  {http://adsabs.harvard.edu/abs/2016PhRvD..93l4046S} {93, 124046}

\bibitem[\protect\citeauthoryear{{Shibata} \& {Taniguchi}}{{Shibata} \&
  {Taniguchi}}{2011}]{2011LRR....14....6S}
{Shibata} M.,  {Taniguchi} K.,  2011, \mn@doi [Living Reviews in Relativity]
  {10.12942/lrr-2011-6}, \href
  {http://adsabs.harvard.edu/abs/2011LRR....14....6S} {14, 6}

\bibitem[\protect\citeauthoryear{{Shibata}, {Sekiguchi}  \&
  {Takahashi}}{{Shibata} et~al.}{2007}]{2007PThPh.118..257S}
{Shibata} M.,  {Sekiguchi} Y.-I.,   {Takahashi} R.,  2007, \mn@doi [Progress of
  Theoretical Physics] {10.1143/PTP.118.257}, \href
  {http://adsabs.harvard.edu/abs/2007PThPh.118..257S} {118, 257}

\bibitem[\protect\citeauthoryear{{Shibata}, {Kiuchi}, {Sekiguchi}  \&
  {Suwa}}{{Shibata} et~al.}{2011}]{2011PThPh.125.1255S}
{Shibata} M.,  {Kiuchi} K.,  {Sekiguchi} Y.,   {Suwa} Y.,  2011, \mn@doi
  [Progress of Theoretical Physics] {10.1143/PTP.125.1255}, \href
  {http://adsabs.harvard.edu/abs/2011PThPh.125.1255S} {125, 1255}

\bibitem[\protect\citeauthoryear{{Shibata}, {Fujibayashi}, {Hotokezaka},
  {Kiuchi}, {Kyutoku}, {Sekiguchi}  \& {Tanaka}}{{Shibata}
  et~al.}{2017}]{2017PhRvD..96l3012S}
{Shibata} M.,  {Fujibayashi} S.,  {Hotokezaka} K.,  {Kiuchi} K.,  {Kyutoku} K.,
   {Sekiguchi} Y.,   {Tanaka} M.,  2017, \mn@doi [\prd]
  {10.1103/PhysRevD.96.123012}, \href
  {http://adsabs.harvard.edu/abs/2017PhRvD..96l3012S} {96, 123012}

\bibitem[\protect\citeauthoryear{{Smartt} et~al.,}{{Smartt}
  et~al.}{2017}]{2017Natur.551...75S}
{Smartt} S.~J.,  et~al., 2017, \mn@doi [\nat] {10.1038/nature24303}, \href
  {http://adsabs.harvard.edu/abs/2017Natur.551...75S} {551, 75}

\bibitem[\protect\citeauthoryear{Steiner, Hempel  \& Fischer}{Steiner
  et~al.}{2013}]{Steiner2013a}
Steiner A.~W.,  Hempel M.,   Fischer T.,  2013, \mn@doi [\apj]
  {10.1088/0004-637X/774/1/17}, 774, 17

\bibitem[\protect\citeauthoryear{{Takahashi}, {El Eid}  \&
  {Hillebrandt}}{{Takahashi} et~al.}{1978}]{1978A&A....67..185T}
{Takahashi} K.,  {El Eid} M.~F.,   {Hillebrandt} W.,  1978, \aap, \href
  {http://adsabs.harvard.edu/abs/1978A%26A....67..185T} {67, 185}

\bibitem[\protect\citeauthoryear{{Thielemann}, {Eichler}, {Panov}  \&
  {Wehmeyer}}{{Thielemann} et~al.}{2017}]{2017ARNPS..67..253T}
{Thielemann} F.-K.,  {Eichler} M.,  {Panov} I.~V.,   {Wehmeyer} B.,  2017,
  \mn@doi [Annual Review of Nuclear and Particle Science]
  {10.1146/annurev-nucl-101916-123246}, \href
  {http://adsabs.harvard.edu/abs/2017ARNPS..67..253T} {67, 253}

\bibitem[\protect\citeauthoryear{{Thompson}, {Burrows}  \&
  {Horvath}}{{Thompson} et~al.}{2000}]{2000PhRvC..62c5802T}
{Thompson} T.~A.,  {Burrows} A.,   {Horvath} J.~E.,  2000, \mn@doi [\prc]
  {10.1103/PhysRevC.62.035802}, \href
  {http://adsabs.harvard.edu/abs/2000PhRvC..62c5802T} {62, 035802}

\bibitem[\protect\citeauthoryear{{Typel}, {R{\"o}pke}, {Kl{\"a}hn}, {Blaschke}
  \& {Wolter}}{{Typel} et~al.}{2010}]{2010PhRvC..81a5803T}
{Typel} S.,  {R{\"o}pke} G.,  {Kl{\"a}hn} T.,  {Blaschke} D.,   {Wolter} H.~H.,
   2010, \mn@doi [\prc] {10.1103/PhysRevC.81.015803}, \href
  {http://adsabs.harvard.edu/abs/2010PhRvC..81a5803T} {81, 015803}

\bibitem[\protect\citeauthoryear{{Wanajo}, {Sekiguchi}, {Nishimura}, {Kiuchi},
  {Kyutoku}  \& {Shibata}}{{Wanajo} et~al.}{2014}]{2014ApJ...789L..39W}
{Wanajo} S.,  {Sekiguchi} Y.,  {Nishimura} N.,  {Kiuchi} K.,  {Kyutoku} K.,
  {Shibata} M.,  2014, \mn@doi [\apjl] {10.1088/2041-8205/789/2/L39}, \href
  {http://adsabs.harvard.edu/abs/2014ApJ...789L..39W} {789, L39}

\bibitem[\protect\citeauthoryear{{Waxman}, {Ofek}, {Kushnir}  \&
  {Gal-Yam}}{{Waxman} et~al.}{2018}]{2018MNRAS.481.3423W}
{Waxman} E.,  {Ofek} E.~O.,  {Kushnir} D.,   {Gal-Yam} A.,  2018, \mn@doi
  [\mnras] {10.1093/mnras/sty2441}, \href
  {http://adsabs.harvard.edu/abs/2018MNRAS.481.3423W} {481, 3423}

\bibitem[\protect\citeauthoryear{{Wilson}, {Couch}, {Cochran}, {Le Blanc}  \&
  {Barkat}}{{Wilson} et~al.}{1975}]{1975NYASA.262...54W}
{Wilson} J.~R.,  {Couch} R.,  {Cochran} S.,  {Le Blanc} J.,   {Barkat} Z.,
  1975, in {Bergman} P.~G.,  {Fenyves} E.~J.,   {Motz} L.,  eds,  Annals of the
  New York Academy of Sciences Vol. 262, Seventh Texas Symposium on
  Relativistic Astrophysics. pp 54--64,
  \mn@doi{10.1111/j.1749-6632.1975.tb31420.x}

\bibitem[\protect\citeauthoryear{{Wilson}, {Mathews}  \& {Marronetti}}{{Wilson}
  et~al.}{1996}]{1996PhRvD..54.1317W}
{Wilson} J.~R.,  {Mathews} G.~J.,   {Marronetti} P.,  1996, \mn@doi [\prd]
  {10.1103/PhysRevD.54.1317}, \href
  {http://adsabs.harvard.edu/abs/1996PhRvD..54.1317W} {54, 1317}

\bibitem[\protect\citeauthoryear{{Wu}, {Tamborra}, {Just}  \& {Janka}}{{Wu}
  et~al.}{2017}]{2017PhRvD..96l3015W}
{Wu} M.-R.,  {Tamborra} I.,  {Just} O.,   {Janka} H.-T.,  2017, \mn@doi [\prd]
  {10.1103/PhysRevD.96.123015}, \href
  {http://adsabs.harvard.edu/abs/2017PhRvD..96l3015W} {96, 123015}

\bibitem[\protect\citeauthoryear{{Zappa}, {Bernuzzi}, {Radice}, {Perego}  \&
  {Dietrich}}{{Zappa} et~al.}{2018}]{2018PhRvL.120k1101Z}
{Zappa} F.,  {Bernuzzi} S.,  {Radice} D.,  {Perego} A.,   {Dietrich} T.,  2018,
  \mn@doi [Physical Review Letters] {10.1103/PhysRevLett.120.111101}, \href
  {http://adsabs.harvard.edu/abs/2018PhRvL.120k1101Z} {120, 111101}

\makeatother
\end{thebibliography}
\appendix
 
\section{Comparative analysis of diffusion time-scale prescriptions used in the literature}\label{appendix:tdiffs} 

 Although leakage schemes have been around for more than two decades, not many comparisons between the different realizations can be found in the literature. Here we want to briefly compare the most common leakage implementations used in the context of neutrino physics in NS mergers, in particular, the schemes from \cite{1996A&A...311..532R} (RJS) and \cite{2003MNRAS.342..673R} (RL). 
 
 There are three main differences between both schemes: the definition of the diffusion time-scale, the energy averaging and the prescription of the neutrino chemical potential to describe the neutrino spectra. 

 As we discussed in section~\ref{sec:model:tdiff}, in a first approximation the diffusion time-scale can be obtained from a dimensional analysis of the diffusion equation as
 \begin{equation}
  t_{\nu_i}^{\mathrm{diff}}\ =\ \frac{3d^2}{c\lambda_{\nu_i}},\label{tdiffdimapendix} 
 \end{equation}
 where $d$ is simply a characteristic length-scale of the system. RJS took the approximation of a homogeneous sphere to define 
 \begin{equation}
 \lambda_{\nu_i}=\frac{d}{\tau_{\nu_i}}. \label{lambda}
 \end{equation}
 By plugging equation~\eqref{lambda} into~\eqref{tdiffdimapendix} we are left with one factor of the length-scale, which is chosen as the integration path for the optical depth, taken as the minimum distance to the neutrinosphere ($r(\tau_{\nu_i}=2/3)$):
 \begin{equation}
  t_{\nu_i}^{\mathrm{diff},\mathrm{RJS}}\ =\ \frac{3d}{c}\tau_{\nu_i}.\label{tdiffRJS} 
 \end{equation}
 Similarly, RL proceeded to further approximate the remaining length-scale as $d\sim \tau_{\nu_i}\lambda_{\nu_i}$ (with $\lambda_{\nu_i}=1/\kappa_{\nu_i}$) to obtain:
 \begin{equation}
  t_{\nu_i}^{\mathrm{diff},\mathrm{RL}}\ =\ \frac{3\lambda_{\nu_i}}{c}\tau_{\nu_i}^2.\label{tdiffRL} 
 \end{equation}
 It is worth noting that other groups have suggested alternative prescriptions for the definition of $d$, such as using the pressure scale-height $d\simeq {P}/{\boldsymbol{\nabla}P}$ \citep{2014MNRAS.441.3444M}. This approach could be generalized to using the scale-height of any convenient scalar quantity that defines the medium in which neutrinos diffuse. 
 
 One of the caveats of grey schemes is the ambiguity associated with the energy averaging of the neutrino quantities. In RJS, the diffusion time-scales are computed from spectrally averaged opacities (and optical depths),
 \begin{equation}
  \bar{\kappa}_{\nu_i}\ =\ \frac{\int_0^{\infty}{\kappa_{\nu_i}(\epsilon)\epsilon^2f(\epsilon;T,\eta_{\nu_i})\mathrm{d}\epsilon}}{\int_0^{\infty}{\epsilon^2f(\epsilon;T,\eta_{\nu_i})}\mathrm{d}\epsilon}. \label{spectralavg}
 \end{equation}
 In RL, on the contrary, the (roughly) $\epsilon^2$ dependence of the opacities is factored out and carried on to the calculation of the ``integrated diffusion rates'', defined as
 \begin{equation}
  R_{\nu_i,\mathrm{diff}}\ =\ \int_0^{\infty}{\frac{E_{\nu_i}^{j=0}(\epsilon)}{t_{\nu_i}^{\mathrm{diff}}(\epsilon)}\mathrm{d}\epsilon}\ =\ \int_0^{\infty}{\frac{E_{\nu_i}^{j=0}(\epsilon)}{t_{\nu_i}^{\mathrm{diff},\mathrm{RL}}\epsilon^2}\mathrm{d}\epsilon}, \label{integralRLnum}
 \end{equation}
 for lepton number diffusion and equivalently for the energy diffusion rate,
 \begin{equation}
  Q_{\nu_i,\mathrm{diff}}\ =\ \int_0^{\infty}{\frac{E_{\nu_i}^{j=1}(\epsilon)}{t_{\nu_i}^{\mathrm{diff}}(\epsilon)}\mathrm{d}\epsilon}\ =\ \int_0^{\infty}{\frac{E_{\nu_i}^{j=1}(\epsilon)}{t_{\nu_i}^{\mathrm{diff},\mathrm{RL}}\epsilon^2}\mathrm{d}\epsilon}. \label{integralRLen}
 \end{equation}
 
 In both schemes, the effective loss term is then calculated as an interpolation between diffusion rates (equations~\ref{integralRLnum} and~\ref{integralRLen}) and production rates (equations~\ref{Rpur} and~\ref{Qpur}):
  \begin{equation}
 R_{\nu_i}^{-}\ =\ R_{\nu_i}\left(1+\frac{R_{\nu_i}}{R_{\nu_i,\mathrm{diff}}}\right)^{-1},\label{Reff2}
 \end{equation}
 and
  \begin{equation}
 Q_{\nu_i}^{-}\ =\ Q_{\nu_i}\left(1+\frac{Q_{\nu_i}}{Q_{\nu_i,\mathrm{diff}}}\right)^{-1},\label{Qeff2}
 \end{equation}
 for number and energy, respectively, which are equivalent to equations~\eqref{Reff} and~\eqref{Qeff} in the case of the RJS averaging.
 
 Moreover, without an actual energy-dependent transport scheme, it is impossible to determine the correct neutrino phase-space distribution. In the optically thick regime, neutrinos remain in $\beta$-equilibrium with the medium, thus their spectrum is a Fermi distribution with the chemical potential being easily obtained from the EoS. RL assume that this behaviour will remain a good approximation even in the optically thin regime, where neutrinos decouple from matter. RJS on the other hand also use a Fermi distribution throughout, but interpolate between the equilibrium chemical potential and an expected value at free streaming conditions of $\mu_{\nu}=0$ (see equation~\ref{etanu}).

\begin{figure}
\begin{center}
\makebox[0pt][c]{%
\minipage{0.5\textwidth}
\includegraphics[width=\textwidth]{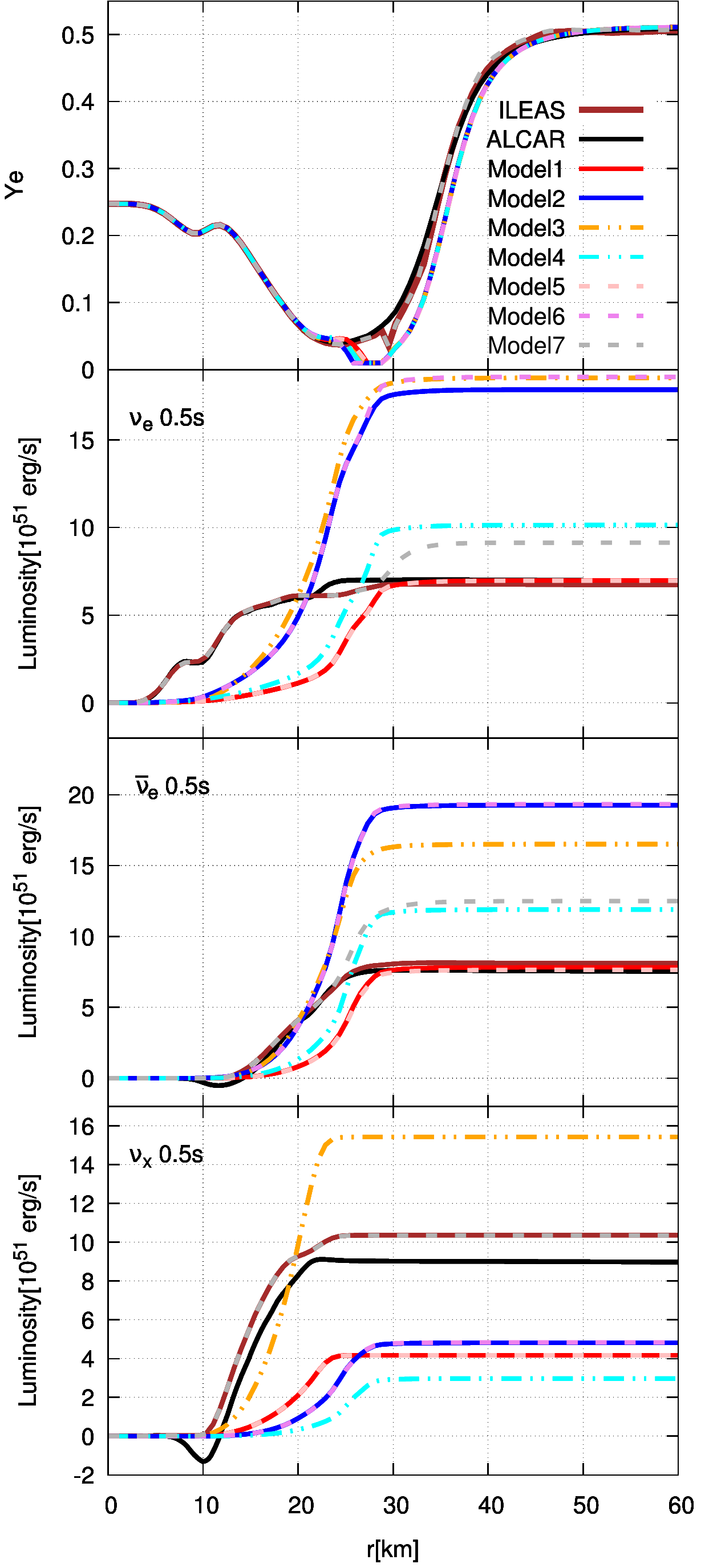}
\endminipage\hfill
}%
\caption{Radial profiles of the electron fraction and the neutrino luminosities of the three neutrino species obtained by the different leakage models summarized in table~\ref{table:tdiffcomp}, applied on a PNS snapshot (relaxed \textsc{ALCAR} background at 0.5~s post-bounce) and relaxed for 5~ms. We did not include absorption or equilibration in any of the numbered models in order to focus on the differences of the leakage module alone. For comparison, we also plot the results obtained by \textsc{ALCAR} and \textsc{ILEAS} (with absorption and equilibration).}\label{Tdiffcomparison}
\end{center}
\end{figure}

 We apply both schemes to one of our PNS cooling snapshots and compare the results to the \textsc{ILEAS} model presented in this work. To further disentangle the contribution of each approximation, we also test different permutations, combining the prescriptions for the diffusion time-scale, the spectral averaging and the neutrino chemical potential out of equilibrium. A list of the models and the prescriptions employed are provided in table~. In order to focus on the impact of the leakage module alone, we let the system relax \textit{without} including equilibration or neutrino absorption in any of the calculations. Relativistic corrections in the diffusion time-scale are omitted as well. Figure~\ref{Tdiffcomparison} shows the radial profiles of the electron fraction and luminosity profiles for all the relaxed models (after 5~ms) together with the ones obtained by \textsc{ALCAR} and \textsc{ILEAS} (including all modules). 
 
 The first aspect that catches the eye is the substantial improvement of our prescription (model 7) with respect to all previous models. For the luminosities this is particularly true in the high-optical-depth regime, where the diffusion time-scale dominates. This is no surprise, as we define $t_{\nu_i}^{\mathrm{diff}}$ directly from the diffusion equation, which encodes much more information about the way neutrinos are transported than the simple ansatz of equation~\eqref{tdiffdimapendix}. Due to the slightly slower increase of RL's $t_{\nu_i}^{\mathrm{diff}}$ with growing optical depth (models 2,3,6) in comparison to RJS's $t_{\nu_i}^{\mathrm{diff}}$ (models 1,4,5), neutrinos escape from further inside the star in the former models. As a consequence, they resemble a bit closer the transport profile, but overproduce neutrinos of all species at lower optical depths. The consequences of the energy averaging are much less straightforward. Differences between a few percent up to a factor 3 can be seen for the different species.
  
 The choice of neutrino chemical potential does not significantly affect the results of the leakage scheme in the chosen snapshot. The differences could become more significant in a scenario where most neutrino luminosities are produced in the semi-transparent region. Based on the current results, however, the interpolated $\mu_{\nu}$ (equation~\ref{etanu}) should be preferred, because it fulfils the correct limit at high optical depth and a well controlled behaviour at low optical depths, thus avoiding an undesirable behaviour of the analytical solutions of the Fermi integrals and their ratios at low optical depth.
 
 In the studied case, the standard formalism of RJS compared to other approaches in the literature yields the best agreement of the outgoing neutrino luminosities for $\nu_e$ and $\bar{\nu}_e$ with \textsc{ALCAR} transport results ($\nu_x$ is worse because of the lack of nucleon-nucleon bremsstrahlung in RJS). Of course, definitive conclusions are not possible on grounds of one test case, and all traditional treatments including the one by RJS fail to reproduce the radial dependences of transport results.
 
 It is worth noticing that, as can be seen in the first panel of figure~\ref{Tdiffcomparison}, all leakage versions produce a similar effect on the electron fraction after relaxation. Namely, the low $Y_e$ trough near the NS surface expands outwards and the matter becomes more neutron-rich. The cause for this effect is simply the inability of any leakage scheme to accurately describe the semi-transparent region, which comes as no surprise being a model constructed as an interpolation between pure diffusion and pure free streaming. A similar effect, albeit to a much smaller extent due to the more accurate diffusion time-scale, can be observed when applying \textsc{ILEAS}. This comparison further highlights the advantages of the scheme presented in this work with respect to some of the leakage versions widely used in the literature.
 
\begin{table*}
  \begin{center}
      \caption{Summary of the prescriptions for the neutrino diffusion time-scale, $t_{\nu_i}^{\mathrm{diff}}$, energy averaging and neutrino chemical potential, $\mu_{\nu}$, employed for all models (including \textsc{ILEAS} and \textsc{ALCAR}) shown in figure~{\protect\ref{Tdiffcomparison}}. Additionally, we present the neutrino luminosities of the three neutrino species obtained by applying them to a PNS snapshot at 0.5~s post-bounce. Models 1--7 do not include the effects of neutrino re-absorption or equilibration, whereas the results obtained by \textsc{ILEAS} and \textsc{ALCAR} do.}
      \label{table:tdiffcomp}
    \begin{threeparttable}
      \begin{tabular}{lcccccc}
      \hline
      \hline
      \noalign{\vskip 2mm}  
      \parbox[c]{0.1\textwidth}{ Model}        & \parbox[c]{0.1\textwidth}{\centering $t_{\nu_i}^{\mathrm{diff}}$} & \parbox[c]{0.1\textwidth}{\centering Energy Avg.} & \parbox[c]{0.1\textwidth}{\centering $\mu_{\nu}$} & \parbox[c]{0.1\textwidth}{\centering $L_{\nu_e}$ ($10^{51}\ \mathrm{erg}\boldsymbol\cdotp\mathrm{s}^{-1}$)}  & \parbox[c]{0.1\textwidth}{\centering $L_{\bar{\nu}_e}$ ($10^{51}\ \mathrm{erg}\boldsymbol\cdotp\mathrm{s}^{-1}$)}  & \parbox[c]{0.1\textwidth}{\centering $L_{\nu_x}$ ($10^{51}\ \mathrm{erg}\boldsymbol\cdotp\mathrm{s}^{-1}$)} \vspace{1mm}  \\ 
      \hline 
      \hline 
      \noalign{\vskip 2mm}                                                                                                                                                                                                                                                                                                                                      
      Model 1        	& RJS\tnote{1}	& RJS     & RJS     & 7.0    & 7.8     & 4.2    \\                                                                                            
      Model 2	    	& RL\tnote{2}	& RL	  & RL	    & 17.9   & 19.3    & 4.8    \\  
      Model 3        	& RL		& RJS     & RJS     & 18.5   & 16.5    & 15.4   \\                                                                                             
      Model 4	    	& RJS		& RL	  & RJS     & 10.1   & 11.9    & 3.0    \\                                                                                                                                                                                                                                                                     
      Model 5        	& RJS		& RJS     & RL	    & 7.0    & 7.6     & 4.2    \\                                                                                              
      Model 6	    	& RL		& RL	  & RJS     & 18.6   & 19.3    & 4.8    \\             
      Model 7	    	& AJJB\tnote{3}	& AJJB	  & AJJB    & 9.1    & 12.5    & 10.4   \\            
      \textsc{ILEAS}	& AJJB	& AJJB	  & AJJB    & 6.7    & 8.1     & 10.4   \\      
      \textsc{ALCAR}	& -		& -	  & -	    & 7.0    & 7.6     & 9.0    \\                            
      \noalign{\vskip 2mm}  
      \hline
      \hline
      \end{tabular}
      \begin{tablenotes}	
       \item[1] Ruffert, Janka \& Sch\"afer {\protect\citep{1996A&A...311..532R}}\ \ \item[2] Rosswog \& Liebend\"orfer {\protect\citep{2003MNRAS.342..673R}} \ \ \item[3] Ardevol, Janka, Just \& Bauswein (this work)
      \end{tablenotes}
    \end{threeparttable}
  \end{center}
\end{table*}

\section{Neutrino reactions}\label{appendix:reac}

 In this appendix we collect the formulae for the different neutrino reactions (opacities and production rates) of all three neutrino species, employed in our scheme. Most reactions and their constants are extracted from \cite{1996A&A...311..532R} and references therein. In this section, unlike in the body of this work, we employ only the superscript of $Q_{\nu_i}^j$ with $j=0,1$ to denote number and energy rates, respectively, for reasons of compactness in the formulation. All production rates and opacities for $\nu_x$ include the contributions of all four species ($\nu_{\mu}$,$\bar{\nu}_{\mu}$,$\nu_{\tau}$ and $\bar{\nu}_{\tau}$). 
 
 \subsection{Opacities for diffusion}\label{appendix:reac:opad}
 
 We define the energy-dependent absorption opacities, $\kappa_{\nu_i,\mathrm{a}}(\epsilon)$, following \cite{1985ApJS...58..771B}, with the correction of stimulated absorption (neutrino phase space blocking) from \cite{2002A&A...396..361R},
 \begin{equation}
  \kappa_{\nu_i,\mathrm{a}}^*(\epsilon)\ =\ \kappa_{\nu_i,\mathrm{a}}(\epsilon)[1-f(\epsilon;T,\eta_{\nu_i}^{\mathrm{eq}})]^{-1}.\label{stimulatedkappa} 
 \end{equation}
 Here, $f(\epsilon;T,\eta_i)=[1+\mathrm{exp}((\epsilon/T)-\eta_i)]^{-1}$ is the distribution function of fermions with degeneracy parameter $\eta_i=\mu_i/T$ and energy $\epsilon$. The superscript `eq', in this case, denotes the usage of the equilibrium neutrino degeneracy instead of the interpolated one (see section~\ref{sec:model:leakage}). The opacity for $\nu_e$ absorption on neutrons, $n$, is given by
 \begin{align}
  \kappa_{\nu_e,\mathrm{a}}^*(\epsilon)\ =\ &\frac{1+3g_A^2}{4}\sigma_0\xi_{np}\frac{[1-f(\epsilon+Q;T,\eta_{e^-})]}{[1-f(\epsilon;T,\eta_{\nu_e}^{\mathrm{eq}})]}\left(\frac{\epsilon+Q}{m_ec^2}\right)^2\nonumber\\
  &\cdotp\left[1-\frac{(m_ec^2)^2}{(\epsilon+Q)^2}\right]^{\frac{1}{2}},\label{kappa_abse_nue}
 \end{align}
 and $\bar{\nu}_e$ absorption on protons, $p$, by
 \begin{align}
  \kappa_{\bar{\nu}_e,\mathrm{a}}^*(\epsilon)\ =\ &\frac{1+3g_A^2}{4}\sigma_0\xi_{pn}\frac{[1-f(\epsilon-Q;T,\eta_{e^+})]}{[1-f(\epsilon;T,\eta_{\bar{\nu}_e}^{\mathrm{eq}})]}\left(\frac{\epsilon-Q}{m_ec^2}\right)^2\nonumber\\
  &\cdotp\left[1-\frac{(m_ec^2)^2}{(\epsilon-Q)^2}\right]^{\frac{1}{2}}\Theta(\epsilon-Q-m_ec^2).\label{kappa_abse_anue}
 \end{align}
 Here $c$ is the speed of light, $g_A\approx1.25$, $\sigma_0=1.76\boldsymbol\cdotp10^{-44}\mathrm{cm}^2$ and $m_e$ the electron mass. $[1-f(\epsilon\pm Q;T,\eta_{e_\mp})]$ are the electron/positron phase space blocking factors and the coefficients $\xi_{np}$ and $\xi_{pn}$ \citep{1985ApJS...58..771B} are related to the nucleon blocking factors $Y_{np}$ and $Y_{pn}$ \citep{1996A&A...311..532R} as
 \begin{equation}
  \xi_{np}\ =\ (n_n+n_p)\boldsymbol\cdotp Y_{np}\ =\ \mathcal{A}\rho\frac{Y_p-Y_n}{\mathrm{e}^{\eta_p-\eta_n}-1},\label{protonblk}
 \end{equation}
 and
 \begin{equation}
  \xi_{pn}\ =\ (n_n+n_p)\boldsymbol\cdotp Y_{pn}\ =\ \mathcal{A}\rho\frac{Y_p-Y_n}{1-\mathrm{e}^{\eta_n-\eta_p}},\label{neutronblk}
 \end{equation}
 where $Y_p$ and $Y_n$ are the proton and neutron number fractions, respectively, and $\mathcal{A}$ is the Avogadro constant \citep{1985ApJS...58..771B}. As we pointed out in section~\ref{sec:model:leakage}, this formulation of the blocking factors assumes nucleons to be well represented by a free Fermi gas. In order to avoid unphysical behaviour, we make use of the free Fermi gas nucleon chemical potentials, which we calculate by inverting the relation \citep{Ramppthesis,Hechtthesis},
 \begin{equation}
  n_N\ =\ \frac{4\pi}{(hc)^3}(2m_Nc^2T)^{3/2}F_{1/2}(\eta_N),\label{nucheminv}
 \end{equation}
 where $N$ refers to the nucleon type, $p$ or $n$.
 
 If one assumes complete dissociation of matter in protons and neutrons, the nucleon fractions can be expressed as $Y_p=Y_e$ and $Y_n=(1-Y_e)$, as in \cite{1996A&A...311..532R}. However, for more consistent comparison to \textsc{ALCAR}, we relaxed this assumption and employed the nucleon number densities obtained from the EoS. The Heaviside step function $\Theta(\epsilon-Q-m_ec^2)$ in equation~\eqref{kappa_abse_anue} ensures that the opacity remains defined and positive, setting the rest-mass difference between particles on both sides of the interaction as the minimum energy for $\bar{\nu}_e$ absorption.
 
 The transport opacities for neutrino-nucleon scattering of all three neutrino species are defined as
 \begin{align}
  \kappa_{\nu_i,\mathrm{s}}(\epsilon)\ =\ C_{N}\sigma_0\xi_{NN}\left(\frac{\epsilon}{m_ec^2}\right)^2,\label{kappa_scte}
 \end{align}
 where $C_{p}=[4(C_V-1)^2+5g_A^2]/24$ and $C_{n}=(1+5g_A^2)/24$ with $C_V=1/2+2\mathrm{sin}^2\theta_W$ and $\mathrm{sin}^2\theta_W=0.23$. We define the nucleon Pauli blocking factor, $Y_{NN}$, following \cite{1993ApJ...405..637M}, as an interpolation between (non-relativistic) degenerate and non-degenerate limits:
 \begin{align}
  &\xi_{NN}\ =\ (n_n+n_p)\boldsymbol\cdotp Y_{NN}\ =\ \mathcal{A}\rho Y_N\frac{\zeta_N}{\sqrt{1+\zeta_N^2}},\\
  &\mathrm{with}\ \ \zeta_N\ =\ \frac{3T}{2E_N^F} ,\label{blksca}
 \end{align}
  where $E_N^F$ is the Fermi energy of nucleon $N$, 
 \begin{align}
  E_N^F\ =\ \frac{h^2}{8\pi^2m_b}\left(3\pi^2n_N\right)^{2/3}. \label{fermie}
 \end{align}

 Similarly, scattering on nuclei of mass number $A$ can be expressed as
 \begin{align}
  \kappa_{\nu_i,\mathrm{s}}(\epsilon;A)\ &=\ \frac{1}{6}A^2\left[C_A-1+\frac{Z}{A}(2-C_A-C_V)\right]^2\nonumber\\
  &\cdotp\sigma_0n_A\left(\frac{\epsilon}{m_ec^2}\right)^2, \label{kappa_scten}
 \end{align}
 where $C_A=1/2$, $Z$ is the proton number of nuclei and $n_A$ the nuclei number density. This equation is used both for scattering on heavy nuclei of average mass and proton numbers $\bar{A}$ and $\bar{Z}$ and for scattering on $\alpha$-particles ($A=4$ and $Z=2$). 
 
 The total opacities for each neutrino species, both for energy and number transport, are simply
 \begin{align}
  \kappa_{\nu_e}(\epsilon)\ =\ &\kappa_{\nu_e,\mathrm{a}}^*(\epsilon)+\kappa_{\nu_e,\mathrm{s}}(\epsilon;n)+\kappa_{\nu_e,\mathrm{s}}(\epsilon;p)+\kappa_{\nu_e,\mathrm{s}}(\epsilon;\alpha)+\nonumber\\
  &\kappa_{\nu_e,\mathrm{s}}(\epsilon;\bar{A}),\nonumber\\
  \kappa_{\bar{\nu}_e}(\epsilon)\ =\ &\kappa_{\bar{\nu}_e,\mathrm{a}}^*(\epsilon)+\kappa_{\bar{\nu}_e,\mathrm{s}}(\epsilon;n)+\kappa_{\bar{\nu}_e,\mathrm{s}}(\epsilon;p)+\label{kappa_totale}\\
  &\kappa_{\bar{\nu}_e,\mathrm{s}}(\epsilon;\alpha)+\kappa_{\bar{\nu}_e,\mathrm{s}}(\epsilon;\bar{A}),\nonumber\\
  \kappa_{\nu_x}(\epsilon)\ =\ &\kappa_{\nu_x,\mathrm{s}}(\epsilon;n)+\kappa_{\nu_x,\mathrm{s}}(\epsilon;p)+\kappa_{\nu_x,\mathrm{s}}(\epsilon;\alpha)+\kappa_{\nu_x,\mathrm{s}}(\epsilon;\bar{A}).\nonumber
 \end{align}

 These opacities are used for the calculation of the diffusion time-scales (equations~\ref{tdiffn} and~\ref{tdiffe}) as explained in section~\ref{sec:model:tdiff}. 
 
 \subsection{Opacities for absorption and optical depth}\label{appendix:reac:opaa}
 
 We use spectrally averaged opacities to estimate the optical depth (equation~\ref{tau}) for the interpolation of the neutrino degeneracies (equation~\ref{etanu}), as well as in the absorption module. For consistency with our production rates, we do not correct these opacities for stimulated absorption (equation~\ref{stimulatedkappa}, see also the discussion in appendix~\ref{appendix:rates}). Following \cite{1996A&A...311..532R}, we average the absorption opacities as
 \begin{align}
  \bar{\kappa}_{\nu_e,\mathrm{a}}^{j}\ &=\ \frac{\int_0^{\infty}{\kappa_{\nu_e,\mathrm{a}}(\epsilon)E_{\nu_e}^j(\epsilon)\mathrm{d}\epsilon}}{\int_0^{\infty}{E_{\nu_e}^j(\epsilon)\mathrm{d}\epsilon}}\nonumber\\
  &=\ \frac{1+3g_A^2}{4(m_ec^2)^2}\sigma_0\xi_{np}\langle1-f(\bar{\epsilon}_{e^-};T,\eta_{e^-})\rangle\nonumber\\
  &\cdotp\frac{T^2F_{4+j}(\eta_{\nu_e})+2QTF_{3+j}(\eta_{\nu_e})+Q^2F_{2+j}(\eta_{\nu_e})}{F_{2+j}(\eta_{\nu_e})},\label{kappa_abs_nue}
 \end{align}
 and
 \begin{align}
  \bar{\kappa}_{\bar{\nu}_e,\mathrm{a}}^{j}\ &=\ \frac{\int_0^{\infty}{\kappa_{\bar{\nu}_e,\mathrm{a}}(\epsilon)E_{\bar{\nu}_e}^j(\epsilon)\Theta(\epsilon-Q)\mathrm{d}\epsilon}}{\int_0^{\infty}{E_{\bar{\nu}_e}^j(\epsilon)\mathrm{d}\epsilon}}\nonumber\\
  &=\ \frac{1+3g_A^2}{4(m_ec^2)^2}\sigma_0\xi_{pn}\langle1-f(\bar{\epsilon}_{e^+};T,\eta_{e^+})\rangle\nonumber\\
  &\cdotp\frac{T^2F_{4+j}(\eta_{\bar{\nu}_e}-Q/T)+(2+j)QTF_{3+j}(\eta_{\bar{\nu}_e}-Q/T)}{F_{2+j}(\eta_{\bar{\nu}_e})}\nonumber\\
  &+\frac{(1+2j)Q^2F_{2+j}(\eta_{\bar{\nu}_e}-Q/T)}{F_{2+j}(\eta_{\bar{\nu}_e})}\nonumber\\
  &+\frac{jQ^3T^{-1}F_{1+j}(\eta_{\bar{\nu}_e}-Q/T)}{F_{2+j}(\eta_{\bar{\nu}_e})},\label{kappa_abs_anue}
 \end{align}
 where $T$ is the matter temperature, $E_{\nu_e}^j(\epsilon)$ is defined as in equation~\eqref{neutrinodensity} and $F_k=\int_0^{\infty}{x^kf(x;T,\eta_{\nu_i})\mathrm{d}x}$ are the Fermi integrals of order $k$ of particle $i$, with $F_{k}(\eta_{\nu_i}- Q/T)$ evaluated including the nucleon rest-mass correction to the lepton energy. In this averaging procedure, we consider the correction of the electron rest mass to the neutrino energy to be negligible. As in \cite{1996A&A...311..532R}, we also approximate the lepton blocking factors, $\langle1-f(\bar{\epsilon}_{i};T,\eta_{i})\rangle$, assuming that the mean electron and positron production energies are equal to those of the absorbed $\nu_e$ and $\bar{\nu}_e$ with an additional correction for the nucleon rest mass difference\footnote{Note that produced $e^-$ will have a minimum energy ${\epsilon}_{e^-}^{\mathrm{min}}=Q$.}
 \begin{align}
 \bar{\epsilon}_{e^-}\ &=\ T\frac{F_5(\eta_{\nu_e})}{F_4(\eta_{\nu_e})}+Q,\label{e-meane}\\ 
 \bar{\epsilon}_{e^+}\ &=\ T\frac{F_5(\eta_{\bar{\nu}_e})}{F_4(\eta_{\bar{\nu}_e})}.\label{e+meane}
 \end{align}
 One can easily recover the results from \cite{1996A&A...311..532R} by assuming the nucleon rest mass to be negligible, $\bar{\epsilon}_{e^-}\approx\bar{\epsilon}_{\nu_e}$ and $\bar{\epsilon}_{e^+}\approx\bar{\epsilon}_{\bar{\nu}_e}$. Note that equations~\eqref{kappa_abs_nue} and~\eqref{kappa_abs_anue} with $j=0$ are only used to calculate the mean neutrino energies for diagnostics (equation~\ref{ethin}). 
 
 Following the same procedure, the spectrally averaged scattering opacities read,
 \begin{align}
  \bar{\kappa}_{\nu_i,\mathrm{s}}^j\ =\ C_{N}\sigma_0\xi_{NN}\left(\frac{T}{m_ec^2}\right)^2\frac{F_{4+j}(\eta_{\nu_i})}{F_{2+j}(\eta_{\nu_i})},\label{kappa_sct}
 \end{align}
 for scattering on nucleons and, 
 \begin{align}
  \bar{\kappa}_{\nu_i,\mathrm{s}}^j(A)\ &=\ \frac{1}{6}A^2\left[C_A-1+\frac{Z}{A}(2-C_A-C_V)\right]^2\nonumber\\
  &\sigma_0n_A\left(\frac{T}{m_ec^2}\right)^2\frac{F_{4+j}(\eta_{\nu_i})}{F_{2+j}(\eta_{\nu_i})}, \label{kappa_sctn}
 \end{align} 
 for scattering on $\alpha$-particles and heavy nuclei. Like for the energy-dependent opacities (equation~\ref{kappa_totale}), we define the total number ($j=0$) and energy ($j=1$) averaged opacities as
 \begin{align}
  \bar{\kappa}_{\nu_e}^j\ =\ &\bar{\kappa}_{\nu_e,\mathrm{a}}^j+\bar{\kappa}_{\nu_e,\mathrm{s}}^j(n)+\bar{\kappa}_{\nu_e,\mathrm{s}}^j(p)+\bar{\kappa}_{\nu_e,\mathrm{s}}^j(\alpha)+\nonumber\\
  &\bar{\kappa}_{\nu_e,\mathrm{s}}^j(\bar{A}),\nonumber\\
  \bar{\kappa}_{\bar{\nu}_e}^j\ =\ &\bar{\kappa}_{\bar{\nu}_e,\mathrm{a}}^j+\bar{\kappa}_{\bar{\nu}_e,\mathrm{s}}^j(n)+\bar{\kappa}_{\bar{\nu}_e,\mathrm{s}}^j(p)+\bar{\kappa}_{\bar{\nu}_e,\mathrm{s}}^j(\alpha)+\label{kappa_total}\\
  &\bar{\kappa}_{\bar{\nu}_e,\mathrm{s}}^j(\bar{A}),\nonumber\\
  \bar{\kappa}_{\nu_x}^j\ =\ &\bar{\kappa}_{\nu_x,\mathrm{s}}^j(n)+\bar{\kappa}_{\nu_x,\mathrm{s}}^j(p)+\bar{\kappa}_{\nu_x,\mathrm{s}}^j(\alpha)+\bar{\kappa}_{\nu_x,\mathrm{s}}^j(\bar{A}).\nonumber
 \end{align}
 
 Finally, in the calculation of the neutrino absorption rates, spectrally averaged absorption opacities are calculated as in equations~\eqref{kappa_abs_nue} and~\eqref{kappa_abs_anue}, but employing the neutrino spectrum from the corresponding ray, as defined in section~\ref{sec:model:abs}. 
 
 \subsection{Production rates}\label{appendix:reac:sourc}
 
 The $\beta$-processes are the main (far dominant) production sources of $\nu_e$ and $\bar{\nu}_e$. From the emissivities obtained by \cite{1985ApJS...58..771B}, we define the corresponding spectrally averaged production rates (including nucleon rest-mass corrections but without electron rest-mass terms) as 
 \begin{align}
  Q_{\nu_e,\beta}^j\ &=\ \frac{1+3g_A^2}{8}\frac{\sigma_0c}{(m_ec^2)^2}\xi_{pn}\langle1-f(\bar{\epsilon}_{\nu_e}^{\beta};T,\eta_{\nu_e})\rangle\nonumber\\
  &\cdotp\frac{8\pi}{(hc)^3}\left[T^{5+j}F_{4+j}(\eta_{e^-}-Q/T)\right.\nonumber\\
  &+2QT^{4+j}F_{3+j}(\eta_{e^-}-Q/T)\nonumber\\
  &\left.+Q^2T^{3+j}F_{2+j}(\eta_{e^-}-Q/T)\right], \label{rate_beta_nue}
 \end{align}
 for $\nu_e$ and
 \begin{align}
  Q_{\bar{\nu}_e,\beta}^j\ &=\ \frac{1+3g_A^2}{8}\frac{\sigma_0c}{(m_ec^2)^2}\xi_{np}\langle1-f(\bar{\epsilon}_{\bar{\nu}_e}^{\beta};T,\eta_{\bar{\nu}_e})\rangle\nonumber\\
  &\cdotp\frac{8\pi}{(hc)^3}\left[T^{5+j}F_{4+j}(\eta_{e^+})\right.\nonumber\\
  &+(2+j)QT^{4+j}F_{3+j}(\eta_{e^+})\nonumber\\
  &+(1+2j)Q^2T^{3+j}F_{2+j}(\eta_{e^+})\nonumber\\
  &\left.+jQ^3T^{2+j}F_{1+j}(\eta_{e^+})\right],\label{rate_beta_anue}
 \end{align}
 for $\bar{\nu}_e$. All quantities and constants are defined as in section~\ref{appendix:reac:opad}. The mean $\nu_e$ and $\bar{\nu}_e$ production energies are approximated in analogy of equations~\eqref{e-meane} and~\eqref{e+meane}, assuming they are equal to the mean energies of the captured electrons and positrons, respectively, with a correction for the nucleon rest mass difference\footnote{Note that electrons must have a minimum energy ${\epsilon}_{e^-}^{\mathrm{min}}=Q$ to be absorbed.}
 \begin{align}
 \bar{\epsilon}_{\nu_e}^{\beta}\ &=\ \mathrm{max}\left(T\frac{F_5(\eta_{e^-})}{F_4(\eta_{e^-})}-Q,0\right),\label{nuemeane}\\
 \bar{\epsilon}_{\bar{\nu}_e}^{\beta}\ &=\ T\frac{F_5(\eta_{e^+})}{F_4(\eta_{e^+})}.\label{anuemeane}
 \end{align}
 Since the neutrino phase space blocking is small in the neutrino production dominated regions, the approximate average value employed in \cite{1996A&A...311..532R}, $[1-f(\epsilon;T,\eta_{\nu_i})]\simeq\langle1-f(\bar{\epsilon};T,\eta_{\nu_i})\rangle$, is very reasonable. We caution the reader that for the tests presented in section~\ref{sec:tests} the $\beta$-production rates were implemented following \cite{2002A&A...396..361R}, as is explained in detail in appendix~\ref{appendix:rates}.
 
 Thermal processes such as electron-positron pair-annihilation are also an important source of neutrino pairs of all three species. Following \cite{1986ApJ...309..653C,1987ApJ...321L.129C}, the $\nu_e$ and $\bar{\nu}_e$ production rates read
 
 \begin{align}
  Q_{\nu_e,\bar{\nu}_e,ee}^j\ &=\ \frac{(C_1+C_2)_{\nu_e\bar{\nu}_e}}{72}\frac{\sigma_0c}{(m_ec^2)^2}\nonumber\\
  &\cdotp\langle1-f(\bar{\epsilon}_{e_i}^{ee};T,\eta_{\nu_e})\rangle\langle1-f(\bar{\epsilon}_{e_i}^{ee};T,\eta_{\bar{\nu}_e})\rangle\nonumber\\
  &\cdotp\left[\frac{8\pi}{(hc)^3}\right]^2\left[T^{4+j}F_{3+j}(\eta_{e^-})T^4F_3(\eta_{e^+})+\right.\nonumber\\
  &\left.T^4F_3(\eta_{e^-} )T^{4+j}F_{3+j}(\eta_{e^+})\right],\label{rate_ee_nue}
  \end{align}  
 where the constants $(C_1+C_2)_{\nu_e\bar{\nu}_e}=(C_V-C_A)^2+(C_V+C_A)^2$, with $C_A$ and $C_V$ as defined in section~\ref{appendix:reac:opad}. Again, the mean neutrino energy in the neutrino phase space blocking is approximated as \cite{1996A&A...311..532R}, 
 \begin{equation}
 \bar{\epsilon}_{e_i}^{ee}\ =\ T\left(\frac{1}{2}\frac{F_4(\eta_{e^-})}{F_3(\eta_{e^-})}+\frac{1}{2}\frac{F_4(\eta_{e^+})}{F_3(\eta_{e^+})}\right).\label{numeane_pair}
 \end{equation}
 It is worth noting that the rates above are for each individual neutrino species. When comparing with the source material \citep{1986ApJ...309..653C,1987ApJ...321L.129C}, one should keep in mind that the energy production rate of $\nu_i$ is half of the energy of the produced pair, whereas the number production rate of $\nu_i$ is the same as the pair production rate. 
 
 For heavy-lepton neutrinos, the production rate via electron-positron annihilation for all 4 $\nu_x$ species in total is expressed as
 \begin{align}
  Q_{\nu_x,ee}^j\ &=\ \frac{(C_1+C_2)_{\nu_x\nu_x}}{18}\frac{\sigma_0c}{(m_ec^2)^2}\left(\langle1-f(\bar{\epsilon}_{e_i}^{ee};T,\eta_{\nu_x})\rangle\right)^2\nonumber\\
  &\cdotp\left[\frac{8\pi}{(hc)^3}\right]^2\left[T^{4+j}F_{3+j}(\eta_{e^-})T^4F_3(\eta_{e^+})+\right.\nonumber\\
  &\left.T^4F_3(\eta_{e^-})T^{4+j}F_{3+j}(\eta_{e^+})\right],\label{rate_ee_nux}
 \end{align}
 with $(C_1+C_2)_{\nu_x\nu_x}=(C_V-C_A)^2+(C_V+C_A-2)^2$.
 
 Transversal plasmon decay will also contribute to the creation of all three neutrino species with a production rate described by
 \begin{align}
  Q_{\nu_e,\bar{\nu}_e,\gamma}^j\ &\approx\ \frac{\pi^3}{3\alpha^*}C_V^2\frac{\sigma_0c}{(m_ec^2)^2}\frac{T^8}{(hc)^6}\gamma^6\mathrm{e}^{-\gamma}(1+\gamma)\nonumber\\
  &\cdotp\langle1-f(\bar{\epsilon}_{e_i}^{\gamma};T,\eta_{\nu_e})\rangle\langle1-f(\bar{\epsilon}_{e_i}^{\gamma};T,\eta_{\bar{\nu}_e})\rangle\nonumber\\
  &\left[\frac{1}{2}T\left(2+\frac{\gamma^2}{1+\gamma}\right)\right]^j, \label{rate_gamma_nue}
 \end{align}
 for $\nu_e$ and $\bar{\nu}_e$, and
 \begin{align}
  Q_{\nu_x,\gamma}^j\ &\approx\ \frac{4\pi^3}{3\alpha^*}(C_V-1)^2\frac{\sigma_0c}{(m_ec^2)^2}\frac{T^8}{(hc)^6}\gamma^6\mathrm{e}^{-\gamma}(1+\gamma)\nonumber\\
  &\left(\langle1-f(\bar{\epsilon}_{e_i}^{\gamma};T,\eta_{\nu_x})\rangle\right)^2\left[\frac{1}{2}T\left(2+\frac{\gamma^2}{1+\gamma}\right)\right]^j,  \label{rate_gamma_nux}
 \end{align}
 for all $\nu_x$ together. $\alpha^*=1/137.036$ is the fine structure constant and $\gamma=5.565\boldsymbol\cdotp 10^{-2}\sqrt{1/3(\pi^2+3\eta_{e^-}^2)}$. The mean energy of neutrinos produced via plasmon decay is taken as
 \begin{equation}
  \bar{\epsilon}_{e_i}^{\gamma}\ =\ \left[\frac{1}{2}T\left(2+\frac{\gamma^2}{1+\gamma}\right)\right].\label{numeane_plasm}
 \end{equation}

 Finally, in addition to the reactions included in \cite{1996A&A...311..532R}, we incorporate the production of $\nu_x$ via nucleon-nucleon bremsstrahlung, which has been shown to have a significant contribution to the production of heavy-lepton neutrinos. For the total energy production rate (four $\nu_x$ species together), we employ the prescription of \cite{2000PhRvC..62c5802T}:
 \begin{align}
  Q_{\nu_x,\mathrm{brems}}^{j=1}\ &=\ 2.08\boldsymbol\cdotp10^{2}\xi_{\mathrm{brems}}\left(Y_n^2+Y_p^2+\frac{28}{3}Y_nY_p\right)\rho T^{5.5}.\label{Q_brems}
 \end{align}
 Following \cite{2006NuPhA.777..356B}, we set the constant $\xi_{\mathrm{brems}}=0.5$, which is its approximate value at the typical neutrinosphere conditions in PNSs. The factor of $2$ higher numerical value of equation~\eqref{Q_brems} compared with \cite{2006NuPhA.777..356B} comes from the fact that we include two neutrino pairs ($\nu_{\mu}\bar{\nu}_{\mu}$ and $\nu_{\tau}\bar{\nu}_{\tau}$) in $\nu_x$. In order to estimate a number production rate, we make an assumption of the average neutrino energy \citep{1998ApJ...507..339H},
 \begin{align}
  \bar{\epsilon}_{\nu_x}^{\mathrm{brems}}\ &\sim 3T, \label{numeane_brems}
 \end{align}
 namely, that all the kinetic energy of the nucleons is transferred to the created neutrinos. Then, the total number production rate is simply,
  \begin{align}
  Q_{\nu_x,\mathrm{brems}}^{j=0}\ &=\ \frac{Q_{\nu_x,\mathrm{brems}}^{j=1}}{\bar{\epsilon}_{\nu_x}^{\mathrm{brems}}}.\label{R_brems}
 \end{align}
 for $\nu_x$ of all kinds in total.
 
 The total neutrino production rates are then written as
 \begin{align}
  R_{\nu_e}\ &=\ Q_{\nu_e,\beta}^{j=0}+Q_{\nu_e,\bar{\nu}_e,ee}^{j=0}+Q_{\nu_e,\bar{\nu}_e,\gamma}^{j=0},\nonumber\\
  R_{\bar{\nu}_e}\ &=\ Q_{\bar{\nu}_e,\beta}^{j=0}+Q_{\nu_e,\bar{\nu}_e,ee}^{j=0}+Q_{\nu_e,\bar{\nu}_e,\gamma}^{j=0}\label{Rpur},\\
  R_{\nu_x}\ &=\ Q_{\nu_x,ee}^{j=0}+Q_{\nu_x,\gamma}^{j=0}+Q_{\nu_x,\mathrm{brems}}^{j=0},\nonumber
 \end{align}
 for numbers of $\nu_e$, $\bar{\nu}_e$ and all kinds of $\nu_x$, respectively, and
 \begin{align}
  Q_{\nu_e}\ &=\ Q_{\nu_e,\beta}^{j=1}+Q_{\nu_e,\bar{\nu}_e,ee}^{j=1}+Q_{\nu_e,\bar{\nu}_e,\gamma}^{j=1},\nonumber\\
  Q_{\bar{\nu}_e}\ &=\ Q_{\bar{\nu}_e,\beta}^{j=1}+Q_{\nu_e,\bar{\nu}_e,ee}^{j=1}+Q_{\nu_e,\bar{\nu}_e,\gamma}^{j=1}\label{Qpur},\\
  Q_{\nu_x}\ &=\ Q_{\nu_x,ee}^{j=1}+Q_{\nu_x,\gamma}^{j=1}+Q_{\nu_x,\mathrm{brems}}^{j=1},\nonumber
 \end{align}
 for energy of $\nu_e$, $\bar{\nu}_e$ and all kinds of $\nu_x$,respectively.

\section{Production rates and opacities used for tests}\label{appendix:rates} 

As we noted in section~\ref{sec:tests:snapPNS}, there exists certain ambiguity in the derivation of the neutrino $\beta$-production rates, which could become a source of uncertainty in our neutrino treatment.

Following the derivation by \cite{1985ApJS...58..771B}, the rate of change of the neutrino distribution function due to $\beta$-interactions (emission and absorption), $Q_{\nu_i}^{\mathrm{net}}$, is proportional to
\begin{align}
 Q_{\nu_i}^{\mathrm{net}}\ \propto\ j_{\nu_i}(\epsilon)\left[1-f(\epsilon;T,\eta_{\nu_i})\right]-f(\epsilon;T,\eta_{\nu_i})\kappa_{\nu_i,\mathrm{a}}(\epsilon).\label{Changerates}
\end{align}
On one hand, one can derive the neutrino production rates from the neutrino emissivities as
\begin{align}
 Q_{\nu_i}^{j}\ =\ \frac{4\pi c}{(hc)^3}\int_0^\infty{\epsilon^{2+j}j_{\nu_i}(\epsilon)[1-f(\epsilon;T,\eta_{\nu_i})]\mathrm{d}\epsilon},\label{leakrates}
\end{align}
where $[1-f(\epsilon;T,\eta_{\nu_i})]$ accounts for the neutrino final state blocking, and the neutrino emissivities are defined as in \cite{1985ApJS...58..771B},
\begin{align}
 j_{\nu_e}(\epsilon)\ &=\ \frac{\sigma_0(1+3g_A^2)}{4m_e^2c^4}\xi_{pn}f(\epsilon+Q;T,\eta_{e^-})(\epsilon+Q)^2\\
 &\cdotp\sqrt{1-\frac{m_e^2c^4}{(\epsilon+Q)^2}},\label{emissivitynue}
\end{align}
for $\nu_e$ and
\begin{align}
 j_{\bar{\nu}_e}(\epsilon)\ &=\ \frac{\sigma_0(1+3g_A^2)}{4m_e^2c^4}\xi_{np}f(\epsilon-Q;T,\eta_{e^+})(\epsilon-Q)^2\\
 &\cdotp\sqrt{1-\frac{m_e^2c^4}{(\epsilon-Q)^2}},\label{emissivityanue}
\end{align}
for $\bar{\nu}_e$. This formulation corresponds to the one adopted for the presented \textsc{ILEAS} scheme, as detailed in appendix~\ref{appendix:reac}.

On the other hand, one can use the Kirchhoff-Planck relation,
\begin{equation}
 \kappa_{\nu_i,\mathrm{a}}f(\epsilon;T,\eta_{\nu_i}^{\mathrm{eq}})\ =\ j_{\nu_i}(\epsilon)\left[1-f(\epsilon;T,\eta_{\nu_i}^{\mathrm{eq}})\right], \label{kirchoffPlank}
\end{equation}
to define a corrected absorption opacity, which includes the effects of stimulated absorption \citep{2002A&A...396..361R}, 
\begin{align}
 \kappa_{\nu_i,\mathrm{a}}^*(\epsilon)\ &=\ \frac{1}{1-f(\epsilon;T,\eta_{\nu_i}^{\mathrm{eq}})}\kappa_{\nu_i,\mathrm{a}}(\epsilon)\\
 &=\ j_{\nu_i}(\epsilon)+\kappa_{\nu_i,\mathrm{a}}(\epsilon),\label{kappa_star}
\end{align}
as we described in equations~\eqref{kappa_abs_nue} and~\eqref{kappa_abs_anue}. Then, equation~\eqref{Changerates} can be rewritten as
\begin{align}
 Q_{\nu_i}^{\mathrm{net}}\ \propto\ \kappa_{\nu_i,\mathrm{a}}^*(\epsilon)\left[f(\epsilon;T,\eta_{\nu_i}^{\mathrm{eq}})-f(\epsilon;T,\eta_{\nu_i})\right].\label{Changerates2}
\end{align}
This expression is \textit{exactly} equivalent to equation~\eqref{Changerates}, and can be interpreted as a redefinition of emission and absorption, with $\kappa_{\nu_i,\mathrm{a}}^*(\epsilon)$ as the new opacity and $j_{\nu_i}^*(\epsilon)=\kappa_{\nu_i,\mathrm{a}}^*(\epsilon)f(\epsilon;T,\eta_{\nu_i}^{\mathrm{eq}})\left[1-f(\epsilon;T,\eta_{\nu_i})\right]^{-1}$ as the new emissivity. The production rates can be calculated accordingly with equation~\eqref{leakrates} employing $j_{\nu_i}^*$ instead of $j_{\nu_i}$ as
\begin{align}
 Q_{\nu_i}^{*,j}\ =\ \frac{4\pi c}{(hc)^3}\int_0^\infty{\epsilon^{2+j}\kappa_{\nu_i,\mathrm{a}}^*(\epsilon)f(\epsilon;T,\eta_{\nu_i}^{\mathrm{eq}})\mathrm{d}\epsilon}.\label{M1rates}
\end{align}
This formulation is adopted by many truncated moment schemes \citep{2002A&A...396..361R,2015MNRAS.453.3386J} because it simplifies the computation of neutrino interactions to $\kappa_{\nu_i,\mathrm{a}}^*(\epsilon)$, instead of calculating the emissivities and opacities separately. Both codes employed in the comparisons presented in section~\ref{sec:tests}, \textsc{ALCAR} and \textsc{VERTEX}, make use of this formulation. For this reason, we have also implemented this formulation of the $\beta$-production rates in \textsc{ILEAS} and used it for such tests. In order to be fully consistent, this reformulation requires to employ the corrected absorption opacities for the neutrino absorption scheme as well. This translates to redefining the opacities in equations~\eqref{kappa_abs_nue} and~\eqref{kappa_abs_anue} as
 \begin{align}
  \bar{\kappa}_{\nu_e,\mathrm{a}}^{*,j}\ &=\ \frac{\int_0^{\infty}{\kappa_{\nu_e,\mathrm{a}}^*(\epsilon)E_{\nu_e}^j(\epsilon)\mathrm{d}\epsilon}}{\int_0^{\infty}{E_{\nu_e}^j(\epsilon)\mathrm{d}\epsilon}}\nonumber\\
  &=\ \frac{1+3g_A^2}{4(m_ec^2)^2}\sigma_0\xi_{np}\frac{\langle1-f(\bar{\epsilon}_{e^-};T,\eta_{e^-})\rangle}{\langle1-f(\bar{\epsilon}_{\nu_e};T,\eta_{\nu_e}^{\mathrm{eq}})\rangle}\nonumber\\
  &\cdotp\frac{T^2F_{4+j}(\eta_{\nu_e})+2QTF_{3+j}(\eta_{\nu_e})+Q^2F_{2+j}(\eta_{\nu_e})}{F_{2+j}(\eta_{\nu_e})},\label{kappa_abs_nue_star}
 \end{align}
 and
 \begin{align}
  \bar{\kappa}_{\bar{\nu}_e,\mathrm{a}}^{*,j}\ &=\ \frac{\int_0^{\infty}{\kappa_{\bar{\nu}_e,\mathrm{a}}^*(\epsilon)E_{\bar{\nu}_e}^j(\epsilon)\Theta(\epsilon-Q)\mathrm{d}\epsilon}}{\int_0^{\infty}{E_{\bar{\nu}_e}^j(\epsilon)\mathrm{d}\epsilon}}\nonumber\\
  &=\ \frac{1+3g_A^2}{4(m_ec^2)^2}\sigma_0\xi_{pn}\frac{\langle1-f(\bar{\epsilon}_{e^+};T,\eta_{e^+})\rangle}{\langle1-f(\bar{\epsilon}_{\bar{\nu}_e};T,\eta_{\bar{\nu}_e}^{\mathrm{eq}})\rangle}\nonumber\\
  &\cdotp\frac{T^2F_{4+j}(\eta_{\bar{\nu}_e}-Q/T)+(2+j)QTF_{3+j}(\eta_{\bar{\nu}_e}-Q/T)}{F_{2+j}(\eta_{\bar{\nu}_e})}\nonumber\\
  &+\frac{(1+2j)Q^2F_{2+j}(\eta_{\bar{\nu}_e}-Q/T)}{F_{2+j}(\eta_{\bar{\nu}_e})}\nonumber\\
  &+\frac{jQ^3T^{-1}F_{1+j}(\eta_{\bar{\nu}_e}-Q/T)}{F_{2+j}(\eta_{\bar{\nu}_e})},\label{kappa_abs_anue_star}
 \end{align}
and using them instead of $\bar{\kappa}_{\nu_i}^\mathrm{a}$ for equations~\eqref{lumiabs} and~\eqref{Q+} in the absorption algorithm. The electron, positron, $\nu_e$ and $\bar{\nu}_e$ mean energies are calculated as in equations~\eqref{e-meane},~\eqref{e+meane},~\eqref{nuemeane} and~\eqref{anuemeane}, respectively.

We remind the reader that for the tests presented in section~\ref{sec:tests} we employed only the reactions included in table~\ref{table:nureaccomp} in order to consistently compare our results with those obtained by the \textsc{ALCAR} and \textsc{VERTEX} codes. The total neutrino production rates used for such tests are,
 \begin{align}
  R_{\nu_e}\ &=\ Q_{\nu_e,\beta}^{*,j=0},\nonumber\\
  R_{\bar{\nu}_e}\ &=\ Q_{\bar{\nu}_e,\beta}^{*,j=0},\label{Rpurtest}\\
  R_{\nu_x}\ &=\ Q_{\nu_x,ee}^{j=0}+Q_{\nu_x,\mathrm{brems}}^{j=0},\nonumber
 \end{align} 
 and
 \begin{align}
  Q_{\nu_e}\ &=\ Q_{\nu_e,\beta}^{*,j=1},\nonumber\\
  Q_{\bar{\nu}_e}\ &=\ Q_{\bar{\nu}_e,\beta}^{*,j=1},\label{Qpurtest}\\
  Q_{\nu_x}\ &=\ Q_{\nu_x,ee}^{j=1}+Q_{\nu_x,\mathrm{brems}}^{j=1},\nonumber
 \end{align}
 for number and energy, respectively, with the $\beta$-production rates described in equation~\eqref{M1rates}.
  
 For a given nucleon species, $n$ or $p$, equation~\eqref{leakrates} and equation~\eqref{M1rates} employ different nucleon blocking factors, either~\eqref{protonblk} or~\eqref{neutronblk}. These two blocking factors obey the relation \citep{1985ApJS...58..771B},
 \begin{equation}
  \xi_{np}\ =\ e^{(\eta_e-\eta_{\nu_i}^{\mathrm{eq}})}\xi_{pn}. \label{xirel}
 \end{equation}
 Employing this equation, and the general mathematical property of Fermi functions, $f(x)=[1+\mathrm{exp}(x)]^{-1}$,
 \begin{equation}
  e^x\ =\ \frac{1-f(x)}{f(x)}, \label{fermirel}
 \end{equation}
 the two definitions of the $\beta$-production rates in equations~\eqref{leakrates} and~\eqref{M1rates} are apparently equivalent except for a factor $\left[1-f(\epsilon;T,\eta_{\nu_i})\right]^{-1}$ in the integrand of equation~\eqref{M1rates}. This factor, which corresponds to the neutrino phase space blocking, amounts to a $1-2$ per cent correction on the local rates (on average $\langle1-f(\bar{\epsilon}_{\nu_e}^{\beta};T,\eta_{\nu_e})\rangle\gtrsim0.98$ with $\bar{\epsilon}_{\nu_e}^{\beta}$ defined as in equation~\ref{nuemeane}). This is accounted for in the corresponding handling of absorption by the correct redefinition of the absorption opacities in equations~\eqref{kappa_abs_nue_star} and~\eqref{kappa_abs_anue_star}. 
 
\begin{figure}
\begin{center}
\makebox[0pt][c]{%
\minipage{0.5\textwidth}
\includegraphics[width=\textwidth]{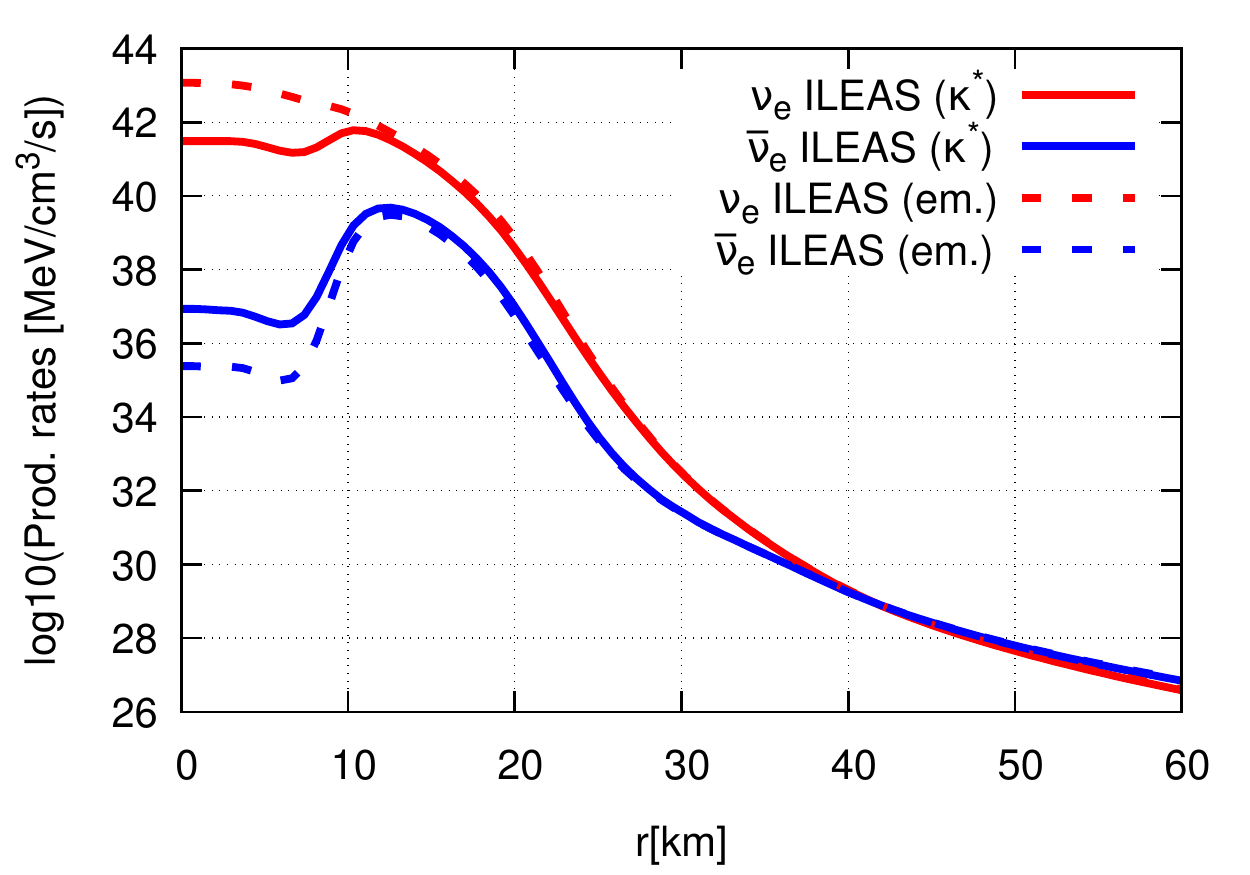}
\endminipage\hfill
}%
\caption{Neutrino $\beta$-production energy rates derived from the modified opacities (lines) and from the emissivities (dashed) for the \textsc{ALCAR} snapshot at 0.5~s post-bounce.}\label{RatecomparisonR}
\end{center}
\end{figure}

 However, figure~\ref{RatecomparisonR} shows a larger difference between the rates described by equations~\eqref{leakrates} (labelled \textit{em.}) and~\eqref{M1rates} (labelled \textit{$\kappa^*$}) in the optically thick region ($r\lesssim20$~km) of a PNS snapshot (\textsc{ALCAR} snapshot at 0.5~s post-bounce). This is due to a more subtle, yet important difference between equations~\eqref{leakrates} and~\eqref{M1rates}. Our definitions of the nucleon phase space blocking (equations~\ref{protonblk} and~\ref{neutronblk}) assume nucleons behave like a free Fermi gas and, accordingly, employ the free Fermi gas nucleon chemical potentials in their calculation (equation~\ref{nucheminv}). As we already pointed out in section~\ref{sec:model:leakage}, it is essential to use the free Fermi gas chemical potentials, instead of the ones provided by the EoS, in order to avoid unphysical behaviour of the blocking factors (becoming negative, bigger than unity or not fulfilling the non-degenerate limits). This inconsistency between the nucleon degeneracies employed in the nucleon blocking factors (free Fermi gas, see equations~\ref{protonblk} and~\ref{neutronblk}) and in the calculation of the neutrino equilibrium degeneracies (EoS, see equations~\ref{etanu} and~\ref{betachem}), is a source of discrepancy between the two approaches. 
 
 Because we employ the free Fermi gas nucleon chemical potentials in the blocking factors, $\eta_{\nu_i}^{\mathrm{eq}}$ in equation~\ref{xirel} is different than the one obtained from the high-density EoS (which considers nucleons as interacting particles) by equation~\eqref{betachem}. In order to be able to recover~\eqref{leakrates} from~\eqref{M1rates} (except for a factor $\left[1-f(\epsilon;T,\eta_{\nu_i})\right]^{-1}$), or vice versa, the same nucleon chemical potentials would need to be used consistently. Because in our formulation this is not the case, there appear notable differences between both rates in the regimes where the nucleon interactions (and thus deviations between the two values of $\eta_{\nu_i}^{\mathrm{eq}}$) become important. 

\begin{figure}
\begin{center}
\makebox[0pt][c]{%
\minipage{0.5\textwidth}
\includegraphics[width=\textwidth]{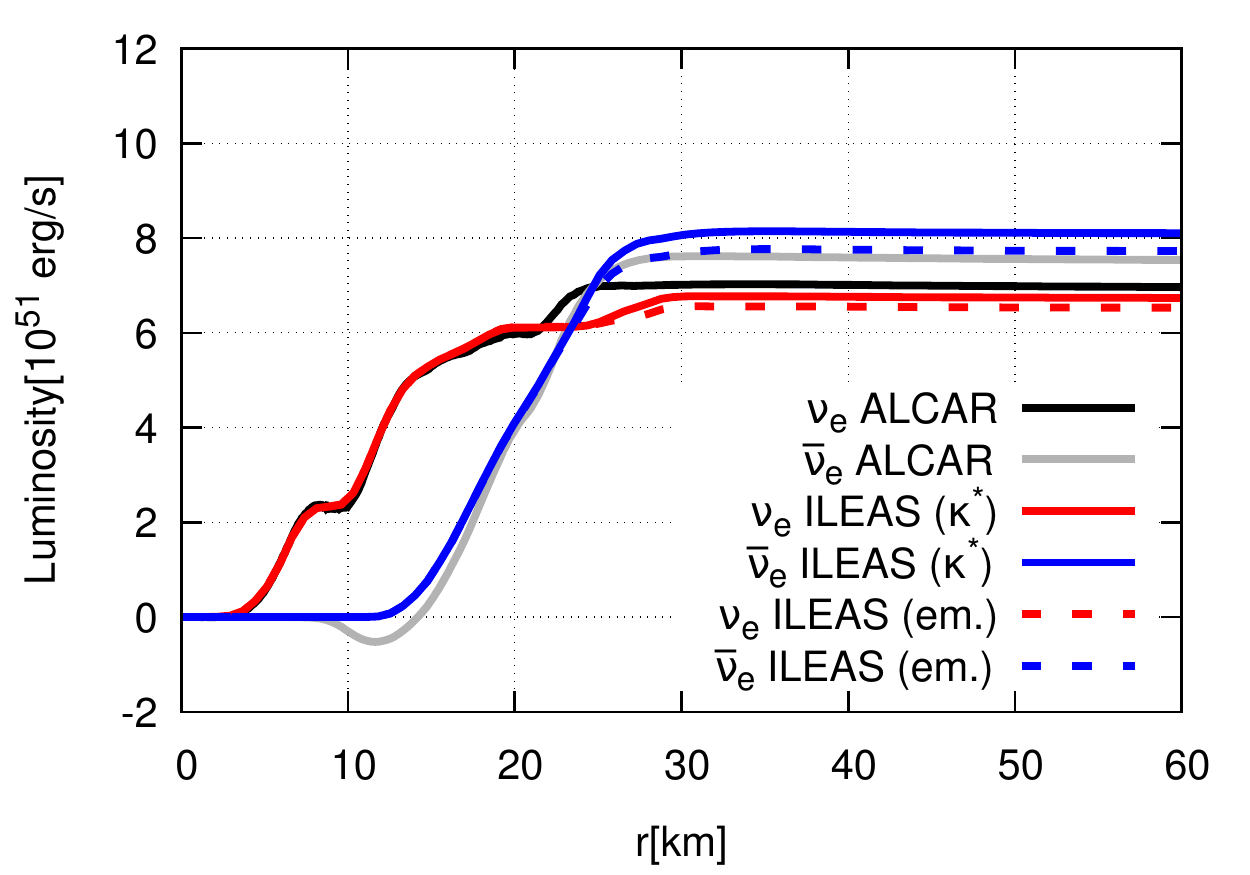}
\endminipage\hfill
}%
\caption{Neutrino luminosity profiles of $\nu_e$ and $\bar{\nu}_e$ obtained by \textsc{ILEAS} after 5~ms relaxation, employing the $\beta$-production rates derived from the modified opacities (solid lines) and from the emissivities (dashed lines), for the \textsc{ALCAR} relaxed snapshot at 0.5~s. The results obtained by \textsc{ALCAR} are also shown for comparison.}\label{RatecomparisonL}
\end{center}
\end{figure}

However, at the high optical depth at which this discrepancy exists, the neutrino transport is dominated by diffusion. As a consequence, even though the production rates differ considerably in the two approaches, neutrino luminosities are vastly produced by the diffusion behaviour, and the impact of the prescription chosen for the production rates is small. In figure~\ref{RatecomparisonL} this small effect in the luminosities (less than 5 per cent) can be inspected. Thus we conclude that the choice of the formulation of the $\beta$-production rates is not significantly relevant for the present work, as its associated uncertainty lies well within the differences between the results obtained by \textsc{ILEAS} in comparison to more sophisticated transport schemes (section~\ref{sec:tests}). If higher accuracy were desired, mean-field effects of the interacting nucleons should be taken into account in the nucleon-neutrino interactions. We refer the reader to \cite{1998PhRvD..58a3009R} for details on the original formulation, and \cite{2012ApJ...755..126R,2012PhRvC..86f5803R,2012PhRvL.109y1104M} for the implementation and application of such corrections in the context of SN and PNS cooling simulations.

\section{Alternative method to compute neutrino-number re-absorption}\label{appendix:nabs} 

\begin{table*}
  \centering
  \caption{Neutrino mean energies obtained by the alternative version of neutrino-number re-absorption in \textsc{ILEAS} (described in appendix~\ref{appendix:nabs}) applied to two of the snapshots of a PNS cooling simulation and one of the BH-torus models discussed in section~\ref{sec:tests}. These data correspond to results after relaxation by 5~ms for PNS cases and 3~ms for BH-torus models. For comparison also the results from transport calculations with ALCAR and VERTEX are listed. The columns labelled \textit{Alternative leakage luminosity} and \textit{Alternative leakage mean energy} provide the neutrino luminosities and mean energies obtained from the alternative version of neutrino-number re-absorption. In addition, the columns labelled \textit{Standard leakage luminosity} and \textit{Standard leakage mean energy} provide the neutrino luminosities and mean energies obtained by the standard ILEAS version copied from table~\ref{table:results}. In both cases the mean energies are calculated by equation~\eqref{eleak}. All values are taken for a local observer in the rest frame of the source at the edge of the grid (100 km).}\label{table:nabs}
    \begin{tabular}{lcccccccc}
      \hline
      \hline
      \noalign{\vskip 2mm}  
      Model        & $\nu$-species & \parbox[c]{1.6cm}{\centering Transport luminosity ($10^{51}\ \mathrm{erg}\boldsymbol\cdotp\mathrm{s}^{-1}$)} &  \parbox[c]{1.6cm}{\centering Alternative leakage luminosity ($10^{51}\ \mathrm{erg}\boldsymbol\cdotp\mathrm{s}^{-1}$)}   &  \parbox[c]{1.6cm}{\centering Standard leakage luminosity ($10^{51}\ \mathrm{erg}\boldsymbol\cdotp\mathrm{s}^{-1}$)}  & \parbox[c]{1.6cm}{\centering Transport mean energy (MeV)} & \parbox[c]{1.6cm}{\centering Alternative leakage\\ mean energy (MeV)} & \parbox[c]{1.6cm}{\centering Standard leakage\\ mean energy (MeV)} & \parbox[c]{1.6cm}{\centering Transport code} \vspace{1mm}  \\ 
      \hline 
      \noalign{\vskip 2mm}                                                                                                                                                                                                                                                          
      PNS 0.5~s        & $\nu_e$     	         & 7.0        & 6.3       & 6.7         & 9.93         & 8.19       & 7.95  & \textsc{ALCAR}  \\
      PNS 0.5~s        & $\bar{\nu}_e$         & 7.6        & 8.6        & 8.1        & 13.32       & 16.00      & 12.62  & \textsc{ALCAR}  \\                                                                                                                     
      PNS 1.2~s        & $\nu_e$     	         & 3.7        & 3.5        & 3.5        & 9.24         & 6.39       & 6.16  & \textsc{VERTEX} \\
      PNS 1.2~s        & $\bar{\nu}_e$         & 3.8        & 3.7        & 3.7        & 11.43       & 14.05      & 12.98  & \textsc{VERTEX} \\                                                                                                                          
      BH-torus 0.1~\Msun & $\nu_e$     	     & 6.5        & 6.1        & 6.5        & 12.02       & 10.83      & 12.69  & \textsc{ALCAR}  \\
      BH-torus 0.1~\Msun & $\bar{\nu}_e$  & 5.2        & 5.2        & 4.8        & 14.20       & 14.29      & 14.50  & \textsc{ALCAR}  \\  
      \noalign{\vskip 2mm}                                                                                                                       
      \hline
      \hline
  \end{tabular}
\end{table*}

\begin{figure*}
\begin{center}
\makebox[0pt][c]{%
\minipage{\textwidth}
\includegraphics[width=\textwidth]{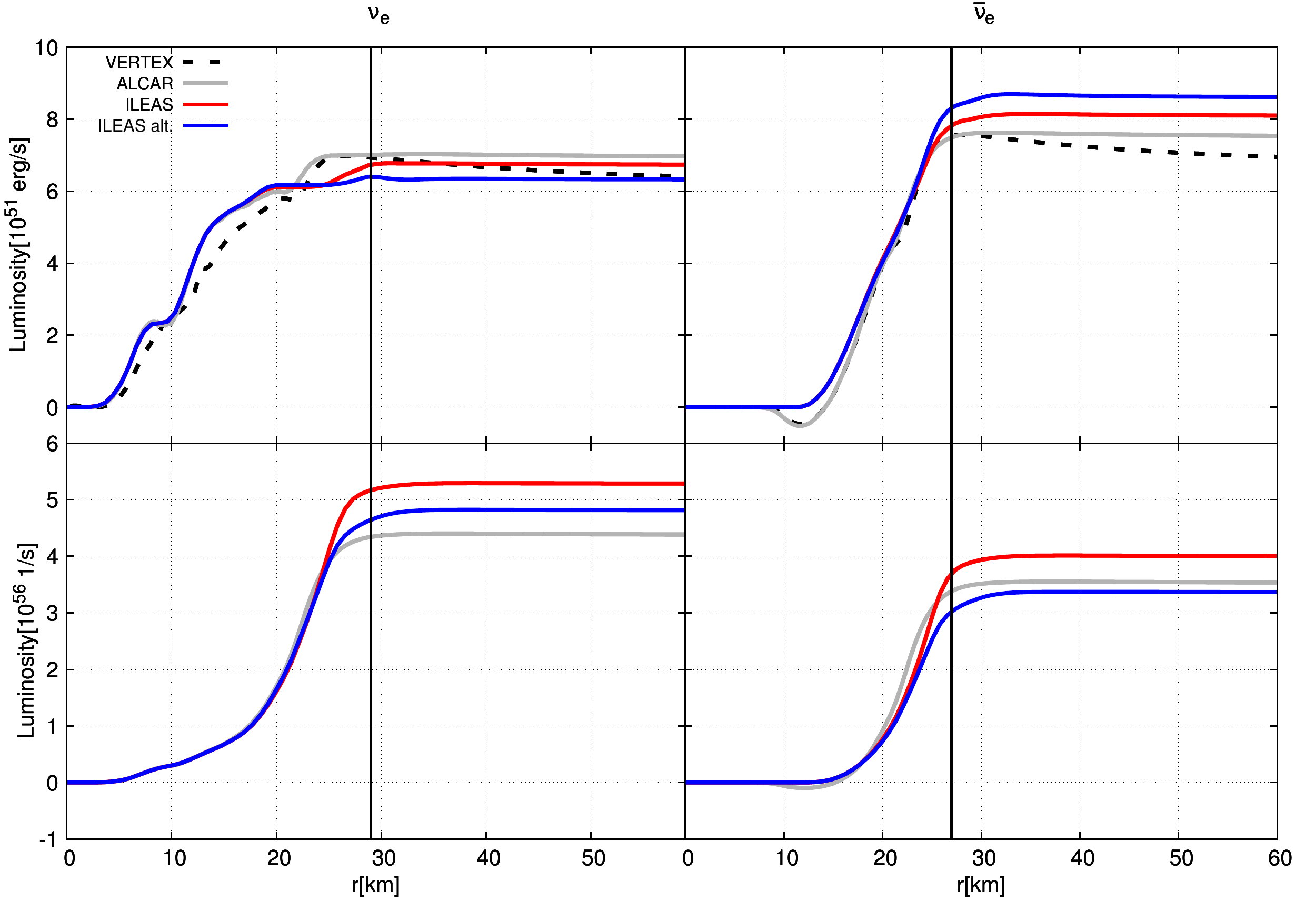}
\endminipage\hfill
}%
\caption{Neutrino energy luminosity (top row) and lepton-number luminosity (bottom row) profiles of $\nu_e$ (left column) and $\bar{\nu}_e$ (right column) obtained by \textsc{ILEAS} after 5~ms relaxation of the \textsc{ALCAR} relaxed snapshot at 0.5~s. Solid red lines display the standard formulation of neutrino re-absorption described in section~\ref{sec:model:abs} and solid blue lines show the alternative version described in appendix~\ref{appendix:nabs}. The results obtained by \textsc{ALCAR} and \textsc{VERTEX} (only neutrino energy luminosities) are also shown for comparison. While the VERTEX calculations take gravitational redshift effects into account, ALCAR and ILEAS do not. The vertical black lines in each panel illustrate the position of the neutrinosphere ($\tau_{\nu_i}=2/3$) for the corresponding neutrino species.}\label{nabslumis}
\end{center}
\end{figure*}

In some of the tests presented in section~\ref{sec:tests} we observe a disagreement of up to $\sim$45 per cent ($\sim$4~MeV difference) for $\nu_e$ and up to $\sim$20 per cent ($\sim$2~MeV) for $\bar{\nu}_e$ when comparing the mean energies of radiated neutrinos obtained by ILEAS (as described by equation~\ref{eleak}) with the ones obtained by the M1 scheme ALCAR. Since it is possible that some approximations introduced in the absorption scheme described in section~\ref{sec:model:abs} (in particular describing neutrino number absorption via equation~\ref{R+}) could have an impact on the neutrino mean energies, we tested an alternative version of the absorption treatment in this appendix.

In a procedure analogous to the one employed in equation~\eqref{Q+} for the energy treatment, we calculated the neutrino number absorption rates, $R_{\nu_i}^+$, independently of the energy absorption rates. Instead of the neutrino energy luminosities and opacities in equation~\eqref{Q+}, we used the neutrino-\textit{number} luminosities,
\begin{equation}
 L_{\nu_i}^{\mathrm{ray,num}}(s)\ \approx\ R_{\nu_i}^{-}(\psi^6V)_{\mathrm{cell,em}},\label{lumapprox2}
\end{equation}
calculated in analogy to equation~\eqref{lumapprox}, with $R_{\nu_i}^{-}$ as defined in equation~\eqref{Reff}, and the spectrally-averaged opacities defined as averages over the number spectra in equations~\eqref{kappa_abs_nue} and~\eqref{kappa_abs_anue} (with $j=0$). In this context the expression of equation~\eqref{Rel} carries only one power of $\alpha(\boldsymbol x_1)/\alpha(\boldsymbol x_2)$ for the number luminosity. The factor $\gamma_{\nu_i,\mathrm{en}}^{\mathrm{eff}}$ (see equation~\ref{gammaeff}) in equation~\eqref{Q+} is also correspondingly replaced by its neutrino-number version $\gamma_{\nu_i,\mathrm{num}}^{\mathrm{eff}}$. The rays used for calculating the number absorption effects were the same as those employed for the energy re-absorption.

While this alternative method only marginally improves the $\nu_e$ leakage mean energies in the tested PNS snapshots by $\lesssim$0.5~MeV compared to ALCAR results, however, the disagreement of the $\bar{\nu}_e$ mean energies increases by $\sim$10--15 per cent ($\sim$1--2.5~MeV). This can be seen by the numbers listed in table~\ref{table:nabs}. In contrast, the results obtained when applying ILEAS on the BH-torus snapshot are less sensitive to the neutrino-number re-absorption model, providing a similar agreement with the ALCAR results for both tested cases.

In figure~\ref{nabslumis}, we plot the radial profiles of the neutrino-energy and number luminosities for the PNS snapshot at 0.5~s and compare the results obtained after 5~ms of relaxation when applying the two different treatments of neutrino-number re-absorption. We find that the results of the alternative method to calculate neutrino-number re-absorption lie within $\sim$10--15 agreement with the ALCAR results. This is compatible with the agreement of transport and standard ILEAS results reported in section~\ref{sec:tests}. Note that changing the neutrino-number absorption in the alternative leakage treatment has indirectly also influence on the neutrino-energy luminosities because of the effects on the $T$ and $Y_e$ profiles of the underlying background model. Comparing the radial luminosity profiles of standard and alternative leakage treatments for the PNS model at 1.2~s shows even better agreement than the PNS snapshot presented in figure~\ref{nabslumis} (see table~\ref{table:nabs}). Also the time evolution of the neutrino emission by BH-torus systems is captured equally well by both ILEAS versions.

It is worth pointing out that the bottom panels in figure~\ref{nabslumis} show that the neutrino-number luminosities are produced in a relatively narrow layer below the neutrinospheres (vertical black lines), i.e. in the semi-transparent region where neutrino losses are governed by the flux-limiting treatment in the diffusion time-scale. The larger discrepancies reported for the neutrino mean energies, especially for $\nu_e$, may thus be understood by the higher sensitivity of neutrino-number losses to the modelling of the semi-transparent region, which constitutes one of the biggest challenges for leakage schemes. Therefore, it is plausible that the alternative version for calculating the neutrino-number re-absorption cannot yield much better results for the mean energies than our standard version.

Since the overall agreement with ALCAR in the neutrino mean energies remains similar, we consider the method described in section~\ref{sec:model:abs} for treating neutrino re-absorption as sufficiently good. Moreover, the comparison presented in this appendix further demonstrates the robustness of our neutrino re-absorption scheme to methodical changes in details.

\label{lastpage}

\end{document}